\documentclass[11pt,oneside]{article} 
\usepackage{tikz}
\usepackage{amsmath,amsthm,graphicx,amssymb,enumerate,color,natbib,bbm,subfig,rotating,tikz,authblk,romannum,array,enumitem,dsfont,booktabs,multirow,mathtools,float}
\usepackage{etoolbox}
\usepackage{fancyhdr}
\usepackage{caption}
\usepackage{dutchcal}
\usepackage[pagewise]{lineno}
\usepackage[font=small]{caption}
\usepackage[mathscr]{euscript}
\usepackage{caption}
\usepackage{enumitem}
\usepackage{tabularx}
\usepackage{enumitem}
\usepackage{setspace}
\usepackage{sansmath}
\usepackage{changepage}
\usepackage{algorithm} 
\usepackage{algpseudocode} 
\usepackage{amsmath}
\usepackage{amssymb}
\usepackage{appendix}
\usepackage{apptools}
\usepackage{natbib}
\usepackage{refcount}
\usepackage{longtable}
\usepackage[top=1in, bottom=1in, left=1in, right=1in]{geometry}
\usetikzlibrary{chains,shapes.multipart,shapes,calc,fit}
\usepackage[percent]{overpic}
\usetikzlibrary{automata,positioning}
\usepackage[bottom]{footmisc}
\usepackage{graphicx}
\usepackage[hidelinks=true]{hyperref}
\usepackage{setspace}
\usepackage[T1]{fontenc}
\usepackage[utf8]{inputenc}
\usepackage[para,online,flushleft]{threeparttable}

\usepackage[all]{xy}
\setcitestyle{round}
\AtAppendix{\renewcommand{\thesection}{Appendix \Alph{section}}}
\newcommand{\rowgroup}[1]{\hspace{-1em}#1}
\DeclareMathAlphabet{\mathpzc}{OT1}{pzc}{m}{it}
\renewcommand{\baselinestretch}{1.3}
\newtheoremstyle{ModifiedStyle}
	{\topsep} 
	{3pt} 
	{} 
	{} 
	{\bfseries} 
	{.} 
	{.5em} 
	{} 
\theoremstyle{ModifiedStyle}

\newtheorem{to-do}{To-Do}

\newcommand{\GradientInverse}{H^\text{\scriptsize -1}}
\newcommand{\E}{\mathbb{E}}
\newcommand{\halfspace}{\hspace{0.2mm}}
\newcommand{\AM}{{\tiny\hspace{-0.7mm} AM \hspace{0.7mm}}}
\newcommand{\PM}{{\tiny\hspace{-0.7mm} PM \hspace{0.7mm}}}
\newcommand{\Wedgesmall}{\text{{\small$\wedge$}}}
\newcommand{\subsubsubsection}[1]{\paragraph{#1}\mbox{}\\}
\newcommand{\Wedgefootnotesize}{\text{{\footnotesize$\wedge$}}}
\newcommand{\Wedgescriptsize}{\text{{\scriptsize$\wedge$}}}
\newcommand{\BSmall}[1]{\mkern-1.7mu\raisebox{-1.2pt}{\scalebox{0.9}{$\scriptscriptstyle B$}}}
\newcommand{\NSmall}[1]{\mkern-1.7mu\raisebox{-1.2pt}{\scalebox{0.9}{$\scriptscriptstyle N$}}}
\newcommand{\Plus}{\raisebox{.4\height}{\scalebox{.6}{+}}}
\newcommand{\Minus}{\raisebox{.4\height}{\scalebox{.8}{-}}}
\def\minus{\texttt{-}}
\def\plus{\texttt{+}}
\DeclareMathOperator*{\maximize}{maximize}
\DeclareMathOperator*{\minimize}{minimize}
\DeclareMathOperator*{\supremum}{supremum}
\DeclareMathOperator*{\infimum}{infimum}
\newcommand*\rot{\rotatebox{90}}
\newcommand{\nquad}{\kern-1em}
\DeclareRobustCommand{\rchi}{{\mathpalette\irchi\relax}}
\newcommand{\irchi}[2]{\raisebox{\depth}{$#1\chi$}} 
\newcommand{\CC}{C\nolinebreak\hspace{-.05em}\raisebox{.4ex}{\tiny\bf +}\nolinebreak\hspace{-.10em}\raisebox{.4ex}{\tiny\bf +}}
\newcommand{\mycite}[1]{\citeauthor{#1}~(\citeyear{#1})}
\makeatletter
\newcommand{\duallabel}[2]{\label{#1}\label{#2}}

\newcommand{\Rom}[1]{\expandafter\@slowromancap\romannumeral #1@}
\newcommand{\Biggg}{\bBigg@{2.5}}
\newcommand{\vast}{\bBigg@{3}}
\newcommand{\Vast}{\bBigg@{3.5}}
\newcommand{\massive}{\bBigg@{4.5}}
\newcommand{\Massive}{\bBigg@{6}}
\newcommand{\thickhline}{%
	\noalign {\ifnum 0=`}\fi \hrule height 1pt
	\futurelet \reserved@a \@xhline}
\newcolumntype{"}{@{\hskip\tabcolsep\vrule width 1pt\hskip\tabcolsep}}
\newcommand{\mycitetalias}[1]{\citeauthor{#1} (\citeyear{#1}, \citealp{#1})}
\newcommand{\customappendixref}[1]{\hyperref[#1]{\Alph{section}}}
\usepackage{cleveref}
\newcommand{\customref}{%
  \hyperref[appendix_computational_methods]{C}%
}
\newcommand{\Blue}[1]{\textcolor{blue}{#1}}
\newcommand{\BlueForOR}[1]{\textcolor{blue}{#1}}
\newcommand{\Red}[1]{\textcolor{red}{#1}}
\newcommand{\Purple}[1]{\textcolor{purple}{#1}}
\newcommand{\Grey}[1]{\textcolor{gray}{#1}}
\newcommand{\Orange}[1]{\textcolor{orange}{#1}}
\newcommand{\Green}[1]{\textcolor{green}{#1}}
\DeclareMathOperator*{\argmin}{argmin}
\DeclareMathOperator*{\argmax}{argmax}
\newcommand{\overbar}[1]{\mkern 1.5mu\overline{\mkern-1.5mu#1\mkern-1.5mu}\mkern 1.5mu}
\newcommand{\PreserveBackslash}[1]{\let\temp=\\#1\let\\=\temp}
\newcolumntype{C}[1]{>{\PreserveBackslash\centering}p{#1}}
\newcolumntype{R}[1]{>{\PreserveBackslash\raggedleft}p{#1}}
\newcolumntype{L}[1]{>{\PreserveBackslash\raggedright}p{#1}}
\newcommand{\fixed@sra}{$\vrule height 2\fontdimen22\textfont2 width 0pt\shortrightarrow$}
\newcommand{\shortarrow}[1]{%
  \mathrel{\text{\rotatebox[origin=c]{\numexpr#1*45}{\fixed@sra}}}
}
\makeatother

\providecommand{\keywords}[1]{\hspace*{-4mm}\small\textbf{\textit{Keywords---}} #1}

\title{\textbf{Dynamic Scheduling of a Multiclass Queue in the Halfin-Whitt Regime: A Computational Approach for High-Dimensional Problems}}

\author{Barış Ata\thanks{\texttt{baris.ata@chicagobooth.edu}} \ and Ebru Kaşıkaralar\thanks{\texttt{ebrukasikaralar@chicagobooth.edu}}}

\date{{\small \vspace{-7mm} \today}}

\makeatletter
\let\LN@align\align
\let\LN@endalign\endalign
\renewcommand{\align}{\linenomath\LN@align}
\renewcommand{\endalign}{\LN@endalign\endlinenomath}
\let\LN@gather\gather
\let\LN@endgather\endgather
\renewcommand{\gather}{\linenomath\LN@gather}
\renewcommand{\endgather}{\LN@endgather\endlinenomath}
\makeatother
\renewcommand\linenumberfont{\normalfont\tiny}
\makeatletter
\AtBeginDocument{%
    \setlength{\abovedisplayskip}{10pt} 
    \setlength{\abovedisplayshortskip}{8pt}
    \setlength{\belowdisplayskip}{10pt} 
    \setlength{\belowdisplayshortskip}{8pt}
}

\begin{document}

\vspace*{-22mm}

{\let\newpage\relax\maketitle}


\vspace*{-9mm}
\begin{abstract}
We consider a multi-class queueing model of a telephone call center, in which a system manager dynamically allocates available servers to customer calls. Calls can terminate through either service completion or customer abandonment, and the manager strives to minimize the expected total of holding costs plus abandonment costs over a finite horizon. Focusing on the Halfin-Whitt heavy traffic regime, we derive an approximating diffusion control problem, and building on earlier work by \citet{beck2021deep}, develop a simulation-based computational method for solution of such problems, one that relies heavily on deep neural network technology. Using this computational method, we propose a policy for the original (pre-limit) call center scheduling problem. Finally, the performance of this policy is assessed using test problems based on publicly available call center data. For the test problems considered so far, our policy does as well as or better than the best benchmark we could find. Moreover, our method is
computationally feasible at least up to dimension 500, that is, for call centers with 500 or more distinct customer classes.

\end{abstract}



\pagenumbering{arabic}
\doublespacing
\setlength{\abovedisplayskip}{8pt}
\setlength{\belowdisplayskip}{8pt}
\setlength{\abovedisplayshortskip}{8pt}
\setlength{\belowdisplayshortskip}{8pt}
\vspace{-6mm}
\section{Introduction}
\vspace{-2mm}
Motivated by call center operations, this paper considers a dynamic scheduling problem for a multiclass queueing system with a single pool of servers. We focus attention on a finite-horizon, nonstationary formulation in order to model the call center operations over a day and develop an effective computational method to solve it in high-dimensional settings, i.e., for systems with many customer classes. To illustrate the effectiveness of our method, we take a data-driven approach whereby we calibrate our model using data from a large US Bank (see Section \ref{data_section}). More specifically, the test problems we study to illustrate the effectiveness of our method are designed and calibrated using the US Bank data set\footnote{Our data set is provided by the Service Engineering Enterprise (SEE) Lab at the Technion, and it is publicly available at https://see-center.iem.technion.ac.il/databases/USBank/. Accessed on August 9, 2023.}.

Call center operations have attracted a significant amount of research attention over the last three decades. The survey papers by \citet{gans2003telephone} and \citet{aksin2007modern} provide an overview through circa 2007; also see \citet{koole2023practice} for a more recent overview. In addition to call center operations, our work blends ideas from three other streams of literature: i) diffusion approximations for queueing systems, ii) stochastic control, and iii) deep learning. 

In what follows, we first derive a diffusion approximation of our dynamic scheduling problem. In doing so, we follow the literature that was initiated by the seminal paper \citet{halfin1981heavy}, later extended by \citet{garnett2002designing}. \citet{halfin1981heavy} consider a single-class queue with many servers in heavy traffic, where the arrival rate and the number of servers grow large, and the traffic intensity of the system approaches one. \citet{garnett2002designing} extend \citet{halfin1981heavy} to incorporate customer abandonments. 

Two important antecedents of our paper in this stream of literature are \citet{harrison2004dynamic} and \citet{atar2004scheduling}. \citet{harrison2004dynamic} is a multiclass extension of \citet{garnett2002designing}, where a system manager makes dynamic server allocation decisions to minimize infinite-horizon discounted costs of customer abandonments and holding costs. The authors consider linear holding and abandonment costs and study this problem in the Halfin-Whitt asymptotic regime. They show that the resulting Hamilton-Jacobi-Bellman (HJB) equation admits a smooth solution and numerically solve a two-dimensional example using the finite element method. \citet{atar2004scheduling} follow a similar approach but consider more general cost structures. In addition to showing the existence and uniqueness of a smooth solution to the HJB equation, the authors prove an asymptotic optimality result for the policy they derive from the (formal) limiting control problem; also see \citet{atar2005scheduling}. 

\begin{figure}[H]
\centering
 \hspace*{-1.5cm}
    \hspace{20mm} \includegraphics[scale = 0.47]{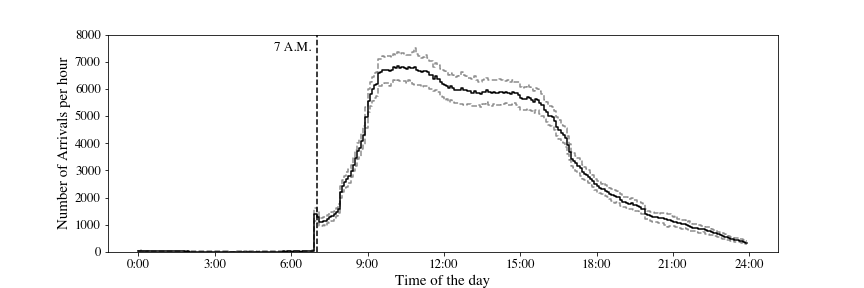}
    \caption{Hourly arrival rate of callers on weekdays during May - July 2003. The resolution of the horizontal axis is five minutes. That is, arrival rates are calculated over five-minute intervals. The solid line shows the hourly average arrival rate across all days. The two dashed lines that enclose it are computed by taking the average plus/minus 2 times the standard deviation of the rates for each five-minute interval across all days.}
    \label{Arrivals_CI}
\end{figure}

Our formulation is a variation of \citet{harrison2004dynamic} and \citet{atar2004scheduling} with the main difference being our focus on a finite-horizon formulation with nonstationary data. As mentioned above, all test problems we consider in our extensive numerical study are constructed using the US Bank data. As one would expect, we observe the following in the data: i) both the arrival rates of callers to various classes and the staffing level vary throughout the day, see Figure \ref{Arrivals_CI} for a plot of the arrival rate over a day; ii) the number of agents working is in the hundreds during most of the day, with the average number of agents being 273; and iii) the average load factor\footnote{The load factor is defined as the ratio of the average amount of work over a specified time to the maximum amount of work that would have been served if the call center had been continuously occupied and it is calculated using Equation (\ref{load_factor}).} of the system is about 90\%. These observations led us to choose a nonstationary, finite-horizon model and to study it in the Halfin-Whitt (many-server) asymptotic regime. 
\vspace{-5mm}
\vspace{-6mm}
Our goal is to develop an effective computational method for solving this problem in high dimensions. To position our work relative to other stochastic systems research, one may recall the framework described by \citet{harrison2003broader} for solving dynamic control problems via diffusion approximations, which we paraphrase as follows:
\vspace{-4mm}
\begin{itemize}
    \setlength\itemsep{-0.5em}
    \item[(a)] formulate a conventional stochastic system model with discrete flow units; 
    \item[(b)] define a notion of heavy traffic and formally (that is, non-rigorously) derive a limiting diffusion control problem appropriate for that regime; 
    \item[(c)] solve the diffusion control problem and propose a policy that implements that solution in the original (that is, pre-limit) problem context;
    \item[(d)] demonstrate the effectiveness of the proposed policy, either by a proof of asymptotic optimality or through a simulation study that compares it against benchmark policies.
\end{itemize}
\vspace{-2mm}
Steps (a) and (b) have been successfully executed by Harrison (\citeyear{harrison1988brownian}, \citeyear{harrison2000}, \citeyear{harrison2003broader})
 for the conventional heavy traffic regime, and by \citet{atar2005scheduling} for the Halfin-Whitt regime. Typically, step (b) relies on a functional central limit theorem (FCLT). Step (d) has also been successfully addressed for various important examples and special cases. For example, \citet{ata2005heavy} provide asymptotic optimality proofs for policies they derive in the conventional heavy traffic regime, and \cite{atar2005scheduling} does the same in the Halfin-Whitt regime, also see \citet{ata2012optimality}. In addition, to illustrate the effectiveness of the policies they derived using diffusion approximations, several authors have used simulation studies; see, for example, \citet{harrison1990scheduling}, \citet{wein1992dynamic}, \citet{chevalier1993scheduling} and \citet{ata2023approximate}.

In general, step (c) has been a major roadblock for successfully implementing the four-step procedure outlined above, because the HJB equation for the approximating diffusion control problem is a nonlinear partial differential equation (PDE) whose dimension can be high. Historically, the state-of-the-art computational method for solving such PDEs has been the finite-element method, which suffers from the curse of dimensionality. Therefore, solving the HJB equation has only been possible for low dimensional problems: for example, \citet{harrison2004dynamic} solved a particular two-dimensional problem, while \citet{kumar2004numerical} and \citet{ata2020dynamic} successfully addressed other examples. In this paper, we solve our limiting diffusion control problem in high dimensions, resolving step (c) for our specific context. Using this solution, we propose a policy for the original call center scheduling problem, and show its effectiveness through a simulation study by comparing its performance to that of the best benchmark we could find (see Section \ref{numerical_results}).

\citet{Ata2024drift} study drift control of a high-dimensional reflected Brownian motion whose state space is an orthant. Those authors develop a computational method based on earlier research by \citet{han2018solving}, and while their work differs significantly from ours, it fills a similar gap in the literature. Specifically, by using their method one can execute step (c) of the four-step framework above for a number of dynamic control problems involving stochastic processing networks in the conventional heavy traffic regime. \citet{ata2025analysis} solves a sequential vehicle routing problem motivated by an eviction enforcement application following a similar approach.

In order to solve the limiting diffusion control problem, we first express it analytically by considering the associated HJB equation, following the approach that is standard in the stochastic control literature; see \citet{fleming2006controlled}. Our HJB equation is a semilinear parabolic partial differential equation, see \citet{gilbarg2001elliptic}. To solve the HJB equation, we build on \citet{beck2021deep}, who developed a method for solving semilinear PDEs. In doing so, we modify their method, taking into account the special structure of our problem. Similar to \citet{beck2021deep}, our method relies heavily on deep neural network technology; see Section \ref{computational_method}. 

\citet{beck2021deep} belongs to a new body of research that has recently emerged on solving partial differential equations (PDEs) using deep neural networks. Broadly speaking, researchers in this field adopt one of two primary approaches: Physics-Informed Neural Networks (PINNs) or methods based on Backward Stochastic Differential Equations (BSDEs).

Physics-Informed Neural Networks (PINNs), introduced by \citet{raissi2019physics}, offer a mesh-free approach to solving partial differential equations (PDEs) by combining training data with the governing physical laws. In this framework, a neural network approximates the target function, and its gradients and Hessians are computed using automatic differentiation. The loss function, which combines errors from the PDE residuals and boundary or initial conditions, is minimized by adjusting the neural network parameters. A significant challenge in this methodology is the computational expense of calculating Hessians via automatic differentiation, especially as the dimensionality of the PDE increases. However, there are several approaches to speed up such computations; see for example, \citet{he2023learning} and \citet{hu2024hutchinson}.

The deep BSDE approach, pioneered by \citet{han2017deep} and \citet{han2018solving}, leverages an equivalence between the semilinear PDEs and BSDEs; see \citet{pardoux1990adapted} and \citet{ma1994solving}. To be specific, \citet{han2018solving} uses neural networks to approximate the (initial) value function and the (space) gradient of the value function over a set of discrete time points, which ultimately yields an approximation of the terminal value function. A natural loss function, defined by the difference between this approximation and the terminal condition, is minimized by adjusting the neural network parameters. This approach offers two key advantages for our purposes: it avoids the need to compute the Hessian of the value function and, with a reference policy, enables efficient sampling from the state space for training. As an aside, \citet{germain2022approximation} provides an approximation error analysis of the deep splitting method\footnote{Although it is not applicable to our setting, for a stationary infinite-horizon version of our problem, one can consider using the reinforcement learning based method of \citet{dai2022queueing} that focuses on an infinite horizon problem formulation under the ergodic cost criterion.}.

However, \citet{han2018solving}'s deep BSDE method results in a large optimization problem for selecting neural network parameters, which can be computationally intensive. \citet{beck2021deep} address this challenge by proposing the deep splitting method, which divides the problem into smaller, more manageable tasks. Using a Feynman-Kac representation, the solution to the PDE is estimated iteratively over small time intervals, resulting in a smaller optimization problem at each step. We build on this method, iteratively solving optimization problems for each interval to approximate the value function and its gradient.

We perform the aforementioned comparison of our proposed policy against benchmarks within the context of several (data-driven) test problems. The US Bank call center data naturally leads to a $17$-dimensional test problem, but this problem is intractable because the associated Markov Decision Process (MDP) suffers from the curse of dimensionality. We also consider two variants of this problem. Additionally, we consider seven high-dimensional test problems: four with dimensions 30, 50, 100 and 500, each designed to have a pathwise optimal solution, and three 100-dimensional test problems that do not admit a pathwise optimal solution. Lastly, we consider a two-dimensional test problem as well as two additional test problems that are three-dimensional. Because they are low dimensional, we can solve the original problem by employing the standard techniques for solving the associated MDP.

To repeat, we consider thirteen test problems. Four of them have pathwise optimal solutions, and we can solve three others using standard techniques for solving MDPs because they are low dimensional. In each of these seven cases, our proposed policy performs as well as the optimal policy in the sense that the difference in their performance is not statistically significant. In the 17-dimensional main test problem and its two variants, the optimal solution is not available. Therefore, we consider a large number of benchmark policies (see Section \ref{computational_benchmarks}) and pick the best one. In these problems too, our policy does as well as the best benchmark policy we could find. Lastly, for the three 100-dimensional problems that do not admit a pathwise optimal solution, the optimal policy is not known. Here, we again pick the best benchmark policy among the ones we consider, and our policy outperforms the best benchmark policy we could find.

The rest of the paper is structured as follows: Section \ref{model_section} presents the model. Section \ref{sec_brownian_control} derives a diffusion approximation in the Halfin-Whitt regime. Section \ref{sect_hjb_equation} defines the HJB equation associated with the diffusion control problem and introduces the proposed policy. Section \ref{computational_method}
 describes our computational method. Section \ref{data} describes the US Bank call center data (Section \ref{data_section}), introduces the test problems (Sections \ref{sect_main_test}-\ref{high_dim_non_pathwise_sect}) and the benchmark policies we use (Section \ref{computational_benchmarks}). All test problems are designed using the US Bank data. Section \ref{numerical_results} presents the computational results, comparing the proposed policy against benchmark policies. 
 Appendices \customappendixref{appendix_a}-G provide further details on the data, the derivation and computation of the benchmark policies, the neural network hyperparameters and the derivation of the algorithm used for our computational method.
\vspace{-6mm}
\section{Model}\label{model_section}
\begin{figure}[h]
    \centering
    \includegraphics[scale=0.25]{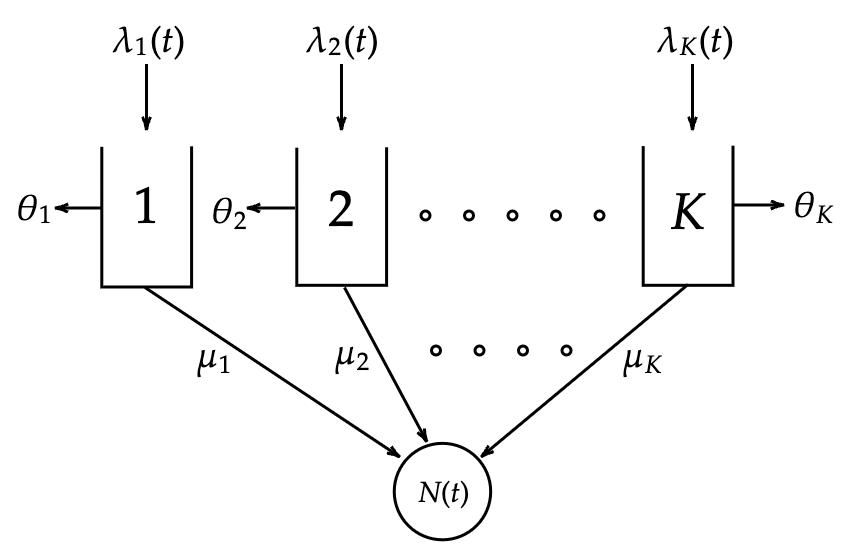}
    \caption{A non-stationary system model.}
    \label{figure_model}
\end{figure}

\vspace{-2mm}
We consider a queueing model of a telephone call center serving $K$ classes of callers. Each day, the call center operates during $[0, T]$, leading to a finite-horizon formulation. Class $k$ callers arrive according to a time-inhomogeneous Poisson process with intensity $\{\lambda_{k}(t): t \in [0,T]\}$. They leave the system either by receiving service or by abandoning while they wait in the queue. Associated with each class $k$ caller are his service and abandonment times. For class $k$, service times form an i.i.d. sequence of exponential random variables with mean $1/\mu_{k} > 0$. Similarly, abandonment times of class $k$ callers are i.i.d. exponential random variables with mean $1/\theta_{k} \geq 0$. The service times, abandonment times, and the caller arrival processes are mutually independent.  A schematic description of the model is given in Figure \ref{figure_model}.

The system state is denoted by $X(t) = \left(X_{1}(t), \ldots, X_{K}(t)\right)^{'}$, where $X_{k}(t)$ denotes the number of class $k$ callers in the system. There is a single pool of homogeneous agents answering calls. Given the nonstationary nature of arrivals to the call center, the number of agents working varies throughout the day. We let $N(t)$ denote the number of agents working at time $t$, and restrict attention to work-conserving policies. That is, an agent does not idle unless all queues are empty.

The system manager decides how to allocate the service effort to callers in the system dynamically. Her control is the $K$-dimensional process $\psi = \{\psi(t): 0 \leq t \leq T \}$, where $\psi_{k}(t)$ denotes the number of class $k$ callers in service at time $t$. Given system state $X(t) = x$, the control must satisfy $\psi(t) \in A(t,x)$, where 
$$A(t,x) = \{a \in \mathbb{R}_{+}^{K}: a \leq x, \, e \cdot a = (e\cdot x) \wedge N(t) \},$$
and $e$ is the $K$-vector of ones. The requirement $a \leq x$ follows because the number of class $k$ callers in service cannot exceed that in the system for $k = 1,\ldots, K.$ The second requirement ensures that the control is work-conserving. That is, the total number of callers in service is the minimum of the number of servers and the total number of callers in the system.

To facilitate the analysis, we also define two auxiliary processes $Y = \{Y(t) \in \mathbb{R}_{+}^{K}: t \in [0,T]\}$ and $Z = \{Z(t) \in \mathbb{R}_{+}: t \in [0,T]\}$ as follows: For $t \in [0,T],$
\begin{align}
    &Y(t) = X(t) - \psi(t), \label{eqn_unscaled_Y_X-psi}\\
    &Z(t) = N(t) - e\cdot\psi(t). \label{eqn_unscaled_Z}
\end{align}
In words, $Y_{k}(t)$ denotes the number of class $k$ callers waiting in the queue at time $t$. Similarly, $Z(t)$ denotes the number of idle servers at time $t$. The work conservation requirement of an admissible policy can equivalently be expressed as follows:
\begin{equation}
    (e \cdot Y(t)) \wedge Z(t) = 0, \quad t \in [0,T]. \label{eqn_work_conserving}
\end{equation}
The system manager strives to minimize the total expected cost that involves holding and abandonment costs
incurred during $[0,T]$ as well as the terminal cost $g(X(T)).$ To be more specific, each class $k$ caller waiting in queue results in a holding cost rate $h_{k}$ per unit of time. Similarly, each abandoning class $k$ caller costs the system $p_{k}$. We define the total cost rate for a class $k$ caller in the queue as follows:
\begin{equation}
c_{k} = h_{k} + \theta_{k}\, p_{k} > 0, \quad k = 1, \ldots, K.   \label{eqn:defn:$c_{k}$:param} 
\end{equation}
Given a control $\psi$, we characterize the system's status at time $t$ by the triple ($X(t), Y(t), Z(t))$. Under suitable assumptions, one can characterize this process as a (controlled) continuous-time Markov chain. The instantaneous cost rate at time $t$ is $c \cdot Y(t)$. Therefore, the total expected cost under policy $\psi$ over the interval $[t,T]$, given that $X(t) = x$, is
\begin{equation}
    J(t,x;\psi) = \mathbb{E}_{x}^{\psi}\left\{\int_{t}^{T} c \cdot Y(s)ds + g(X(T))\right\}, \label{eqn_$c_{k}$_of_policy}
\end{equation}
where $\mathbb{E}_{x}^{\psi}$ denotes the conditional expectation starting in state $x$ under policy $\psi$. The terminal cost corresponds to the overtime pay associated with serving the callers in the queue at the end of the day. We model it as follows:
\begin{equation}
    g(x) = \bar{c} \, \left(e \cdot x - N(t)\right)^{+}, \quad x \in \mathbb{R}_{+}^{K},\label{eqn_terminal_$c_{k}$}
\end{equation}
where $\bar{c}$ is the overtime pay for an agent to serve a caller in the queue. 

The problem described in this section is a continuous-time MDP. In principle, it can be solved using standard dynamic programming techniques. However, that approach is not computationally tractable for problems with high-dimensional state vectors, which is our focus. Therefore, we take a different approach. As a preliminary, we first derive a diffusion approximation to this problem formally (i.e., nonrigorously). We then study the resulting Brownian control problem using a novel computational approach that uses deep neural network approximations.
\vspace{-6mm}
\section{A Brownian control problem in the Halfin-Whitt regime}\label{sec_brownian_control}
\vspace{-2mm}
We consider a sequence of systems, each having the structure described in Section \ref{model_section}, indexed by $r = 1,2, \ldots.$ A superscript of $r$ is attached to various quantities of interest to emphasize their dependence on $r.$ To be specific, we assume the arrival, service, and abandonment rates vary with $r$ as follows:
For $k = 1,\ldots, K$,
\begin{align}
    &\lambda_{k}^{r}(t) = r \lambda_{k}(t) + \sqrt{r} \zeta_{k}(t) + o(\sqrt{r}), \quad t \in [0,T], \label{eqn_scaled_lambda}\\
    &\mu_{k}^{r} = \mu_{k}  \text{ and } \theta_{k}^{r} = \theta_{k}, \label{eqn_scaled_mu_theta}
\end{align}
where $\lambda_{k}(\cdot)$ and $\zeta_{k}(\cdot)$ are given functions. We estimate them from data in Section \ref{data_section} for our test problems. Similarly, the number of agents varies with $r$ as follows:
\begin{equation}
    N^{r}(t) = r N(t), \quad t \in [0,T]. \label{eqn_no_agents_in_nth_system}   
\end{equation}
Crucially, we assume that the system primitives satisfy the following heavy traffic assumption:
\begin{equation}
    \sum_{k=1}^{K} \frac{\lambda_{k}(t)}{\mu_{k}} = N(t), \quad t \in [0,T]. \label{eqn_HT_condition}
\end{equation}
Combining (\ref{eqn_scaled_lambda}) and (\ref{eqn_HT_condition}), the difference between the arriving work and service capacity at each time $t$ determines the uncontrollable drift-rate terms $\zeta_{k}(t)$ for $k = 1,\ldots, K$. The state descriptor will have two additional drift-rate terms that are controllable; see Equations (\ref{eqn_infinitesimal_drift}) and (\ref{eqn_state_process}). 

Under the foregoing assumptions, the system is a balanced, high-volume system. When the system parameter $r$ is large, the heavy traffic assumption can be interpreted as follows: Focusing on the first-order or fluid-scale terms, that is, terms of order $r$, the workload arriving at the system equals the service capacity at all times (see Equation (\ref{eqn_HT_condition})). On the other hand, the second-order terms in Equation (\ref{eqn_scaled_lambda}) contribute to the drift of the system state (see Equation (\ref{eqn_state_process}) below). Defining the nominal number of class $k$
 callers in service (on the fluid scale) as 
 \begin{equation*}
     \psi_{k}^{*}(t) = \frac{\lambda_{k}(t)}{\mu_{k}}, \quad k = 1,\ldots, K \text{ and } t \in [0,T],
 \end{equation*}
we now introduce the scaled state, control, queue length, and idleness processes $\hat{X}^{r}, \hat{\psi}^{r}, \hat{Y}^{r}, \hat{Z}^{r}$, respectively, as follows: For $k = 1,\ldots, K$, $t \in [0,T]$ and $r \geq 1,$
\begin{align*}
    &\hat{X}_{k}^{r}(t) = \frac{X_{k}^{r}(t) - r\psi_{k}^{*}(t)}{\sqrt{r}},\\
    & \hat{\psi}_{k}^{r}(t) = \frac{\psi_{k}^{r}(t) - r \psi_{k}^{*}(t)}{\sqrt{r}},\\
    & \hat{Y}_{k}^{r}(t) = \frac{Y_{k}^{r}(t)}{\sqrt{r}},\\
    &\hat{Z}^{r}(t) = \frac{Z^{r}(t)}{\sqrt{r}}.
\end{align*}
The scaled control $\hat{\psi}^{r}$ satisfies 
\begin{equation*}
    e \cdot \hat{\psi}^{r}(t) \leq 0 \text{ and } \hat{\psi}^{r}(t) \leq \hat{X}^{r}(t).
\end{equation*}
The first inequality follows from the natural requirement that $e \cdot \psi^{r}(t) \leq N^{r}(t)$ and Equations (\ref{eqn_no_agents_in_nth_system}) and (\ref{eqn_HT_condition}), whereas the second inequality is immediate from $\psi^{r}(t) \leq X^{r}(t)$, which holds because $\psi^{r}$ is admissible for the $r^{th}$ system. Similarly, Equations (\ref{eqn_scaled_Y_X-psi}) - (\ref{eqn_scaled_Z}) below follow from Equations (\ref{eqn_unscaled_Y_X-psi}) - (\ref{eqn_unscaled_Z}):
\vspace{-7mm}
\begin{align}
    &\hat{Y}^{r}(t) = \hat{X}^{r}(t) - \hat{\psi}^{r}(t), \label{eqn_scaled_Y_X-psi}\\
    &\hat{Z}^{r}(t) = - e \cdot \hat{\psi}^{r}(t). \label{eqn_scaled_Z}
\end{align}
In addition, the scaled queue length and idleness processes inherit the following work conservation property from Equation (\ref{eqn_work_conserving}):
\vspace{-5mm}
\begin{equation}
    (e \cdot \hat{Y}^{r}(t)) \wedge \hat{Z}^{r}(t) = 0. \label{eqn_nth_system_work_conserving}
    \vspace{-5mm}
\end{equation}
In what follows, we restrict attention to control policies that satisfy the following:
\vspace{2mm}
\begin{equation}
    \psi_{k}^{r}(t) = \frac{\lambda_{k}(t)}{\mu_{k}}r + \sqrt{r}\psi_{k}(t) + o(\sqrt{r}). \label{eqn_policies_of_interest}
\end{equation}
The leading term on the right-hand side reflects a flow-balance condition. Without it, the system manager incurs costs of order $r.$ However, for policies that satisfy (\ref{eqn_policies_of_interest}), we expect the costs to be of order $\sqrt{r}.$ The scaling of the second term on the right-hand side is natural under the diffusion scaling we consider.

For such policies, we now derive the infinitesimal drift and covariance of the scaled state process $\hat{X}^{r}$ to facilitate our formal derivation of its diffusion limit. In particular, for $t \in [0,T],\, k = 1, \ldots, K$,\, 
$j \neq k$, and small $h > 0$, the following holds:
\begin{align}
    &\mathbb{E}\left[\hat{X}_{k}^{r}(t + h) - \hat{X}_{k}^{r}(t) \mid \hat{X}^{r}(t)\right] = \left[\zeta_{k}(t) - \mu_{k}\hat{\psi}_{k}(t) - \theta_{k} \hat{Y}_{k}(t)\right]h + o(h), \label{eqn_infinitesimal_drift}\\
    &\mathbb{E}\left[(\hat{X}_{k}^{r}(t + h) - \hat{X}_{k}^{r}(t))^{2} \mid \hat{X}^{r}(t)\right] = 2\lambda_{k}(t)h + o(h), \label{eqn_infinitesimal_variance}\\
    &\mathbb{E}\left[(\hat{X}_{k}^{r}(t+h) - \hat{X}_{k}^{r}(t))(\hat{X}_{j}^{r}(t+h) - \hat{X}_{j}^{r}(t)) \mid \hat{X}^{r}(t)\right] = o(h). \label{eqn_infinitesimal_covariance}
\end{align}
Passing to the limit formally as $r \rightarrow \infty$ and denoting the weak limit of $(\hat{X}^{r}, \hat{\psi}^{r}, \hat{Y}^{r}, \hat{Z}^{r})$ by $(\hat{X}, \hat{\psi}, \hat{Y}, \hat{Z}),$ we deduce from Equations (\ref{eqn_infinitesimal_drift}) - (\ref{eqn_infinitesimal_covariance}) that the limiting state process $\hat{X}$ satisfies the following\footnote{Our treatment is purely formal and is meant to motivate our computational method. A rigorous weak convergence result justifying our approximation is left for future research.}: For $k = 1,\ldots, K$ and $t \in [0,T],$ 
\begin{equation}
    d \hat{X}_{k}(t) = \left(\zeta_{k}(t) - \mu_{k}\hat{\psi}_{k}(t) - \theta_{k}\hat{Y}_{k}(t)\right)dt + \sqrt{2\lambda_{k}(t)}\,dB_{k}(t), \label{eqn_state_process}
\end{equation}
where $B(t) = \left(B_{k}(t)\right)$ is a $K$-dimensional standard Brownian motion.

Also, the limiting control, queue length, and idleness processes, $\hat{\psi}, \hat{Y}$, and $\hat{Z}$, respectively, satisfy the following: 
\vspace{-6mm}
\begin{align}
    &\hat{Y}(t) = \hat{X}(t) - \hat{\psi}(t), \label{eqn_limiting_Y}\\
    &\hat{Z}(t) = -e \cdot \hat{\psi}(t),  \\
    &(e \cdot \hat{Y}(t)) \wedge \hat{Z}(t) = 0,
\end{align}
which follow from Equations (\ref{eqn_scaled_Y_X-psi}) - (\ref{eqn_nth_system_work_conserving}), respectively. 

To minimize technical complexity, we restrict attention to Markov controls. Moreover, following \citet{atar2004scheduling}, we adopt a choice of control that is more convenient mathematically. To be specific, letting $\mathcal{U} = \{u \in \mathbb{R}_{+}^{K}: e \cdot u = 1\}$, the system manager chooses a policy $u:[0,T] \times \mathbb{R}^{K} \rightarrow \mathcal{U}$, where $u_{k}(t,x)$ denotes the fraction of total backlog kept in class $k$ at time $t$ when the system state is $x$. We let $\hat{X}^{u}$ denote the state process under policy $u$. Similarly, the superscript $u$ will be attached to various processes to emphasize their dependence on policy $u$ as needed for clarity.

Given a policy $u$, one can represent the corresponding (limiting) queue length and idleness processes $\hat{Y}^{u},\,\hat{Z}^{u}$ as follows:
\begin{align}
    &\hat{Y}^{u}_{k}(t) = (e \cdot \hat{X}^{u}(t))^{+}u_{k}(t, \hat{X}^{u}(t)), \quad k = 1,\ldots, K, \label{eqn_limiting_Yi_under_control_u}\\
    &\hat{Z}^{u}(t) = (e \cdot \hat{X}^{u}(t))^{-}.
\end{align}
Combining Equations (\ref{eqn_limiting_Y}) and (\ref{eqn_limiting_Yi_under_control_u}) yields 
\begin{equation*}
    \vspace{-2mm}
    \hat{\psi}_{k}(t) = \hat{X}^{u}_{k}(t) - (e \cdot \hat{X}^{u}(t))^{+}u_{k}(t, \hat{X}^{u}(t)), \quad k = 1,\ldots, K.
\end{equation*}
Then for $k = 1,\ldots, K,$ defining
\begin{align}
&b_{k}(t,x,u) = \zeta_{k}(t) - \mu_{k}x_{k} + (e\cdot x)^{+}(\mu_{k} - \theta_{k})u_{k}(t,x), \label{eqn_drift}\\
&\sigma_{k}(t) = \sqrt{2\lambda_{k}(t)}, \label{eqn_sigma} 
\end{align}
and letting $b(\cdot) = (b_{k}(\cdot))$ and $\sigma(t) = \operatorname{diag}(\sigma_{1}(t), \ldots, \sigma_{K}(t))$, the following controlled diffusion process describes the limiting state process:
\begin{equation}
    d\hat{X}(t) = b\big(t, \hat{X}(t), u(t,\hat{X}(t))\big)dt + \sigma(t)\,dB(t), \quad t\in [0,T]. \label{eqn_x_limit}
\end{equation}
Given a control $u$ and the limiting system state $\hat{X}(t) = x$, the instantaneous expected cost rate at time $t$ is 
\vspace{-4mm}
\begin{equation}
    c\,(t,x,u) = (e \cdot x)^{+}\sum_{k=1}^{K}c_{k}u_{k}.
\end{equation}
Therefore, the total expected cost under a policy $u$ starting in state $x$ at time $t$, denoted by $\hat{J}(t,x;u)$, is given as follows\footnote{One can show that $\hat{J}$ is the formal limit of the scaled cost process $J^{r}/\sqrt{r}$.}:
\begin{equation}
    \hat{J}(t,x;u) = \mathbb{E}_{x}^{u} \left\{\int_{t}^{T} c\left(s,\hat{X}^{u}(s), u(s, \hat{X}^{u}(s))\right)ds + \hat{g}(\hat{X}^{u}(T))\right\},  \label{eqn_j_definition}
\end{equation}
where $\hat{g}(x) = \bar{c} \, (e \cdot x)^+$ and $\mathbb{E}_{x}^{u}$ denotes the conditional expectation starting in state $x$ under policy $u$. We now define the optimal value function as 
\vspace{-1mm}
\begin{equation*}
    V(t,x) = \inf \hat{J}(t,x;u),
    \vspace{-4mm}
\end{equation*}
where the infimum is taken over the class of admissible policies.

\section{HJB equation}\label{sect_hjb_equation}
\vspace{-2mm}
The HJB equation provides the means for solving the control problem analytically; see \citet{fleming2006controlled}. As a preliminary to introducing the HJB equation, we define the differential operator $\mathcal{L}$ and the auxiliary function $\mathcal{H}$ as follows:
\begin{align*}
    &\mathcal{L} = \sum_{k=1}^{K} \lambda_{k}(t)\frac{\partial^{2}}{\partial x_{k}^{2}}, \\
    & \mathcal{H}(t,x,p) = \inf_{u \in \mathcal{U}}\left[b(t,x,u)\cdot p + c\,(t,x,u)\right].
\end{align*}
\vspace{-2mm}
Then the HJB equation involves finding a sufficiently smooth function $V(t,x)$ that solves the following partial differential equation: For $(t,x) \in [0,T] \times \mathbb{R}_{+}^{K},$
\begin{align*}
    &\frac{\partial V}{\partial t}(t,x) + \mathcal{L} V(t,x) + \mathcal{H}(t,x, \nabla_{x}V(t,x)) = 0,\\
    &V(T,x) = \hat{g}(x),
\end{align*}
where $\nabla_{x}$ denotes the gradient operator with respect to variable $x$. Rewriting the function $\mathcal{H}$ more explicitly yields the following HJB equation: For $(t,x) \in [0,T] \times \mathbb{R}_{+}^{K}$,
\begin{align}
    \begin{split}
        &\frac{\partial V}{\partial t} (t,x) + \sum_{k=1}^{K} \lambda_{k}(t) \frac{\partial^{2}V}{\partial x_{k}^{2}}(t,x) \\
        &\qquad \qquad + \sum_{k=1}^{K}(\zeta_{k}(t) - \mu_{k}x_{k})\frac{\partial V}{\partial x_{k}}(t,x) + (e \cdot x)^{+}\min_{k=1,\ldots,K}\left[c_{k} + (\mu_{k} - \theta_{k})\frac{\partial V}{\partial x_{k}}(t,x)\right] = 0, \label{eqn_HJB_PDE}
    \end{split}\\
     &V(T,x) = \hat{g}(x). \label{eqn_HJB_terminal_condition}
\end{align}
\noindent\textbf{Proposed policy and its interpretation in the pre-limit system.} The minimization operation in the HJB equation suggests a natural way to propose an effective policy; see the last term of the left-hand side of Equation (\ref{eqn_HJB_PDE}). To do so, we define a time-varying effective holding cost function as follows:
\begin{equation}
    \,\,\,\,\,\, \phi_{k}(t,x) = c_{k} + (\mu_{k} - \theta_{k}) \frac{\partial V}{\partial x_{k}}(t,x), \,\, \text{ for } \,\, (t,x) \in [0,T] \times \mathbb{R}^{K}. \label{eq_effective_$c_{k}$}
\end{equation}
Then we define a permutation $i(t,x) = (i_{1}(t,x), \ldots, i_{K}(t,x))$ of $(1,\ldots,K)$ such that\footnote{Ties are broken by maintaining the order induced by the original class indices.}
\begin{align*}
\phi_{{i}_{1}}(t,x) \geq \cdots \geq \phi_{{i}_{K}}(t,x) \ \ \textrm{for} \ (t,x) \in [0,T] \times \mathbb{R}^K.
\end{align*}
Class $i_1(t,x)$ is the most expensive class whereas class $i_K(t,x)$ is the cheapest one to keep the backlog in at time $t$ when the system state is $x$. Thus, the proposed policy, denoted by $u^*(t,x)$, keeps the backlog in the cheapest class at all times. That is, for each $(t,x) \in [0,T] \times \mathbb{R}^K$, it satisfies
\vspace{-4mm}
\begin{align*}
    u^{\ast }_{i_K(t,x)}(t,x) = 1 \ \textrm{and} \ u^{*}_i(t,x)= 0 \ \textrm{for} \ i \neq i_K(t,x).
\end{align*}
Implementing this policy in the pre-limit system literally is not always possible. Thus, we propose a natural modification of it instead. The proposed policy keeps the backlog in the cheapest buffers, that is, the higher indexed classes under permutation $i(t,x)$, as much as possible. Because we restrict attention to work-conserving policies, this is equivalent to prioritizing lower-indexed classes when assigning servers to callers. To be specific, servers are first assigned to class $i_{1}$, then class $i_{2}$ and so on.  That is, given the system state $x$ at time $t$, we determine the number of class $k$ jobs in service as follows:
\begin{align*}
    &\psi_{{i}_{1}}(t) = x_{{i}_{1}} \wedge N^{r}(t),\\ 
    &\psi_{{i}_{k}}(t) = x_{{i}_{k}} \wedge \left[N^{r}(t) -  \sum_{l=1}^{k-1} \psi_{i_{l}}(t)\right], \quad k = 2,\ldots, K.
\end{align*}
\section{Computational method}\label{computational_method}
\vspace{-2mm}
Our computational method adopts the deep splitting method proposed by \citet{beck2021deep} for solving parabolic PDEs to our problem setting.  

To begin, we specify a \textit{reference policy} to generate sample paths of the system state to be visited during the training process. Loosely speaking, we aim to choose a reference policy so that its paths tend to occupy the parts of the state space that we expect the optimal policy to visit frequently. We denote our reference policy by $\tilde{u}$. Then, the corresponding \textit{reference process}, denoted by $\tilde{X}$, satisfies the following: For $t \in [0,T]$ and $k = 1,\ldots,K$, 
\begin{equation}
    d \tilde{X}_{k}(t) = b_{k}\big(t, \tilde{X}(t), \tilde{u}(t,\tilde{X}(t))\big)dt + \sigma_{k}(t)dB_{k}(t), 
\end{equation}
where $b_{k}(t,x,u)$ and $\sigma(t)$ are defined as in Equations (\ref{eqn_drift}) and (\ref{eqn_sigma}), respectively. The sample paths of the \textit{reference process} serve as the training data for our computational method.

Next, we derive a stochastic identity from the HJB equation (\ref{eqn_HJB_PDE}) by evaluating it at $(t,\tilde{X}(t))$ for $t \in [0,T]$ and by adding and subtracting the following term:
$$\sum_{k=1}^{K}(e\cdot \tilde{X}(t))^{+}(\mu_{k}-\theta_{k})\, \tilde{u}_{k}(t,\tilde{X}(t))\frac{\partial V}{\partial x_{k}}(t,\tilde{X}(t)),$$ 
which yields the stochastic identity
\begin{align}
    \frac{\partial V}{\partial t}(t,\tilde{X}(t)) + \sum_{k=1}^{K}\lambda_{k}(t)\frac{\partial^{2}V}{\partial x_{k}^{2}}(t,\tilde{X}(t)) &+ \sum_{k=1}^{K}b_{k}\big(t, \tilde{X}(t), \tilde{u}(t,\tilde{X}(t))\big)\frac{\partial V}{\partial x_{k}}(t,\tilde{X}(t)) \nonumber \\
    &- F(t,\tilde{X}(t),\nabla_{x}V(t,\tilde{X}(t))) = 0, \quad t \in [0,T],
    \label{modified_HJB}
\end{align}
where for $(t,x,v) \in [0,T] \times \mathbb{R}^{K} \times \mathbb{R}^{K},$ the auxiliary function $F$ is defined as follows:
\begin{flalign}
    F(t,x,v) = (e \cdot x)^{+}\left(\sum_{k=1}^{K}v_{k}\,(\mu_{k} - \theta_{k})\,\tilde{u}_{k}(t,x) - \min_{k=1,\ldots, K}\Big[c_{k} + (\mu_{k} - \theta_{k})\,v_{k}\Big]\right). \label{eqn_f_function}
\end{flalign}

We proceed by fixing a sufficiently fine partition $0 = t_{0} < t_{1} < \ldots < t_{N} = T$ of the time horizon $[0,T]$. Then, 
given a reference policy $\tilde{u}$, we generate the discretized state-pair samples $(\tilde{X}(t_{n}), \tilde{X}(t_{n+1}))$ for $n = 0,1,\ldots,N-1$ by first initializing $\tilde{X}(t_{n})$ using the empirical distribution estimated in Subroutine 1 and then applying Equation (\ref{path2}) to generate $\tilde{X}(t_{n+1})$; see Subroutine 2. In Equation (\ref{modified_HJB}), the term involving the auxiliary function $F$ is nonlinear whereas the rest of the terms constitute a linear PDE if one ignores the last term. The idea behind the deep splitting method is to approximate the nonlinear term in Equation (\ref{modified_HJB}) over each subinterval $[t_{n},t_{n+1}]$ so as to arrive at a linear PDE that approximates the original PDE, given in Equation (\ref{modified_HJB}). One then proceeds with a derivation of a Feynman-Kac representation of that linear PDE to propose a solution to the original PDE on that subinterval. In our context, this ultimately gives rise to the following relationship for $n = 0,1,\ldots,N-1$:
\begin{equation}
    \mathbb{E}\left[V\big(t_{n+1}, \tilde{X}(t_{n+1})\big) \Bigm| \tilde{X}(t_{n}) \right] = V\big(t_{n}, \tilde{X}(t_{n})\big) + F\big(t_{n}, \tilde{X}(t_{n}), \nabla_{x}V(t_{n}, \tilde{X}(t_{n}))\big)(t_{n+1} - t_{n}). \label{key_identity}
\end{equation}
See \ref{alg_derivation_appendix} for a formal derivation.

Next, for each $n = 0, 1,\ldots, N-1$, we approximate the value function $V(t_{n}, \cdot)$ with a neural network $H^{n}(\,\cdot\,; \omega_n)$ and the gradient function $\nabla_{x}V(t_{n}, \cdot)$ with a neural network $G^{n}(\,\cdot\,; \nu_n)$, where $\omega_{n}$ and $\nu_{n}$ are parameter vectors. Given the neural network approximations $H^{n}(\,\cdot \,;\omega_{n})$ and $G^{n}(\,\cdot \, ;\nu_{n})$ for $n = 0,1,\ldots,N-1$, the deep splitting method proceeds backward in time, starting with $n = N$. Using the terminal cost function $\hat{g}(\cdot)$, we set $V(t_{N}, \tilde{X}(t_{N})) = \hat{g}(\tilde{X}(t_{N}))$.

Because the condition expectation in (\ref{key_identity}) can be viewed as a projection operation, it corresponds to minimizing a certain $L^{2}$ norm. Thus, to determine the optimal neural network parameters $\omega_{N-1}^{\star}, \, \nu_{N-1}^{\star}$, we define the loss function 
\begin{align*}
      \ell_{N-1}(\omega_{N-1}, \nu_{N-1}) = \mathbb{E}\Big[\Big(\hat{g}(\tilde{X}(t_{N})) &- H^{N-1}(\tilde{X}(t_{N-1})\,;\,\omega_{N-1}) \\
      & - F\big(t_{N-1}, \tilde{X}(t_{N-1}),  G^{N-1}(\tilde{X}(t_{N-1})\,;\,\nu_{N-1})\big)\Delta t_{N-1} \Big)^{2}\Big],
\end{align*}
and we set $(\omega_{N-1}^{\star}, \, \nu_{N-1}^{\star}) \in \operatorname{argmin} \ell_{N-1}(\omega_{N-1}, \nu_{N-1})$. 

Proceeding similarly for $n = N-2, \ldots,1,0$, we first define the loss function $\ell_{n}(\omega_{n}, \nu_{n})$ recursively as follows: 
\begin{align*}
      \ell_{n}(\omega_{n}, \nu_{n}) = \mathbb{E}\Big[\Big( H^{n+1}(\tilde{X}(t_{n+1})\,;\,\omega^{\star}_{n+1}) - H^{n}(\tilde{X}(t_{n})\,;\,\omega_{n}) 
      - F\big(t_{n}, \tilde{X}(t_{n}),  G^{n}(\tilde{X}(t_{n})\,;\,\nu_{n})\big)\Delta t_{n} \Big)^{2}\Big].
\end{align*}
We then set the optimal neural network parameters as $(\omega_{n}^{\star}, \nu_{n}^{\star}) \in \operatorname{argmin} \ell_{n}(\omega_{n}, \nu_{n}).$

We summarize our computational method in Algorithm \ref{ds_alg}, where we also replace the loss functions with their empirical estimates. That is, we replace the expectations with the corresponding sample averages using the samples generated by Subroutine 2. More specifically, letting $\{\tilde{X}^{(s)}(t_{n}), \tilde{X}^{(s)}(t_{n+1})\}_{s = 1}^{S}$ for $n = 0,1,\ldots,N-1$, we define the empirical loss functions, denoted by $\hat{\ell}_{n}(\omega, \nu)$ for $n = 0,1,\ldots,N-1$, recursively, as follows:
\vspace{-1mm}
\begin{align*}
    \hat{\ell}_{N-1}(\omega_{N-1}, \nu_{N-1}) = \frac{1}{S} \sum_{s = 1}^{S} \Big(\hat{g}(\tilde{X}^{(s)}&(t_{N})) - H^{N-1}(\tilde{X}^{(s)}(t_{N-1});\omega_{N-1})\\
    & - F\big(t_{N-1}, \tilde{X}^{(s)}(t_{N-1}), G^{N-1}(\tilde{X}^{(s)}(t_{N-1}); \nu_{N-1})\big)\Delta t_{N-1}\Big)^{2},
\end{align*}
\vspace{-3mm}
and for $n = 0,1,\ldots,N-2$, given $\omega_{n+1}^{\star}$ from the previous optimization step, we let 
\begin{align*}
    \hat{\ell}_{n}(\omega_{n}, \nu_{n}) = \frac{1}{S} \sum_{s = 1}^{S} \Big(H^{n+1}(\tilde{X}^{(s)}(t_{n+1}); \omega^{\star}_{n+1}) &- H^{n}(\tilde{X}^{(s)}(t_{n});\omega_{n})\\
    & - F\big(t_{n}, \tilde{X}^{(s)}(t_{n}), G^{n}(\tilde{X}^{(s)}(t_{n}); \nu_{n})\big)\Delta t_{n}\Big)^{2}.
\end{align*}
\begin{algorithm}[!htb]
        \floatname{algorithm}{Subroutine}
        \label{subroutine_1}
	\caption*{\textbf{Subroutine 1} Distribution generator}
        \begin{minipage}{\textwidth} 
	\begin{algorithmic}[1]
            \Statex \textbf{Input:} A fixed reference policy $\tilde{u}(\cdot)$, the drift term $b(\cdot, \tilde{X}(\cdot), \tilde{u}(\cdot))$, the variance term $\sigma^{2}(\cdot),$ the time horizon $T$, the number of intervals $N$, a discretization step-size $\Delta t_{n}$ (for simplicity, we assume $\Delta t_{n} \triangleq T/N$), an initial distribution $\Gamma_{0}$,  and a random initial state $x_{0} \sim \Gamma_{0}$, $J$ sample path replications.
            \Statex \textbf{Output:} Empirical distribution $\Gamma_{n} = (\Gamma_{n}^{(k)})$ for $k = 1,\ldots, K$ and $n = 0,1,\ldots, N-1$.
		\Function{GENERATOR}{$T, \Delta t_{n}, x, J$}
            \State For time interval $[0,T]$ and $N = T/\Delta t_{n}$, construct the partition $0 = t_{0} < \ldots < t_{N} = T,$
            \Statex  \quad \, where $\Delta t_{n} = t_{n+1} - t_{n}$ for $n = 0,\ldots,N-1.$ 
            \For {$j \leftarrow 1$ to $J$}
            \State Generate $N$ i.i.d. $K$-dimensional Gaussian random vectors $\Delta B^{(j)}(t_{n}) = (\Delta B^{(j)}_{k}(t_{n}))$ 
            \Statex \quad \quad \quad with mean zero and covariance matrix $\Delta t_{n}I$ for $n = 0,1,\ldots, N-1$.
		\For {$n \leftarrow 0$ to $N-1$}
            \For {$k \leftarrow 1$ to $K$}
            \vspace{1mm}
		\State \hspace{-3mm} $\tilde{X}^{(j)}_{k}(t_{n+1}) \leftarrow \tilde{X}^{(j)}_{k}(t_{n}) + b_{k}\left(t_{n}, \tilde{X}^{(j)}(t_{n}), \tilde{u}(t_{n}, \tilde{X}^{(j)}(t_{n}))\right)\Delta t_{n} + \sigma_{k}(t_{n})\Delta B^{(j)}_{k}(t_{n})$
		\EndFor
		\EndFor
            \EndFor
            \For {$n \leftarrow 0$ to $N-1$}
            \For {$k \leftarrow 1$ to $K$}
            \State Estimate mean $\mu_{k}(\tilde{X}_{k}(t_{n})) = \frac{1}{J}\sum_{j=1}^{J}\tilde{X}_{k}^{(j)}(t_{n})$.
            \State Estimate standard deviation $\sigma_{k}(\tilde{X}_{k}(t_{n})) = \sqrt{\frac{1}{J}\sum_{j=1}^{J}\left(\tilde{X}_{k}^{(j)}(t_{n}) - \mu_{k}
            (\tilde{X}_{k}(t_{n}))\right)^{2}}$.
            \State Fit the empirical distribution for class $k$ as $\Gamma_{n}^{(k)} := \mathscr{N}(\mu_{k}(\tilde{X}_{k}(t_{n}), \sigma_{k}(\tilde{X}_{k}(t_{n}))).$
            \EndFor
            \EndFor
		\State \textbf{return} $\Gamma_{n} = (\Gamma_{n}^{(k)})$ for $k = 1,\ldots, K$ and $n = 0,1\ldots, N-1$.
        \EndFunction
	\end{algorithmic} 
 \end{minipage}
\end{algorithm}
\begin{algorithm}[H]
        \floatname{algorithm}{Subroutine}
        \label{subroutine_1}
	\caption*{\textbf{Subroutine 2} Sampling Process}
        \begin{minipage}{\textwidth} 
	\begin{algorithmic}[1]
            \Statex \textbf{Input:} A fixed reference policy $\tilde{u}(\cdot)$, the drift term $b(\cdot, \tilde{X}(\cdot), \tilde{u}(\cdot))$, the variance term $\sigma^{2}(\cdot),$  a batch size $S$, a discretization step-size $\Delta t_{n}$ (for simplicity, we assume $\Delta t_{n} \triangleq T/N$), a distribution $\Gamma_{n}$ (see Subroutine 1 for the GENERATOR of $\Gamma_{n}$).
            \Statex \textbf{Output:} State pair sample $\{\tilde{X}^{(s)}(t_{n}), \tilde{X}^{(s)}(t_{n+1})\}_{s=1}^{S}$ 
		\Function{SAMPLING}{$t_n, \Delta t_{n}, \Gamma_n$}
             \State Generate one $K$-dimensional Gaussian random vector $\Delta B(t_{n})$ with mean zero and
            \Statex \quad \, covariance matrix $\Delta t_{n}I$.
            \State Initialize a random state $\tilde{X}^{(s)}(t_{n}) \sim \Gamma_{n}$ for $s = 1,\ldots,S$.
            \For {$k \leftarrow 1$ to $K$}
		\State \begin{equation}\tilde{X}^{(s)}_{k}(t_{n+1}) \leftarrow \tilde{X}^{(s)}_{k}(t_{n}) + b_{k}\left(t_{n}, \tilde{X}^{(s)}(t_{n}), \tilde{u}(t_{n}, \tilde{X}^{(s)}(t_{n}))\right)\Delta t_{n} + \sigma_{k}(t_{n})\Delta B_{k}(t_{n}). \label{path2} \end{equation}
            \vspace{1mm}
		\EndFor
		\State \textbf{return} $\{\tilde{X}^{(s)}(t_{n}), \, \tilde{X}^{(s)}(t_{n+1})\}_{s =1}^{S}$
        \EndFunction
	\end{algorithmic} 
 \end{minipage}
\end{algorithm}
\newpage
Given the approximation of the gradient function $G^{n}(\,\cdot\,;\,\nu^{\star}_{n})$ for $n = 0,\ldots,N-1$, we approximate the effective holding cost function $\phi$ defined in Equation (\ref{eq_effective_$c_{k}$}) as follows: For $t \in [t_{n}, t_{n+1})$, $n = 0,1,\ldots, N-1$ and $k = 1,\ldots, K$, we let 
\begin{equation*} 
    \tilde{\phi}_{k}(t, x) = c_{k} + (\mu_{k} - \theta_{k})\,G^{n}_{k}(x;\nu_{n}^{\star}).
\end{equation*}
Following the approach detailed in Section \ref{sect_hjb_equation}, we use the approximate effective holding cost function $\tilde{\phi}$ to order classes from the most expensive to the cheapest. Given this ordering, we propose keeping the backlog in the cheapest buffers as much as possible, as described in Section \ref{sect_hjb_equation}. Equivalently, this corresponds to assigning the servers to the ``most expensive'' class first, then to the second most expensive class, and so on. 

In our numerical study, we consider the following five reference policies: i) evenly split, ii) weighted split, iii) minimal, iv) randomly split, and v) static priority. The evenly split reference policy distributes the backlog evenly across classes. That is, we set $\tilde{u}_{k}(t_{n},\cdot) = 1/K$ for all $k,n$. The weighted-split reference policy generalizes the evenly-split reference policy slightly. It divides the classes $1,\ldots,K$ into two sets: $\mathcal{C}$ and $\{1,\ldots,K \} \textbackslash \mathcal{C}$. Then it sets
\begin{align*} 
	\tilde{u}_{k} (t_n,\cdot)  \,=\, \left \{\!\! \begin{array}{ll} 
		w_1, & \text{if } k \in \mathcal{C}, \\
		w_2, & \text{otherwise,}
	\end{array} \right.
\end{align*}

\noindent where $w_1$ and $w_2$ satisfy $\sum_{k=1}^K \tilde{u}_k(t_n,\cdot) =1$. If $w_1=w_2$, then the weighted-split reference policy reduces to the evenly-split reference policy. \ref{appendix_computational_method} describes another particular weighted-split reference policy, where set $\mathcal{C}$ and weights $w_1$ and $w_2$ are chosen by taking into consideration the call volume of each class. The minimal reference policy sets $\tilde{u}_{k}(t_{n}, \cdot) = 0$ for all $k,n$. Under the randomly split reference policy, the backlog is distributed uniformly over the unit simplex. To be more specific, we let $\tilde{u}(t_{n},x)$ be a sequence of i.i.d. Dirichlet\,($1,\ldots,1$) random vectors for $n = 0,\ldots, N-1$ and $x \in \mathbb{R}^{K}$. Lastly, the static priority reference policy puts all backlog in one class. Namely, we pick a class, say $k^{\star}$ and for all $k,n$ and $x$, set the control as follows:

\begin{singlespace}
    \begin{equation*}
    \tilde{u}_{k} (t_n,x) = \begin{cases}
			1 & \text{ if } k = k^{\star},\\
                0 & \text{ otherwise.}
		 \end{cases}
    \end{equation*}
\end{singlespace}
\begin{algorithm}[!htb] 
    \caption{Deep Splitting}
    \label{ds_alg}
    \begin{algorithmic}[1]
            \Statex \textcolor{black}{\textbf{Input:} A batch size $S$, a time horizon $T$, the number of intervals $N$, the number of subnetworks $2N$, a discretization step-size $\Delta t_{n}$ (for simplicity, we assume $\Delta t_{n} \triangleq T/N$), a fixed reference policy $\tilde{u}(\cdot)$, and an optimization solver (SGD, ADAM, RMSProp, etc).} 
        \Statex \textbf{Output:} The approximation of the value function $H^{n}(\,\cdot\,;\,\omega^{\star}_n)$, the approximation of the gradient function $G^{n}(\,\cdot\,;\,\nu^{\star}_{n})$, and the neural network weights $(\omega_{n}^{\star}, \nu_{n}^{\star})$ for $n = 0,\ldots, N-1$.
        \State Set $V(t_{N}, \cdot) = \hat{g}(\cdot)$.
        \State Generate $S$ state pairs $\{\tilde{X}^{(s)}(t_{N-1}), \tilde{X}^{(s)}(t_{N})\}_{s=1}^{S}$ by invoking SAMPLING$(t_{N-1}, \Delta t_{N-1})$.
        \State Initialize the neural network weights $(\omega_{N-1},\nu_{N-1})$ with Kaiming or Xavier initialization.
        \State Compute the minimizer of the empirical loss $\hat{\ell}_{N-1}(\omega_{N-1},\nu_{N-1})$
        \Statex \begin{align*}
            \hat{\ell}_{N-1}(\omega_{N-1}, \nu_{N-1}) = \frac{1}{S} \sum_{s = 1}^{S} \Big(\hat{g}&(\tilde{X}^{(s)}(t_{N})) - H^{N-1}(\tilde{X}^{(s)}(t_{N-1});\omega_{N-1}) \\
            &- F\big(t_{N-1}, \tilde{X}^{(s)}(t_{N-1}), G^{N-1}(\tilde{X}^{(s)}(t_{N-1}); \nu_{N-1})\big)\Delta t_{N-1}\Big)^{2}.
        \end{align*}
        \State Update $(\omega_{N-1}^{\star},\nu_{N-1}^{\star}) \in \operatorname{argmin}\ell_{N-1}(\omega_{N-1},\nu_{N-1})$.
        \For {$n \leftarrow N-2$ to $0$}
            \State Generate $S$ state pair samples $\{\tilde{X}^{(s)}(t_{n}), \tilde{X}^{(s)}(t_{n+1})\}$ by invoking SAMPLING$(t_{n}, \Delta t_{n})$.
            \State Initialize the neural networks weights as  $(\omega_{n},\nu_{n}) = (\omega^{\star}_{n+1}, \nu^{\star}_{n+1})$.
            \State Compute the minimizer of the empirical loss $\hat{\ell}_{n}(\omega_{n},\nu_{n})$
            \Statex \begin{align*}
    \hat{\ell}_{n}(\omega_{n}, \nu_{n}) = \frac{1}{S} \sum_{s = 1}^{S} \Big(H^{n+1}(\tilde{X}^{(s)}(t_{n+1})\,;\,& \omega^{\star}_{n+1}) - H^{n}(\tilde{X}^{(s)}(t_{n})\,;\,\omega_{n})\\
    & - F\big(t_{n}, \tilde{X}^{(s)}(t_{n}), G^{n}(\tilde{X}^{(s)}(t_{n}); \nu_{n})\big)\Delta t_{n}\Big)^{2}.
\end{align*}
        \State Update $(\omega_{n}^{\star},\nu_{n}^{\star}) \in \operatorname{argmin}\ell_{n}(\omega_{n},\nu_{n})$.
        \EndFor
        \State \textbf{return} Functions $H^{n}(\cdot\,;\,\omega^{\star}_{n})$ and $G^{n}(\, \cdot \,;\nu^{\star}_{n})$ for $n = 0,1,\ldots, N-1$.
    \end{algorithmic} 
\end{algorithm}
\section{Data, test problems, and benchmark policies}\label{data}
\subsection{Data} \label{data_section}
We use the publicly available data set of a US bank call center that is provided by the Service Enterprise Engineering Lab at the Technion\footnote{Available at https://see-center.iem.technion.ac.il/databases/USBank/. Accessed on August 9, 2023. }. The data contains records of agent activities and individual calls between March 2001 and October 2003. The call center operates 24 hours a day, seven days a week, across four sites located in New York, Pennsylvania, Rhode Island, and Massachusetts. Agents are divided into six groups that are referred to as nodes. There isn't a one-to-one correspondence between nodes and sites. In particular, each node may include agents who are housed in different sites. Similarly, a site may house agents who belong to different nodes. The six nodes are numbered 1, 2, 3, 5, 6, and 7. The call center receives up to 330,000-350,000 calls a day on weekdays and 170,000-190,000 calls a day on weekends. Over a thousand agents work on weekdays and a few hundred on weekends, unevenly distributed among the six different nodes.

Customers enter the system through the voice response unit (VRU), an automated system allowing them to complete transactions independently. Most customers leave the system after performing some self-service transaction via the VRU, but around 55,000-65,000 calls, about 20\% of daily arrivals, speak with an agent. We only focus on these callers for our analysis. The VRU forwards these calls to the agents capable of performing the desired service. Over time, the call center stopped some of its services offered in 2001 and started additional services after November 2002. Thus, to focus on the calls that show similar arrival patterns and request the same services, we restrict our analysis to the calls arriving between May and July 2003. During this period, the call center serves 15 different types of customers; see Table \ref{stats}. 

Each call is divided into one or more subcalls that trace its activities in the system from entry to exit. Table \ref{characteristics} lists call characteristics observed in the data. To remove outliers, we focus our analysis on the calls having normal termination, transfer, short abandonment, and abandonment as an outcome, consisting of more than 99\% of the observations. Also, we restrict attention to the first subcall, which starts when the customer first joins the queue to speak with an agent and ends when the first service is completed. The total duration of the first subcalls accounts for about 70\% of the total talk time of all calls.

\begin{figure}[H]
\centering
    \includegraphics[scale = 0.45]{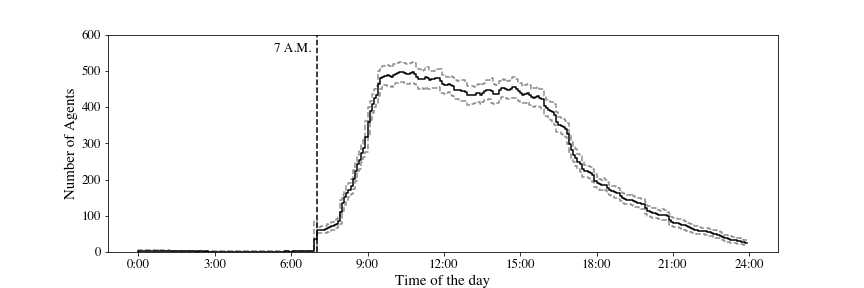}
    \caption{The number of agents (after the aforementioned service capacity adjustment) on weekdays during May-July 2003. The resolution of the horizontal axis is five minutes. The solid line depicts the average staffing level observed throughout all days. The two dashed lines that enclose it are derived by calculating the average staffing level plus or minus twice the standard deviation for each five-minute interval across all days. }
    \label{Agents_CI}
\end{figure}

The call volume is significantly higher on weekdays than it is on weekends. We focus attention on weekday calls between 7 AM and midnight. Figure \ref{Arrivals_CI} shows the call arrival rate during a weekday. The data provides codes for the state of agents for every second of a shift (see Table \ref{states} in \ref{data_specifics}). We divide the day into five-minute intervals to find the number of active agents in each interval using this information. We consider an agent to be available in an interval if she is already logged in and has not yet logged out or taken a break. This allows us to determine the number of agents working throughout the day. Since we focus on the first subcalls, we adjust the service capacity allocated to those subcalls according to the time spent on them relative to the total duration of the calls. Figure \ref{Agents_CI} displays the average number of agents over a day after this adjustment.

As alluded to above, customers who cannot be served immediately after exiting VRU are placed in a queue. These customers can abandon the queue before entering the service. The time to abandon is observed for the callers who abandon. For other callers, the time to abandon is censored by their wait time until they enter service. Only 2\% of the callers abandon, leading to heavy censoring of the abandonment times. Such heavy censoring can lead to biased estimates; see \citet{brown2005statistical} and \citet{akcsin2013structural}. Therefore, we use the bias-corrected Kaplan-Meier estimator proposed by~\citet{stute1994jackknife} to find the average abandonment times for each service class; see the last column of Table \ref{stats}.
\vspace{-4mm}
\vspace{-4mm}
Finally, to further eliminate outliers, we focus our analysis on the calls with a service time shorter than 30 minutes and a waiting time shorter than 15 minutes, constituting 99.5\% of our observations. The Retail class calls that have arrived at nodes 5, 6, and 7 to be served by an agent consist of less than 0.01\% of all calls. Thus, we remove these observations from our analysis. Consequently, we focus on three nodes, numbered 1, 2, and 3 for Retail calls. Because the Retail class alone constitutes more than half of all calls, we split it into three different classes corresponding to nodes 1, 2, and 3, where the calls are handled; see the first column of Table \ref{stats}.

In summary, we focus on 4,016,560 calls from any of the 15 service classes offered by the call center, received on weekdays during May-July 2003, excluding holidays, between 7 AM and midnight, having joined the queue to speak with an agent. To repeat, we further restrict attention to their first subcalls. The summary statistics for those calls are given in Table  \ref{stats}.

We supplement the US Bank data by estimating the holding and abandonment costs using other data sources; see Table \ref{cost1}. To do so, we assume the opportunity cost of an hour spent waiting equals the foregone hourly wage of the caller. US Bureau of Labor Statistics\footnote{Available at https://www.bls.gov/news.release/empsit.t19.htm. Accessed on August 11, 2023.} reports \$24 as the average hourly wage for the retail industry. 
Thus, we set the hourly holding cost rate of the Retail class as \$24. Then, we divide the remaining classes into two groups based on their perceived importance/priority relative to the Retail class. Namely, we put classes Premier, Business, Platinum, and Priority Service into one group, and the rest in another group. We further divide the lower priority group into two according to their call volume. For those classes whose arrival rate is less than 1\% of the total arrival rate, we set their holding cost rate as \$20, the lowest value we use. For the rest of the classes in that group, we set their hourly holding cost rate as \$22. Lastly, for those classes in the higher priority group, we set their hourly holding cost rates as shown\footnote{Note that the weighted average of the holding cost rates is about \$24, the holding cost rate of the Retail class.} in Table \ref{cost1} with Priority Service and Platinum classes having the highest value whereas the Premier class having the lowest value within that group.
\vspace{2mm}
\begin{table}[H]
	\centering
	\setlength\tabcolsep{4pt} 
	{\small 
             \scalebox{0.87}{
		\begin{tabular}{lccccccc}
			\toprule
			Class & Number of & Arrival & Average service & Average abandonment & $\mu$ & $\theta$\\
			\rule{0pt}{3.5ex} &  observations& percentage (\%) & time (sec.) & time (sec.) & (per hr.) & (per hr.) \\
			\midrule
			Retail (Node: 1) & 618,077 & 15.39 & 209.04 & 594.30 & 17.22 & 6.06\\
                Retail (Node: 2) & 916,473 & 22.82 & 208.67 & 460.94 & 17.25 & 7.81\\
                Retail (Node: 3) & 622,495 & 15.50 & 208.69 &  689.63 & 17.25 & 5.22\\
			Premier & 138,815 & 3.46 & 273.70 & 367.66 & 13.15 & 9.79\\
			Business & 193,564 & 4.82 & 217.35 & 419.41 & 16.56 & 8.58\\
			Platinum & 13,784 & 0.34 & 209.32 & 480.31 & 17.20 & 7.50\\
			Consumer Loans & 277,930 & 6.92 & 236.95 & 739.10 & 15.19 & 4.87\\
			Online Banking & 106,145 & 2.64 & 339.75 & 644.75 & 10.60 & 5.58\\
			EBO & 28,730 & 0.72 & 364.59 & 437.02 & 9.87 & 8.24\\
			Telesales & 251,513 & 6.26 & 374.20 & 400.25  & 9.62 & 8.99\\
			Subanco & 20,576 & 0.51 & 305.24 & 563.06 & 11.79 & 6.39\\
			Case Quality & 33,652 & 0.84 & 362.71 & 388.48 & 9.93 & 9.27\\
			Priority Service & 58,991 & 1.47 & 347.94 & 393.99 & 10.35 & 9.14\\
			AST & 137,437 & 3.42 & 287.47 & 480.11 & 12.52 & 7.50\\
			CCO & 335,051 & 8.34 & 236.79 & 506.98 & 15.20 & 7.10\\
			Brokerage & 232,338 & 5.78 & 285.30 & 522.39 & 12.62 & 6.89\\
			BPS &  30,989 & 0.77 & 265.27 & 608.49 & 13.57 & 5.92\\
			\bottomrule 
		\end{tabular}
	}
 }
        \caption{Summary Statistics for the Data Used in the Analysis.}
	\label{stats}
\end{table}
\begin{table}[H]
	\centering
        \scalebox{0.8}{
		\begin{tabular}{lccccccccc}\toprule
			Call ID and ID and the service group of the agent who answered the call\\
			Type of service received by the caller\\
			Date and time (in seconds) of entering and exiting the queue\\
			Date and time (in seconds) of entering and exiting the service\\
			The outcome of the call (handled/transferred/abandoned/disconnected/error)\\
			Queue time - time a call spends in the queue before entering the service\\
            Node - the identifier of the site where
            the call is being processed\\
			\bottomrule
		\end{tabular}
	}
        \caption{Call characteristics observable in the data.}
	\label{characteristics}
\end{table}
To set the abandonment penalties, we view a caller's value from service as a proxy for his abandonment penalty. We use the caller's value of the time spent in service as a proxy for his value for the service. This, in turn, corresponds to his hourly waiting cost divided by 12, because the servers can handle about 12 calls per hour. While admittedly a crude approach, it results in abandonment penalties that are ordered in the same way as the holding costs are ordered, e.g., abandonment penalty for the Platinum class is higher than that for the Retail class.

ZipRecruiter, a popular job listing platform, tracks thousands of salary reports across different industrial sectors. According to their data\footnote{Available at https://www.ziprecruiter.com/Salaries/Call-Center-Salary-per-Hour. Accessed on August 12, 2023.}, the average hourly salary of a call center representative in the U.S. is about \$17. Therefore, we use an overtime rate  $1.5 \times 17 = \$25.5$ per hour, which corresponds to an overtime rate of $\bar{c} = \$25.5/12 = \$2.12$ per call.

\begin{table}[H]
	\centering
        \small
	\setlength\tabcolsep{4pt} 
	{\small 
            \scalebox{0.87}{
		\begin{tabular}{lccccccccc}
			\toprule
  Class & Arrival & \textbf{$p$} & \,\,  \textbf{$h$}  \\ 
  & percentage (\%) & (per job) &  (per hour) &\\
  \midrule
    Subanco & 0.51 & \$1.667 & \$20\\
    EBO & 0.72 & \$1.667 & \$20\\
    BPS & 0.77 & \$1.667 & \$20\\
    Case Quality & 0.84 & \$1.667 & \$20\\
    \midrule
    Online Banking & 2.64 & \$1.833 & \$22\\
    AST & 3.42 & \$1.833 & \$22\\
    Brokerage & 5.78 & \$1.833 & \$22\\
    Telesales & 6.26 & \$1.833 & \$22\\
    Consumer Loans & 6.92 & \$1.833 & \$22\\
    CCO & 8.34 & \$1.833 & \$22\\ 
    \midrule
    Retail (Node: 1, 2, 3) & 53.71 & \$2.000 & \$24\\
    Premier & 3.46 & \$2.167 & \$26\\
    \midrule
    Business & 4.82 & \$2.500 & \$30\\
    Platinum & 0.34 & \$2.667 & \$32\\
    Priority Service & 1.47 & \$2.667 & \$32\\
			\bottomrule 
		\end{tabular}
	}
 }
        \caption{Abandonment and Holding Cost Rates.}
	\label{cost1}
\end{table}

\subsection{Main test problem and its variants}\label{sect_main_test}
For the main test problem, we set the values of the problem primitives using the US Bank data. Recall that the call center offers 15 different classes of service. Moreover, we split the Retail class into three different classes based on the nodes where the calls are served. Thus, we set the number of classes to $K=17$ and the length of the planning horizon to $T=17$ hours. 

Recall that time horizon $[0,T]$ is partitioned as $0 = t_{0} < t_{1} < \ldots < t_{N} = T$. We let $\lambda_{k}^{r}(t_n)$ denote the arrival rate of class $k$ during the $n^{\text{th}}$ time interval $[t_{n}, t_{n+1})$ for $n = 0,1,\ldots, N-1$, which we estimate from the data; see Figures \ref{graph_arrival_class1_17dim_node1} - \ref{graph_arrival_15} of \ref{appdx_data_main_test}. We also estimate the mean service and abandonment times, denoted by $m_k$ and $1/\theta_k$ respectively, directly from the data; see fourth and fifth columns of Table \ref{stats}. These yield the limiting service and abandonment rates; see Equation (\ref{eqn_scaled_mu_theta}). The number of agents working throughout the day estimated from data for each period $[t_{n},t_{n+1})$, denoted by $\{N^{r}(t_n): n = 0,1,\ldots, N-1\}$, are displayed in Figure \ref{Agents_CI}. In what follows, we let $\Delta t = T/N$ and let $t_{n+1} - t_{n} = \Delta t$ for all $n$.

Additionally, we set the system parameter $r=400$, reflecting the order of magnitude for the staffing level, and use it to calculate various limiting parameters that are used crucially in our computational method. To be specific, having determined the system parameter $r$, we define the limiting staffing levels as follows:
\begin{equation}\label{eqn_limit_staff}
    N(t_n) = \frac{N^{r}(t_n)}{r}, \quad n = 0,1,\ldots N-1,
\end{equation}
see Figure \ref{limiting_agents} in  \ref{appdx_data_main_test} for the limiting staffing levels throughout the day. In addition,  we set the limiting arrival rate $\lambda_{k}(\cdot)$ for $k = 1,\ldots, K$ (see Figures \ref{graph_arrival_class1_17dim_limit} - \ref{graph_arrival_class17_17dim_limit} in \ref{appdx_data_main_test}) as follows:
\begin{equation}\label{eqn_limit_arrival}
    \lambda_{k}(t_n) = q_{k}\frac{N(t_n)}{\sum_{k=1}^{K}\frac{q_{k}}{\mu_{k}}}, \quad n = 0,1\ldots, N-1,
\end{equation}
where $q_{k}$ denotes the fraction of class $k$ customers (see the third column of Table \ref{stats}, also see Equation (\ref{eqn_HT_condition})). We then set the second-order terms $\zeta_{k}(\cdot)$ using Equation (\ref{eqn_scaled_lambda}) as follows:
\begin{equation}\label{eqn_limit_zeta}
\zeta_{k}(t_n) = \frac{1}{\sqrt{r}}\left(\lambda_{k}^{r}(t_n) - r \lambda_{k}(t_n)\right), \quad k=1,\ldots, K, \,\, n = 0,1,\ldots, N-1,
\end{equation}
see Figures \ref{main_test_zeta_class1} - \ref{main_test_zeta_class15} in \ref{appdx_data_main_test} for the resulting $\zeta_{k}(\cdot)$ functions for $k=1,\ldots, K.$

Given these primitives of the Brownian control problem, we compute the proposed policy using our method and compare its performance against the benchmark policies introduced in Section \ref{computational_benchmarks}. In doing so, we observe that the $c\mu/\theta$ rule stands out among the benchmark policies we considered for the main test problem. In light of this observation, we further investigate the system dynamics throughout the day. 

\begin{figure}[H]
    \centering
    \includegraphics[scale=0.23]{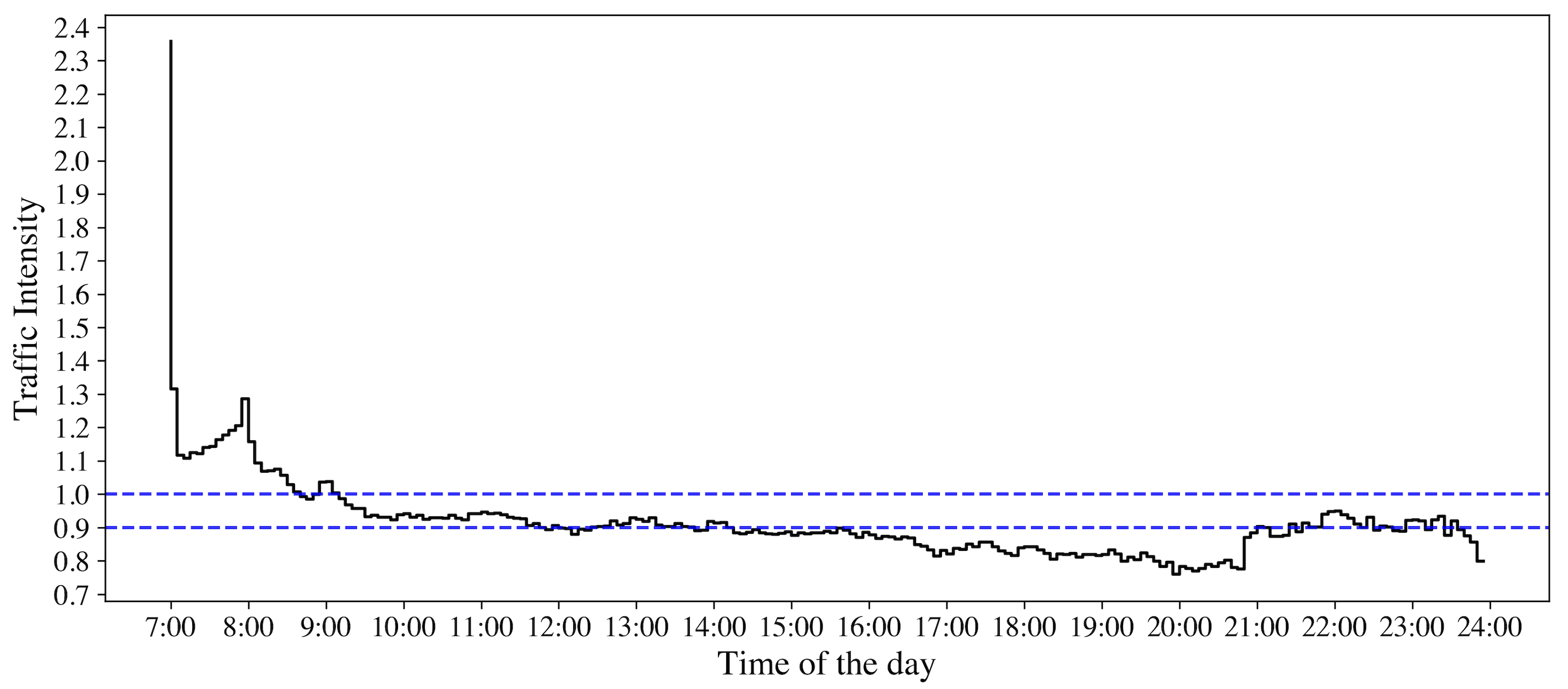}
    \caption{The average traffic intensity observed throughout the day. The resolution of the horizontal axis is five minutes. That is, traffic intensity is calculated over five-minute intervals.}
    \label{fig:traffic-intensity}
\end{figure}

\begin{figure}[H]
\centering
 \includegraphics[scale=0.38]{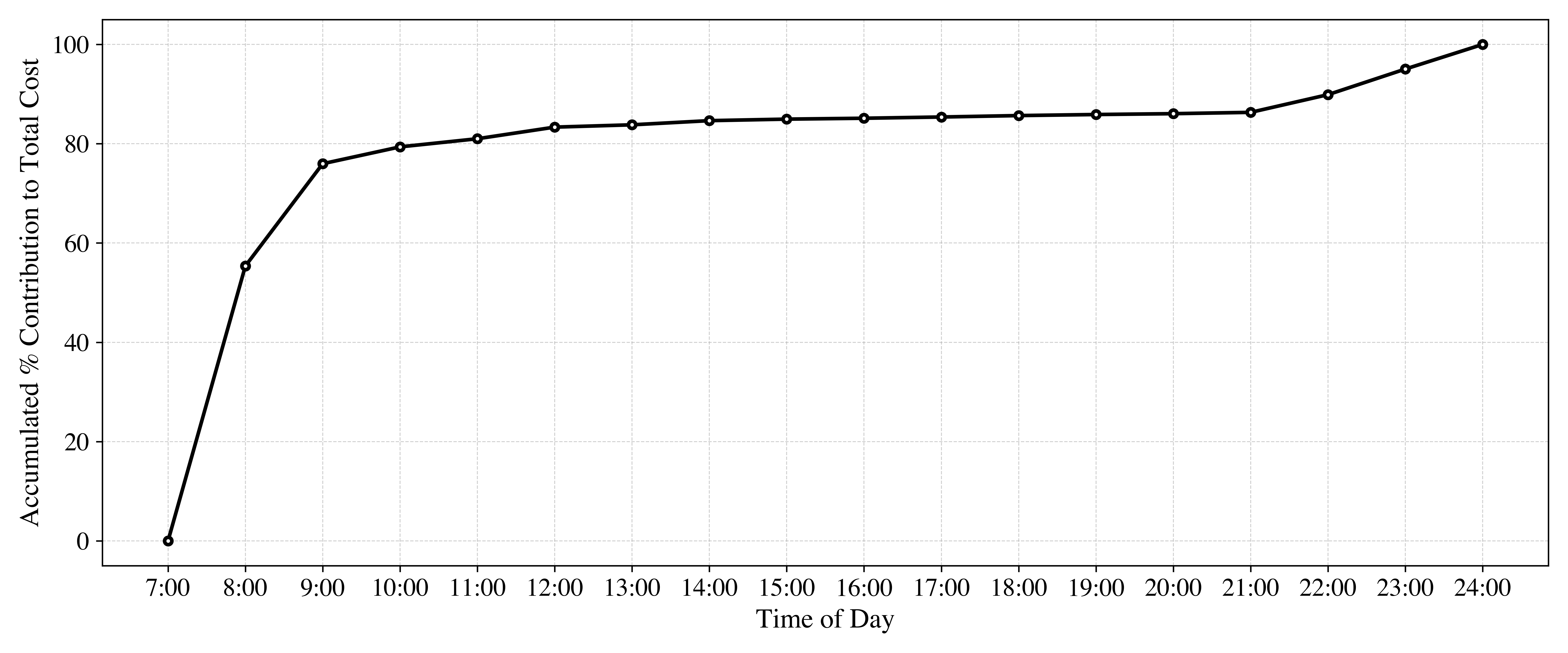}
    \caption{Hourly accumulation of the system cost.}
    \label{fig:grh3}
\end{figure}

Figure \ref{fig:traffic-intensity} shows the traffic intensity throughout the day. Notably, the system is overloaded between 7 AM and 9 AM, with the traffic intensity exceeding 1 during this period. For the remainder of the day, the traffic intensity remains below 1. Furthermore, 76\% of the total queueing costs occur during the overloaded period from 7 AM to 9 AM, as shown in Figure \ref{fig:grh3}.  \citet{atar2010cmu} proves that the $c\mu/\theta$ rule is asymptotically optimal for overloaded systems. Therefore, it is natural to expect the $c\mu/\theta$ rule to perform well for this problem instance.
\vspace{-1.5mm}
\vspace{-5mm}
\vspace{-7.5mm}
Next, we consider two variations of our main test problem to explore if the $c\mu/\theta$ policy performs well more broadly. For those two test problems, we adjust the staffing levels so that the traffic intensity is 0.95 throughout the day, i.e., the system is critically loaded; see Figure \ref{fig:critically_loaded_agents} in \ref{data_variants}. 

A numerical study in \ref{two_dim_numerical_study} that utilizes the approximations developed in \citet{garnett2002designing} leads to the following insight: As the abandonment rate of a class increases, its expected queue length decreases as one would expect. However, it decreases slowly enough that the product abandonment rate $\times$ queue length increases as the abandonment rate increases. Building on this insight, consider a two-class queueing system whose classes are labeled by their $c\mu/\theta$ ranking and modify the system parameters as follows: First, increase the abandonment rate of class 2 (the lower-ranked class under the $c\mu/\theta$ rule) while simultaneously increasing its holding cost rate to preserve its ranking under the $c\mu/\theta$ rule. Second, decrease the abandonment rate of class 1 (the higher-ranked class under the $c\mu/\theta$ rule) while simultaneously decreasing its holding cost rate to maintain its priority under the $c\mu/\theta$ rule. Under the updated parameters, if one reverses the priorities, i.e., uses the static rule that prioritizes class 2 over class 1, the system performance improves relative to the $c\mu/\theta$ rule. We build on this insight to design the two variants described below; see \ref{variants_design} for further details.

To do so, we consider the classes that are expected to have larger queues under the $c\mu/\theta$ rule and divide the classes into two groups accordingly: high-priority, denoted by $\mathcal{G}_{H}$, and low-priority, denoted by $\mathcal{G}_{L}$, based on their rank order with respect to the $c\mu/\theta$ rule. Intuitively, we expect the queue length to be negligible for classes in $\mathcal{G}_{H}.$ We further divide the low-priority group into two subgroups\footnote{These subgroups are determined such that the total call volume and mean service rates of the classes in $\mathcal{G}_{L}^{\alpha}$ and $\mathcal{G}_{L}^{\beta}$ are similar.}, denoted by $\mathcal{G}_{L}^{\alpha}$ and $\mathcal{G}_{L}^{\beta}$, as shown in Table \ref{division}.
\begin{table}[H]
	\centering
	\setlength\tabcolsep{4pt} 
	{\footnotesize 
             \scalebox{1.0}{
		\begin{tabular}{llllcccccc}
			\toprule
			Group &    Classes \\
			\midrule
                 $\mathcal{G}_H$ & Platinum, Retail (Node: 3), Retail (Node: 1), Business, Consumer Loans, Retail (Node: 2)\\
                 \midrule
                 $\mathcal{G}^{\alpha}_{L}$ & CCO, BPS, Priority Service, Brokerage\\
                 $\mathcal{G}^{\beta}_{L}$ & Premier, Online Banking, AST, Subanco, Telesales, EBO, Case Quality \\
			\bottomrule
		\end{tabular}
	}
 }
        \caption{The combination of original classes into high-priority group and low-priority subgroups. Within each subgroup, classes are listed from left to right in descending $c\mu/\theta$ order.}
	\label{division}
\end{table}
We then build on the observations described in the preceding paragraph to design the two variants of the test problem. More specifically, we adjust the parameters of the low-priority classes, i.e., those in $\mathcal{G}_{L}^{\alpha}$ and $\mathcal{G}_{L}^{\beta}$, while preserving their original $c\mu/\theta$ values. Table \ref{updated_stats} summarizes these adjustments for any $(\alpha, \beta)$ pair, ensuring that the original $c\mu/\theta$ values remain unchanged.

\noindent \textbf{First variant of the main test problem.} We scale the abandonment rates $\theta_{k}$ and holding cost rates $h_{k}$ by $\alpha = 0.7$ for $k \in \mathcal{G}_{L}^{\alpha}$ and by $\beta = 1.3$ for $k \in \mathcal{G}_{L}^{\beta}$. Under these changes, the $c\mu/\theta$ rule no longer performs well compared to other benchmark policies; see Table \ref{comp_benchmark} in \ref{appendix_static_priority}. Specifically, in this case, the alternative priority policy, which reverses the priority order of the low-priority subgroups $\mathcal{G}_{L}^{\alpha}$ and $\mathcal{G}_{L}^{\beta}$, outperforms the $c\mu/\theta$ rule by 10\%. To be specific, while the $c\mu/\theta$ rule prioritizes subgroups as $\mathcal{G}_{H} > \mathcal{G}_{L}^{\alpha} > \mathcal{G}_{L}^{\beta}$, the alternative policy reverses the order of the low-priority subgroups, prioritizing them as $\mathcal{G}_{H} > \mathcal{G}_{L}^{\beta} > \mathcal{G}_{L}^{\alpha}$. The resulting parameters for the first variant are shown in Table \ref{first_variant_data} in \ref{data_variants}.\\
\\
\noindent \textbf{Second variant of the main test problem.} We apply the same parameter changes as in the first variant example, but with $\alpha = 0.6$ and $\beta = 1.4$. Once again, the $c\mu/\theta$ proves suboptimal, as shown in Table \ref{comp_benchmark} in \ref{appendix_static_priority}. In this case, the alternative priority policy, which reverses the priority order of the low-priority subgroups $\mathcal{G}_{L}^{\alpha}$ and $\mathcal{G}_{L}^{\beta}$ outperforms the $c\mu/\theta$ rule by 20\%. The resulting parameters for the second variant are shown in Table \ref{second_variant_data} in \ref{data_variants}.\\

\begin{table}[H]
	\centering
	\setlength\tabcolsep{4pt} 
	{\small 
             \scalebox{0.87}{
		\begin{tabular}{lccccccc}
			\toprule
			Class & Arrival & $\mu$ & $\theta$ & $h$ & $p$ & $c$ & $c\mu/\theta$\\
			\rule{0pt}{3.5ex} & percentage (\%) & (per hr.) & (per hr.) & (per hr.) & (per job) & (per hr.) &  \\
			\midrule
			Platinum & 0.34 & 17.20 & 7.50 & \$32.00 & \$2.667 & \$51.99 & 119.230\\
                Retail (Node: 3) & 15.50 & 17.25 & 5.22 & \$24.00 & \$2.000 & \$34.44 & 113.810\\
                Retail (Node: 1) & 15.39 & 17.22 & 6.06 & \$24.00 & \$2.000 & \$36.12 & 102.638\\
			Business & 4.82 & 16.56 & 8.58 & \$30.00 & \$2.500 & \$51.46 & 99.321\\
			Consumer Loans & 6.92 & 15.19 & 4.87 & \$22.00 & \$1.833 & \$30.93 & 96.474\\
			Retail (Node: 2) & 22.82 & 17.25 & 7.81 & \$24.00 & \$2.000 & \$39.62 & 87.509\\
            \midrule
		CCO & 8.34 & 15.20 & 7.10\,$\alpha$ & \$22.00\,$\alpha$ & \$1.833 & \$35.02\,$\alpha$ & 74.972\\
		BPS & 0.77 & 13.57 & 5.92\,$\alpha$ & \$20.00\,$\alpha$ & \$1.667 & \$29.86\,$\alpha$ & 68.446\\
            Priority Service & 1.47 & 10.35 & 9.14\,$\alpha$ & \$32.00\,$\alpha$ & \$2.667 & \$56.37\,$\alpha$ & 63.833\\
             Brokerage & 5.78 & 12.62 & 6.89\,$\alpha$ & \$22.00\,$\alpha$ & \$1.833 & \$34.63\,$\alpha$ & 63.429\\
             \midrule
            Premier & 3.46 & 13.15 & 9.79\,$\beta$ & \$26.00\,$\beta$ & \$2.167 & \$47.22\,$\beta$ & 63.426\\
            Online Banking & 2.64 & 10.60 & 5.58\,$\beta$ & \$22.00\,$\beta$ & \$1.833 & \$32.24\,$\beta$ & 61.244\\
            AST & 3.42 & 12.52 & 7.50\,$\beta$ & \$22.00\,$\beta$ & \$1.833 & \$35.75\,$\beta$ & 59.679\\
		Subanco & 0.51 & 11.79 & 6.39\,$\beta$ & \$20.00\,$\beta$ & \$1.667 & \$30.66\,$\beta$ & 56.569\\
		Telesales & 6.26 & 9.62 & 8.99\,$\beta$ & \$22.00\,$\beta$ & \$1.833 & \$38.49\,$\beta$ & 41.187\\
		EBO & 0.72 & 9.87 & 8.24\,$\beta$ & \$20.00\,$\beta$ & \$1.667 & \$33.73\,$\beta$ & 40.402\\
		Case Quality & 0.83 & 9.93 & 9.27\,$\beta$ & \$20.00\,$\beta$ & \$1.667 & \$35.44\,$\beta$ & 37.963\\
			\bottomrule 
		\end{tabular}
	}
 }
        \caption{The system parameters adjusted with coefficients $\alpha \in (0,1]$ and $\beta \in [1,\infty)$ for low-priority subgroups $\mathcal{G}_{L}^{\alpha}$ and $\mathcal{G}_{L}^{\beta}$, respectively.}
	\label{updated_stats}
\end{table}
\color{black}
\subsection{Low dimensional test problems}\label{lower_dimensional}
This section introduces test problems of dimensions $K = 2,\,3$. It is computationally feasible to solve such problems using standard MDP techniques. As such, their optimal policies provide natural benchmarks for our proposed policy. To design these test problems, we simply partition the 17 classes into $K \,(= 2 \text{ or } 3)$ and combine the classes in each group to define a new class; see Table \ref{class_division_2dim} for $K = 2$ and Table \ref{class_division_3dim} for $K = 3$ for details.

The resulting arrival rates are given in Figure \ref{graph_arrival_2dim} of \ref{appendix_2dim} for the 2-dimensional test problem and in Figure \ref{graph_arrival_3dim} of \ref{appendix_3dim} for the 3-dimensional test problem. The mean service and abandonment rates are calculated by taking the weighted average\footnote{The weight of a class is the percentage of arrivals to that class within its group when combining the classes of the main test example.} of the rates associated with corresponding classes in the main test example; see Tables \ref{table_2dim} - \ref{table_3dim}. Similarly, we set the hourly holding cost rates $h_{k}$, the abandonment cost rates $p_k$, and the total cost rates $c_k$ by taking the weighted average of the cost rates corresponding classes in the main test example; see the last three columns of the Tables \ref{table_2dim} - \ref{table_3dim}. Also, we use an overtime rate $\bar{c}$ of \$2.12 per call, as done in the main test example.
\begin{table}[!htb]
	\centering
	\vspace{0mm}
	\setlength\tabcolsep{4pt} 
	{\footnotesize 
             \scalebox{0.9}{
		\begin{tabular}{llllcccccc}
			\toprule
			Class &  &  Names of the Combined Classes \\
			\midrule
			1 & & Retail (Node: 2), Business, Telesales, Consumer Loans,\\
   & &Online Banking, CCO\\
			2 & & Retail (Node: 1,\,3), Premier, Platinum, EBO, Subanco, Case Quality,\\
			&& Priority Service, AST, Brokerage, BPS\\
			\bottomrule 
		\end{tabular}
	}
 }
        \caption{The combination of original classes into two new classes.}
	\label{class_division_2dim}
\end{table}
\vspace{1mm}
\begin{table}[!htb]
	\centering
	\vspace{0mm}
	\setlength\tabcolsep{4pt} 
	{\footnotesize 
             \scalebox{0.9}{
		\begin{tabular}{llllcccccc}
			\toprule
			Class &  &  Names of the Combined Classes \\
			\midrule
			 1 & & Retail (Node: 2), Business, Telesales\\
			 2 & & Retail (Node: 1), Consumer Loans, Online Banking, CCO\\
			 3 & & Retail (Node: 3), Premier, Platinum, EBO, Subanco, Case Quality,\\
                & & Priority Service, AST, Brokerage, BPS\\
			\bottomrule 
		\end{tabular}
	}
 }
        \caption{The combination of original classes into three new classes.}
	\label{class_division_3dim}
\end{table}

The derivation of the limiting quantities follows the same steps as in Section \ref{data_section}; see the second columns of Tables \ref{table_2dim} - \ref{table_3dim} for the updated arrival percentages $q_{k}$. Figure \ref{graph_arrival_limit_2dim} in \ref{appendix_2dim} and Figure \ref{graph_arrival_limit_3dim} in \ref{appendix_3dim} display the limiting arrival rates $\lambda_k(\cdot)$ for the 2 and 3-dimensional test problems, respectively. Similarly, Figure \ref{graph_zeta_2dim} in \ref{appendix_2dim} and Figure \ref{graph_zeta_limit_3dim} in \ref{appendix_3dim} display functions $\zeta_{k}(\cdot)$ of the 2 and 3-dimensional test problems, respectively.
\begin{table}[h]
	\centering
	\vspace{0mm}
	\setlength\tabcolsep{4pt} 
	{\footnotesize 
             \scalebox{0.9}{
		\begin{tabular}{lccccccccc}
			\toprule
			Class & Arrival & $\mu$ & $\theta$ & $p$ & $h$ & $c$ \\
			\rule{0pt}{3.5ex} & percentage (\%) &  (per hr) & (per hr) & (per job) & (per hr) & (per hr)\\
			\midrule
			 1 & 51.80 & 15.32 & 7.40 & \$1.97 & \$23.63 & \$38.20\\
                2 & 48.20 & 15.49 &  6.45 & \$1.99 &  \$23.83 & \$36.64\\
			\bottomrule 
		\end{tabular}
	}
 }
        \caption{Summary statistics for the 2-dimensional example.}
	\label{table_2dim}
\end{table}
\begin{table}[!htb]
	\centering
	\vspace{0mm}
	\setlength\tabcolsep{4pt} 
	{\footnotesize 
        \scalebox{0.9}{
		\begin{tabular}{lccccccccc}
			\toprule
			Class &  Arrival & $\mu$ & $\theta$ & $p$  & $h$ & $c$ \\
			\rule{0pt}{3.5ex} & percentage (\%) & (per hr)& (per hr) & (per job) & (per hr) & (per hr) \\
                \midrule
			 1  & 33.90 & 15.74 & 8.14 & \$2.04 & \$24.48 & \$41.09 \\
			 2  & 33.29 & 15.77 & 6.03 & \$1.91 & \$22.92 & \$34.45 \\
                 3 & 32.81 & 14.68 & 6.64 & \$1.98 & \$23.75 & \$36.88\\
			\bottomrule 
		\end{tabular}
	}
 }
        \caption{Summary statistics for the 3-dimensional problem.}
	\label{table_3dim}
\end{table}

A simulation study that considered the six possible static priority policies for the 3-dimensional example revealed that the performance difference between the best and worst static policies was modest (7.91\%). Thus, we consider an additional 3-dimensional test problem to demonstrate the robustness of our proposed algorithm when the range of expected costs for different policies is larger. Specifically, for the new test problem, all problem primitives of the main 3-dimensional test example remain the same, except for the abandonment penalty $p_{k}$, the holding cost rate $h_{k}$, and the cost rate $c_{k}$ for Class 2. As shown in the last three columns of Table \ref{table_3dim_extend} in \ref{3dim_variant_data}, these are half of those in the main 3-dimensional test example, cf. Table \ref{table_3dim}. For this test problem, the performance difference between the best and worst static policies is 83.45\%.

\subsection{High dimensional test problems that have pathwise optimal policies}\label{higher_dimensional}
To illustrate the scalability of our method, we introduce four additional test problems of dimensions 30, 50, 100 and 500. We use $J > K = 17$ to distinguish the number of classes of the higher dimensional test problems from that of the main test problem. We also put a $\sim$ on various quantities associated with the new system with $J$ classes.

Crucially, we design the high-dimensional test problems $(J = 30, 50, 100 \text{ and } 500)$ so that they admit pathwise optimal policies. For these test problems, different customer classes differ from each other only in their arrival rates, holding cost rates and total cost rates. The remaining problem primitives, that is, service times, abandonment rates, and abandonment penalties are the same across different classes. Of course, these observations follow from our design choices.

To be specific, for each class $j=1,\ldots, J$, we set the arrival rate process $\tilde{\lambda}_{j}^{\tilde{r}}(\cdot)$ by drawing randomly with replacement from the original arrival rate processes $\{\lambda_{k}^{r}(\cdot), \,\, k=1,\ldots,K\}$. The original classes used to determine the arrival rate process $\{\tilde{\lambda}_{j}^{\tilde{r}}(\cdot),\,j = 1,\ldots, J\}$ are shown in the second column of Tables \ref{stats_30dim_path}, \ref{stats_50dim_path}, \ref{stats_100dim_path_part1} and \ref{stats_500dim_path} in \ref{high_dimensional_supplementary_tables} for the respective test problems. Then we use the resulting arrival rate processes $\tilde{\lambda}_{j}^{\tilde{r}}(\cdot)$ to calculate the fraction of class $j$ customers in the new system, denoted by $\tilde{q}_{j}$, as follows\footnote{For simplicity, we set $\Delta t_{n} = T/N$ for $n =0,1\ldots,N-1$ and the $\Delta t $ terms in the following equation cancel out.}:
\begin{equation}
\tilde{q}_{j} = \frac{\sum_{n=0}^{N-1}\tilde{\lambda}_{j}^{\tilde{r}}(t_n)}{\sum_{j=1}^{J}\sum_{n=0}^{N-1}\tilde{\lambda}_{j}^{\tilde{r}}(t_{n})}, \quad j = 1,\ldots,J.
\end{equation}

Similarly, we independently draw $J$ random samples with replacement from the mean service times $\{m_{k}, \, k = 1, \ldots, K\}$ and abandonment rates $\{\theta_{k}, \, k = 1, \ldots, K\}$; see Table \ref{stats} for $m_{k}$ and $\theta_{k}$ for $k = 1, \ldots, K$. The sampled service times and abandonment rates are denoted by $\{\tilde{m}_{j}, \, j = 1, \ldots, J\}$ and $\{\tilde{\theta}_{j}, \, j = 1, \ldots, J\}$, respectively. The original classes used to define $\tilde{m}_{j}$ and $\tilde{\theta}_{j}$ are listed in the fourth and fifth columns of Tables \ref{stats_30dim_path}, \ref{stats_50dim_path}, \ref{stats_100dim_path_part1}, and \ref{stats_500dim_path} in \ref{high_dimensional_supplementary_tables} for the test problems with dimensions 30, 50, 100, and 500, respectively. To determine the common service times $\tilde{m}$ and abandonment rates $\tilde{\theta}$ across the $J$ classes, we take a weighted average of the $J$ randomly drawn service times $\tilde{m}_{j}$ and abandonment rates $\tilde{\theta}_{j}$, using $\tilde{q}_{j}$ as weights. The resulting common service times and abandonment rates for each test problem are shown in Table \ref{sampled_common_service_abandonment_rates}.

\begin{table}[H]
	\centering
	\setlength\tabcolsep{4pt}
	{\footnotesize 
             \scalebox{1.0}{
		\begin{tabular}{lccccccccc}
			\toprule
			Problem Dimension  & $\tilde{m}$ & $\tilde{\theta}$\\
			\rule{0pt}{3.5ex} &  (hours) & (per hr)\\
			\midrule
			  30 & 0.07107 & 7.80\\[0.2em]
                50 & 0.07304 & 7.14\\[0.2em]
                100 & 0.07215 & 7.35\\[0.2em]
                500 & 0.07273 & 7.37\\
			\bottomrule 
		\end{tabular}
	}
 }
\caption{The common service times and abandonment rates across all class for each of the high-dimensional test example.}
\label{sampled_common_service_abandonment_rates}
\end{table}
Recall $\{N^{r}(t_{n}): n = 1,\ldots,N-1\}$ denotes the staffing level throughout the day, estimated directly from the data (see Figure \ref{Agents_CI}). Then an aggregate daily utilization of the system for the main test problem can be calculated directly from the data. To be specific, the daily utilization $\rho^{r}$ of the $K$-dimensional main test problem is given as follows:
\begin{equation}
    \rho^{r} = \frac{\sum_{k=1}^{K}m_{k}\sum_{n=0}^{N-1}\lambda_{k}^{r}(t_{n})\,\Delta t}{\sum_{n=0}^{N-1}N^{r}(t_{n}) \, \Delta t}. \label{load_factor}
\end{equation}
In order to define the remaining parameters of the high-dimensional system, we begin with the system parameter $\tilde{r}$. Recall that the system parameter corresponds to the scale of the system, i.e., the total volume of arriving work and the total service capacity. We expect the scale of the high-dimensional system to be of order $J/K$ times that of the main test problem. Thus, we set
\begin{equation}
\tilde{r} = \lceil r \, J/K \rceil.  \label{scaling_parameter}  
\end{equation} 
We then define the utilization $\tilde{\rho}^{\tilde{r}}$ for the  $J$-dimensional system so that $\sqrt{\tilde{r}}\,(1-\tilde{\rho}^{\tilde{r}}) = \sqrt{r}\,(1-\rho^{r})$. Thus, the utilization $\tilde{\rho}^{\tilde{r}}$ is given by: 
\begin{equation}
    \tilde{\rho}^{\tilde{r}} = 1 - (1-\rho^{r})/\sqrt{J/K}.
\end{equation}
The rationale for this choice is that the larger the system is, the larger the utilization can be without sacrificing system performance due to the statistical economies of scale. In particular, this choice of the system utilization for the large system leads to drift terms of similar magnitudes in the approximating Brownian control problems for the two systems. Then what remains to be determined are the staffing levels $\tilde{N}^{\tilde{r}}(t_{n})$ for $n = 0,1,\ldots,N-1$. To this end, we first note that the following must hold by definition:
\begin{align}
    \sum_{n=0}^{N-1}\tilde{N}^{\tilde{r}}(t_{n})\Delta t  = \frac{\sum_{j=1}^{J}\tilde{m}\sum_{n=0}^{N-1}\tilde{\lambda}^{\tilde{r}}(t_{n})\Delta t}{\tilde{\rho}^{\tilde{r}}}, \label{staffing_equation}
\end{align}
which determines the total service capacity in a day. Next, we set $\tilde{N}^{\tilde{r}}(t_{n})$ for each $n$. To do so, we assume that the staffing profiles for the main test problem and the high-dimensional test problem are identical in the following sense: 
\begin{align}
    \frac{\tilde{N}^{\tilde{r}}(t_{n})}{\sum_{n=0}^{N-1}\tilde{N}^{\tilde{r}}(t_{n})} \approx \frac{N^{r}(t_{n})}{\sum_{n=0}^{N-1}N^{r}(t_{n})} = \frac{N(t_{n})}{\sum_{n=0}^{N-1}N(t_{n})}, \quad n = 0,1,\ldots,N-1. \label{staffing_ratio}
\end{align}
Combining (\ref{staffing_equation}) - (\ref{staffing_ratio}) leads us to the setting
\begin{align}
    \tilde{N}^{\tilde{r}}(t_{n}) = \Bigg\lceil \frac{\sum_{j=1}^{J} \tilde{m} \sum_{n=0}^{N-1}\lambda_{k}^{r}(t_{n})}{\tilde{\rho}^{\tilde{r}}}\cdot \frac{N(t_{n})}{\sum_{n=0}^{N-1} N(t_{n})}\Bigg\rceil, \label{staffing_high_dim_equation}
\end{align}
where we use $\lceil \cdot \rceil$ to ensure the number of agents is integer valued. This completes our definition of the parameters of the (pre-limit) high-dimensional system.

Next, we define the limiting quantities associated with the high-dimensional system following the logical progression of ideas used in Equations (\ref{eqn_limit_staff}) - (\ref{eqn_limit_zeta}). In particular, the limiting staffing levels are given as follows:
\begin{equation}
    \tilde{N}(t_{n}) = \frac{\tilde{N}^{\tilde{r}}(t_{n})}{\tilde{r}}, \quad n = 0,1,\ldots,N-1. \label{adjusted_agents}
\end{equation}
In addition, the limiting arrival rates $\tilde{\lambda}_{j}(\cdot)$ for $j = 1,\ldots, J$ are given as follows:
\begin{equation}
\tilde{\lambda}_{j}(t_{n}) = \tilde{q}_{j}\frac{\tilde{N}(t_{n})}{\sum_{j=1}^{J}\tilde{q}_{j}\tilde{m}} = \tilde{q}_{j}\frac{\tilde{N}(t_{n})}{\tilde{m}}, \quad j = 1,\ldots, J, \quad n = 0,1,\ldots,N-1. \label{lim_lambda_high} 
\end{equation}
Note that the choices made in (\ref{adjusted_agents}) - (\ref{lim_lambda_high}) satisfy the heavy traffic condition, cf. Equation (\ref{eqn_HT_condition}). Then we compute the second-order terms $\tilde{\zeta}_{j}(\cdot)$ for $j = 1,\ldots,J$ as done in Equation (\ref{eqn_limit_zeta}). That is, we set
\begin{equation}
    \tilde{\zeta}_{j}(t_{n}) = \frac{1}{\tilde{r}}\left(\tilde{\lambda}^{\tilde{r}}_{j}(t_{n}) - \tilde{r}\tilde{\lambda}_{j}(t_{n})\right), \quad j = 1,\ldots,J, \,\, n = 0,1,\ldots,N-1. \label{high_dim_zeta}
\end{equation}
Lastly, we generate a uniform grid of hourly holding cost rates between \$14 and \$34 with a fixed grid size $\Delta \tilde{h}$. The grid size $\Delta \tilde{h}$ is determined so that the number of possible hourly holding cost rates within this range is greater than or equal to the dimension $(J)$ of the test example\footnote{To be specific, we use $\Delta \tilde{h}$ values of 0.5, 0.25, 0.125 and 0.04 for $J$ values of 30, 50, 100, and 500, respectively.}. Then, we randomly draw $J$ samples without replacement from this range to determine the hourly holding cost rates $\tilde{h}_{j}$. The resulting cost parameters for the $J = 30, 50, 100$, and $500$-dimensional test examples are shown in Tables \ref{stats_30dim_path}, \ref{stats_50dim_path}, \ref{stats_100dim_path_part1}, and \ref{stats_500dim_path} in \ref{high_dimensional_supplementary_tables}, respectively.

\begin{figure}[H]
    \centering
    \includegraphics[scale=0.4]{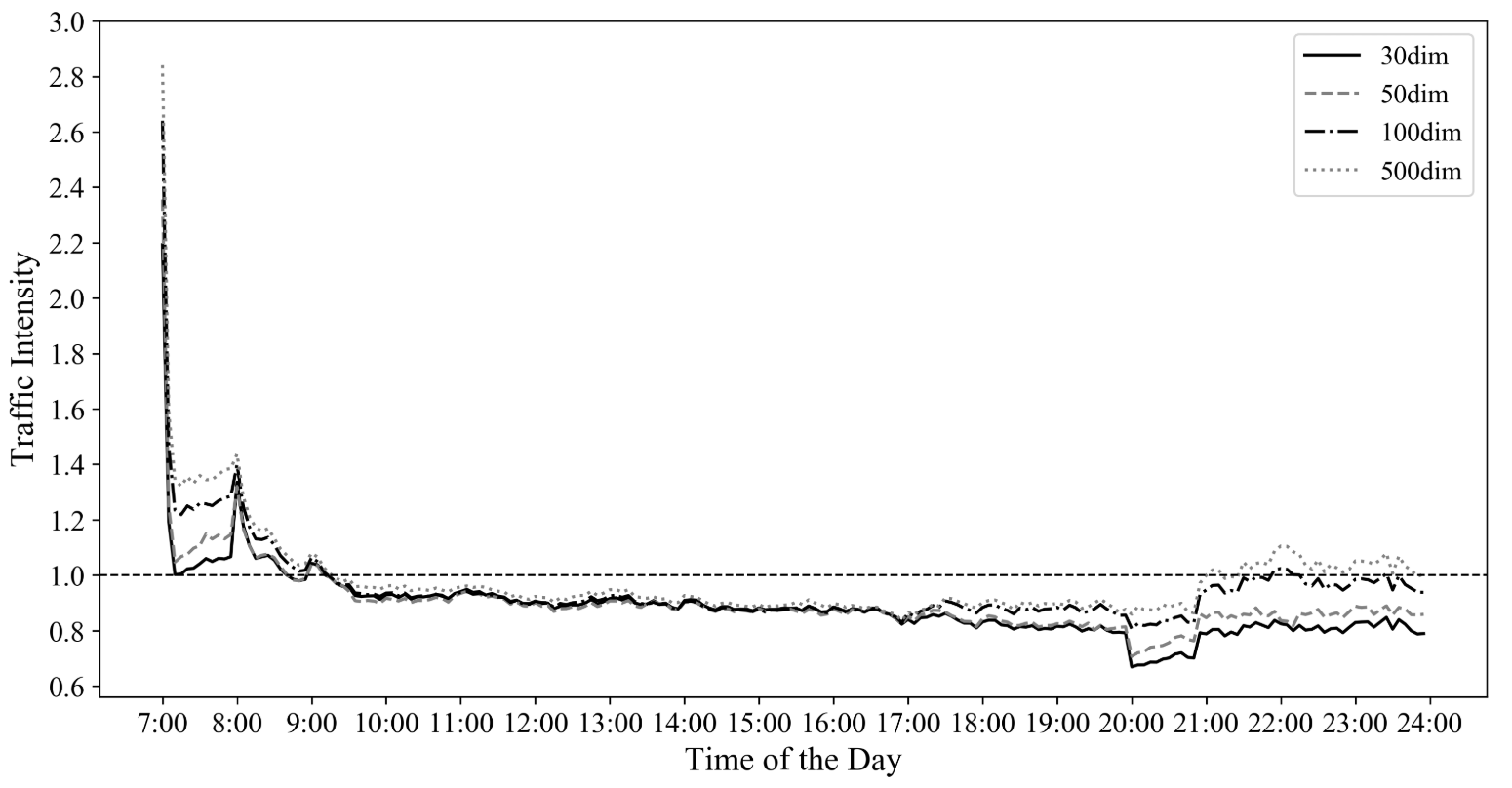}
    \caption{The average traffic intensity observed throughout the day for the high dimensional test problems. The resolution of the horizontal axis is five minutes. That is, traffic intensity is calculated over five-minute intervals.}
    \label{fig:high_dim_utilization}
\end{figure}

Because we bootstrap the parameters of the high-dimensional problem from those of the main test problem, our choice of staffing given in Equation (\ref{staffing_high_dim_equation}) leads to a similar utilization profile throughout the day as in the main test problem; see Figure \ref{fig:high_dim_utilization}. In particular,
the system starts overloaded and most queueing costs are incurred 7 AM to 9 AM.
\vspace{-5mm}
Recall that we chose common abandonment and service rate parameters for all classes in these test problems so that the static priority policy that ranks classes according to their cost rates is pathwise optimal. For more general service and abandonment rate parameters, one expects the $c\mu/\theta$ rule to be near optimal because of the overloaded nature of the system, as shown in Figure \ref{fig:high_dim_utilization}. Incidentally, the $c\mu/\theta$ rule reduces to the pathwise optimal policy we identify for these test problems because the service and abandonment rates are common across classes.

In the next subsection, we consider high-dimensional test problems whose service and abandonment rate parameters vary across classes. Thus, the policy that is optimal for the four test problems discussed in this section is no longer optimal for those test problems. In addition, we set the system utilization to 95\% throughout the day to avoid the overloaded scenario described above,  in which case one expects the $c\mu/\theta$ rule to be near optimal as discussed earlier.

\subsection{High dimensional test problems that do not admit a pathwise optimal policy}
\label{high_dim_non_pathwise_sect}
In this section, we introduce three additional test problems, each building on one of the three 17-dimensional test problems from Section \ref{sect_main_test}—the main problem and its two variants. Following a similar approach to Section \ref{higher_dimensional}, we begin by generating
the arrival rate process $\tilde{\lambda}_{j}^{\tilde{r}}(\cdot)$ for each class $j = 1, \dots, 100$ by randomly sampling them with replacement from the original arrival rate processes $\{\tilde{\lambda}_{k}^{\tilde{r}}(\cdot),\, k=1, \dots, K\}$. The original classes corresponding to these arrival rates are listed in the second column of Tables \ref{stats_100dim_non_path_main}, \ref{stats_100dim_non_path_var1} and \ref{stats_100dim_non_path_var2} in \ref{high_dimensional_supplementary_tables} for each of these three test problems.

Next, we independently sample 100 values with replacement from the mean service times $\{m_{k}, \,k=1,\ldots,K\}$ to determine the mean service times $\tilde{m}_{j}$ for $j = 1,\ldots,100$. Similarly, we determine the abandonment rates $\tilde{\theta}_{j}$ by sampling 100 values from the abandonment rates $\{\theta_{k}, \, k=1, \dots, K\}$, corresponding to those of the main test problem, the first variant, and the second variant described in Section \ref{sect_main_test}, generating three distinct test problem scenarios. The original classes used to define $\tilde{m}_{j}$ and $\tilde{\theta}_{j}$ are listed in the fourth and fifth columns of Tables \ref{stats_100dim_non_path_main}, \ref{stats_100dim_non_path_var1} and \ref{stats_100dim_non_path_var2} in \ref{high_dimensional_supplementary_tables} for each of these three test problems.

We then set the staffing levels $\tilde{N}^{\tilde{r}}(\cdot)$ so that the traffic intensity is 0.95 throughout the day, i.e., the system is critically loaded. The limiting staffing levels $\tilde{N}(\cdot)$, arrival rates $\tilde{\lambda}_{j}(\cdot)$ and the second-order terms $\tilde{\zeta}_{j}(\cdot)$ for $j = 1,\ldots,100$ are calculated using Equation (\ref{adjusted_agents}), (\ref{lim_lambda_high}), and (\ref{high_dim_zeta}), respectively. Lastly, following the approach used in the test problems of Section \ref{higher_dimensional}, we construct a uniform grid of hourly holding cost rates ranging from \$14 to \$34, with a fixed step size of 0.125, for the 100-dimensional test problem. From this grid, we then randomly select 100 values without replacement to determine hourly holding cost rates $\tilde{h}_{j}$. The abandonment rates $\tilde{p}_{j}$ are then determined by dividing the hourly holding cost rates $\tilde{h}_{j}$ by 12, consistent with previous test examples. The resulting cost parameters are shown in Tables \ref{stats_100dim_non_path_main}, \ref{stats_100dim_non_path_var1} and \ref{stats_100dim_non_path_var2} in \ref{high_dimensional_supplementary_tables} for each of these three test problems. To the best of our knowledge, they do not admit a pathwise optimal policy.

\subsection{Benchmark policies}\label{computational_benchmarks}
This section introduces the benchmark policies we consider to assess the performance of the proposed policy in each of the thirteen test problems; see Section \ref{numerical_results} for the computational results.

\noindent \textbf{Policies that are pathwise optimal.} Recall that four of the thirteen test problems admit pathwise optimal solutions. Specifically, the four high dimensional $(J = 30, 50, 100, \text{ and } 500)$ cases described in Section \ref{higher_dimensional} have pathwise optimal solutions. For these test problems, the optimal policy is the static priority policy that prioritizes classes based on their total cost rates.

\noindent \textbf{Optimal policies for MDP formulations in the low dimensional cases.}
For the three low dimensional $(K = 2, \,3)$ test cases, introduced in Section \ref{lower_dimensional}, it is computationally feasible to find the optimal policies using standard dynamic programming techniques, which constitute natural benchmarks; see \ref{appendix_mdp_low_dim} for further details. It is worth mentioning that for the 3-dimensional variant test problem described in Table \ref{table_3dim_extend} in \ref{3dim_variant_data}, the optimal policy is the static priority policy that prioritizes classes based on their total cost rates.
\vspace{-3mm}
\subsubsection{Benchmark policies for the main test problem and its variants}
\label{static_benchmarks_main_variant}
We consider the following benchmark policies for the main test problem and its two variants: 

\noindent \textbf{Static Priority Rules.} Because it is infeasible to search over all possible (17!) static priority policies, we consider the following ones as building blocks:
\vspace{-8mm}
\begin{singlespace}
\begin{enumerate}
    \item The $c\mu/\theta$ rule proposed by \citet{atar2010cmu},
    \item The $c\mu$ rule proposed by \citet{cox1961queues},
    \item The static priority rule based on cost rates $c_{k}$,
    \item The static priority rule based on $\mu_{k} - \theta_{k}$,
    \item The static priority rule based on $c_{k}\,(\mu_{k} - \theta_{k})$.
\end{enumerate}
\end{singlespace}
The $c\mu/\theta$ and $c\mu$ rules are well-studied policies and optimal for different models; see \citet{atar2010cmu} and \citet{cox1961queues}, respectively. The static priority rule based on the cost rates is the pathwise optimal policy for four of our test problems. The last two static priority policies are derived using the definition of the effective holding cost function in Equation (\ref{eq_effective_$c_{k}$}). To be specific, because we expect $\partial V$/$\partial x_{k} \geq 0$, the higher $(\mu_{k} - \theta_{k})$, the higher the effective holding cost is. Thus, we prioritize the classes with higher $(\mu_k - \theta_k),$ overlooking the fact that their cost rate $c_{k}$ and $\partial V/\partial x_{k}$ differs across different classes. The last static rule above also considers the cost rate $c_{k}$ and ranks classes according to the combined effect, i.e., $c_{k}(\mu_{k} - \theta_{k}).$ We also use these five static priority rules as benchmark policies for the three 100-dimensional test problems introduced in Section \ref{high_dim_non_pathwise_sect}, which do not admit a pathwise optimal solution.

\noindent \noindent \textbf{Extended static priority rules.} Recall that, to design the variant examples in Section \ref{sect_main_test}, we divide the classes into three subgroups: one high-priority subgroup, $\mathcal{G}_{H}$, and two low-priority subgroups, $\mathcal{G}_{L}^{\alpha}$ and $\mathcal{G}_{L}^{\beta}$. Building on this design choice, we extend the aforementioned five static policies. Specifically, we propose two additional sets of static priority policies. The first set ranks the subgroups such that $\mathcal{G}_{H} > \mathcal{G}_{L}^{\alpha} > \mathcal{G}_{L}^{\beta}$, while the second set reverses the order of the low-priority subgroups, ranking them as $\mathcal{G}_{H} > \mathcal{G}_{L}^{\beta} > \mathcal{G}_{L}^{\alpha}$. For both sets, we apply the five static priority policies within each subgroup—$\mathcal{G}_{H}$, $\mathcal{G}_{L}^{\alpha}$, and $\mathcal{G}_{L}^{\beta}$—generating 120 additional static policies\footnote{Within $\mathcal{G}_{H}$, the order induced by $c\mu$ and $c_k$ are the same, resulting in 4 distinct prioritization schemes. In $\mathcal{G}_{L}^{\alpha}$, $c\mu$ and $c_k$ induce the same order, and the order of $\mu_k - \theta_k$ is equivalent to that of $c_k(\mu_k - \theta_k)$, leading to 3 distinct schemes. In $\mathcal{G}_{L}^{\beta}$, there are 5 distinct prioritization schemes. This gives a total of 60 distinct prioritization schemes for $\mathcal{G}_{H} > \mathcal{G}_{L}^{\alpha} > \mathcal{G}_{L}^{\beta}$. Since we also consider $\mathcal{G}_{H} > \mathcal{G}_{L}^{\beta} > \mathcal{G}_{L}^{\alpha}$, the total becomes 120.
} for the main test problem and its two variants. See \ref{appendix_additional_static_benchmarks} for further details.

\noindent \textbf{A class of dynamic policies building on an auxiliary 3-dimensional MDP.} We also design an auxiliary, low-dimensional MDP that focuses specifically on ``low-priority'' classes. To do this, we first divide the classes into two groups: high-priority, denoted by $\mathcal{G}_{H}$, and low-priority, denoted by $\mathcal{G}_{L}$. The low-priority group is further subdivided into three subgroups: $\mathcal{G}_{L}^{1},\, \mathcal{G}_{L}^{2}, \,\mathcal{G}_{L}^{3}$. We assume that the classes in the high-priority subgroup $\mathcal{G}_{H}$ are always given the highest priority. Because we expect their queue lengths to be small, we delete those classes from the problem for computational simplicity and focus attention on the remaining low-priority classes. Next, we solve a 3-dimensional MDP focusing on these three low-priority subgroups. Finally, to establish a priority rule within each subgroup, we apply the five static priority policies described earlier. This leads to 240 additional benchmark policies\footnote{In $\mathcal{G}_{H}$, $c\mu$ and $c_k$ maintain the same order, resulting in 4 unique prioritization schemes. Similarly, in $\mathcal{G}_{L}^{1}$, $c\mu$ and $c_k$ induce the same order, and since the order of $\mu_k - \theta_k$ is equivalent to that of $c_k(\mu_k - \theta_k)$, this leads to 3 distinct schemes. Meanwhile, $\mathcal{G}_{L}^{2}$ has 5 distinct prioritization schemes. Lastly, in $\mathcal{G}_{L}^{3}$, the order of $c\mu$ is the same as that of $c_{k}$, leading to 4 distinct prioritization schemes. Altogether, this accounts for 240 distinct prioritization schemes.}
 for the main test problem and its two variants; see \ref{appendix_mdp_high_dim}.

\noindent \textbf{Heuristic dynamic index policies.} Lastly, we consider three dynamic index policy heuristics that use alternative neural network approximations to estimate the effective holding cost function in Equation (\ref{eq_effective_$c_{k}$}). For details, see \ref{appendix_dynamic_benchmarks}. 

\noindent See \ref{appendix_static_priority} for the performance comparison of all benchmark policies.
\vspace{-5mm}
\section{Computational results}\label{numerical_results}
\vspace{-2mm}
This section compares our proposed policy, derived using our computational method (see Section \ref{computational_method}), to the benchmark policies introduced in Section \ref{computational_benchmarks}. We report the results using 99\% confidence intervals throughout this section.  However, we also conducted the same analyses using 95\% confidence intervals and confirmed that our conclusions remain consistent across both levels of confidence; see \ref{comparison_95}. Our proposed policy performs on par with the best benchmark for the low-dimensional problems, the main test problem and its variants, and the four high-dimensional problems that admit a pathwise optimal solution. In addition, it outperforms the best benchmark in the 100-dimensional instance that does not admit a pathwise optimal solution.
\vspace{-4mm}
\subsection{Results for the low dimensional test problems}
\begin{figure}[H]
     \centering
     \captionsetup{position=bottom} 
     \captionsetup[subfigure]{labelformat=empty}
        \subfloat[Our Policy]{\hspace{-4mm}\includegraphics[width=0.5\linewidth]{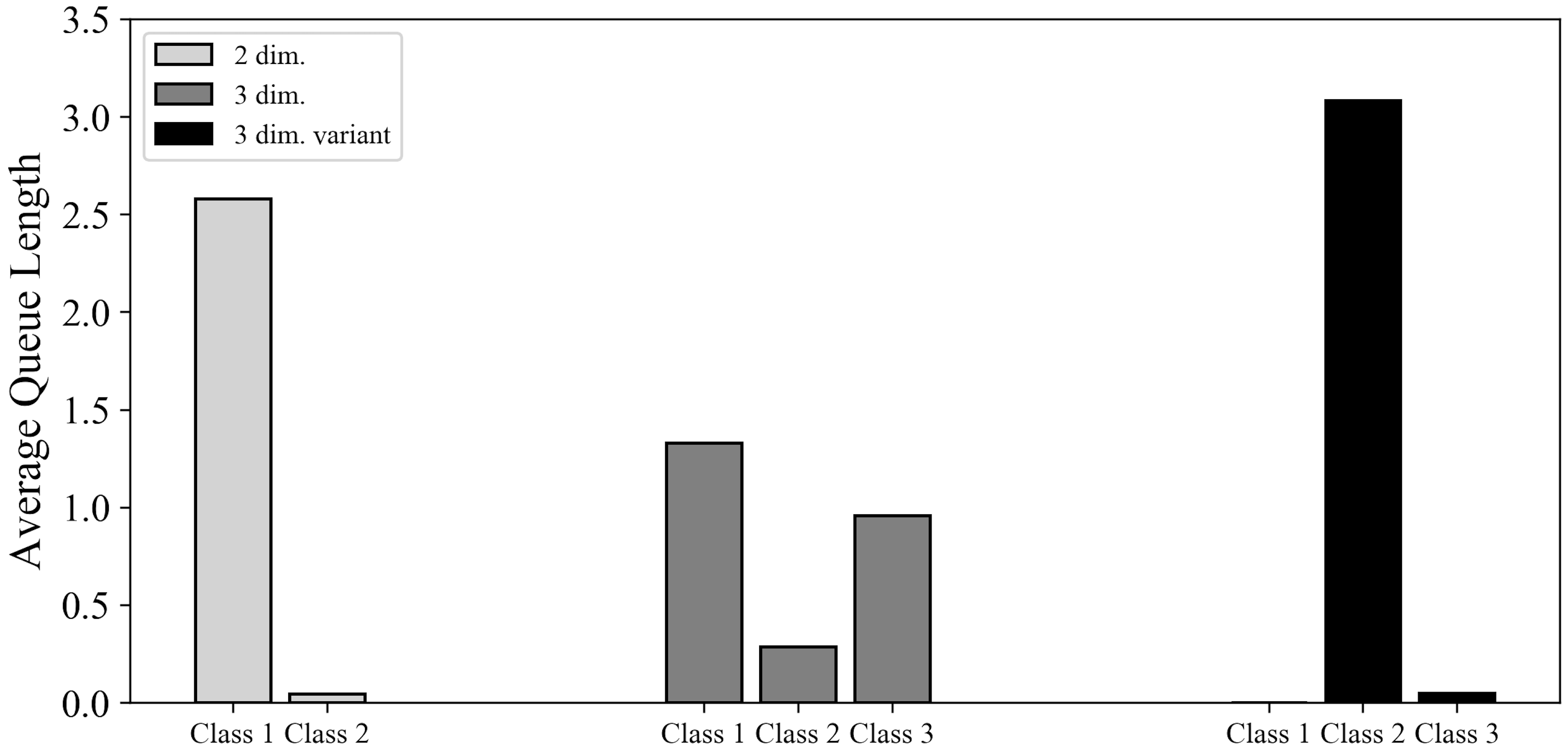}}
    \hfill
    \subfloat[Benchmark Policy]{\includegraphics[width=0.5\linewidth]{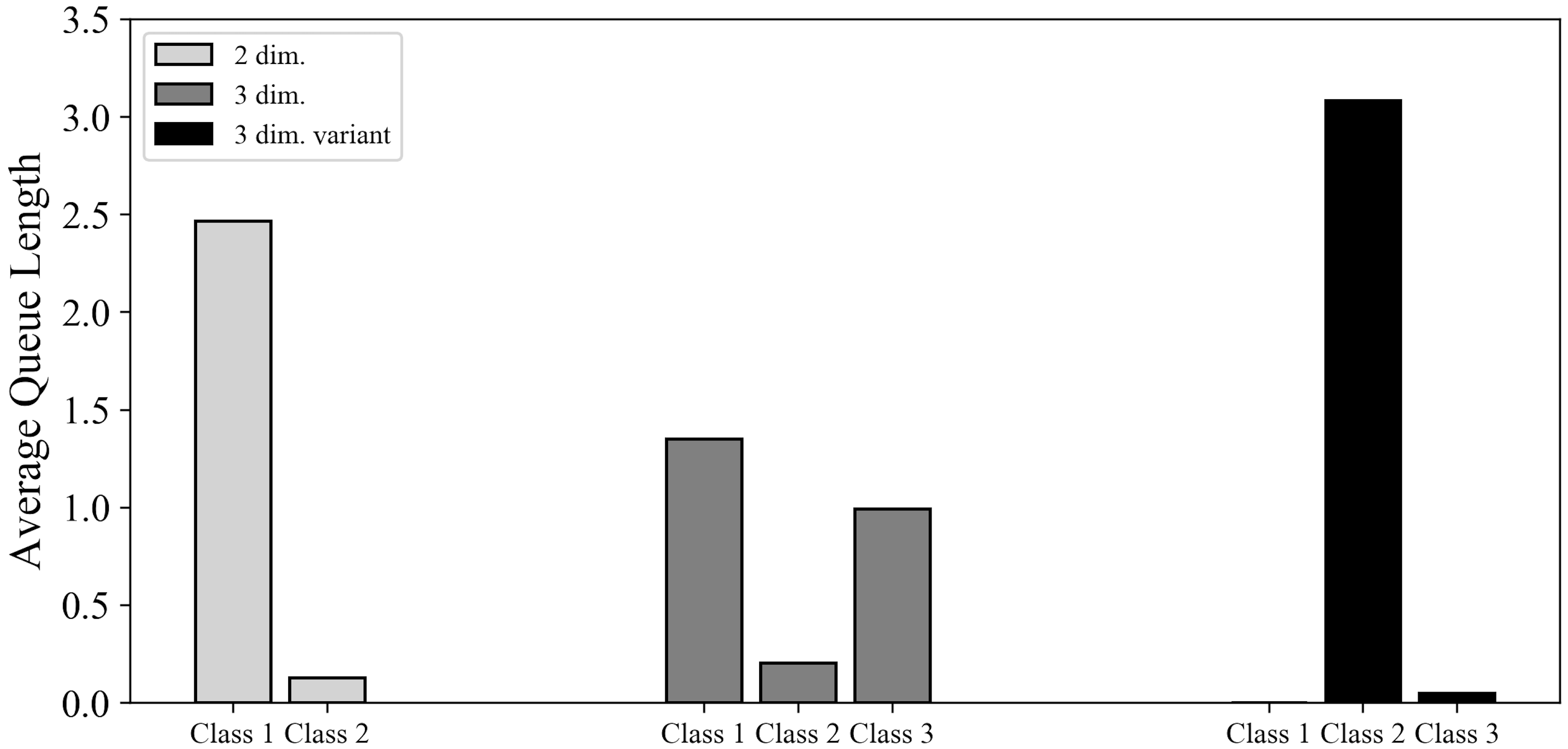}}
     \caption{Graphical representation of the average queue length under the policy learned from neural networks and the benchmark policy for the low dimensional test problems.} \label{queue_length_distribution}
\end{figure}

\begin{figure}[H]
     \centering
     \captionsetup{position=bottom} 
     \captionsetup[subfigure]{labelformat=empty}
    \subfloat[Our Policy]{\hspace{-4mm}\includegraphics[width=0.5\linewidth]{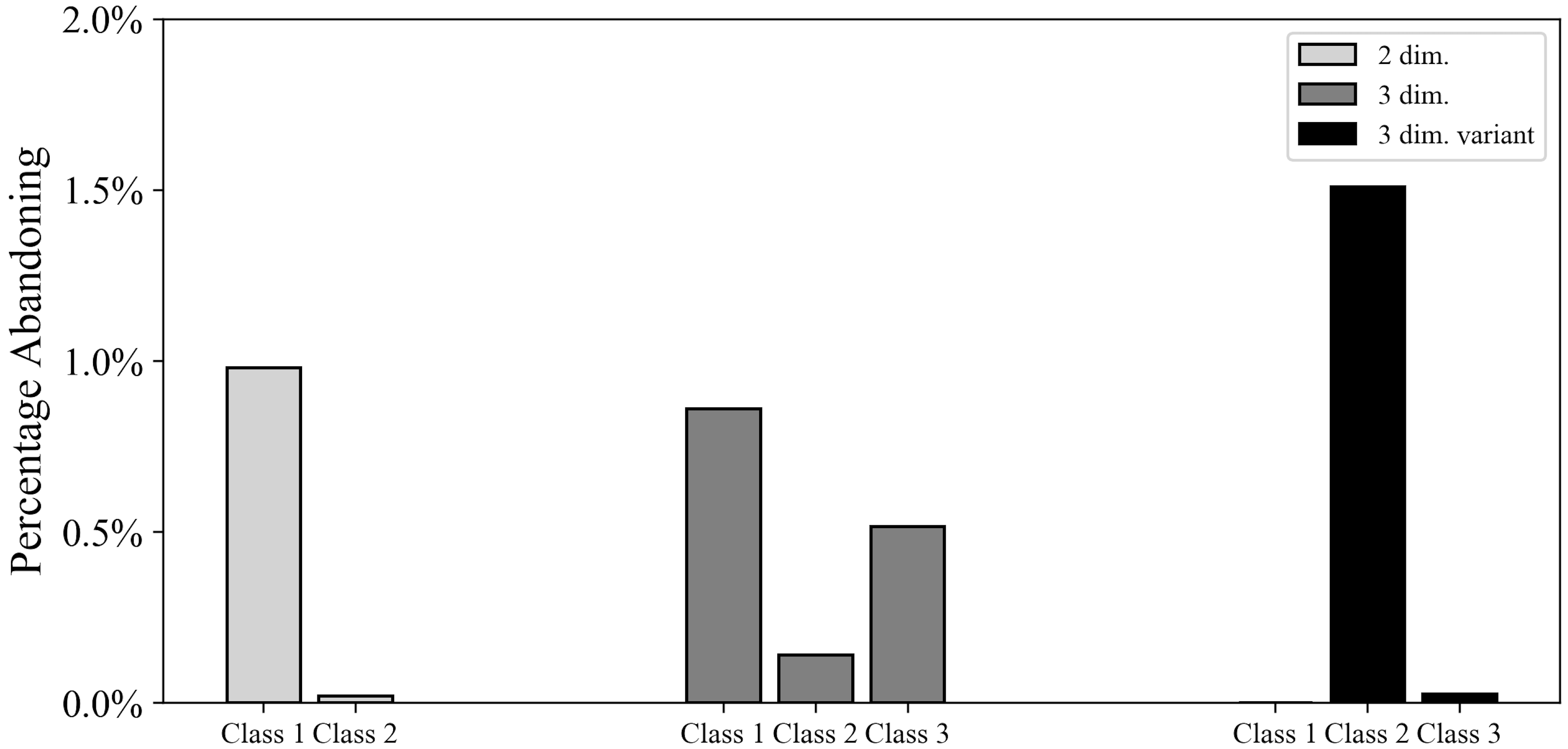}}\hfill
    \subfloat[Benchmark Policy]{\includegraphics[width=0.5\linewidth]{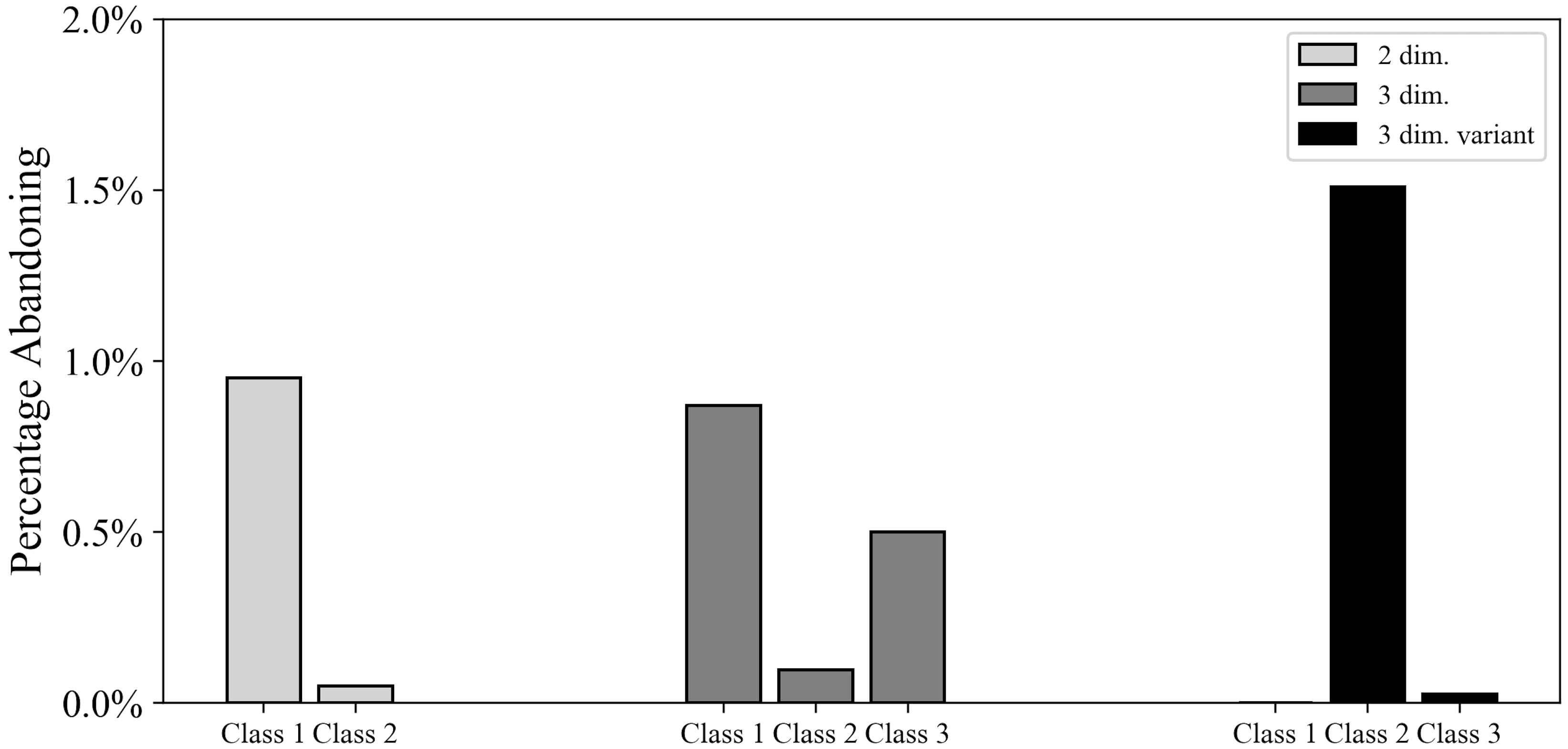}}
    \caption{Graphical representation of the average fraction of abandonments over arrivals under the policy learned from neural networks and the benchmark policy for the low dimensional test problems. } 
    \label{abandonments_lower_dimensional}
\end{figure}

For the low dimensional test problems ($K = 2,3$), the benchmark policy is the optimal policy computed using standard dynamic programming techniques;  see \ref{appendix_mdp_low_dim} for details. Table \ref{results_lower_test1} reports the average total costs obtained in a simulation study\footnote{We use the same random seed for each simulation study with 10,000 replications. All the performance figures reported are subject to simulation and discretization errors.} along with the percentage optimality gap between the optimal policy and our proposed policy. They have similar performance. For the last test problem shown in Table \ref{results_lower_test1} (the 3-dimensional variant), the optimal policy is a static priority policy that ranks class 1 highest, class 3 second, and class 2 lowest. Our method learns this policy. Because we use common random numbers for comparison, our proposed policy and the benchmark policy have the same performance in all simulation runs done for this test problem. 
\vspace{2mm}
\begin{table}[H]
    \centering
    \renewcommand{\arraystretch}{1.2}
    \setlength\tabcolsep{4pt} 
	{\small 
             \scalebox{0.9}{
    \begin{tabular}{lcccccccccccccccc}
        \toprule
        Method &&&&& 2-Dimensional &&&&& 3-Dimensional &&&&& 3-Dimensional variant\\
        \midrule
        Our Policy &&&&& 1682.84 $\pm$ 8.25  &&&&& 1690.25 $\pm$ 8.39 &&&&& 934.02 $\pm$ 4.73 \\
        Benchmark &&&&& 1681.19 $\pm$ 8.23  &&&&& 1682.67 $\pm$ 8.30  &&&&& 934.02 $\pm$ 4.73 \\
        \midrule
        Optimality Gap &&&&&  0.10\% $\pm$ 0.69\% &&&&& 0.45\% $\pm$ 0.70\% &&&&& 0.00\% $\pm$ 0.72\%\\
        \bottomrule
    \end{tabular}
    }
    }
    \caption{Performance comparison of our proposed policy with the benchmark policy in the low dimensional test problems. The first two rows show the total cost $\pm$ half-length of the 99\% confidence interval. Similarly, the last row shows the percentage optimality gap $\pm$ half-length of the 99\% confidence interval.}
    \label{results_lower_test1}
\end{table}
\vspace{-5mm}
Figures \ref{queue_length_distribution} - \ref{abandonments_lower_dimensional} display how the backlog and the percentage of arrivals that abandon, respectively, are distributed across different classes for the two policies. They show that our proposed policy is similar to the benchmark policy\footnote{Because the abandonment rates are low, the allocation of the service effort is nearly proportional to the offered load of that class under both the benchmark and proposed policies. Thus, the two policies allocate the service effort across different classes similarly.}.
\vspace{-7mm}
\vspace{-9mm}
\subsection{Results for the main test problem and its variants}

\begin{figure}[H]
  \captionsetup{position=bottom} 
  \captionsetup[subfigure]{labelformat=empty}
  \subfloat[Our Policy (Main)]{\hspace{-4mm}\includegraphics[width=0.5\linewidth]{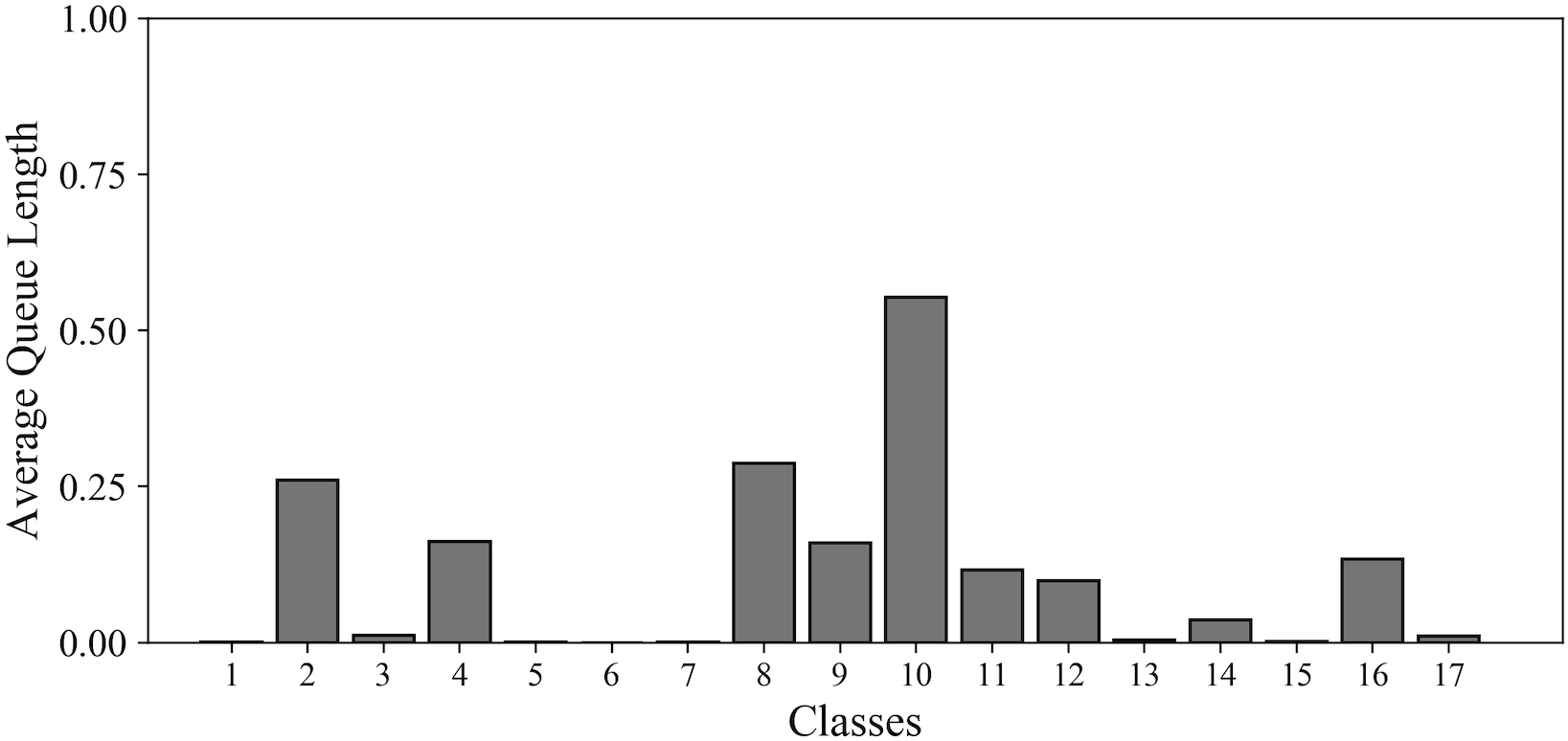}}\hfill
  \subfloat[Benchmark Policy (Main)]{\includegraphics[width=0.5\linewidth]{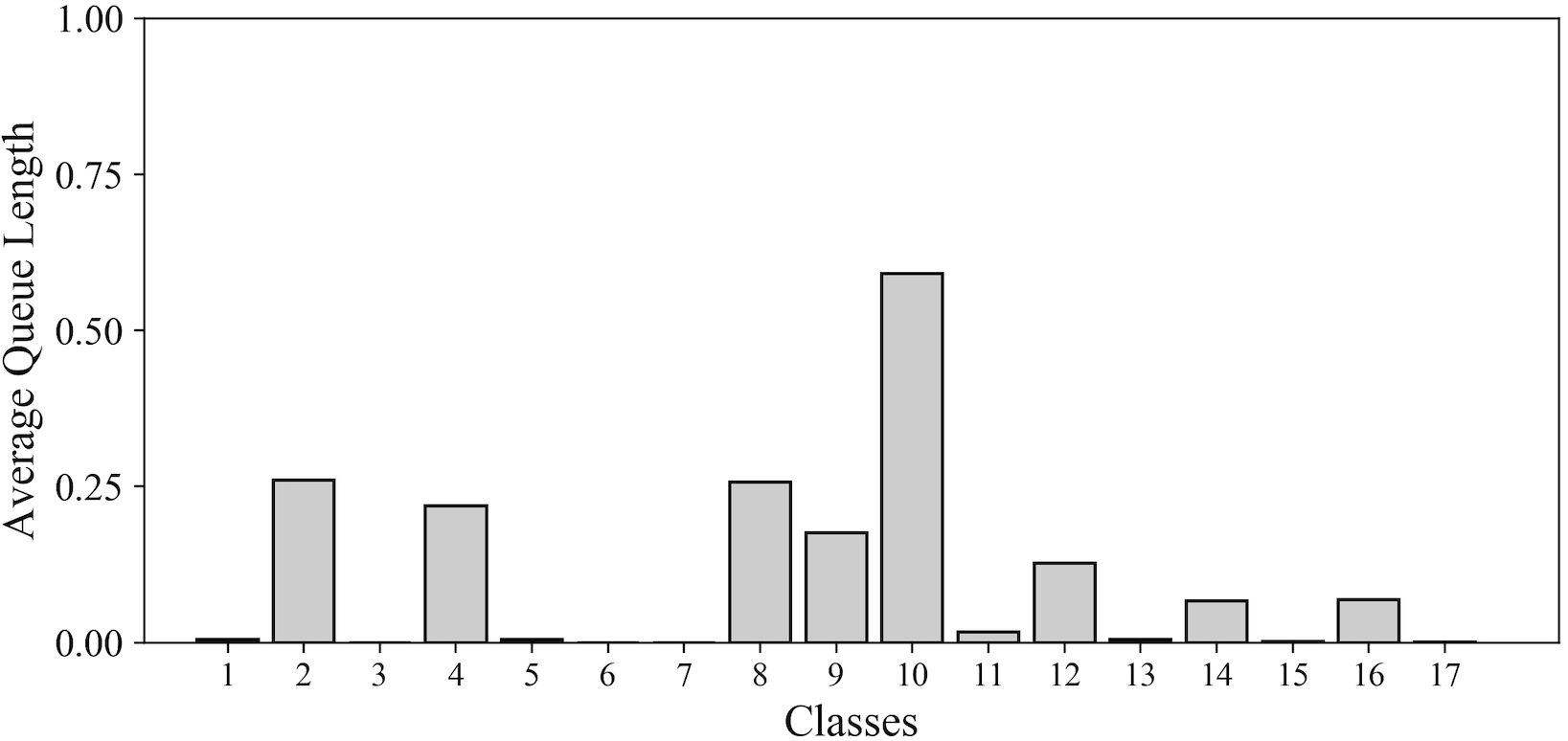}}\\
  \subfloat[Our Policy (First variant)]{\hspace{-4mm} \includegraphics[width=0.5\linewidth]{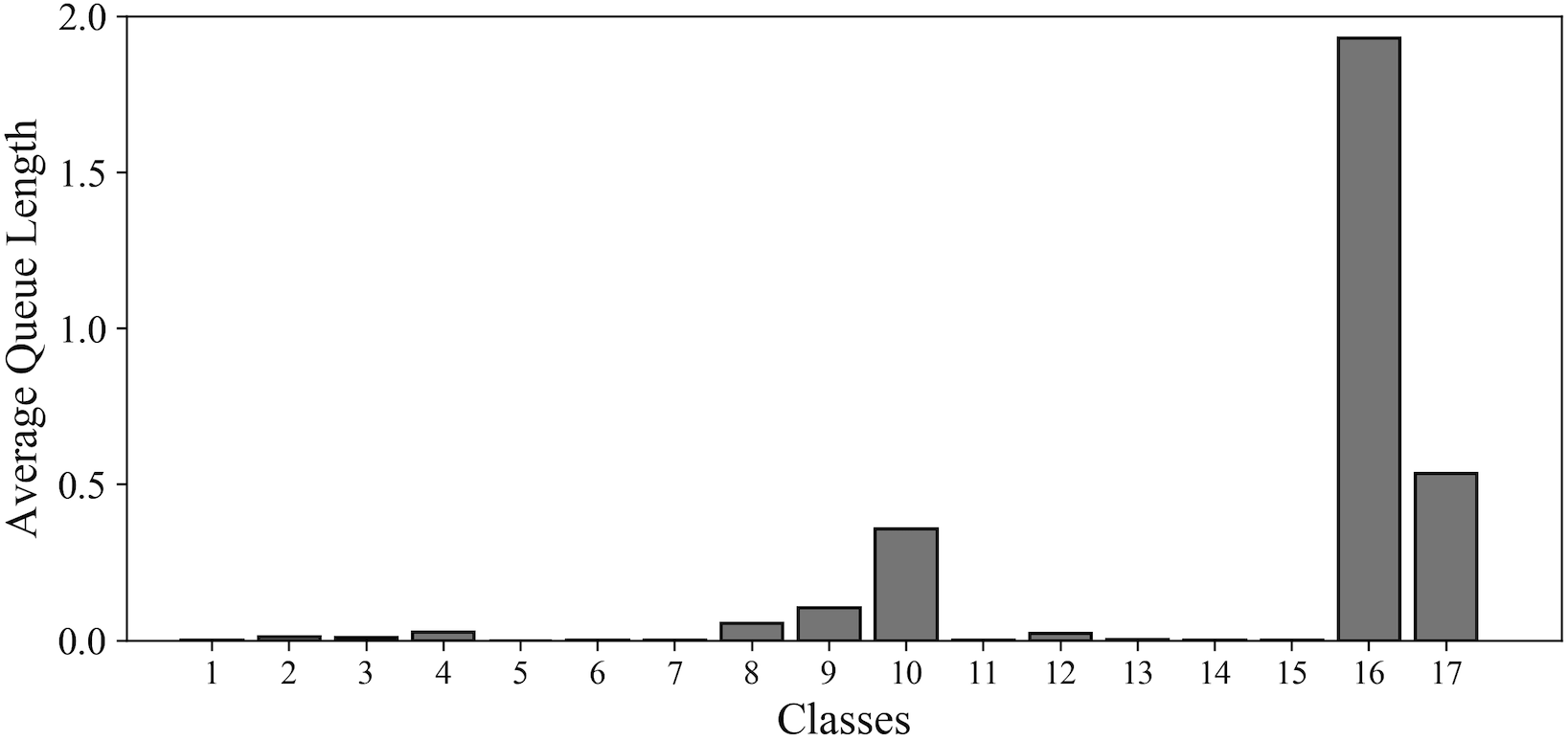}}\hfill
  \subfloat[Benchmark Policy (First variant)]{\includegraphics[width=0.5\linewidth]{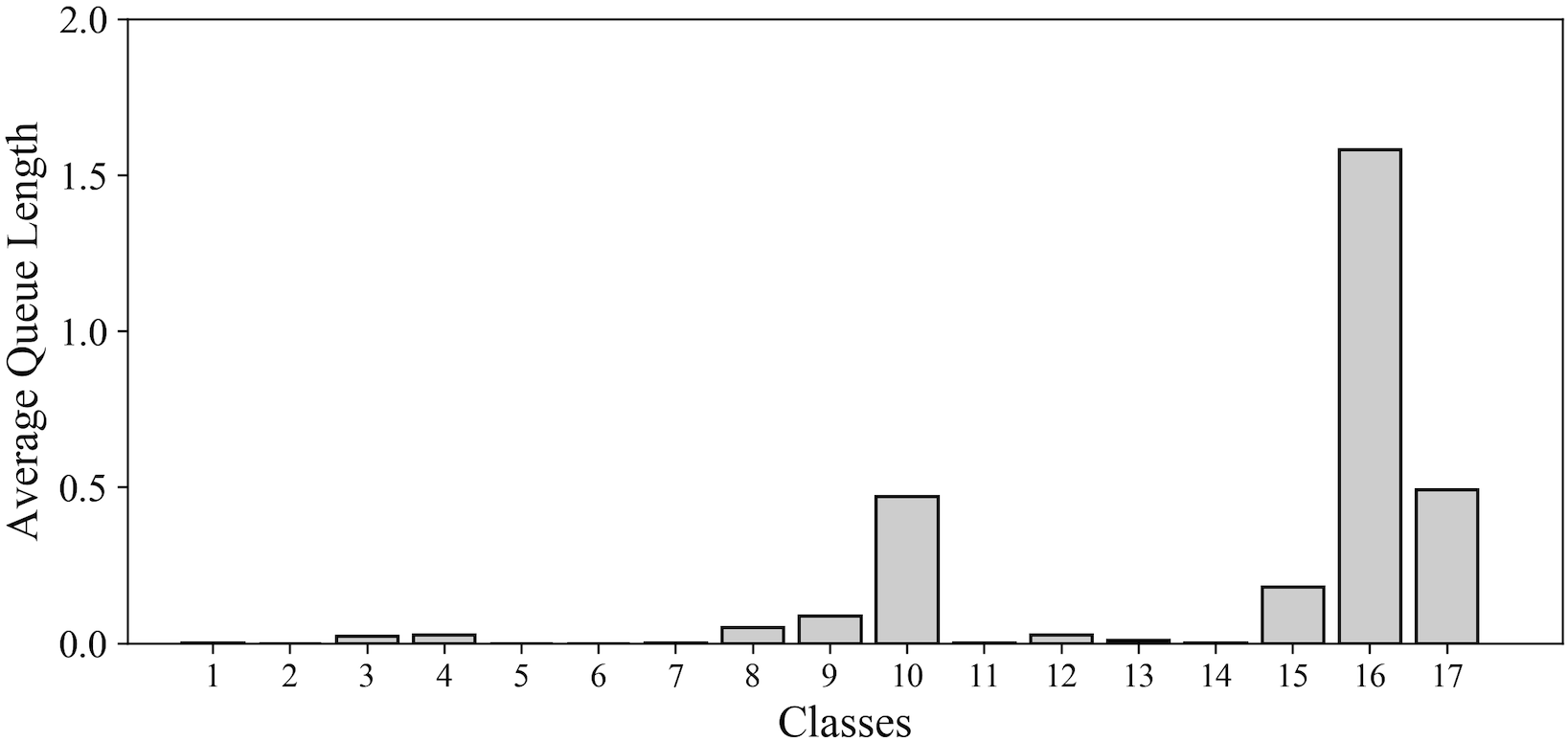}}\\
  \subfloat[Our Policy (Second variant)]{\hspace{-4mm} \includegraphics[width=0.5\linewidth]{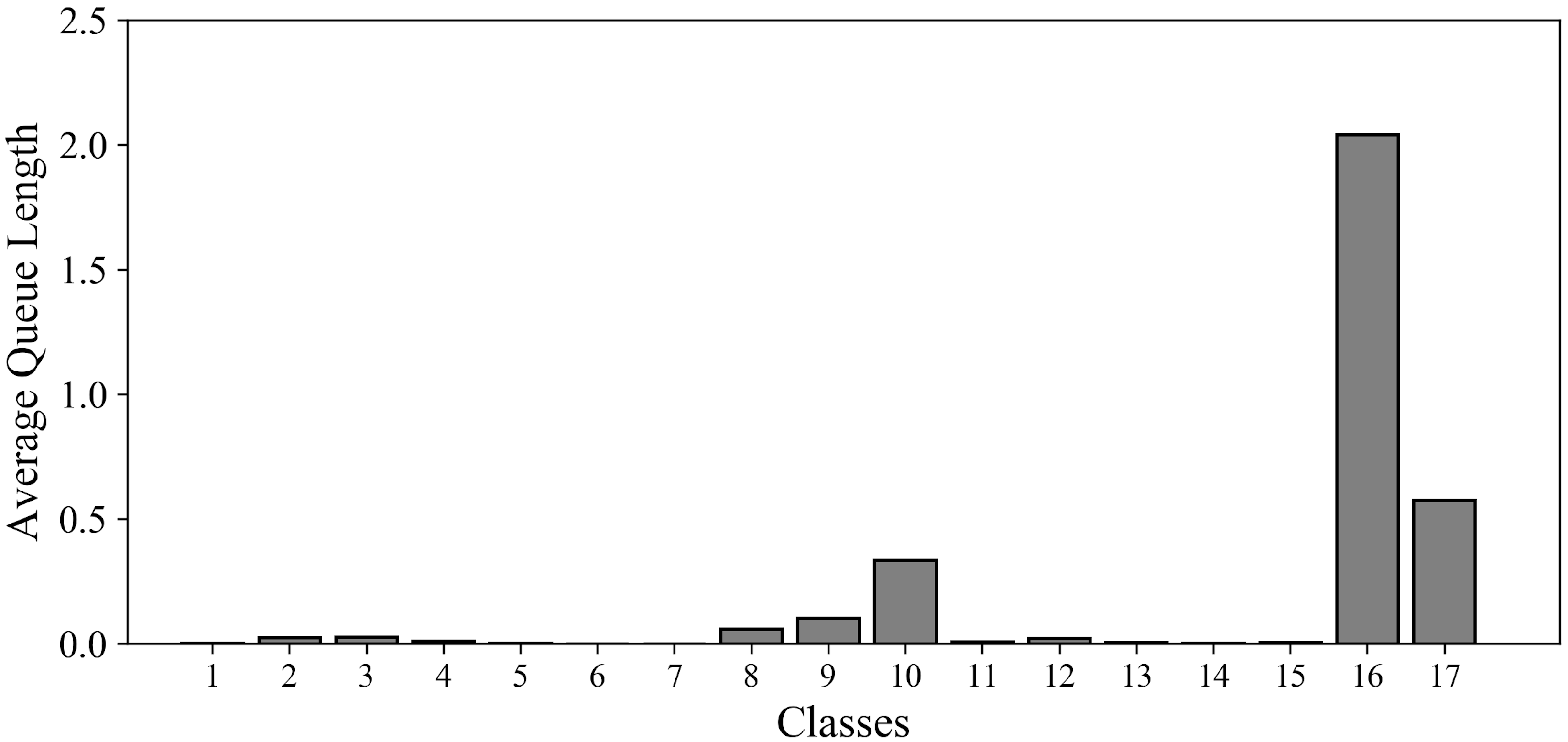}}\hfill
  \subfloat[Benchmark Policy (Second variant)]{\includegraphics[width=0.5\linewidth]{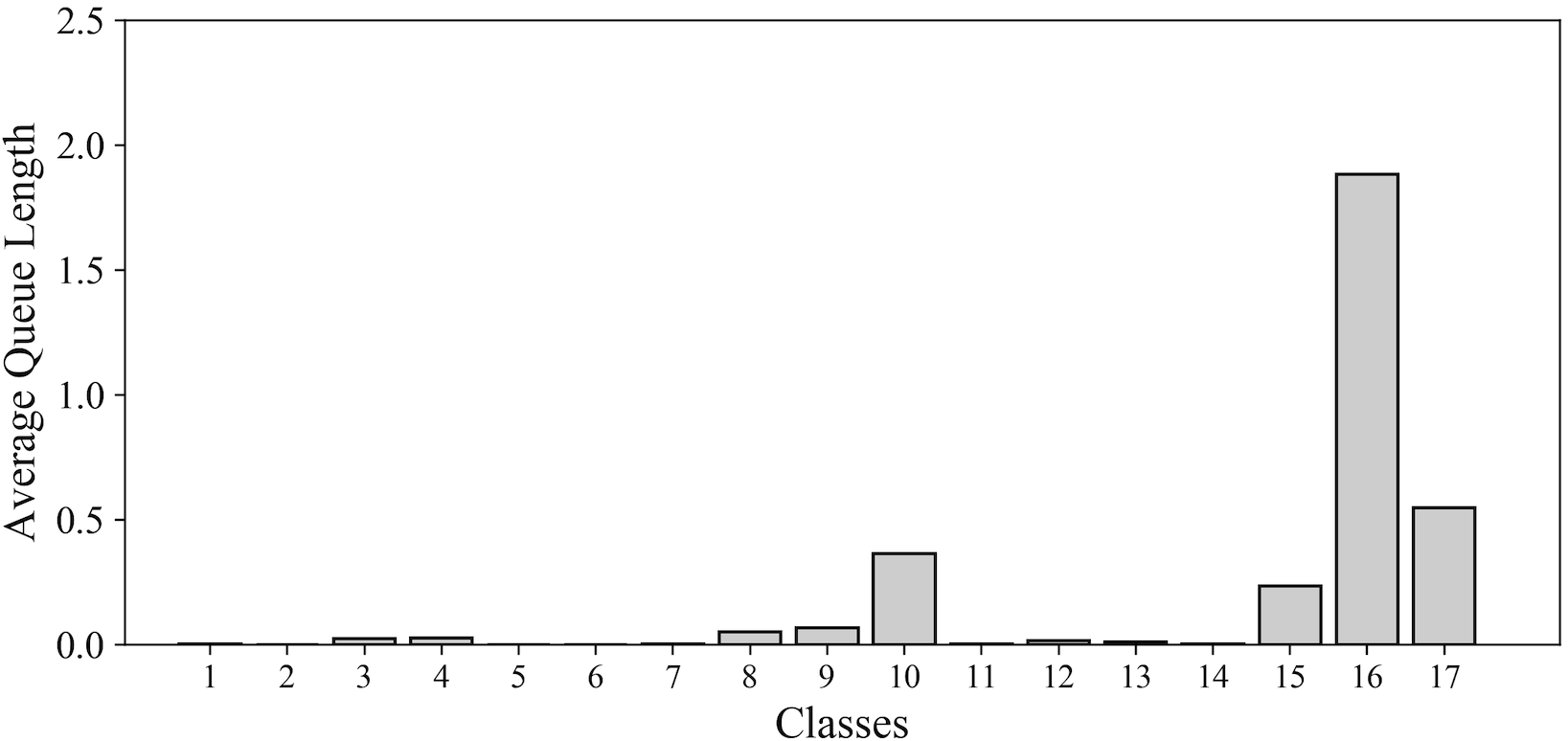}}
  \caption{Graphical representation of the average queue length under the policy learned from neural networks and the benchmark policy for the main test problem and its variations.}
  \label{17dim_main_queue}
\end{figure}

\begin{figure}[H]
  \captionsetup{position=bottom} 
  \captionsetup[subfigure]{labelformat=empty}
  \subfloat[Our Policy (Main)]{\hspace{-4mm} \includegraphics[width=0.5\linewidth]{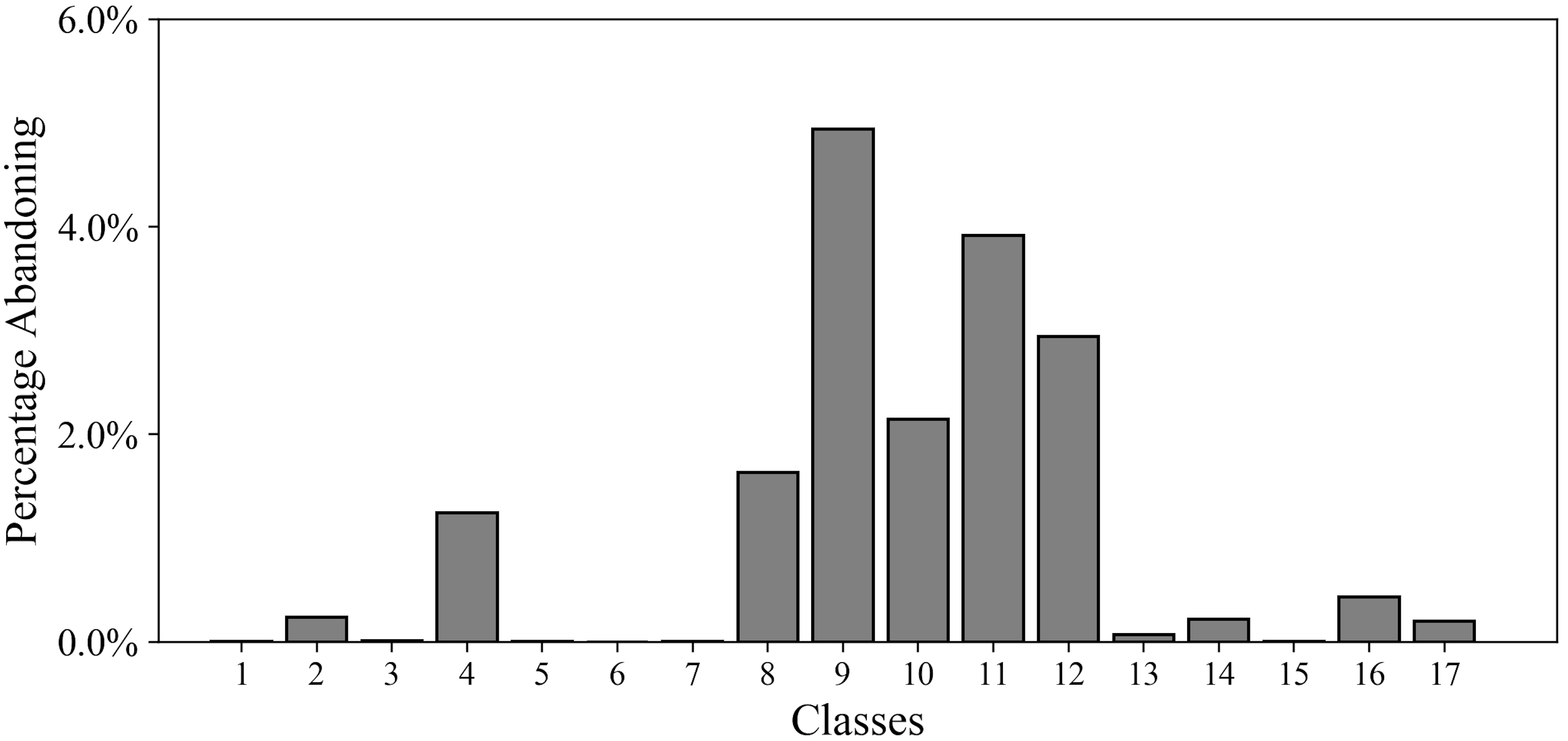}}\hfill
  \subfloat[Benchmark Policy (Main)]{\includegraphics[width=0.5\linewidth]{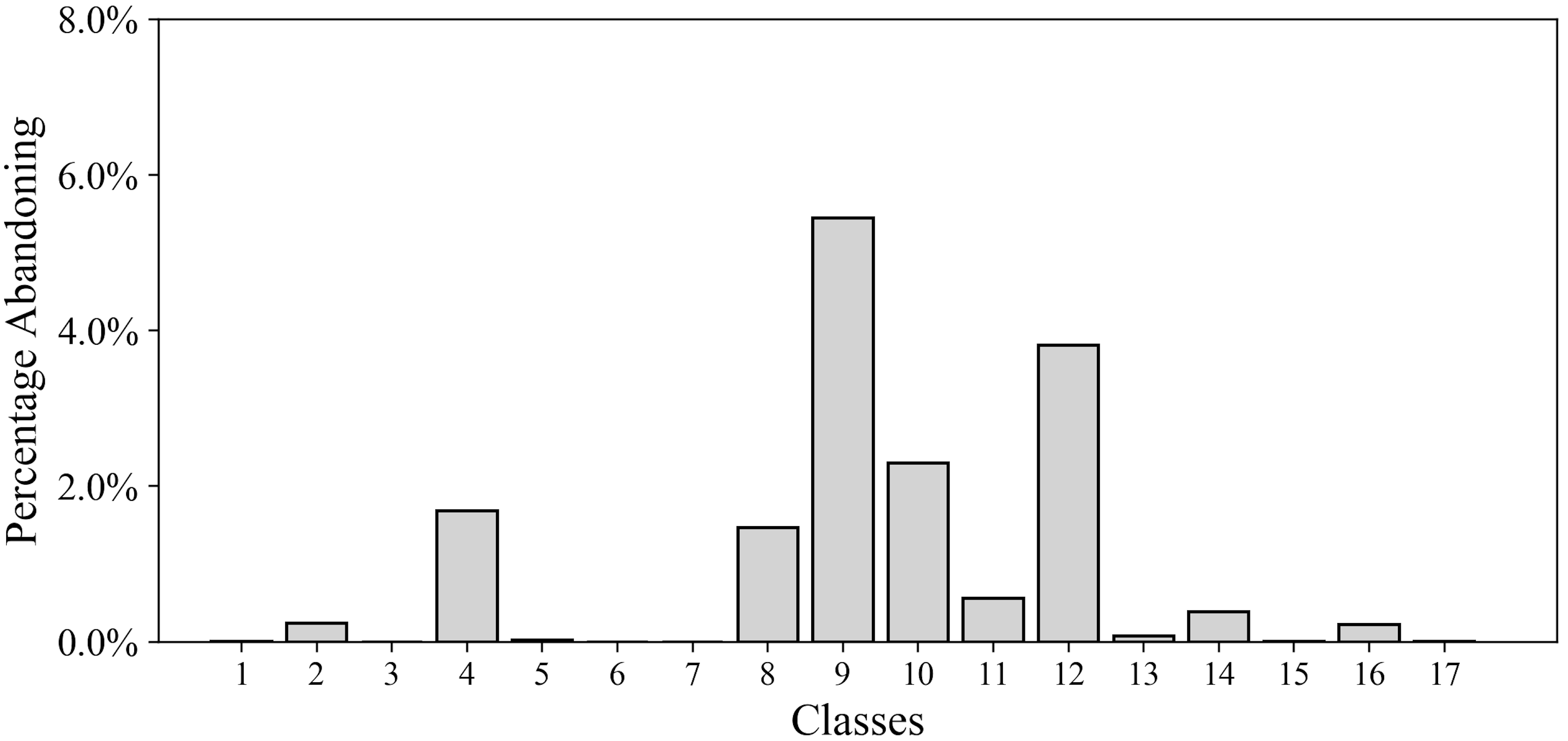}}\\
  \subfloat[Our Policy (First variant)]{\hspace{-4mm}\includegraphics[width=0.5\linewidth]{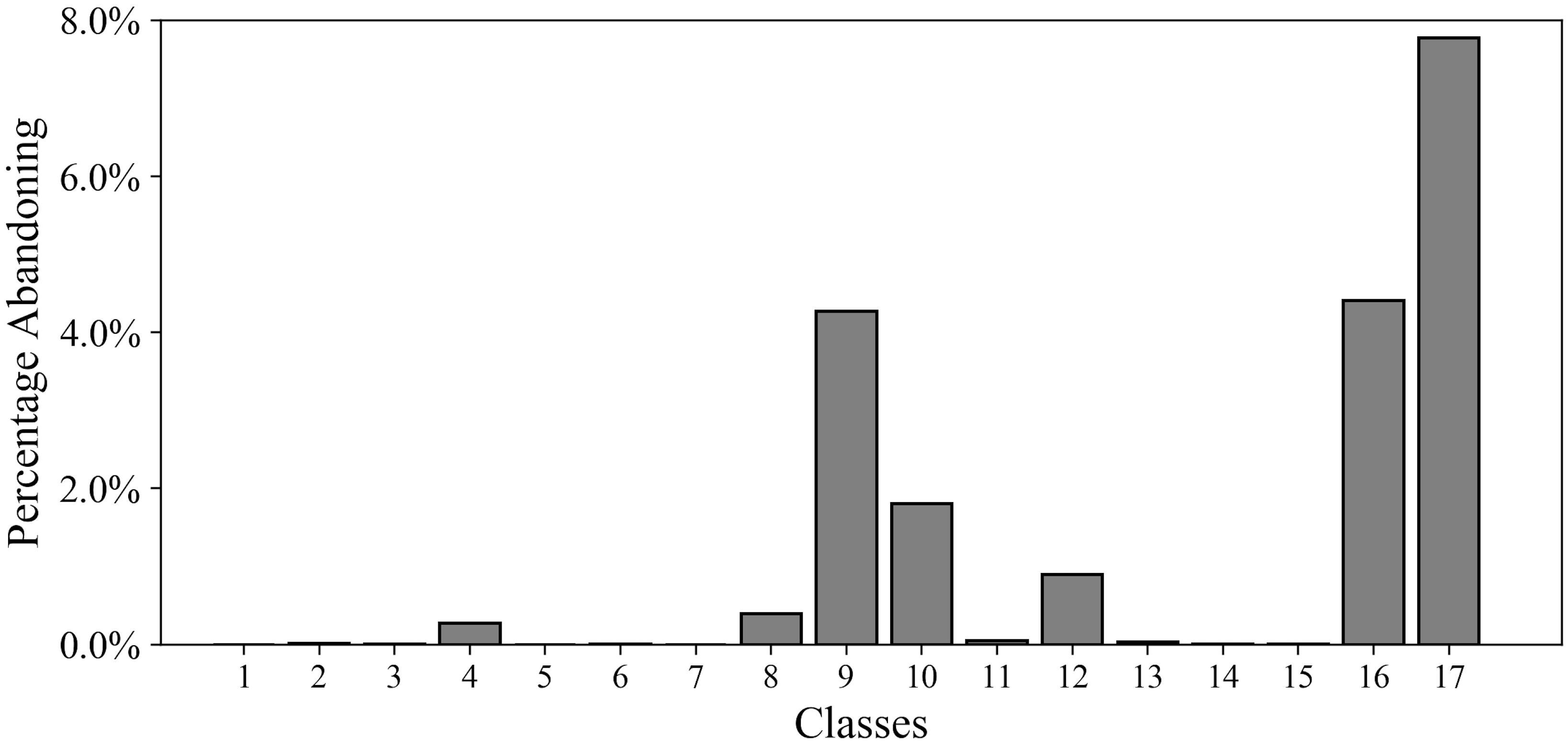}}\hfill
  \subfloat[Benchmark Policy (First variant)]{\includegraphics[width=0.5\linewidth]{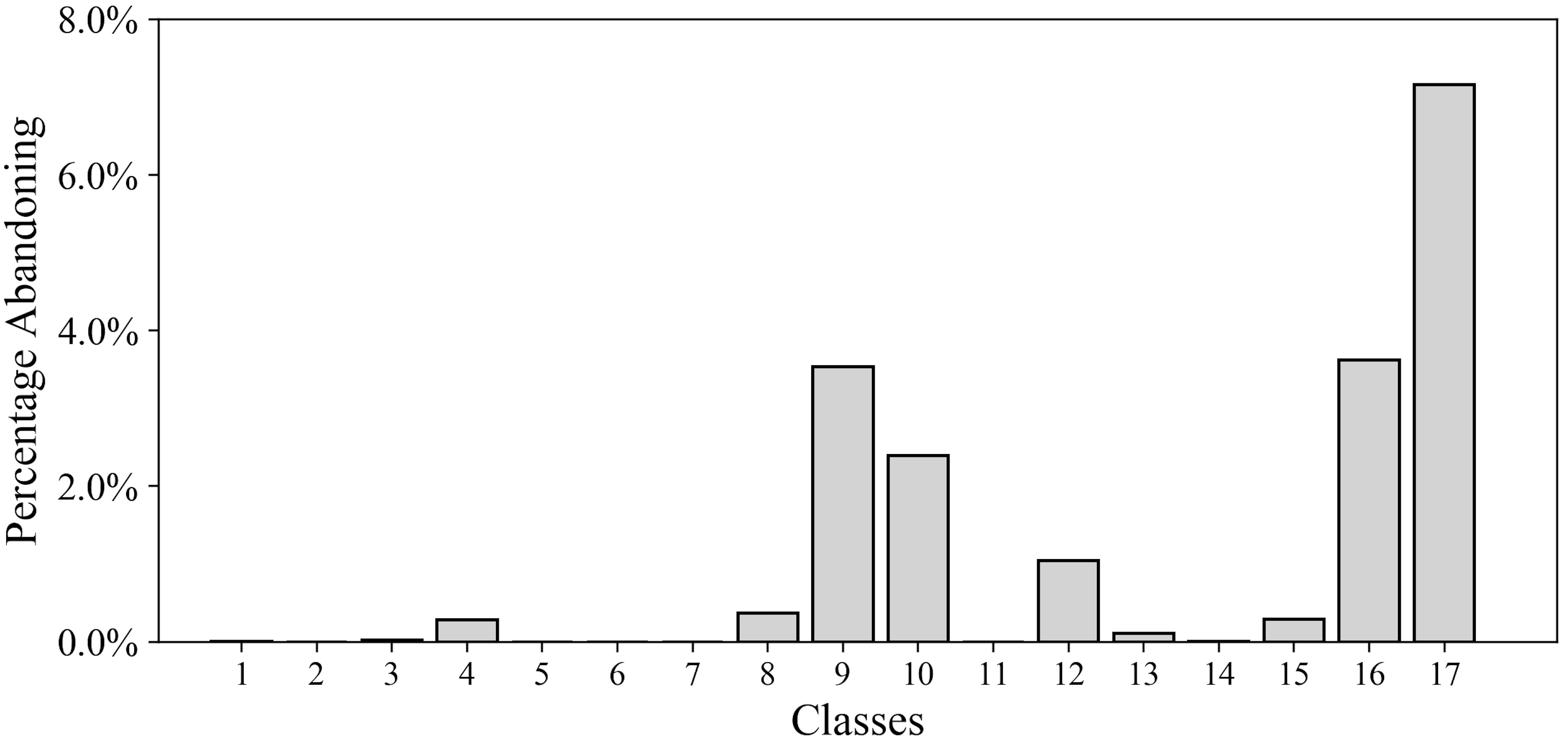}}\\
  \subfloat[Our Policy (Second variant)]{\hspace{-4mm} \includegraphics[width=0.5\linewidth]{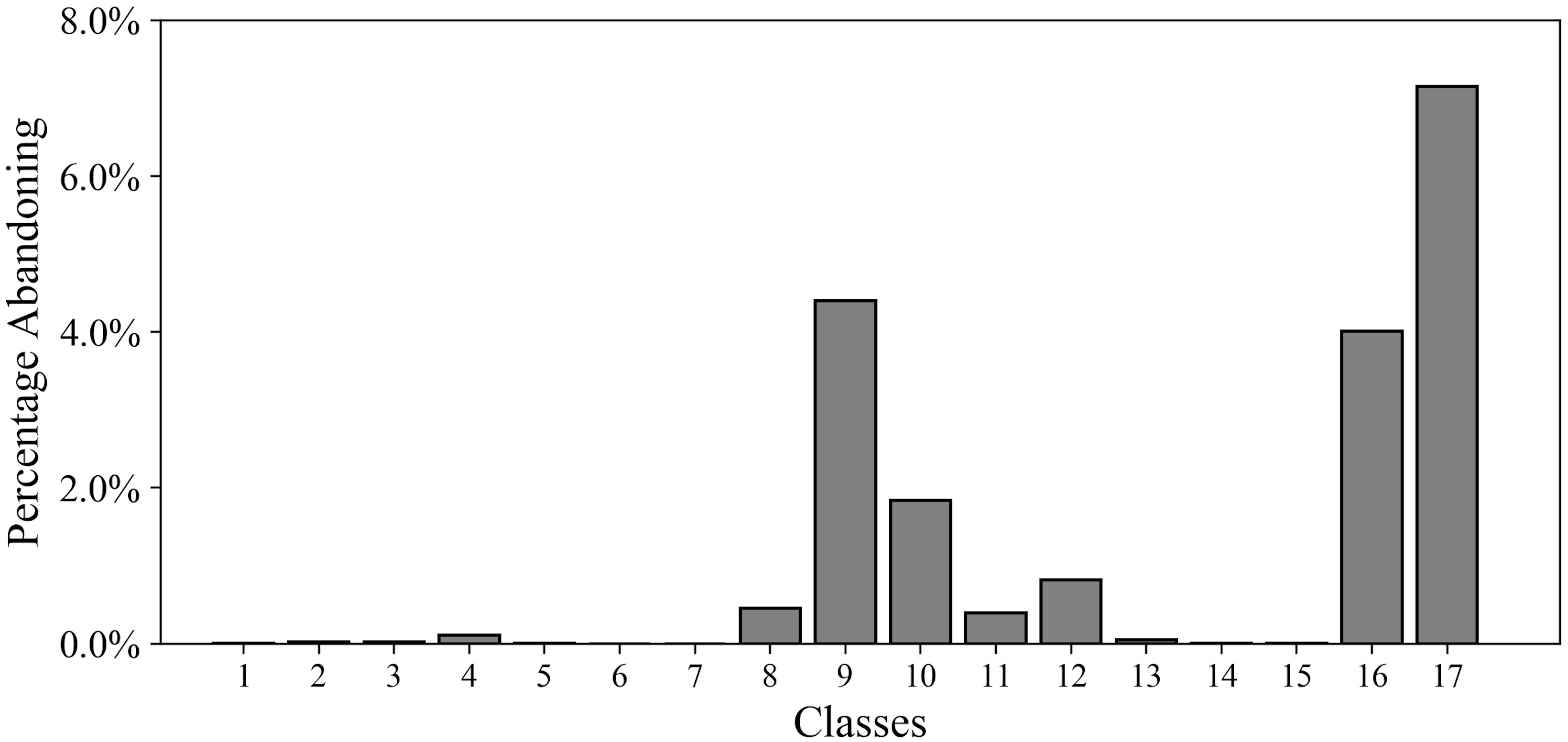}}\hfill
  \subfloat[Benchmark Policy (Second variant)]{\includegraphics[width=0.5\linewidth]{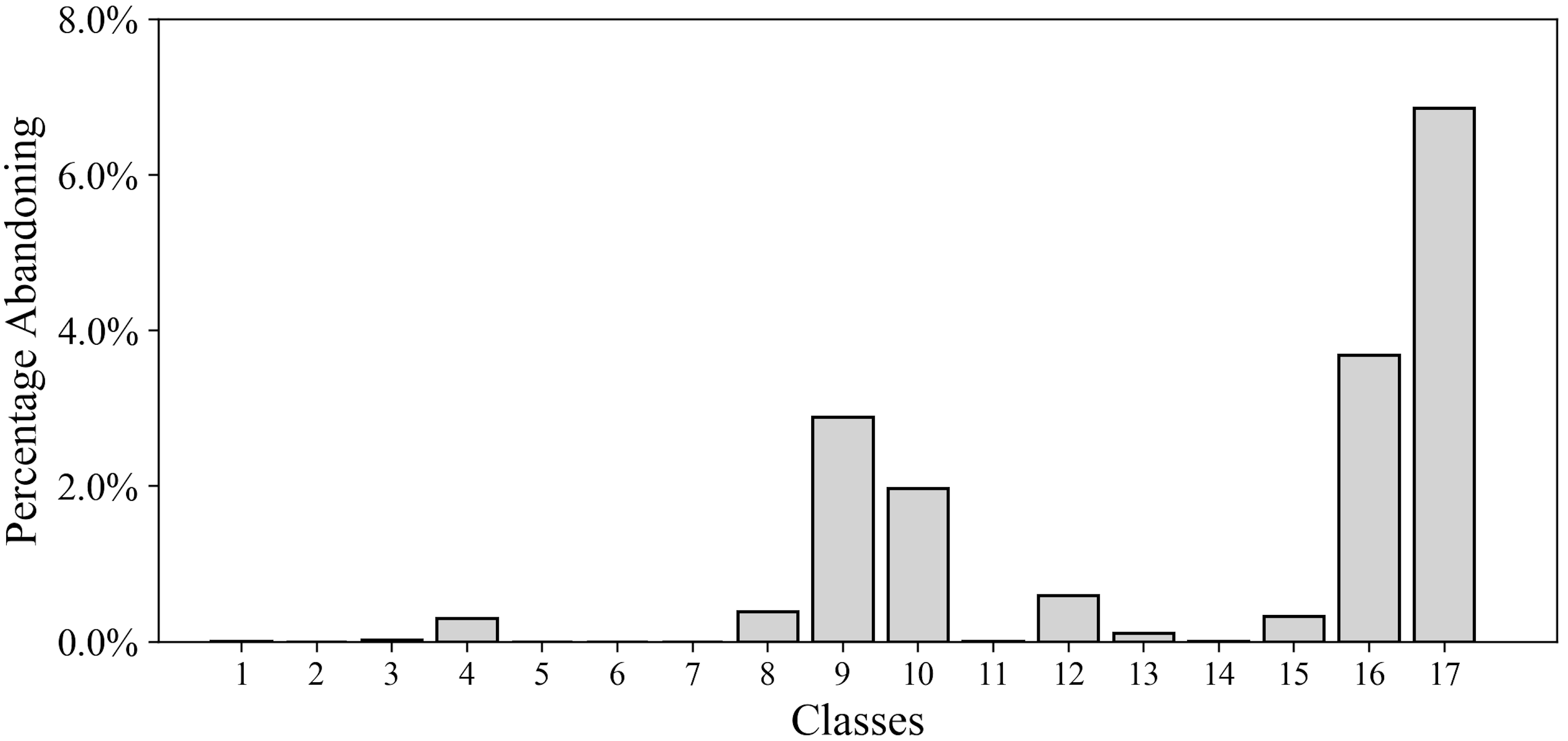}}
  \caption{Graphical representation of the average fraction of abandonments over arrivals under the policy learned from neural networks and the benchmark policy for the main test problem and its variations.}
  \label{17dim_main_abandonments}
\end{figure}

For our main test problem and its two variants (see Section \ref{sect_main_test}), the optimal policy is not known, but we consider various benchmark policies (see Section \ref{computational_benchmarks}) and pick the best one to assess the performance of our proposed policy. Table \ref{results_main_test} presents the average total costs derived from a simulation study\footnote{We use the same random seed for each simulation study with 10,000 replications. All the performance figures reported are subject to simulation and discretization errors.}, alongside the percentage performance gap between the best benchmark policy and our proposed policy.
\begin{table}[H]
    \centering
    \renewcommand{\arraystretch}{1.2}
    \setlength\tabcolsep{4pt} 
	{\small 
             \scalebox{0.9}{
    \begin{tabular}{lcccccccccccccccc}
        \toprule
          Method &&&&& Main &&&&& First variant &&&&& Second variant\\
        \midrule
          Our Policy &&&&& 1154.41 $\pm$ 5.91 &&&&& 1470.06 $\pm$ 7.86 &&&&& 1396.31 $\pm$ 7.79\\
          Benchmark &&&&& 1157.23 $\pm$ 5.88 &&&&& 1467.59 $\pm$ 7.70 &&&&& 1394.72 $\pm$ 7.68\\
          \midrule
          Performance Gap &&&&& -0.24\% $\pm$ 0.72\% &&&&& 0.17\% $\pm$ 0.75\% &&&&& 0.11\% $\pm$ 0.78\%\\
        \bottomrule
    \end{tabular}
    }
    \caption{Performance comparison of our proposed policy with the benchmark policy in the main test problem and its two variants. The first two rows show the total cost $\pm$ half-length of the 99\% confidence interval for our proposed policy and the best benchmark policy. The last row shows the percentage performance gap $\pm$ half-length of the 99\% confidence interval.}
    \label{results_main_test}
    }
\end{table}
\vspace{-6mm}
Figures \ref{17dim_main_queue} - \ref{17dim_main_abandonments} display how the queue lengths and the percentage of arrivals that abandon, respectively, vary across different classes for the two policies. The numerical results and the figures show that our proposed policy is similar to the benchmark policy for the main test problem and its two variants.
\vspace{-6mm}

Our computational method significantly reduces the runtime for the test problems where identifying good benchmarks is computationally intensive. These include low dimensional test problems, where the benchmark policy is the optimal MDP solution, as well as the main test problem and its two variants, where the benchmark policy is a dynamic policy derived by combining the auxiliary low dimensional MDP solution with the five static priority policies introduced in Section \ref{static_benchmarks_main_variant}.  Solving the MDP takes one day for 2-dimensional instance and six days for 3-dimensional instances. Similarly, solving the auxiliary low dimensional MDP\footnote{Although the MDP that focuses solely on low-priority classes is also three-dimensional—since we remove high-priority classes from the system—the resulting truncated state space is smaller than that of the original three-dimensional problem. Consequently, the auxiliary MDP for low-priority classes has a shorter runtime compared to solving the MDP for the full three-dimensional problem instance.} and searching through static priority policies for the main test problem and its variants takes two days. In stark contrast, our computational method solves low dimensional problems under 15 minutes and the main test problem and its variants under one hour.

\subsection{Results for the high dimensional test problems that have pathwise optimal policies}

\begin{figure}[H]
  \captionsetup{position=bottom} 
  \captionsetup[subfigure]{labelformat=empty}
  \subfloat[Our Policy ($K = 30$)]{\hspace{-4mm}\includegraphics[width=0.5\linewidth]{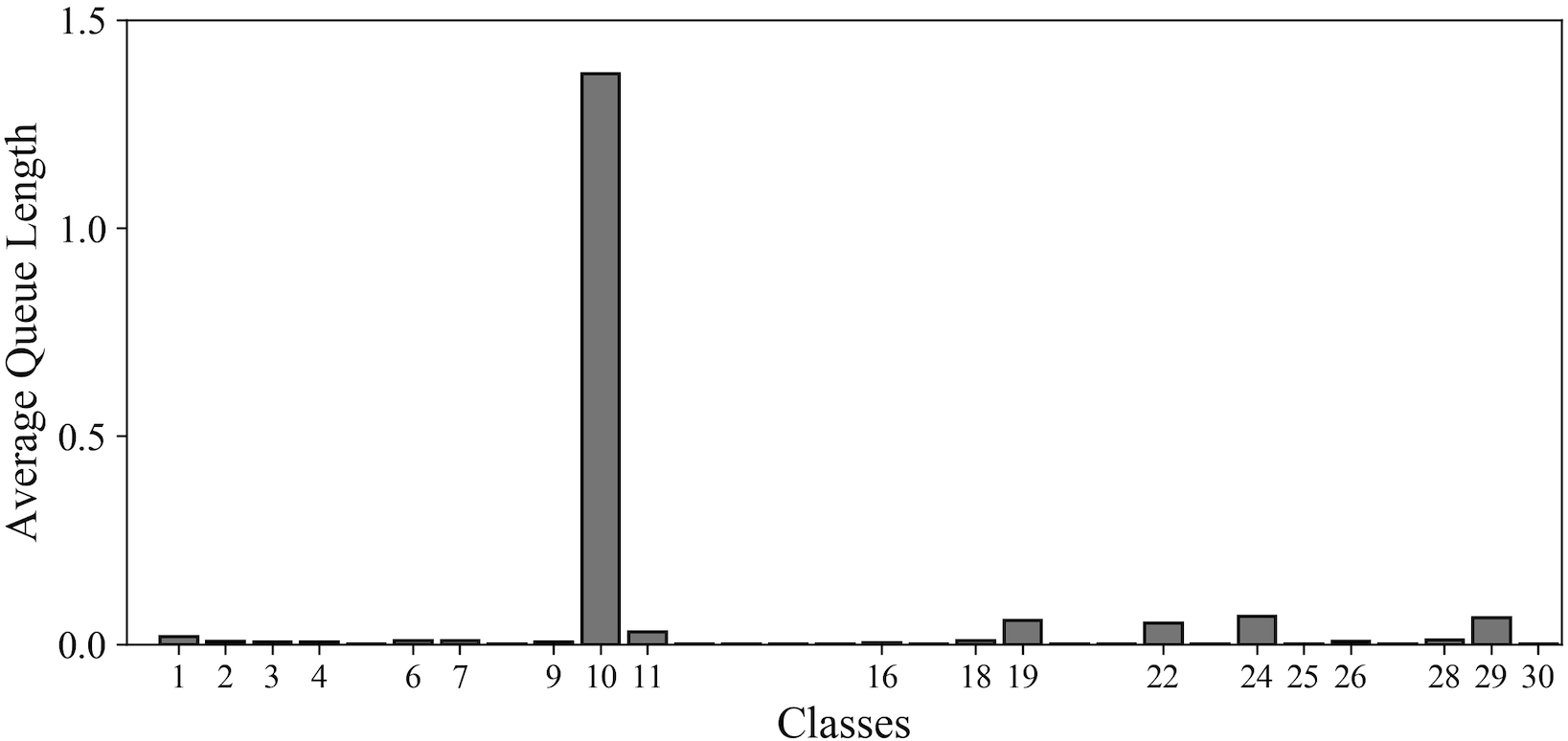}}\hfill
  \subfloat[Benchmark Policy ($K = 30$)]{\includegraphics[width=0.5\linewidth]{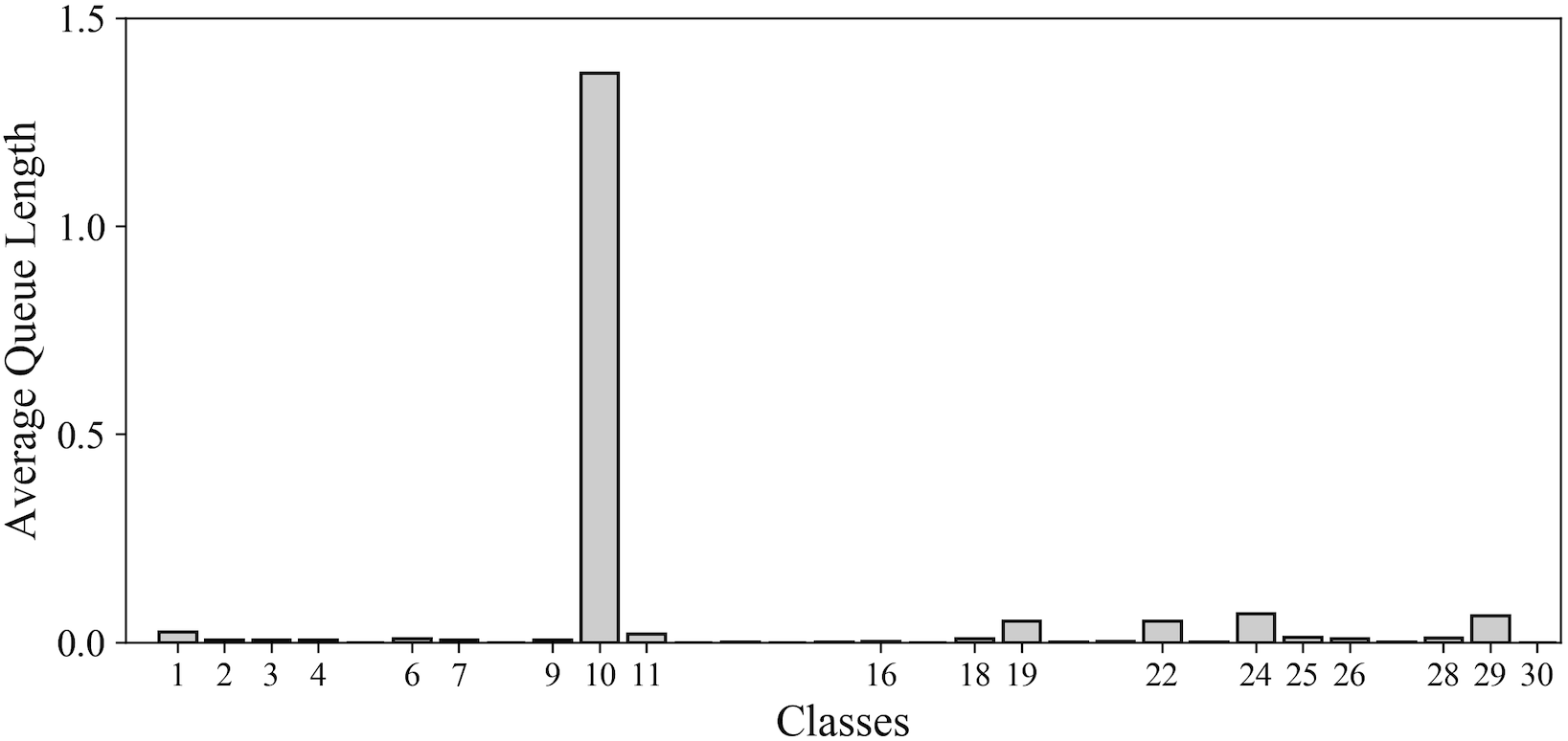}}\\
  \subfloat[Our Policy ($K = 50$)]{\hspace{-4mm}\includegraphics[width=0.5\linewidth]{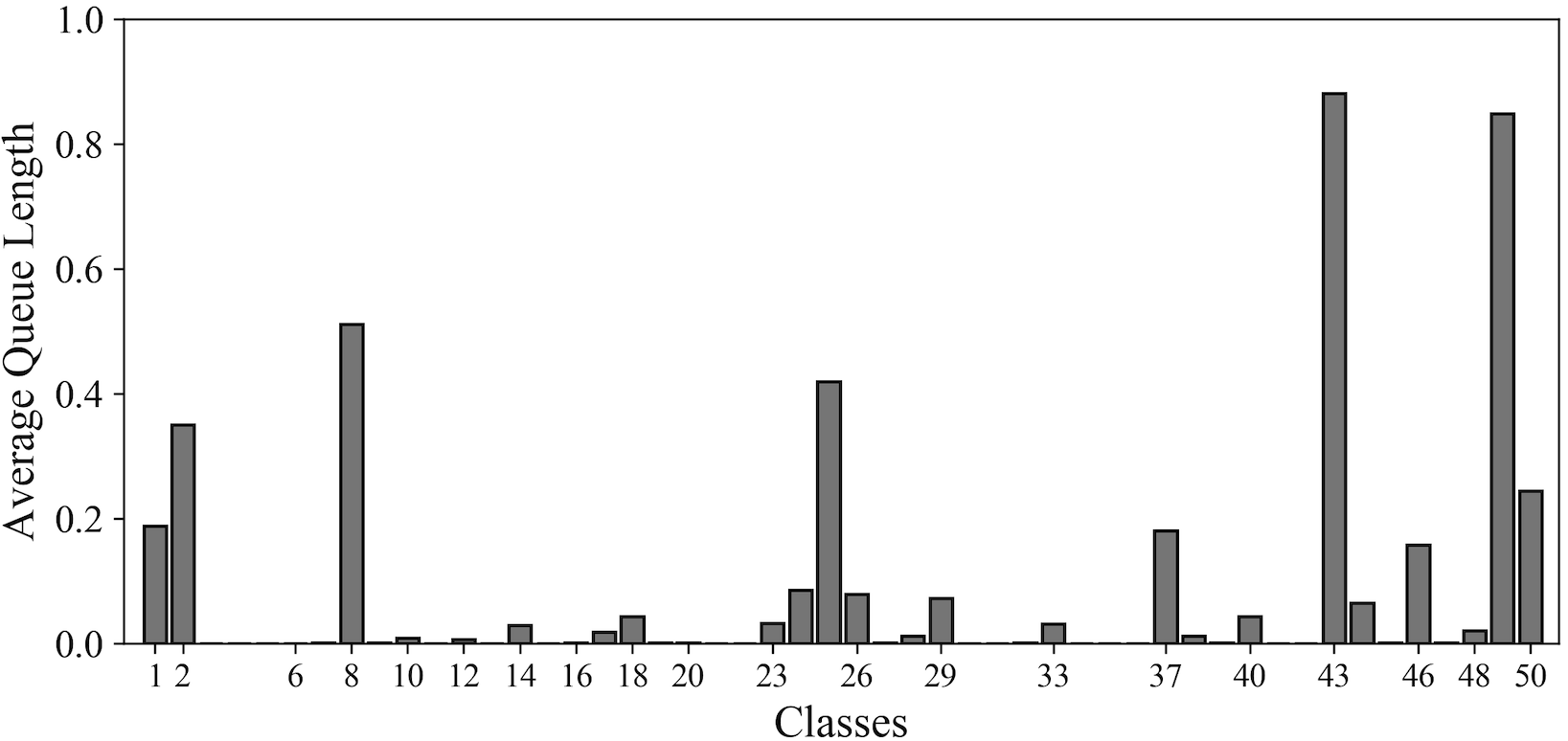}}\hfill
  \subfloat[Benchmark Policy ($K = 50$)]{\includegraphics[width=0.5\linewidth]{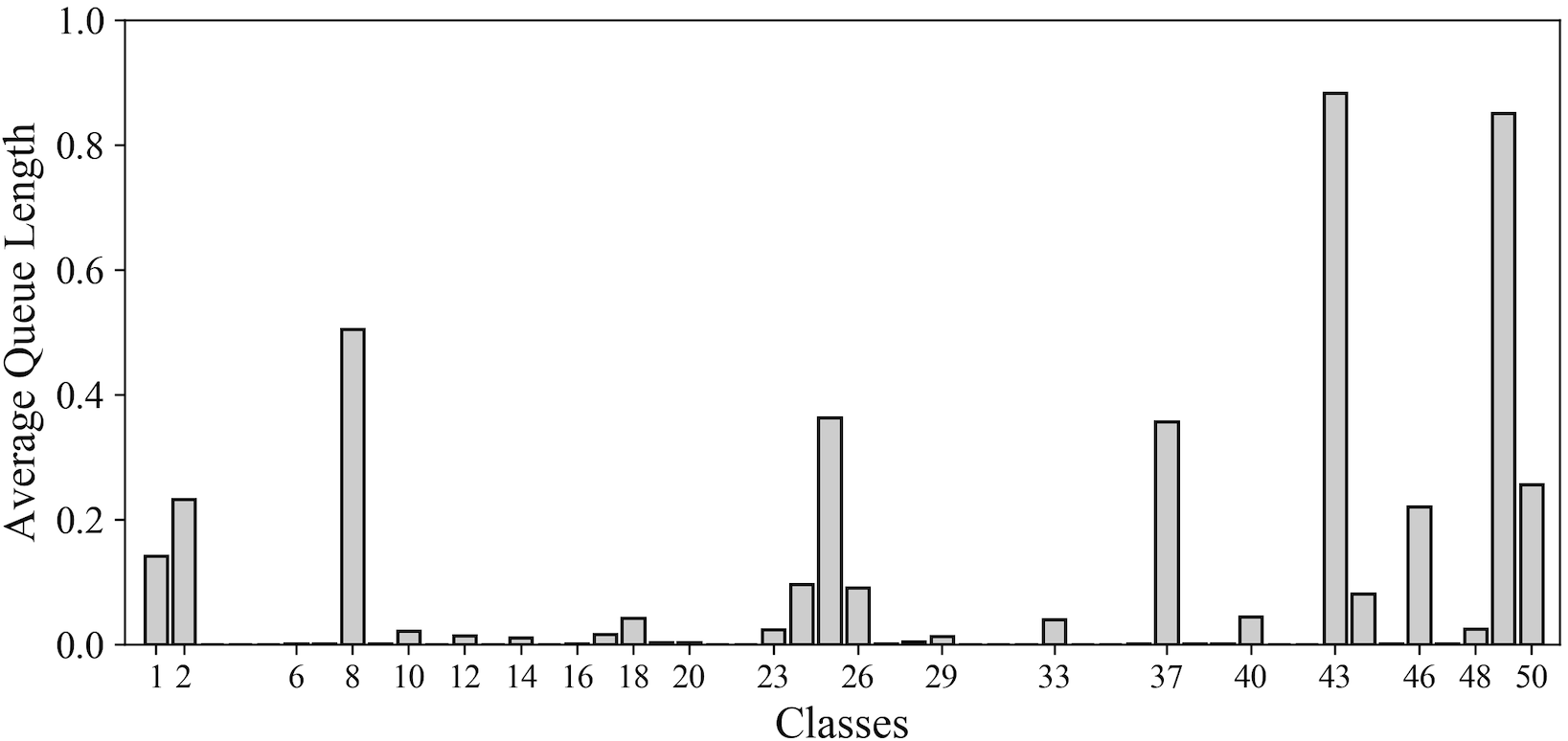}}\\
  \subfloat[Our Policy ($K = 100$)]{\hspace{-4mm}\includegraphics[width=0.5\linewidth]{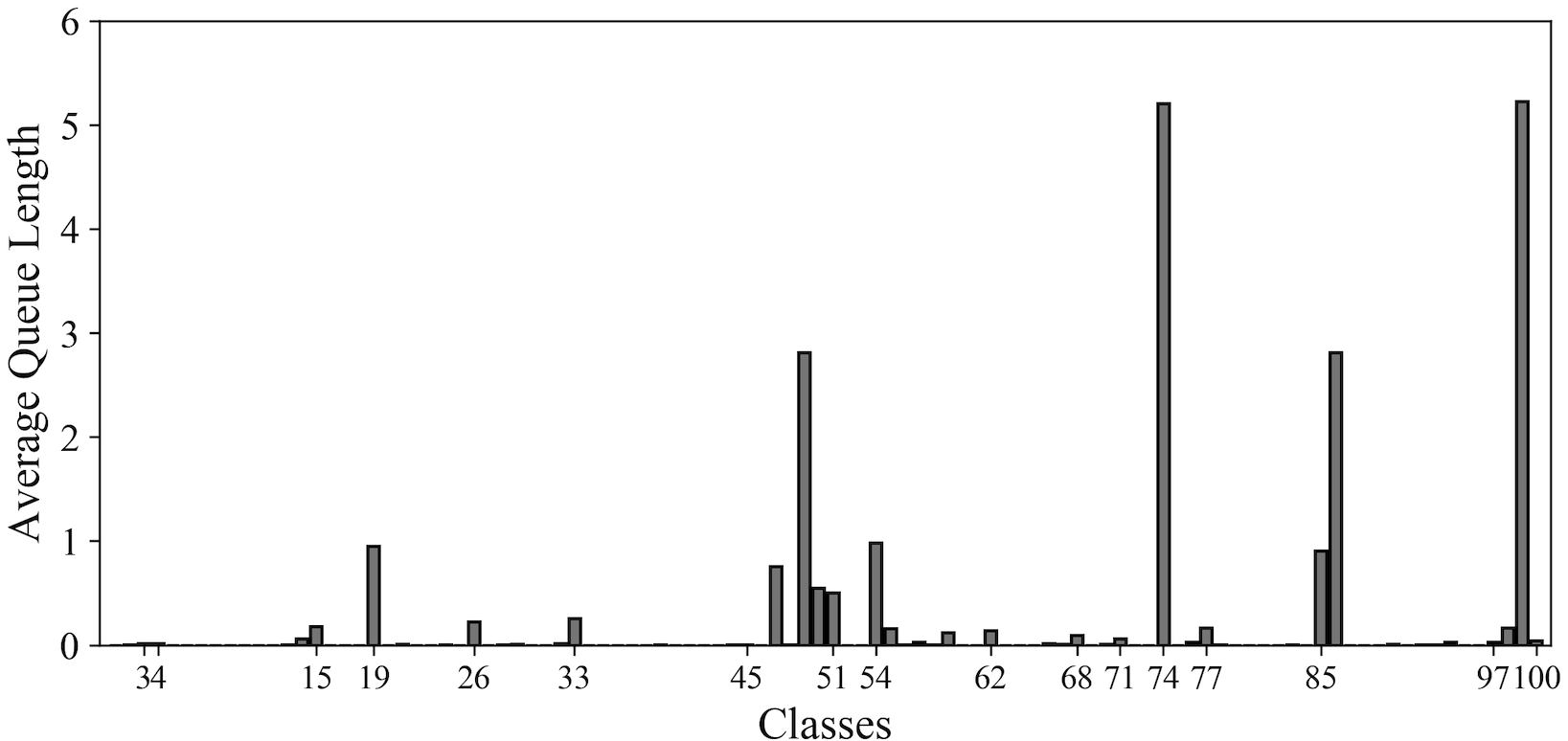}}\hfill
  \subfloat[Benchmark Policy ($K = 100$)]{\includegraphics[width=0.5\linewidth]{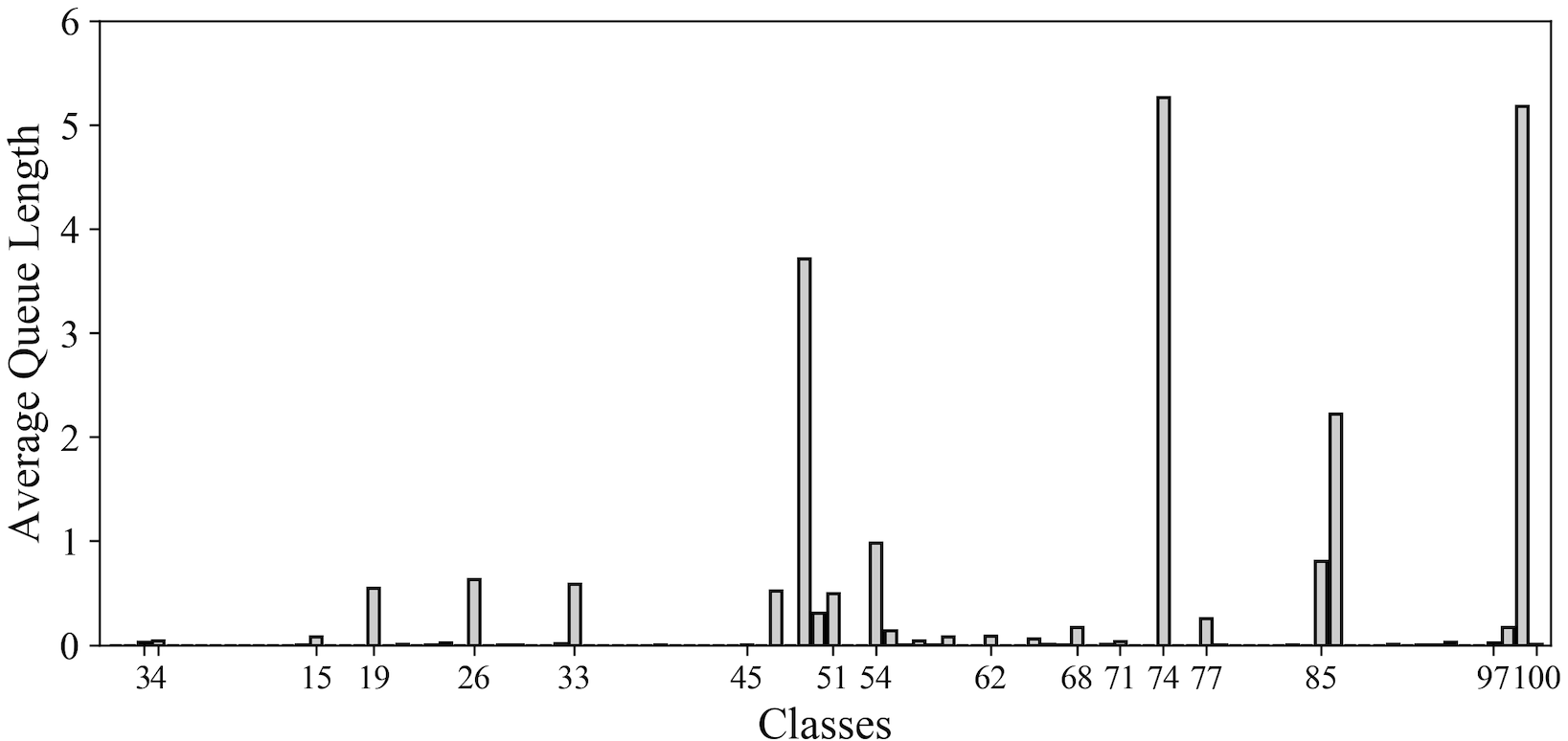}}\\
  \subfloat[Our Policy ($K = 500$)]{\hspace{-5mm}\includegraphics[width=0.514\linewidth]{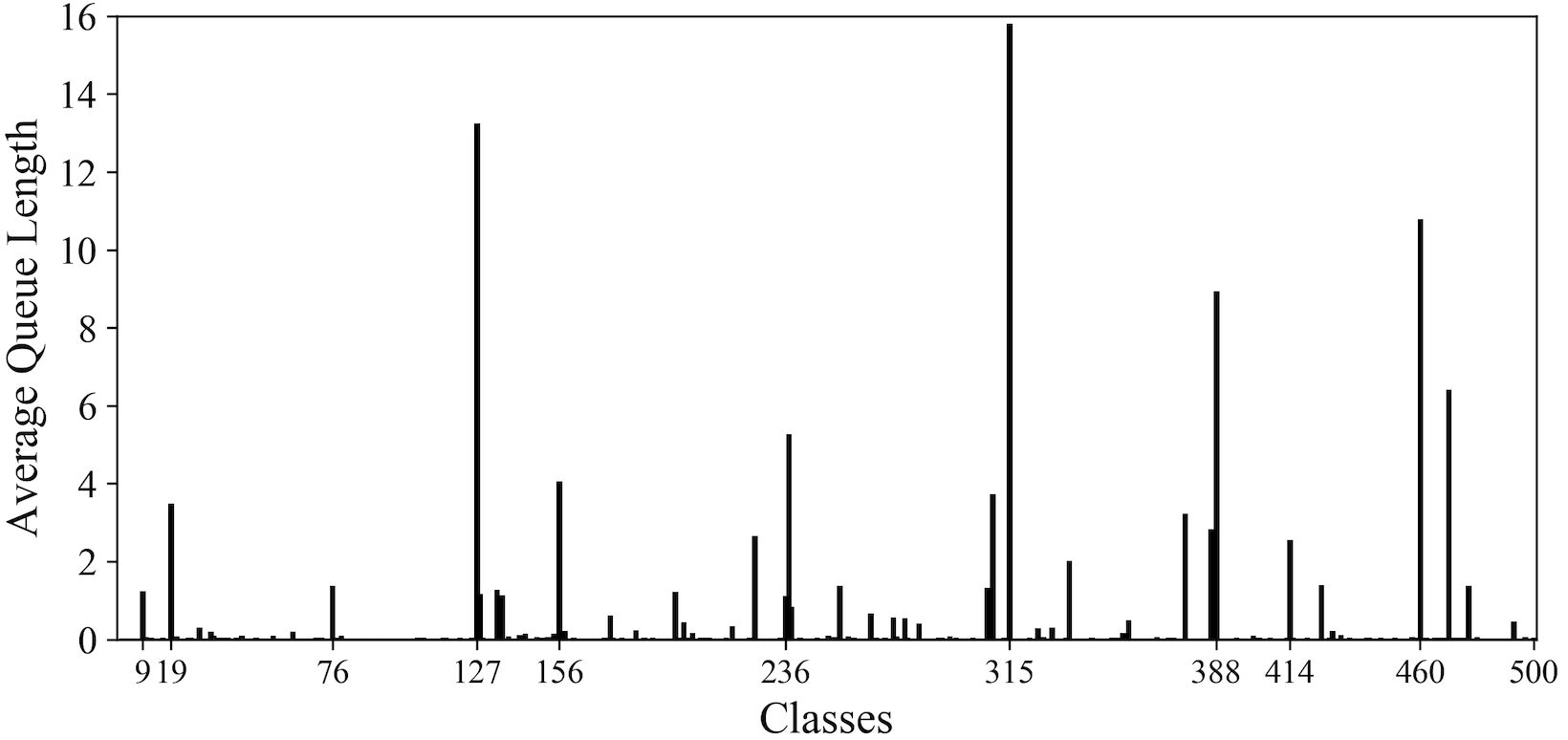}}\hfill
  \subfloat[Benchmark Policy ($K = 500$)]{\hspace{1.5mm}\includegraphics[width=0.514\linewidth]{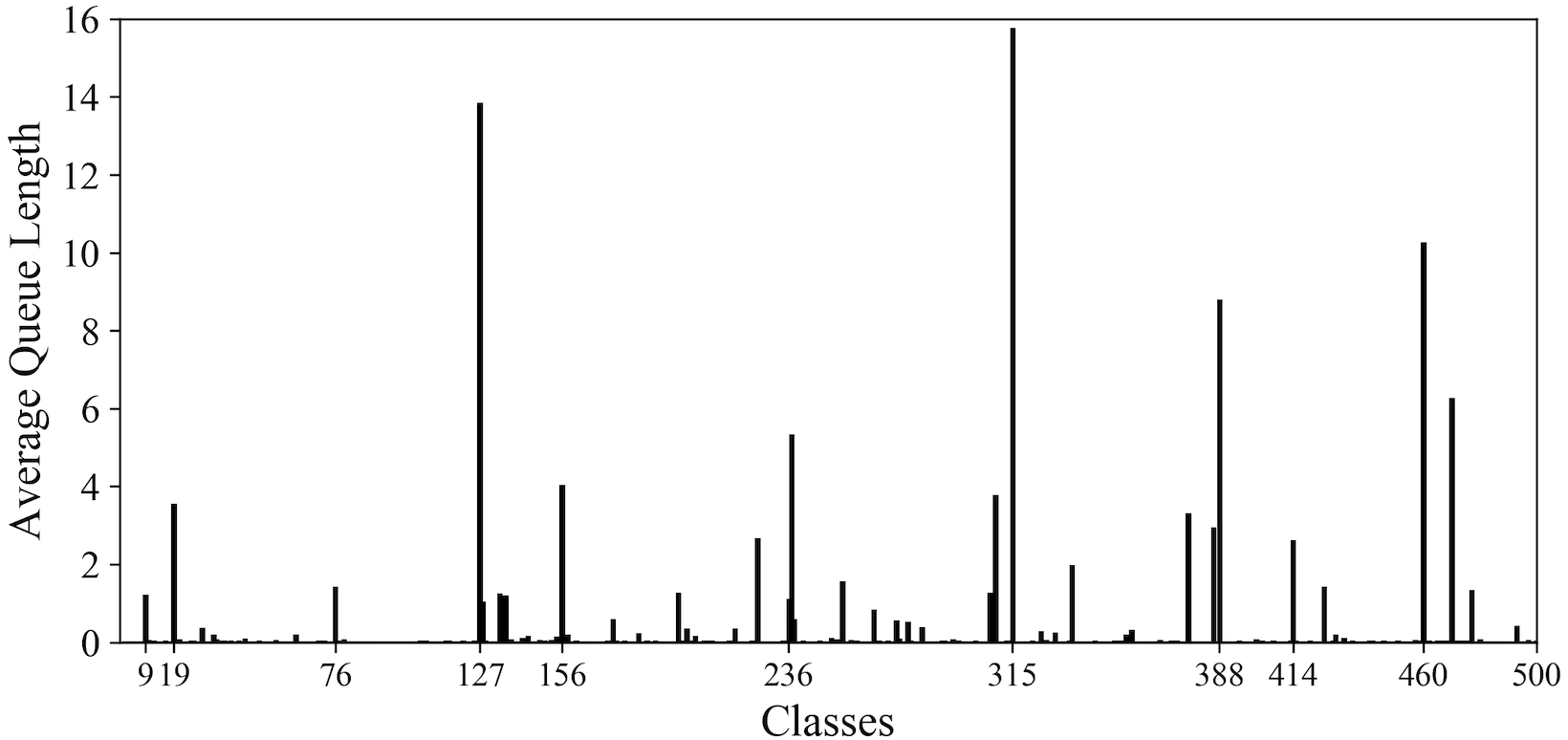}}
  \caption{Graphical representation of the average queue length under the policy learned from neural networks and the benchmark policy for the 30, 50, 100, and 500-dimensional test problems that admit a pathwise optimal solution.}
  \label{highdim_queue}
\end{figure}

\begin{figure}[H]
  \captionsetup{position=bottom} 
  \captionsetup[subfigure]{labelformat=empty}
  \subfloat[Our Policy ($K = 30$)]{\hspace{-4mm} \includegraphics[width=0.5\linewidth]{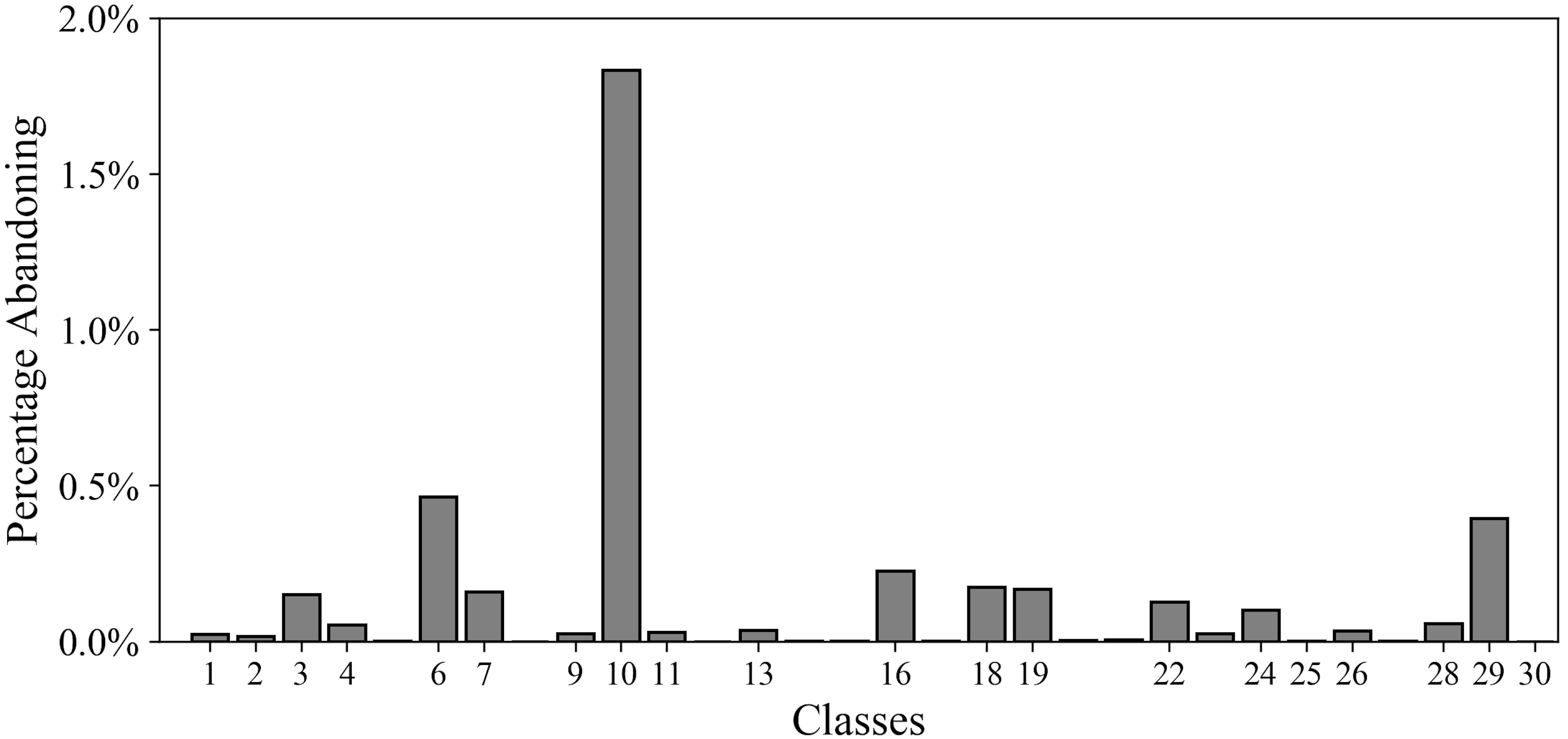}}\hfill
  \subfloat[Benchmark Policy ($K = 30$)]{\includegraphics[width=0.5\linewidth]{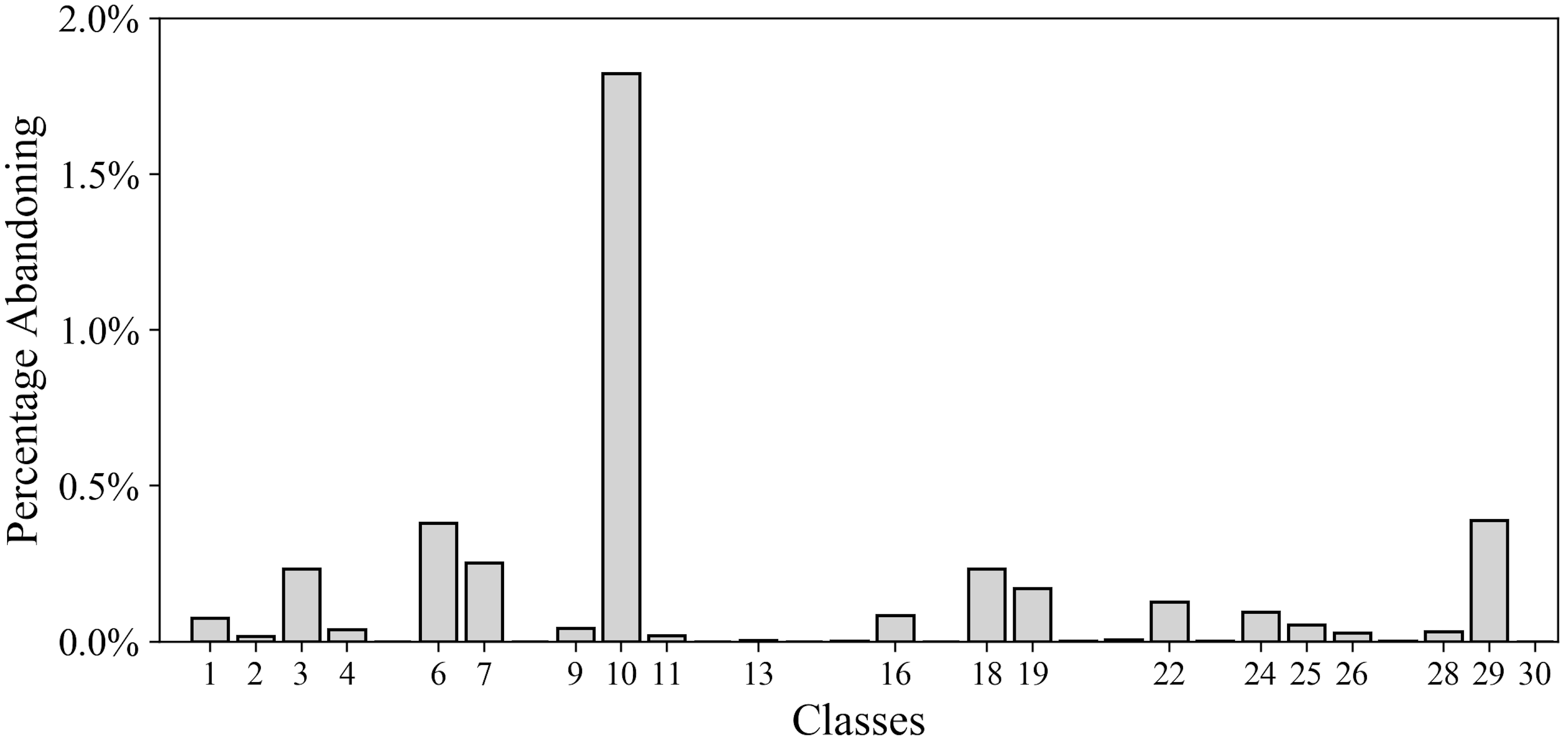}}\\
  \subfloat[Our Policy ($K = 50$)]{\hspace{-4mm}\includegraphics[width=0.5\linewidth]{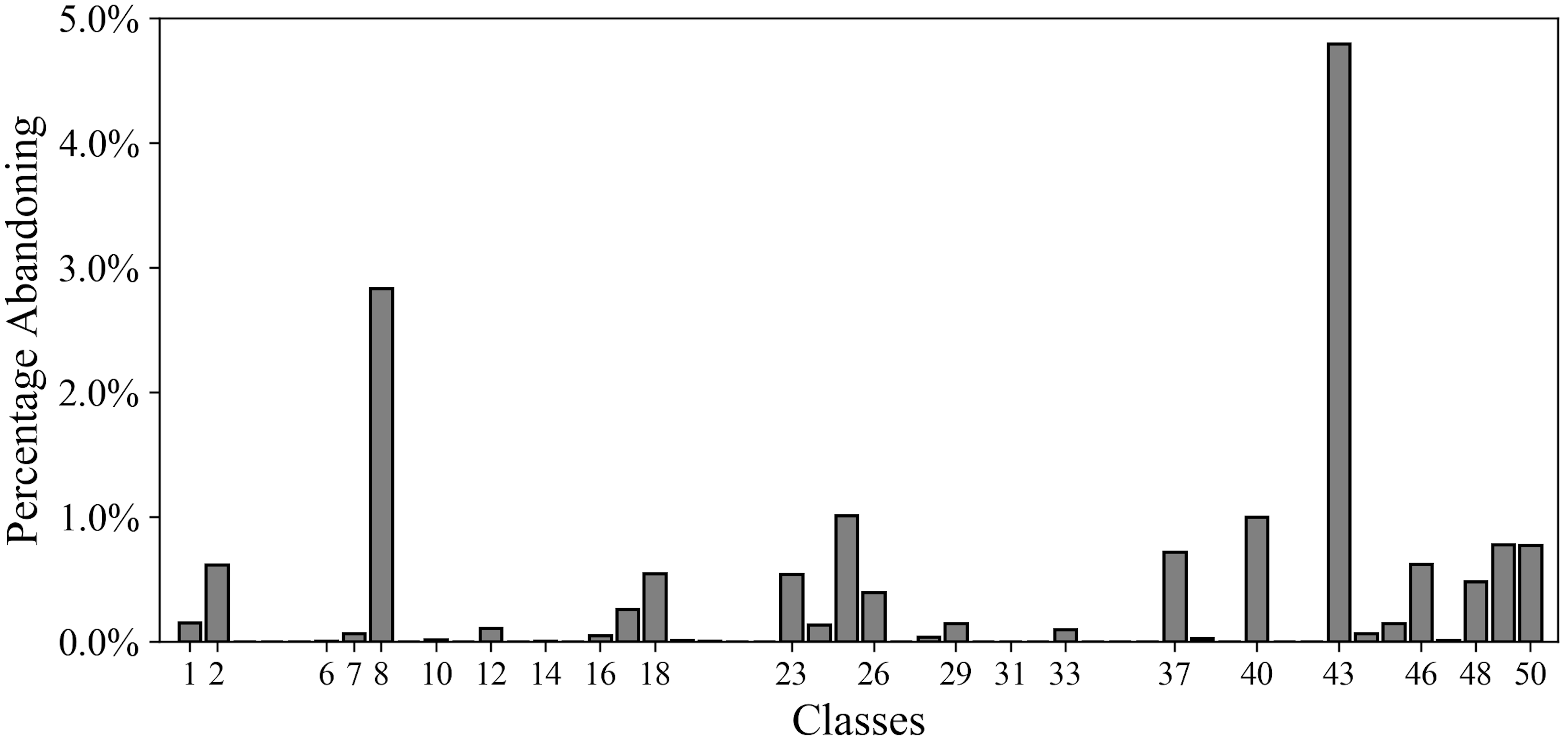}}\hfill
  \subfloat[Benchmark Policy ($K = 50$)]{\includegraphics[width=0.5\linewidth]{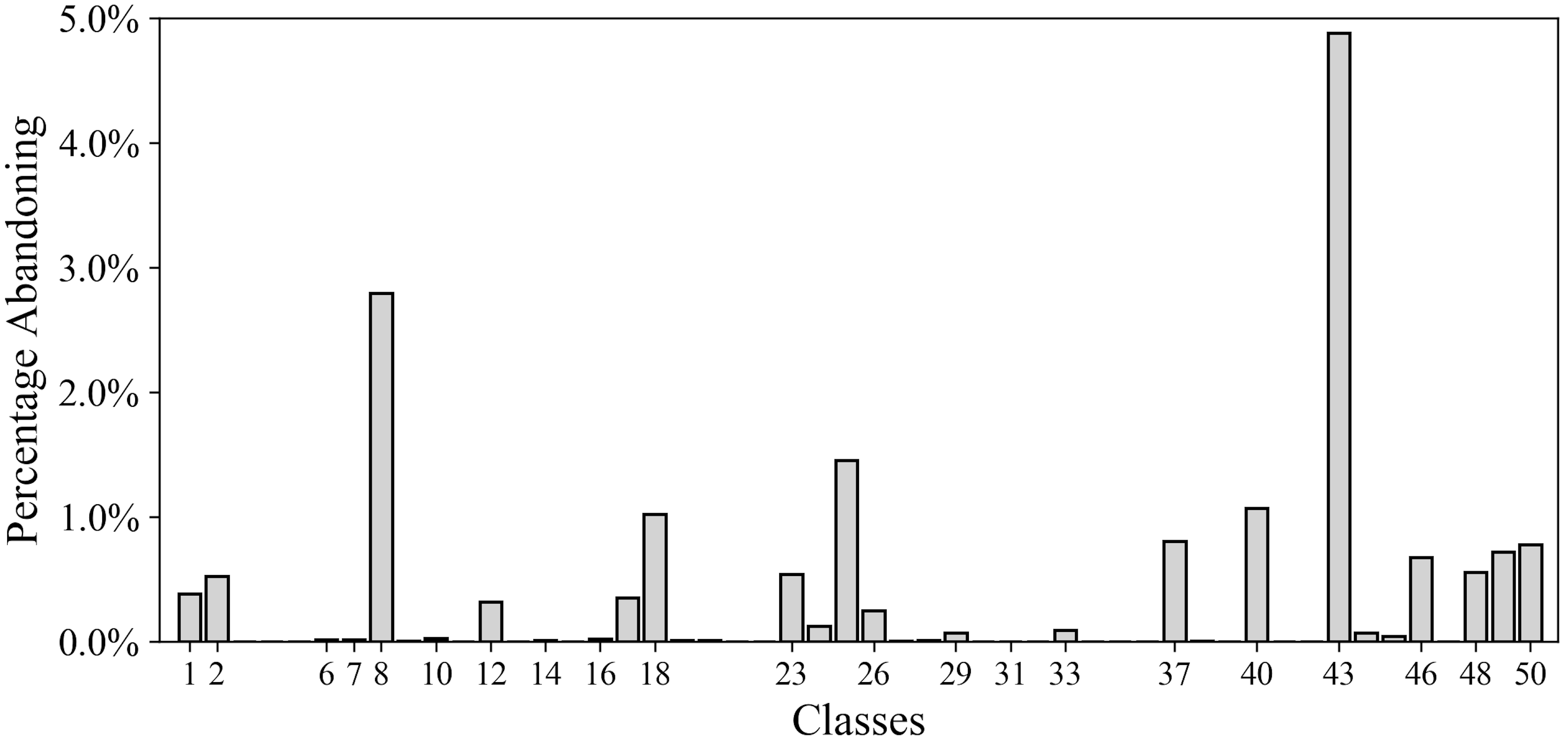}}\\
  \subfloat[Our Policy ($K = 100$)]{\hspace{-4mm}\includegraphics[width=0.5\linewidth]{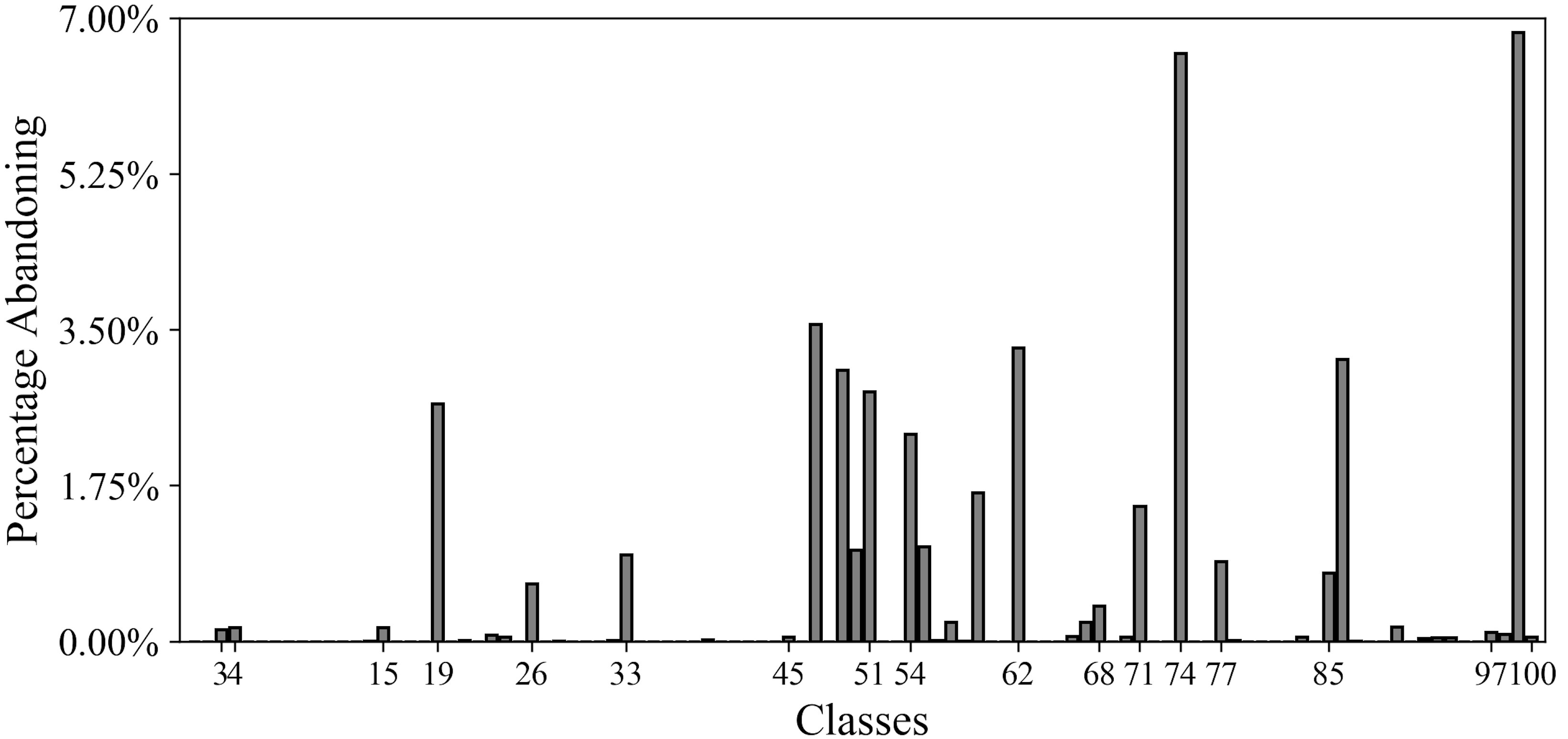}}\hfill
  \subfloat[Benchmark Policy ($K = 100$)]{\includegraphics[width=0.5\linewidth]{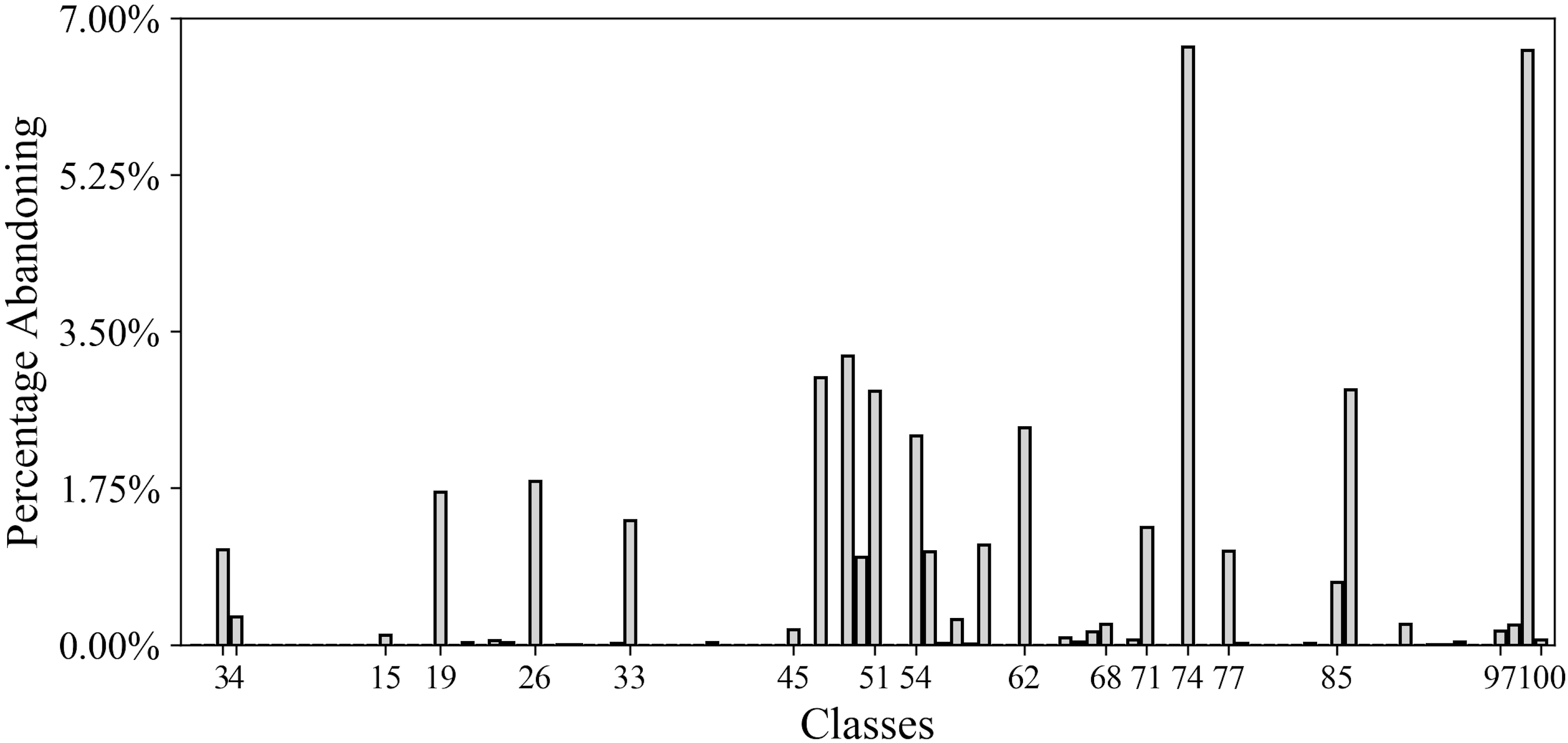}}
  \\
  \subfloat[Our Policy ($K = 500$)]{\hspace{-4mm}\includegraphics[width=0.5\linewidth]{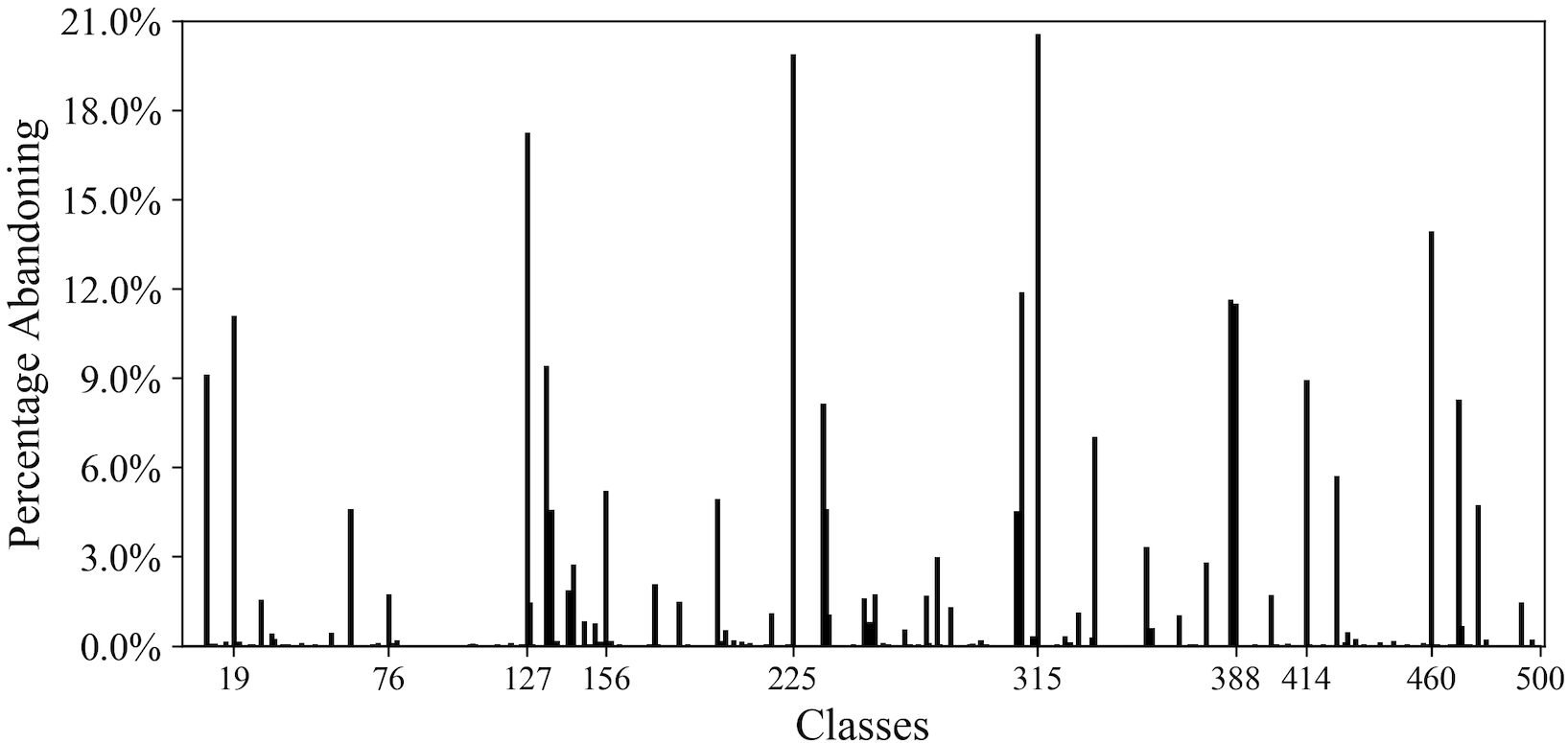}}\hfill
  \subfloat[Benchmark Policy ($K = 500$)]{\includegraphics[width=0.5\linewidth]{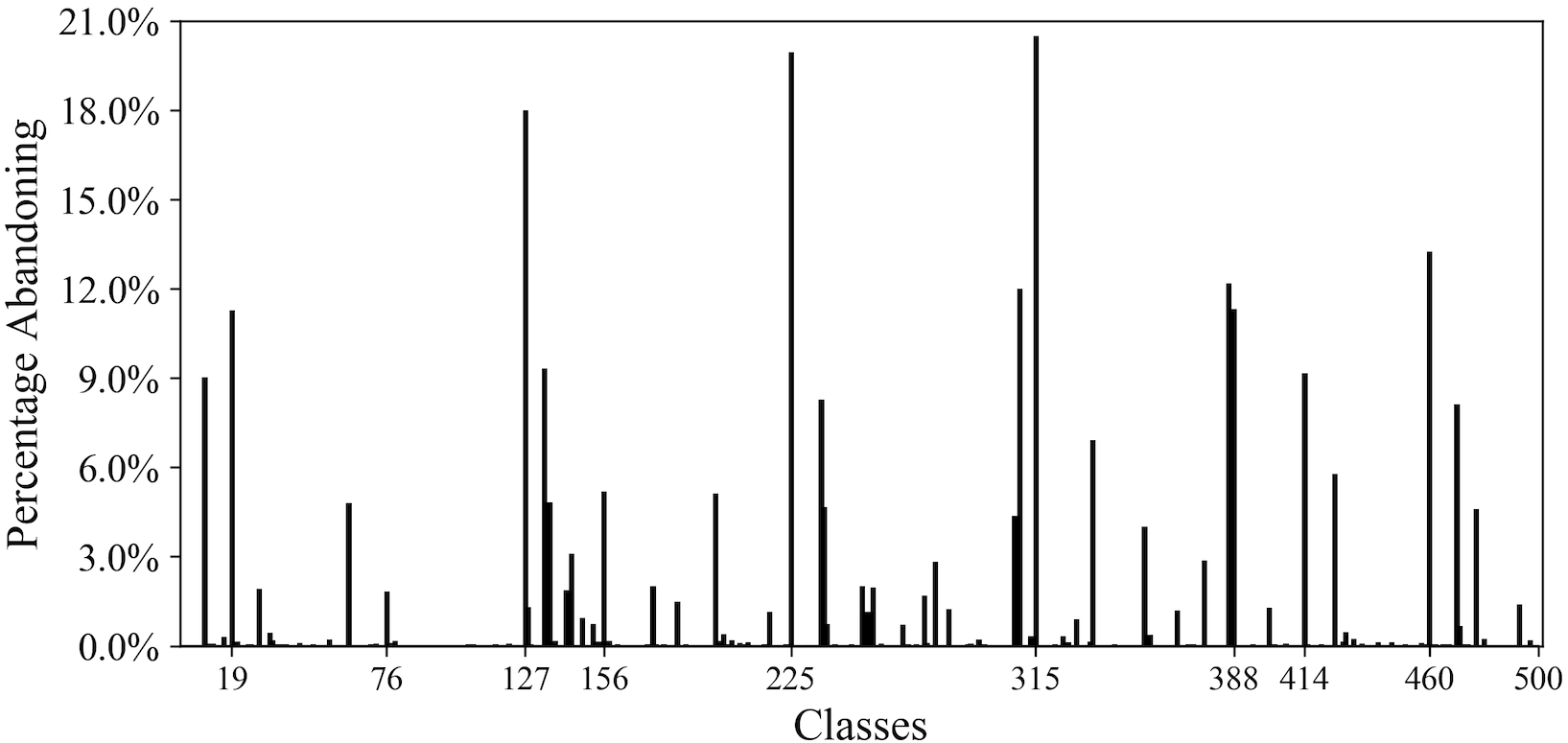}}
  \caption{Graphical representation of the average fraction of abandonments over arrivals under the policy learned from neural networks and the benchmark policy for the $30$, $50$, $100$ and 500-dimensional test problems that admit a pathwise optimal solution. Note that we use different scales on the vertical axes to improve visibility.}
\label{high_dim_abandons}
\end{figure}

The high dimensional test problems introduced in Section \ref{higher_dimensional} ($K = 30, 50, 100$ \text{ and } $500$) admit pathwise optimal solutions. To be specific, the optimal policy is a static priority policy in each case, and we use it as the benchmark policy. Table \ref{results_high_test1} reports the simulated performance (average total costs) of our proposed policy and the benchmark policy along with the percentage optimality gap between the two policies. The results show our proposed policy is near optimal. Specifically, the difference in performance compared to the optimal policy is not statistically significant at the 99\% confidence level.
\begin{table}[H]
    \centering
    \renewcommand{\arraystretch}{1.2}
    \setlength\tabcolsep{4pt} 
	{\small 
             \scalebox{0.9}{
    \begin{tabular}{lccccccccccccccccccccc}
        \toprule
        Method &&&& 30-Dimensional &&&& 50-Dimensional &&&& 100-Dimensional &&&& 500-Dimensional\\
        \midrule
        Our Policy &&&& 911.23 $\pm$ 6.41 &&&& 2428.70 $\pm$ 12.24 &&&& 11067.63 $\pm$ 21.70 &&&& 53340.77 $\pm$ 79.09 \\
        Benchmark &&&& 908.05 $\pm$ 6.40 &&&& 2422.18 $\pm$ 12.21 &&&& 11062.99 $\pm$ 21.90 &&&& 53309.68 $\pm$ 78.20\\
        \midrule
        Optimality Gap &&&&  0.35\% $\pm$ 1.00\% &&&& 0.27\% $\pm$ 0.71\% &&&& 0.04\% $\pm$ 0.28\% &&&& 0.06\% $\pm$ 0.21\%\\
        \bottomrule
    \end{tabular}
    }
    \caption{Performance comparison of our proposed policy with the benchmark policy in the high dimensional test problems that admit a pathwise optimal solution. The first two rows show the total cost $\pm$ half-length of the 99\% confidence interval for each case. Similarly, the last row shows the percentage optimality gap $\pm$ half-length of the 99\% confidence interval for each case.}
    \label{results_high_test1}
    }
\end{table}
Figures \ref{highdim_queue} - \ref{high_dim_abandons} display how the queue lengths and the percentage of calls that abandon, respectively, vary across different classes for the two policies. They show that our proposed policy is similar to the benchmark policy for each of the test problems considered here. 
\newpage
\subsection{Results for the high dimensional test problems that do not admit a pathwise optimal policy}
For the three 100-dimensional test problems introduced in Section \ref{high_dim_non_pathwise_sect}, the optimal policy remains unknown and, to the best of our knowledge, they do not admit a pathwise optimal solution. In addition, due to their high dimensionality and lack of special structure, we were unable to design simple yet effective benchmark policies, as we did for the main test problem and its variants (see Section \ref{static_benchmarks_main_variant}). In that case, the special structure of the test problems that we identified allowed us to develop hundreds of benchmark policies based on five basic static priority policies. In contrast, for these test problems, we rely only on the same five static priority policies as benchmarks; see Table \ref{results_100dim_non_pathwise}. Table \ref{results_100dim_non_pathwise} presents the average total costs derived from a simulation study, along with the percentage performance gap between the best benchmark policy and our proposed policy. Our proposed policy significantly outperforms the best benchmark policy for each of these test problems. 
\begin{table}[H]
    \centering
    \renewcommand{\arraystretch}{1.2}
    \setlength\tabcolsep{4pt} 
	{\small 
             \scalebox{0.85}{
    \begin{tabular}{lcccccccccccccc}
        \toprule
          Method &&&& 100-Dimensional Main &&&& 100-Dimensional First Variant &&&& 100-Dimensional Second Variant\\
        \midrule
            Our Policy &&&& 1382.14 $\pm$ 8.89 &&&& 1342.77 $\pm$ 8.37 &&&& 1337.13 $\pm$ 8.19\\
            $c\mu/\theta$ rule  &&&& 1490.02 $\pm$ 8.83 &&&& 1458.34 $\pm$ 8.24 &&&& 1454.49 $\pm$ 8.10\\
            $c\mu$ rule  &&&& 1409.68 $\pm$ 9.27 &&&& 1415.07 $\pm$ 8.96 &&&& 1466.97 $\pm$ 9.61 \\
            $c_{k}$ rule &&&&  1441.79 $\pm$ 9.56 &&&& \, 1647.21 $\pm$ 11.40 &&&& \, 1812.00 $\pm$ 12.92\\
            $\mu_{k} - \theta_{k}$ rule &&&&  \,\, 2009.27 $\pm$ 11.81 &&&& \, 1958.73 $\pm$ 11.05 &&&& \, 1953.13 $\pm$ 10.86  \\
            $c_{k}(\mu_{k} - \theta_{k})$ rule &&&& 1537.11 $\pm$ 9.13 &&&& \, 2308.48 $\pm$ 12.71 &&&& \, 2298.48 $\pm$ 12.54 \\
            \midrule
            Performance Gap &&&& -1.95\% $\pm$ 0.90\% &&&& -5.11\% $\pm$ 0.84\% &&&& -8.07\% $\pm$ 0.76\% \\
            \bottomrule
    \end{tabular}
    }
    \caption{Performance comparison of our proposed policy and the benchmark policies considered for three 100-dimensional problem that do not admit a pathwise optimal solution. The first two rows show the total cost $\pm$ half-length of the 99\% confidence interval for our proposed policy and the best benchmark policy. The last row shows the percentage performance gap $\pm$ half-length of the 99\% confidence interval.}
    \label{results_100dim_non_pathwise}
    }
\end{table}
\setlength{\bibsep}{0.0pt}
\setstretch{1.0}
\bibliographystyle{apalike}
\bibliography{bibfile}

\appendix
\renewcommand{\thesection}{Appendix \Alph{section}} 
\newpage
\section{Data used for the test problems}\label{appendix_a} 
\fancyhf{}  
\fancyhead[C]{\Large \textbf{NOT INTENDED FOR PRINT PUBLICATION}}
\fancyfoot[C]{\thepage}
\thispagestyle{fancy} 

\doublespacing
For the graphs shown in this section, the resolution of the horizontal axis is 5 minutes. For the main test problem, the graphs follow the order of the class names listed in the first column of Table \ref{stats}.

\subsection{Main test problem}\label{appdx_data_main_test}

\begin{figure}[H]
     \centering
     \begin{minipage}{0.48\textwidth}
     \centering
         \includegraphics[scale = 0.45]{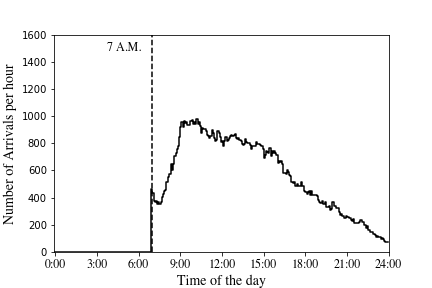} 
        \captionsetup{font={footnotesize}}
        \caption{Hourly Arrival Rates for Retail (Node: 1)}
\label{graph_arrival_class1_17dim_node1}
    \label{graph_one}
     \end{minipage}
     \hfill
     \begin{minipage}{0.48\textwidth}
         \centering
\includegraphics[scale = 0.45]{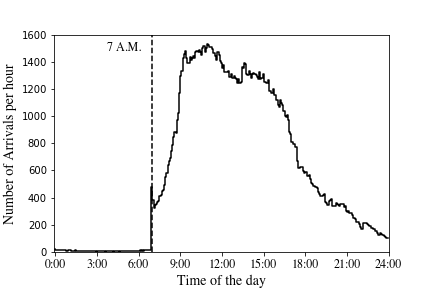} \captionsetup{font={footnotesize}}
        \caption{Hourly Arrival Rates for Retail (Node: 2)}
\label{graph_arrival_class2_17dim_node1}
     \end{minipage}
\end{figure}

\begin{figure}[H]
     \begin{minipage}{0.48\textwidth}
         \centering
         \includegraphics[scale = 0.45]{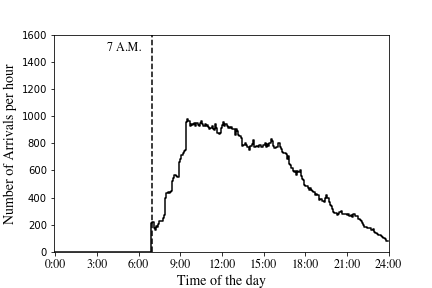} 
        \captionsetup{font={footnotesize}}
        \caption{Hourly Arrival Rates for Retail (Node: 3)}
        \label{graph_arrival_class3_17dim_node3}
         \end{minipage}
    \hfill
    \begin{minipage}{0.48\textwidth}
         \centering
        \includegraphics[scale = 0.45]{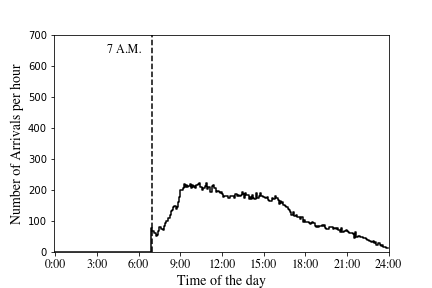}
         \captionsetup{font={footnotesize}}
         \caption{Hourly Arrival Rates for Premier}
     \end{minipage}
\end{figure}

\begin{figure}[H]
\centering
    \hspace{20mm}\includegraphics[scale = 0.45]{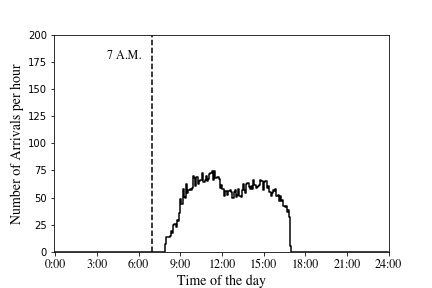}
    \captionsetup{font={footnotesize}}
    \hspace{2cm} \caption{Hourly Arrival Rates for BPS}
    \label{graph_arrival_15}
\end{figure}

\textbf{Graphs of the prelimit arrival rates $\lambda^{r}_{k}(\cdot)$.} Figures \ref{graph_one} - \ref{graph_arrival_15} display the average hourly arrival pattern for weekday calls of each class during May - July 2003. To improve visibility, we scale the vertical axes of the graphs according to the call volumes\footnote{Using the arrival percentages shown in Table \ref{stats}, we observe that Retail is the class with the largest call volume. Thus, we set the $y$-axis of Retail class arrivals from 0 to 1600 for each node. For the classes with a medium call volume, we set the $y$-axes from 0 to 700. Lastly, for the classes with the smallest call volume, i.e., the classes with less than 1\% aggregate arrival percentage, we set the $y$-axes of the graphs from 0 to 200.}.

\begin{figure}[H]
     \centering
     \begin{minipage}{0.48\textwidth}
         \centering
        \includegraphics[scale = 0.45]{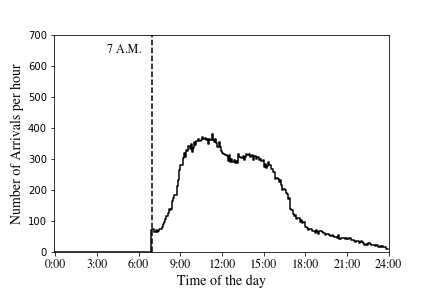}
         \captionsetup{font={footnotesize}}
         \caption{Hourly Arrival Rates for Business}
     \end{minipage}
     \hfill
     \begin{minipage}{0.48\textwidth}
         \centering
         \includegraphics[scale = 0.45]{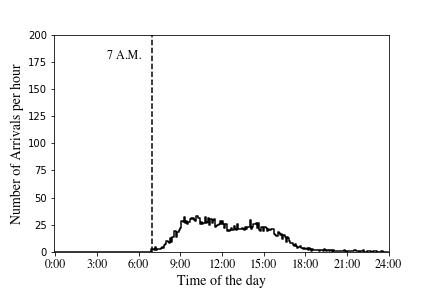}
         \captionsetup{font={footnotesize}}
         \caption{Hourly Arrival Rates for Platinum}
     \end{minipage}
\end{figure}
\vspace{-5mm}
\begin{figure}[H]
     \centering
     \begin{minipage}{0.48\textwidth}
         \centering
         \includegraphics[scale = 0.45]{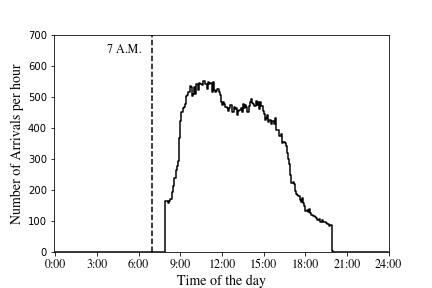}
         \captionsetup{font={footnotesize}}
         \caption{Hourly Arrival Rates for Consumer Loans}
     \end{minipage}
     \hfill
     \begin{minipage}{0.48\textwidth}
         \centering
         \includegraphics[scale = 0.45]{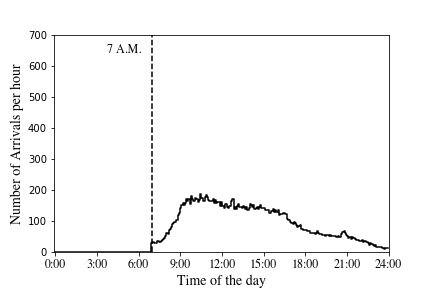}
         \captionsetup{font={footnotesize}}
         \caption{Hourly Arrival Rates for Online Banking}
     \end{minipage}
\end{figure}
\vspace{-5mm}
\begin{figure}[H]
     \centering
     \begin{minipage}{0.48\textwidth}
         \centering
         \includegraphics[scale = 0.45]{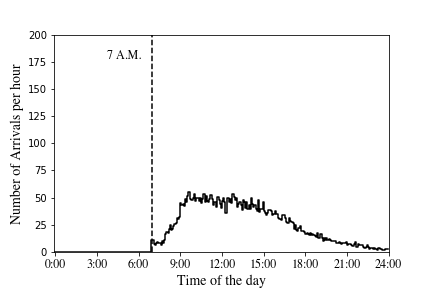}
         \captionsetup{font={footnotesize}}
         \caption{Hourly Arrival Rates for EBO}
     \end{minipage}
     \hfill
     \begin{minipage}{0.48\textwidth}
         \centering
         \includegraphics[scale = 0.45]{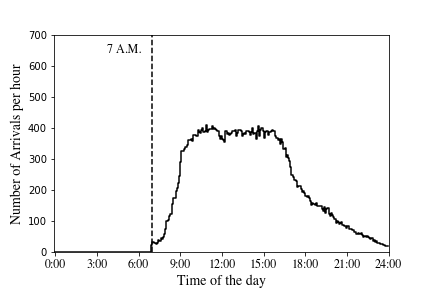}
         \captionsetup{font={footnotesize}}
         \caption{Hourly Arrival Rates for Telesales}
     \end{minipage}
\end{figure}
\vspace{-5mm}
\begin{figure}[H]
     \centering
     \begin{minipage}{0.48\textwidth}
         \centering
         \includegraphics[scale = 0.45]{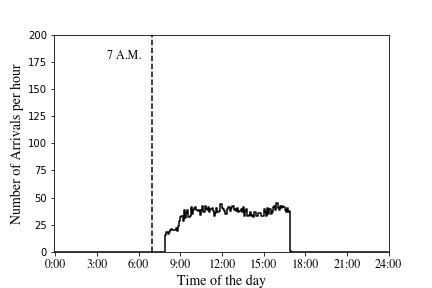}
         \captionsetup{font={footnotesize}}
         \caption{Hourly Arrival Rates for Subanco}
     \end{minipage}
     \hfill
     \begin{minipage}{0.48\textwidth}
         \centering
         \includegraphics[scale = 0.45]{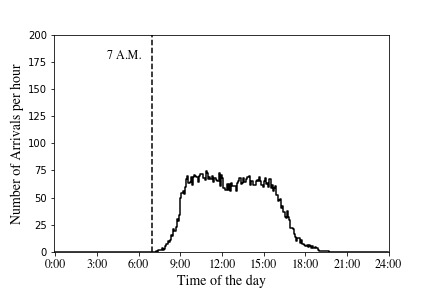}
         \captionsetup{font={footnotesize}}
         \caption{Hourly Arrival Rates for Case Quality}
     \end{minipage}
\end{figure}
\begin{figure}[H]
     \centering
     \begin{minipage}{0.48\textwidth}
         \centering
         \includegraphics[scale = 0.45]{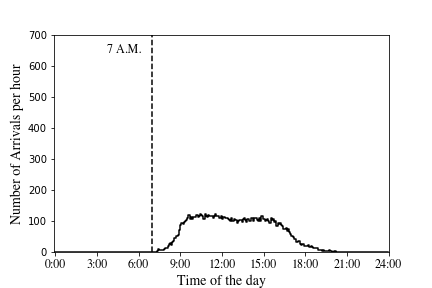}
         \captionsetup{font={footnotesize}}
         \caption{Hourly Arrival Rates for Priority Service}
     \end{minipage}
     \hfill
     \begin{minipage}{0.48\textwidth}
         \centering
         \includegraphics[scale = 0.45]{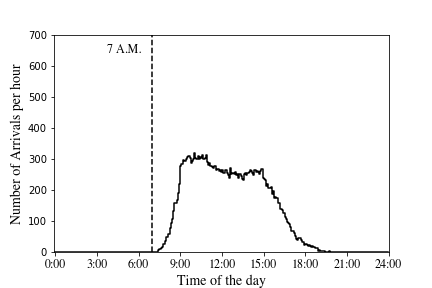}
         \captionsetup{font={footnotesize}}
         \caption{Hourly Arrival Rates for AST}
     \end{minipage}
\end{figure}

\begin{figure}[H]
     \centering
     \begin{minipage}{0.48\textwidth}
         \centering
         \includegraphics[scale = 0.45]{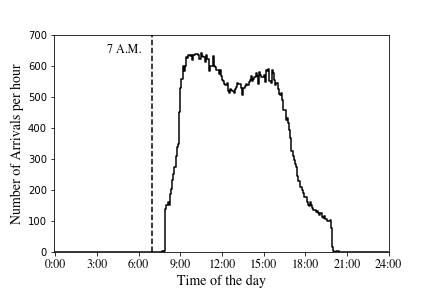}
         \captionsetup{font={footnotesize}}
         \caption{Hourly Arrival Rates for CCO}
     \end{minipage}
     \hfill
     \begin{minipage}{0.48\textwidth}
         \centering
         \includegraphics[scale = 0.45]{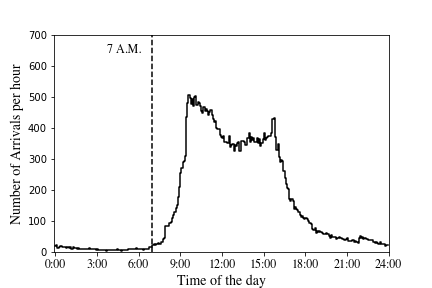}
         \captionsetup{font={footnotesize}}
         \caption{Hourly Arrival Rates for Brokerage}
     \end{minipage}
\end{figure}

\begin{figure}[H]
     \centering
     \begin{minipage}{0.48\textwidth}
     \centering
         \includegraphics[scale = 0.45]{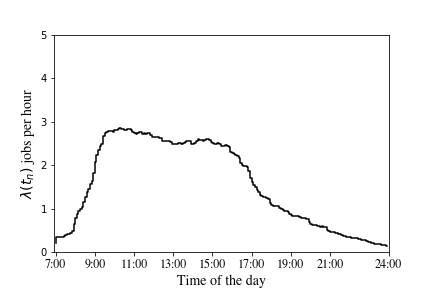}
        \captionsetup{font={footnotesize}}
        \caption{Hourly $\lambda(t)$ terms for Retail (Node: 1)}
\label{graph_arrival_class1_17dim_limit}
     \end{minipage}
     \hfill
     \begin{minipage}{0.48\textwidth}
         \centering
         \includegraphics[scale = 0.45]{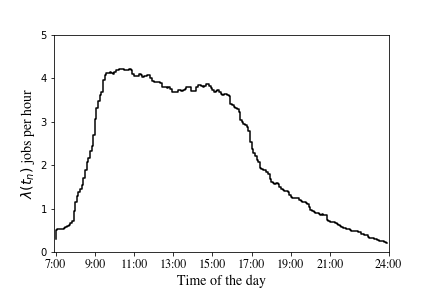}
        \captionsetup{font={footnotesize}}
        \caption{Hourly $\lambda(\cdot)$ terms for Retail (Node: 2)}
    \label{graph_arrival_class2_17dim_limit}
     \end{minipage}
\end{figure}

\begin{figure}[H]
     \begin{minipage}{0.48\textwidth}
         \centering
         \includegraphics[scale = 0.45]{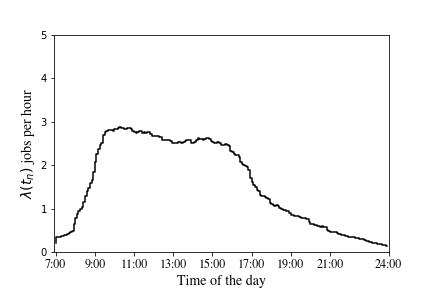}
        \captionsetup{font={footnotesize}}
        \caption{Hourly $\lambda(\cdot)$ terms for Retail (Node: 3)}
\label{graph_arrival_class3_17dim_limit}
     \end{minipage}
        \hfill
     \begin{minipage}{0.48\textwidth}
         \centering
         \includegraphics[scale = 0.45]{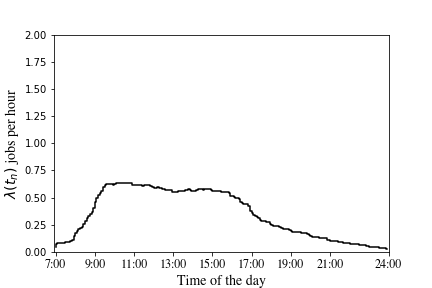}
         \captionsetup{font={footnotesize}}
         \caption{Hourly $\lambda(\cdot)$ terms for Premier}
     \end{minipage}
\end{figure}

\begin{figure}[H]
         \centering
         \includegraphics[scale = 0.45]{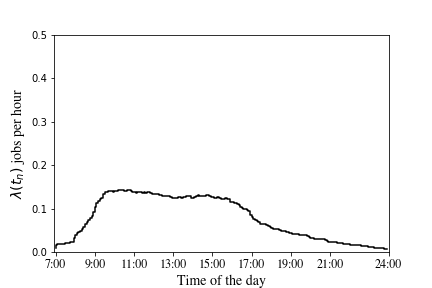}
         \captionsetup{font={footnotesize}}
         \caption{Hourly $\lambda(\cdot)$ terms for BPS}
\label{graph_arrival_class17_17dim_limit}
\end{figure}

\vspace{2mm}
\noindent \textbf{Graphs of the limiting arrival rates $\lambda_k(\cdot)$.} Figures \ref{graph_arrival_class1_17dim_limit} - \ref{graph_arrival_class17_17dim_limit} display the hourly limiting arrival rates for each class $k$, denoted by $\lambda_k(\cdot)$, which are calculated using Equation (\ref{eqn_limit_arrival}). To improve visibility, the vertical axes have been scaled relative to the magnitude of the limiting arrival rates\footnote{Specifically, for classes with the highest limiting arrival rates, the y-axes range from 0 to 5. For those with the lowest limiting arrival rates, we set the y-axes from 0 to 0.5. Lastly, for the remaining classes, the y-axes of the graphs are set between 0 and 2.}.

\vspace{-5mm}
\vspace{-5mm}
\begin{figure}[H]
     \begin{minipage}{0.48\textwidth}
         \centering
         \includegraphics[scale = 0.45]{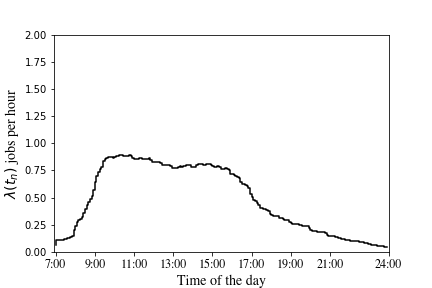}
         \captionsetup{font={footnotesize}}
         \caption{Hourly $\lambda(\cdot)$ terms for Business}
     \end{minipage}
        \hfill
     \begin{minipage}{0.48\textwidth}
         \centering
         \includegraphics[scale = 0.45]
{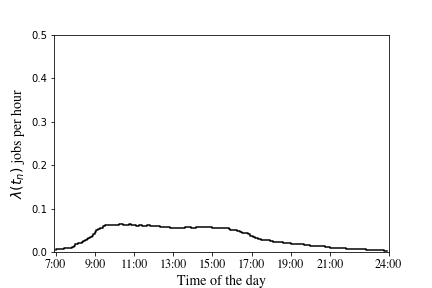}
         \captionsetup{font={footnotesize}}
         \caption{Hourly $\lambda(\cdot)$ terms for Platinum}
     \end{minipage}
\end{figure}
\vspace{-5mm}
\begin{figure}[H]
     \begin{minipage}{0.48\textwidth}
         \centering
         \includegraphics[scale = 0.45]{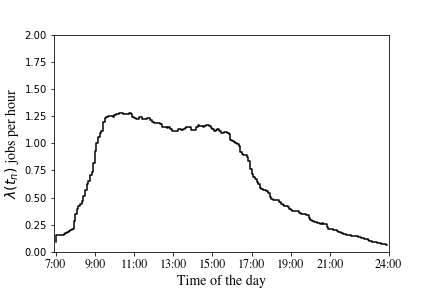}
         \captionsetup{font={footnotesize}}
         \caption{Hourly $\lambda(\cdot)$ terms for Consumer Loans}
     \end{minipage}
     \hfill
     \begin{minipage}{0.48\textwidth}
         \centering
         \includegraphics[scale = 0.45]{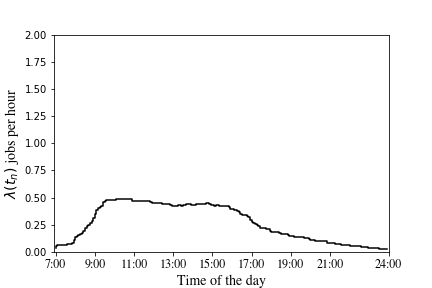}
         \captionsetup{font={footnotesize}}
         \caption{Hourly $\lambda(\cdot)$ terms for Online Banking}
     \end{minipage}
\end{figure}

\begin{figure}[H]
     \begin{minipage}{0.48\textwidth}
         \centering
         \includegraphics[scale = 0.45]{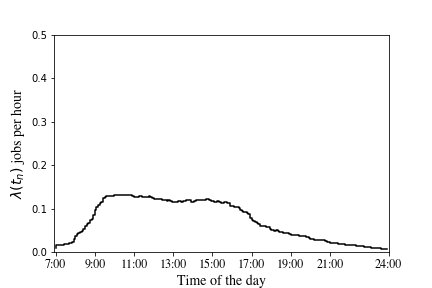}
         \captionsetup{font={footnotesize}}
         \caption{Hourly $\lambda(\cdot)$ terms for EBO}
     \end{minipage}
    \hfill
     \begin{minipage}{0.48\textwidth}
         \centering
         \includegraphics[scale = 0.45]{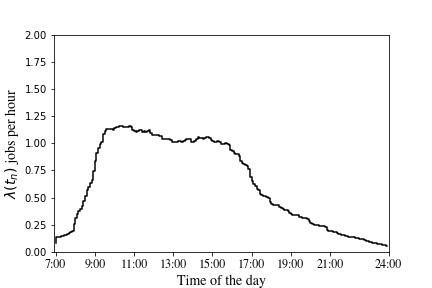}
         \captionsetup{font={footnotesize}}
         \caption{Hourly $\lambda(\cdot)$ terms for Telesales}
     \end{minipage}
 \end{figure}
\vspace{-5mm}
 \begin{figure}[H]
     \begin{minipage}{0.48\textwidth}
         \centering
         \includegraphics[scale = 0.45] {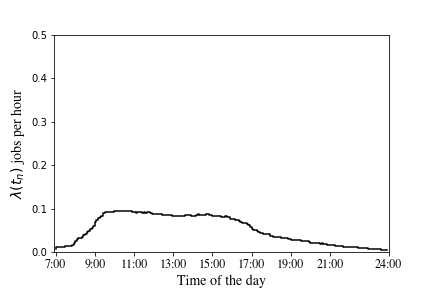}
         \captionsetup{font={footnotesize}}
         \caption{Hourly $\lambda(\cdot)$ terms for Subanco}
     \end{minipage}
    \hfill
     \begin{minipage}{0.48\textwidth}
         \centering
         \includegraphics[scale = 0.45]{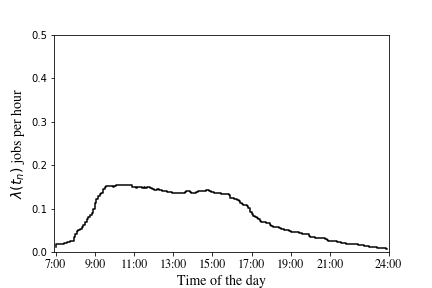}
         \captionsetup{font={footnotesize}}
         \caption{Hourly $\lambda(\cdot)$ terms for Case Quality}
     \end{minipage}
  \end{figure}
\vspace{-5mm}
  \begin{figure}[H]
     \begin{minipage}{0.48\textwidth}
         \centering
         \includegraphics[scale = 0.45]{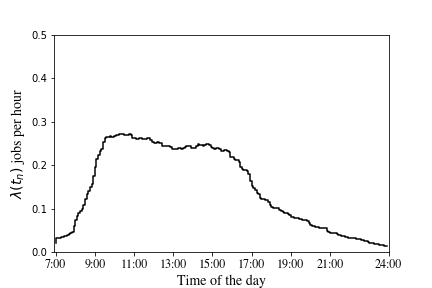}
         \captionsetup{font={footnotesize}}
         \caption{Hourly $\lambda(\cdot)$ terms for Priority Service}
     \end{minipage}
    \hfill
     \begin{minipage}{0.48\textwidth}
         \centering
         \includegraphics[scale = 0.45]{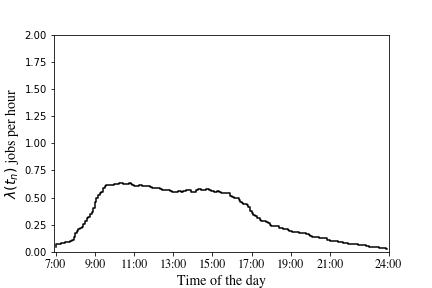}
         \captionsetup{font={footnotesize}}
         \caption{Hourly $\lambda(\cdot)$ terms for AST}
     \end{minipage}
   \end{figure}
  \vspace{-5mm}
   \begin{figure}[H]
     \begin{minipage}{0.48\textwidth}
         \centering
         \includegraphics[scale = 0.45]{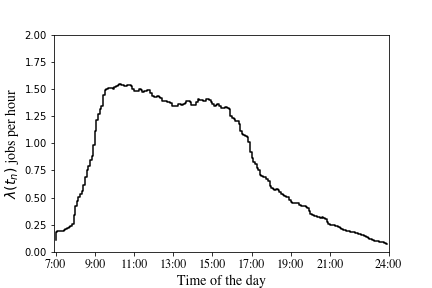}
         \captionsetup{font={footnotesize}}
         \caption{Hourly $\lambda(\cdot)$ terms for CCO}
     \end{minipage}
     \hfill
     \begin{minipage}{0.48\textwidth}
         \centering
         \includegraphics[scale = 0.45]{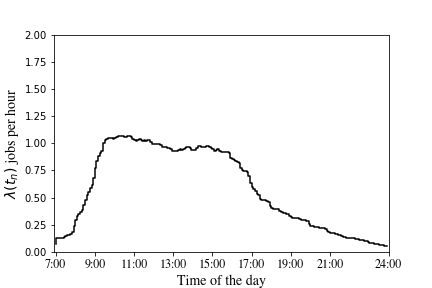}
         \captionsetup{font={footnotesize}}
         \caption{Hourly $\lambda(\cdot)$ terms for Brokerage}
     \end{minipage}
   \end{figure}

\begin{figure}[H]
     \centering
     \begin{minipage}{0.48\textwidth}
     \centering
         \includegraphics[scale = 0.45]{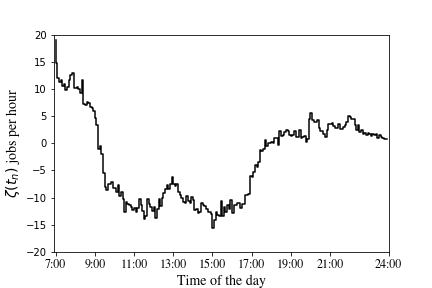} 
        \captionsetup{font={footnotesize}}
        \caption{Hourly $\zeta(\cdot)$ terms for Retail (Node: 1)}
        \label{main_test_zeta_class1}
     \end{minipage}
     \hfill
     \begin{minipage}{0.48\textwidth}
         \centering
         \includegraphics[scale = 0.45]{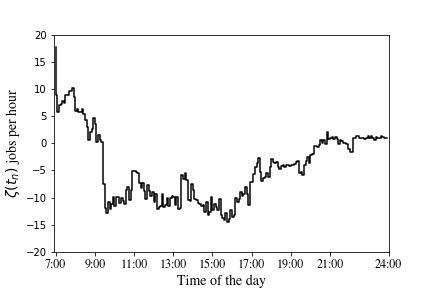} 
        \captionsetup{font={footnotesize}}
        \caption{Hourly $\zeta(\cdot)$ terms for Retail (Node: 2)}
        \label{graph_zeta_class2_17dim_limit}
     \end{minipage}
\end{figure}

\begin{figure}[H]
     \begin{minipage}{0.48\textwidth}
         \centering
         \includegraphics[scale = 0.45]{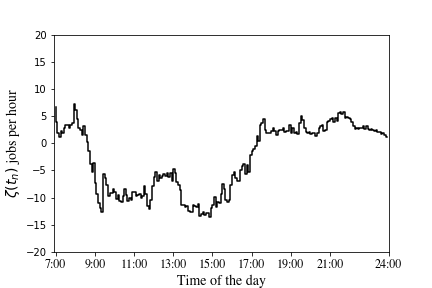}  
        \captionsetup{font={footnotesize}}
        \caption{Hourly $\zeta(\cdot)$ terms for Retail (Node: 3)}
        \label{graph_zeta_class3_17dim_limit}
     \end{minipage}
    \hfill
     \begin{minipage}{0.48\textwidth}
         \centering
         \includegraphics[scale = 0.45]{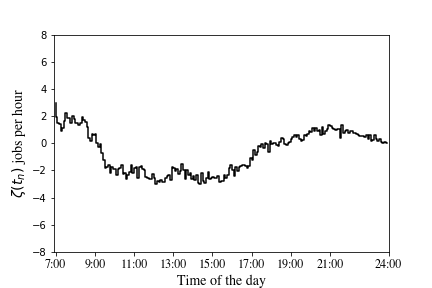}
         \captionsetup{font={footnotesize}}
         \caption{Hourly $\zeta(\cdot)$ terms for Premier}
     \end{minipage}
\end{figure}

\begin{figure}[H]
         \centering
         \includegraphics[scale = 0.45]{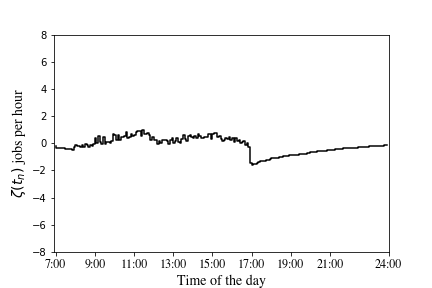}
         \captionsetup{font={footnotesize}}
         \caption{Hourly $\zeta(\cdot)$ terms for BPS}
         \label{main_test_zeta_class15}
\end{figure}

\textbf{\hspace{-6mm}\textbf{Graphs of the second-order terms $\zeta_k(\cdot)$.}} Figures \ref{main_test_zeta_class1} - \ref{main_test_zeta_class15} show the hourly second-order terms $\zeta_{k}(\cdot)$, which are calculated using Equation (\ref{eqn_limit_zeta}). To improve visibility, the vertical axes have been adjusted to reflect the magnitude of the variance of the $\zeta_k(\cdot)$ terms\footnote{To be specific, for the classes with the largest variance in the second-order terms $\zeta_{k}(\cdot)$, we set the y-axes from -20 to 20. For those with the smallest variance in the second-order terms $\zeta_{k}(\cdot)$, the y-axes of the graphs are set from -8 to 8. For $\zeta_{k}(\cdot)$ graphs of the remaining classes, the y-axes are set between -1 to 1.}. 

\begin{figure}[H]
     \begin{minipage}{0.48\textwidth}
         \centering
         \includegraphics[scale = 0.45]{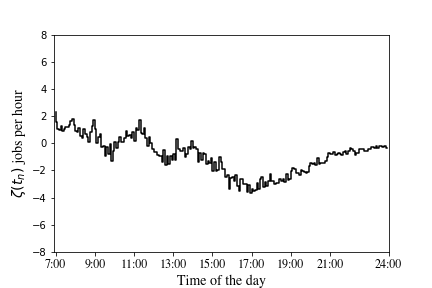}
         \captionsetup{font={footnotesize}}
         \caption{Hourly $\zeta(\cdot)$ terms for Business}
     \end{minipage}
     \hfill
     \begin{minipage}{0.48\textwidth}
         \centering
         \includegraphics[scale = 0.45]
         {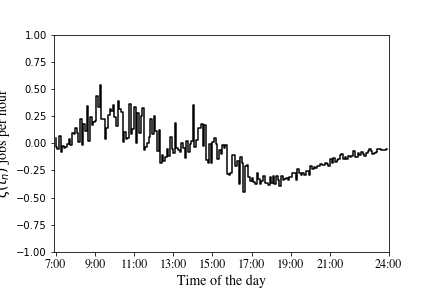}
         \captionsetup{font={footnotesize}}
         \caption{Hourly $\zeta(\cdot)$ terms for Platinum}
     \end{minipage}
\end{figure}
\vspace{-5mm}
\begin{figure}[H]
     \begin{minipage}{0.48\textwidth}
         \centering
         \includegraphics[scale = 0.45]{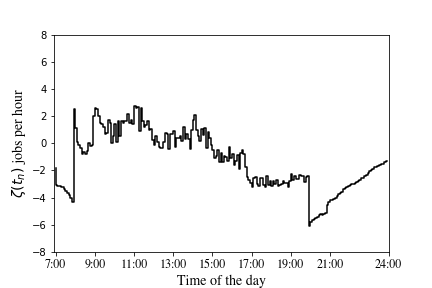}
         \captionsetup{font={footnotesize}}
         \caption{Hourly $\zeta(\cdot)$ terms for Consumer Loans}
     \end{minipage}
     \hfill
     \begin{minipage}{0.48\textwidth}
         \centering
         \includegraphics[scale = 0.45]{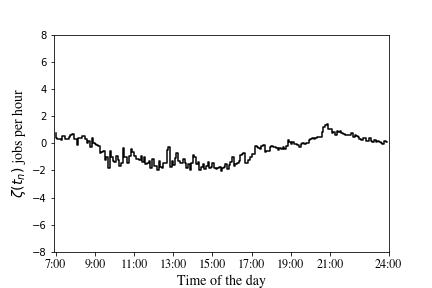}
         \captionsetup{font={footnotesize}}
         \caption{Hourly $\zeta(\cdot)$ terms for Online Banking}
     \end{minipage}
\end{figure}
\vspace{-5mm}
\begin{figure}[H]
     \begin{minipage}{0.48\textwidth}
         \centering
         \includegraphics[scale = 0.45]{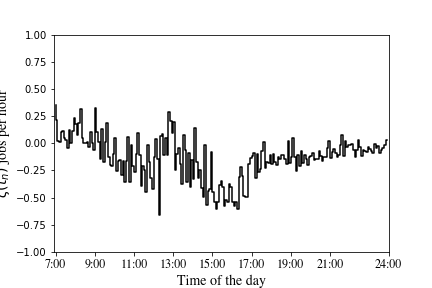}
         \captionsetup{font={footnotesize}}
         \caption{Hourly $\zeta(\cdot)$ terms for EBO}
     \end{minipage}
     \hfill
     \begin{minipage}{0.48\textwidth}
         \centering
         \includegraphics[scale = 0.45]{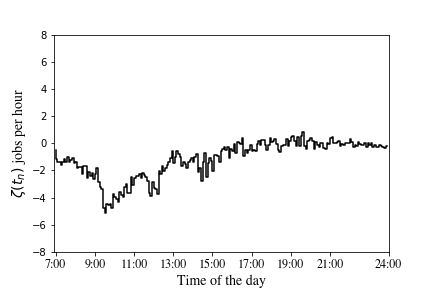}
         \captionsetup{font={footnotesize}}
         \caption{Hourly $\zeta(\cdot)$ terms for Telesales}
     \end{minipage}
\end{figure}
\vspace{-5mm}
\begin{figure}[H]
     \begin{minipage}{0.48\textwidth}
         \centering
         \includegraphics[scale = 0.45]{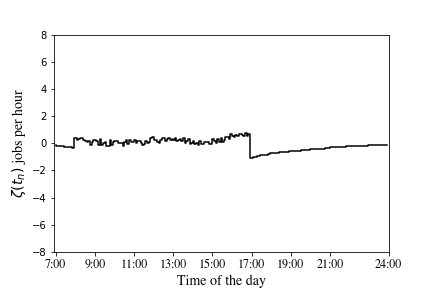}
         \captionsetup{font={footnotesize}}
         \caption{Hourly $\zeta(\cdot)$ terms for Subanco}
     \end{minipage}
    \hfill
     \begin{minipage}{0.48\textwidth}
         \centering
         \includegraphics[scale = 0.45]{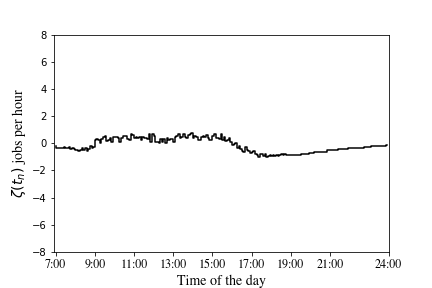}
         \captionsetup{font={footnotesize}}
         \caption{Hourly $\zeta(\cdot)$ terms for Case Quality}
     \end{minipage}
\end{figure}

\begin{figure}[H]
     \begin{minipage}{0.48\textwidth}
         \centering
         \includegraphics[scale = 0.45]{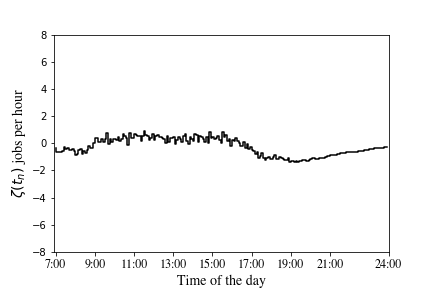}
         \captionsetup{font={footnotesize}}
         \caption{Hourly $\zeta(\cdot)$ terms for Priority Service}
     \end{minipage}
     \hfill
     \begin{minipage}{0.48\textwidth}
         \centering
         \includegraphics[scale = 0.45]{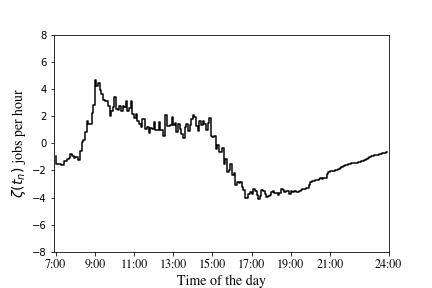}
         \captionsetup{font={footnotesize}}
         \caption{Hourly $\zeta(\cdot)$ terms for AST}
     \end{minipage}
\end{figure}
\vspace{-7mm}
\begin{figure}[H]
     \begin{minipage}{0.48\textwidth}
         \centering
         \includegraphics[scale = 0.45]{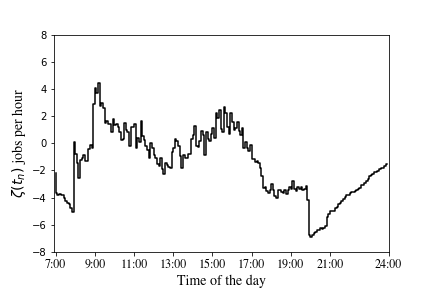}
         \captionsetup{font={footnotesize}}
         \caption{Hourly $\zeta(\cdot)$ terms for CCO}
     \end{minipage}
     \hfill
     \begin{minipage}{0.48\textwidth}
         \centering
         \includegraphics[scale = 0.45]{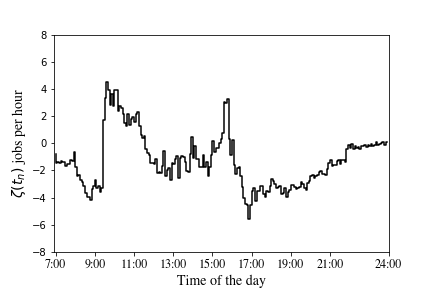}
         \captionsetup{font={footnotesize}}
         \caption{Hourly $\zeta(\cdot)$ terms for Brokerage}
     \end{minipage}
\end{figure}
\vspace{-8mm}
\begin{figure}[H]
    \centering
    \includegraphics[scale=0.45]{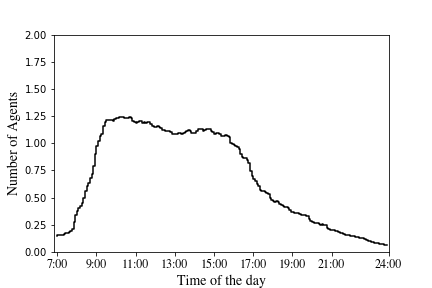}
    \caption{The limiting staffing levels $N(\cdot)$ throughout the day}
    \label{limiting_agents}
\end{figure}

\vspace{-6mm}
\noindent \textbf{The limiting staffing levels $N(\cdot)$.} Substituting the system parameter $r = 400$, and the number of agents $N^{r}(t_n)$, shown in Figure \ref{Agents_CI}, into Equation (\ref{eqn_limit_staff}), we find the limiting staffing levels as shown in Figure \ref{limiting_agents}. 
\vspace{-12mm}
\subsubsection{Possible agent states observed in the US Bank call center data} \label{data_specifics}

\begin{table}[H]
	\centering
            \scalebox{0.75}{
		\begin{tabular}{lrrrrr}\toprule
		Code &&&& State\\
		\midrule
		1 &&&& Incoming Call\\
		2 &&&& Outgoing Call\\
		20-21 &&&& Sign-on\\
		30-31 &&&& Sign-off\\
		40-49 &&&& Idle \\
		50 &&&& Available\\
		60-62 &&&& Break\\
			\bottomrule
		\end{tabular}
	}
        \caption{The agent states.}
	\label{states}
\end{table}
\vspace{-5mm}
\subsubsection{Data used for the main test problem and its variants}\label{data_variants}
\begin{table}[H]
	\centering
	\vspace{0mm}
	\setlength\tabcolsep{4pt} 
	{\small 
             \scalebox{0.8}{
		\begin{tabular}{lccccccccc}
			\toprule
			Class & Arrival & $\mu$ & $\theta$ & $p$ & $h$ & $c$\\
			\rule{0pt}{3.5ex} & percentage (\%)  & (per hr) & (per hr) & (per job) & (per hr) & (per hr)\\
			\midrule
			Retail (Node: 1) & 15.39 & 17.22 & 6.06 & \$2.000 & \$24.00 & \$36.12\\
                Retail (Node: 2) & 22.82 & 17.25 & 7.81 & \$2.000 & \$24.00 & \$39.62\\
                Retail (Node: 3) & 15.50 & 17.25 &  5.22 &\$2.000 & \$24.00 & \$34.44\\
			Premier & 3.46 & 13.15 & 9.79& \$2.167  & \$26.00 & \$47.22\\
			Business  & 4.82 & 16.56 & 8.58 & \$2.500  & \$30.00 & \$51.46\\
			Platinum  & 0.34 & 17.20 & 7.50 & \$2.667  & \$32.00 & \$51.99\\
			Consumer Loans  & 6.92 & 15.19 & 4.87 & \$1.833 & \$22.00 & \$30.93\\
			Online Banking  & 2.64 & 10.60 & 5.58 & \$1.833  & \$22.00 & \$32.24\\
			EBO  & 0.72 & 9.87 & 8.24 & \$1.667  & \$20.00 & \$33.73\\
			Telesales  & 6.26 & 9.62 & 8.99 & \$1.833  & \$22.00 & \$38.49\\
			Subanco  & 0.51 & 11.79 & 6.39 & \$1.667  & \$20.00 & \$30.66\\
			Case Quality  & 0.84 & 9.93 & 9.27 & \$1.667  & \$20.00 & \$35.44\\
			Priority Service & 1.47 & 10.35 & 9.14 & \$2.667  & \$32.00 & \$56.37\\
			AST  & 3.42 & 12.52 & 7.50 & \$1.833  & \$22.00 & \$35.75\\
			CCO  & 8.34 & 15.20 & 7.10 & \$1.833  & \$22.00 & \$35.02\\
			Brokerage  & 5.78 & 12.62 & 6.89 & \$1.833  & \$22.00 & \$34.63\\
			BPS & 0.77 & 13.57 & 5.92 & \$1.667  & \$20.00 & \$29.86\\
			\bottomrule 
		\end{tabular}
	}
 }
        \caption{Summary statistics for the data used in the main test problem.}
	\label{stats_17dim_path}
\end{table}
\begin{table}[H]
	\centering
	\vspace{0mm}
	\setlength\tabcolsep{4pt} 
	{\small 
             \scalebox{0.8}{
		\begin{tabular}{lccccccccc}
			\toprule
			Class & Arrival & $\mu$ & $\theta$ & $p$ & $h$ & $c$\\
			\rule{0pt}{3.5ex} & percentage (\%)  & (per hr) & (per hr) & (per job) & (per hr) & (per hr)\\
			\midrule
			Retail (Node: 1) & 15.39 & 17.22 & 6.06 & \$2.000 & \$24.00 & \$36.12\\
                Retail (Node: 2) & 22.82 & 17.25 & 7.81 & \$2.000 & \$24.00 & \$39.62\\
                Retail (Node: 3) & 15.50 & 17.25 &  5.22 &\$2.000 & \$24.00 & \$34.44\\
			Premier & 3.46 & 13.15 & 12.73 & \$2.167  & \$33.80 & \$61.39\\
			Business  & 4.82 & 16.56 & 8.58 & \$2.500  & \$30.00 & \$51.46\\
			Platinum  & 0.34 & 17.20 & 7.50 & \$2.667  & \$32.00 & \$51.99\\
			Consumer Loans  & 6.92 & 15.19 & 4.87 & \$1.833 & \$22.00 & \$30.93\\
			Online Banking  & 2.64 & 10.60 & 7.25 & \$1.833  & \$28.60 & \$41.91\\
			EBO  & 0.72 & 9.87 & 10.71 & \$1.667  & \$26.00 & \$43.85\\
			Telesales  & 6.26 & 9.62 & 11.69 & \$1.833  & \$28.60 & \$50.04\\
			Subanco  & 0.51 & 11.79 & 8.31 & \$1.667  & \$26.00 & \$39.86\\
			Case Quality  & 0.84 & 9.93 & 12.05 & \$1.667  & \$26.00 & \$46.07\\
			Priority Service & 1.47 & 10.35 & 6.40 & \$2.667  & \$22.40 & \$39.46\\
			AST  & 3.42 & 12.52 & 9.75 & \$1.833  & \$28.60 & \$46.48\\
			CCO  & 8.34 & 15.20 & 4.97 & \$1.833  & \$15.40 & \$24.51\\
			Brokerage  & 5.78 & 12.62 & 4.82 & \$1.833  & \$15.40 & \$24.24\\
			BPS & 0.77 & 13.57 & 4.14 & \$1.667  & \$14.00 & \$20.90\\
			\bottomrule 
		\end{tabular}
	}
 }
        \caption{Summary statistics for the data used in the first variant test problem.}
	\label{first_variant_data}
\end{table}
\begin{table}[H]
	\centering
	\vspace{0mm}
	\setlength\tabcolsep{4pt} 
	{\small 
             \scalebox{0.8}{
		\begin{tabular}{lccccccccc}
			\toprule
			Class & Arrival & $\mu$ & $\theta$ & $p$ & $h$ & $c$\\
			\rule{0pt}{3.5ex} & percentage (\%)  & (per hr) & (per hr) & (per job) & (per hr) & (per hr)\\
			\midrule
			Retail (Node: 1) & 15.39 & 17.22 & 6.06 & \$2.000 & \$24.00 & \$36.12\\
                Retail (Node: 2) & 22.82 & 17.25 & 7.81 & \$2.000 & \$24.00 & \$39.62\\
                Retail (Node: 3) & 15.50 & 17.25 &  5.22 &\$2.000 & \$24.00 & \$34.44\\
			Premier & 3.46 & 13.15 & 13.71 & \$2.167  & \$36.40 & \$66.11\\
			Business  & 4.82 & 16.56 & 8.58 & \$2.500  & \$30.00 & \$51.46\\
			Platinum  & 0.34 & 17.20 & 7.50 & \$2.667  & \$32.00 & \$51.99\\
			Consumer Loans  & 6.92 & 15.19 & 4.87 & \$1.833 & \$22.00 & \$30.93\\
			Online Banking  & 2.64 & 10.60 & 7.81 & \$1.833  & \$30.80 & \$45.14\\
			EBO  & 0.72 & 9.87 & 11.54 & \$1.667  & \$28.00 & \$47.22\\
			Telesales  & 6.26 & 9.62 & 12.59 & \$1.833  & \$30.80 & \$53.89\\
			Subanco  & 0.51 & 11.79 & 8.95 & \$1.667  & \$28.00 & \$42.92\\
			Case Quality  & 0.84 & 9.93 & 12.98 & \$1.667  & \$28.00 & \$49.62\\
			Priority Service & 1.47 & 10.35 & 5.48 & \$2.667  & \$19.20 & \$33.82\\
			AST  & 3.42 & 12.52 & 10.50 & \$1.833  & \$30.80 & \$50.05\\
			CCO  & 8.34 & 15.20 & 4.26 & \$1.833  & \$13.20 & \$21.01\\
			Brokerage  & 5.78 & 12.62 & 4.13 & \$1.833  & \$13.20 & \$20.78\\
			BPS & 0.77 & 13.57 & 3.55 & \$1.667  & \$12.00 & \$17.92\\
			\bottomrule 
		\end{tabular}
	}
 }
        \caption{Summary statistics for the data used in the second variant test problem.}
	\label{second_variant_data}
\end{table}
\begin{figure}[H]
    \centering
    \includegraphics[width=0.85\linewidth]{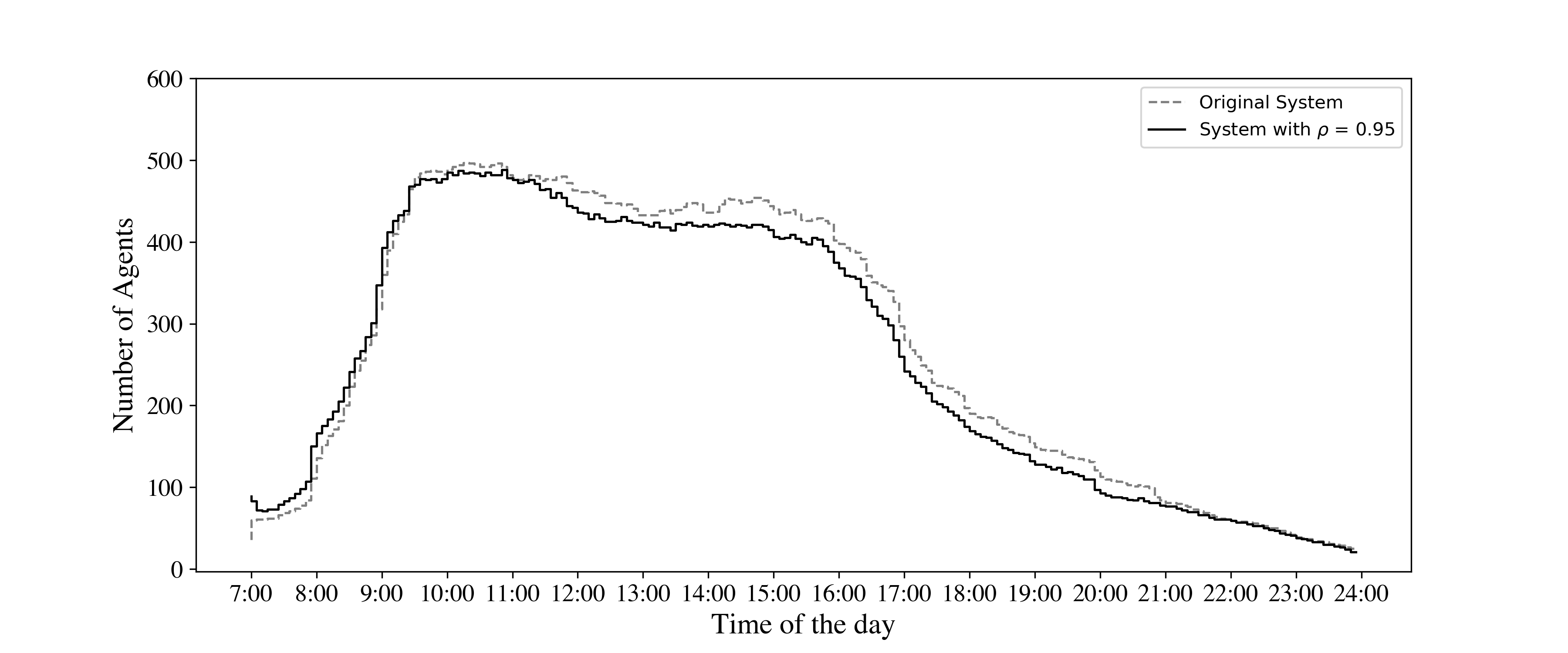}
    \caption{The number of agents for the target system utilization of 0.95 in comparison to the original system. The resolution of the horizontal axis is 5 minutes. That is, the number of agents is calculated over five-minute intervals.}
\label{fig:critically_loaded_agents}
\end{figure}
\subsection{Low dimensional test problems}\label{appendix_lower}
\subsubsection{The 2-dimensional test problem}\label{appendix_2dim}
\begin{figure}[H]
     \begin{minipage}{0.48\textwidth}
     \centering
         \includegraphics[scale = 0.45]{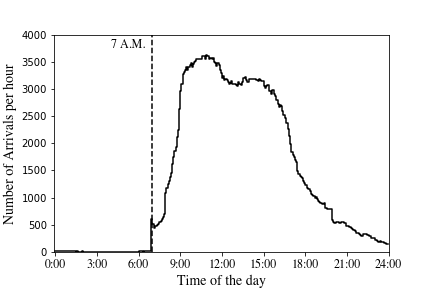}
         \captionsetup{font={footnotesize}}
        \caption*{(a) Class 1}
     \end{minipage}
     \hfill
     \begin{minipage}{0.48\textwidth}
         \centering
         \includegraphics[scale = 0.45]{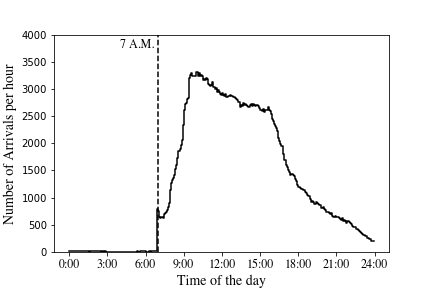}
         \captionsetup{font={footnotesize}}
         \caption*{(b) Class 2}
     \end{minipage} 
     \caption{Hourly arrival rates $\lambda^{r}_{k}(\cdot)$}
     \label{graph_arrival_2dim}
\end{figure}

\begin{figure}[H]
     \centering
     \begin{minipage}{0.48\textwidth}
     \centering
         \includegraphics[scale = 0.45]{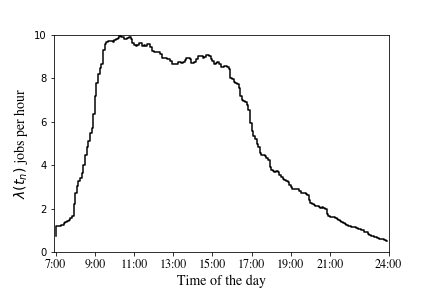}
         \captionsetup{font={footnotesize}}
        \caption*{(a) Class 1}
 \label{graph_arrival_class1_2dim_limit}
     \end{minipage}
     \hfill
     \begin{minipage}{0.48\textwidth}
         \centering
         \includegraphics[scale = 0.45]{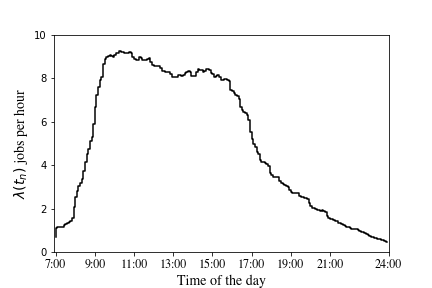}
         \captionsetup{font={footnotesize}}
         \caption*{(b) Class 2}
     \end{minipage}
     \caption{Hourly limiting arrival rates $\lambda_{k}(\cdot)$}
\label{graph_arrival_limit_2dim}
\end{figure}

\begin{figure}[H]
     \begin{minipage}{0.48\textwidth}
     \centering
         \includegraphics[scale = 0.45]{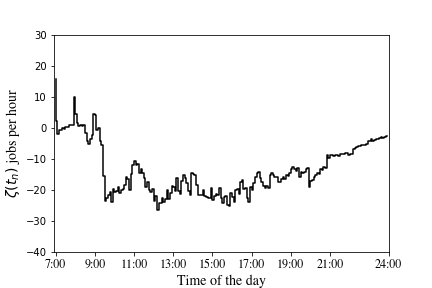}
            \captionsetup{font={footnotesize}}
             \caption*{(a) Class 1}
     \end{minipage}
     \begin{minipage}{0.48\textwidth}
         \centering
         \includegraphics[scale = 0.45]{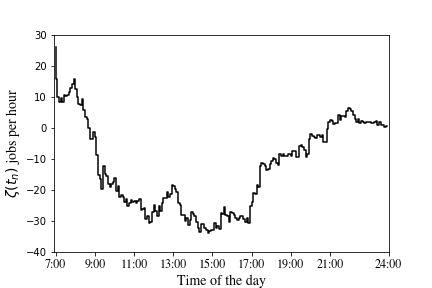}
         \captionsetup{font={footnotesize}}
        \caption*{(b) Class 2}
     \end{minipage}
     \caption{Hourly $\zeta_{k}(\cdot)$ terms}
      \label{graph_zeta_2dim}
\end{figure}

\textbf{Graphs of the prelimit arrival rates $\lambda^{r}_{k}(\cdot)$.} To determine the hourly arrival rates for the classes of the 2-dimensional test problem, we aggregate the hourly arrival rates $\lambda^{r}_{k}(\cdot)$ of the classes of the main test problem 
combined into the two new classes as shown in Table \ref{class_division_2dim}. Figure \ref{graph_arrival_2dim} displays the hourly arrival pattern for each of the two classes.
\vspace{-5mm}
\noindent \textbf{Graphs of the limiting arrival rates $\lambda_k(\cdot)$.} Figure \ref{graph_arrival_limit_2dim} shows the hourly limiting arrival rates for each class $k$, denoted by $\lambda_{k}(\cdot)$, calculated using Equation (\ref{eqn_limit_arrival}). 
\vspace{-5mm}
\noindent \textbf{Graphs of the second-order terms $\zeta_k(\cdot)$.} Figure \ref{graph_zeta_2dim} shows the hourly second-order terms, denoted by $\zeta_{k}(\cdot)$, which are calculated using Equation (\ref{eqn_limit_zeta}). 
\vspace{-5mm}
\subsubsection{3-dimensional test problems} \label{appendix_3dim}
\begin{figure}[H]
    \centering
     \hspace{-8mm} \begin{minipage}[t]{0.33\textwidth} 
        \centering
        \includegraphics[scale=0.4]{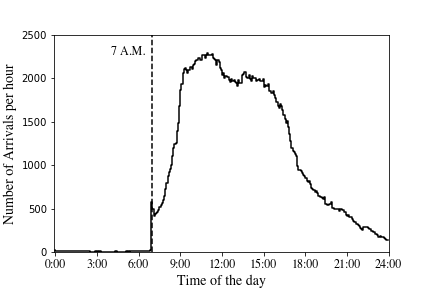}
        \captionsetup{font={footnotesize}}
        \caption*{(a) Class 1}
    \end{minipage}%
    \hfill
    \hspace{-6mm} \begin{minipage}[t]{0.33\textwidth}
        \centering
        \includegraphics[scale=0.4]{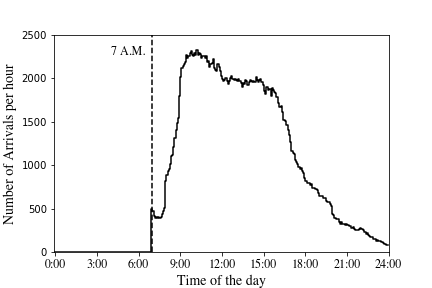}
        \captionsetup{font={footnotesize}}
        \caption*{(b) Class 2}
    \end{minipage}%
    \hfill
    \hspace{-6mm} \begin{minipage}[t]{0.33\textwidth}
        \centering
        \includegraphics[scale=0.4]{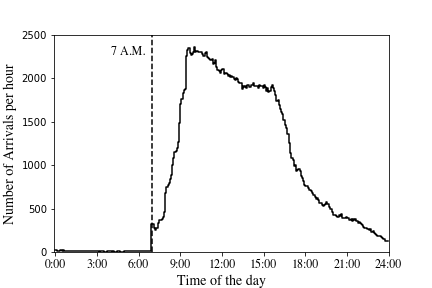}
        \captionsetup{font={footnotesize}}
        \caption*{(c) Class 3}
    \end{minipage}
    \caption{Hourly arrival rates $\lambda^{r}_{k}(\cdot)$}
     \label{graph_arrival_3dim}
\end{figure}

\begin{figure}[H]
     \centering
     \hspace{-6mm}\begin{minipage}{0.33\textwidth}
     \centering
         \includegraphics[scale = 0.4]{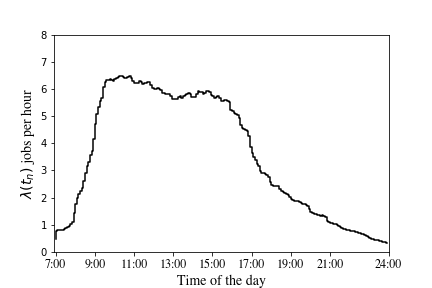}
         \captionsetup{font={footnotesize}}
        \caption*{(a) Class 1}
     \end{minipage}
     \hfill
     \hspace{-6mm}\begin{minipage}{0.33\textwidth}
         \centering
         \includegraphics[scale = 0.4]{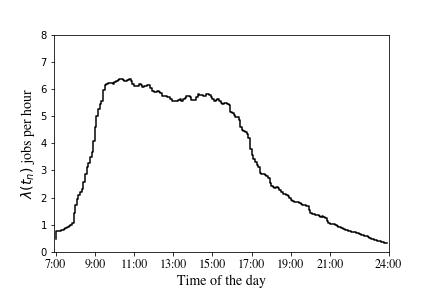}
         \captionsetup{font={footnotesize}}
         \caption*{(b) Class 2}
     \end{minipage}
     \hfill
     \hspace{-6mm}\begin{minipage}{0.33\textwidth}
         \centering
         \includegraphics[scale = 0.4]{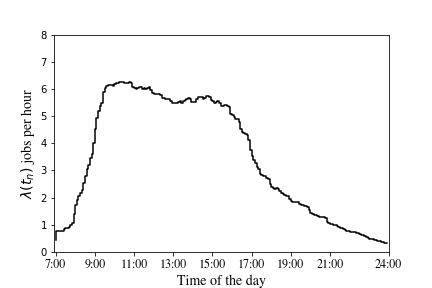}
         \captionsetup{font={footnotesize}}
         \caption*{(c) Class 3}
     \end{minipage}
     \caption{Hourly limiting arrival rates $\lambda_{k}(\cdot)$}
     \label{graph_arrival_limit_3dim}
\end{figure}

Recall that the two test problems we consider (that are three-dimensional) differ only in their cost parameters.\\
\textbf{Graphs of the prelimit arrival rates $\lambda^{r}_{k}(\cdot)$.} To set the hourly arrival rates $\lambda^{r}_{k}(\cdot)$ of the classes for the 3-dimensional test problem, we aggregate the hourly arrival rates $\lambda^{r}_{k}(\cdot)$ of the classes of the main test problem according to the class definitions provided in Table \ref{class_division_3dim}. Figure \ref{graph_arrival_3dim} illustrates the hourly arrival patterns for each of these three classes.
\noindent \textbf{Graphs of the limiting arrival rates $\lambda_k(\cdot)$.} Figure \ref{graph_arrival_limit_3dim} shows the hourly limiting arrival rates for each class $k$, denoted by $\lambda_{k}(\cdot)$, calculated using Equation (\ref{eqn_limit_arrival}). 

\begin{figure}[H]
     \centering
     \hspace{-6mm}\begin{minipage}{0.33\textwidth}
     \centering
         \includegraphics[scale = 0.4]{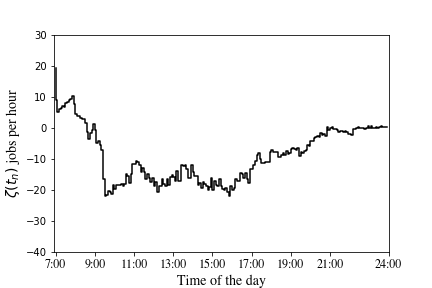}
         \captionsetup{font={footnotesize}}
        \caption*{(a) Class 1}
     \end{minipage}
     \hfill
     \hspace{-6mm}\begin{minipage}{0.33\textwidth}
         \centering
         \includegraphics[scale = 0.4]{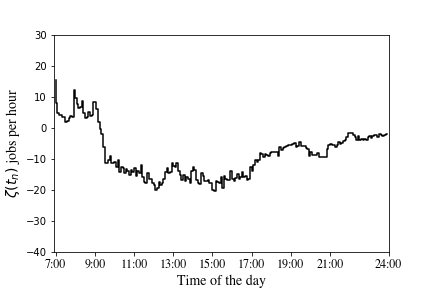}
         \captionsetup{font={footnotesize}}
         \caption*{(b) Class 2}
     \end{minipage}
     \hfill
     \hspace{-6mm}\begin{minipage}{0.33\textwidth}
         \centering
         \includegraphics[scale = 0.4]{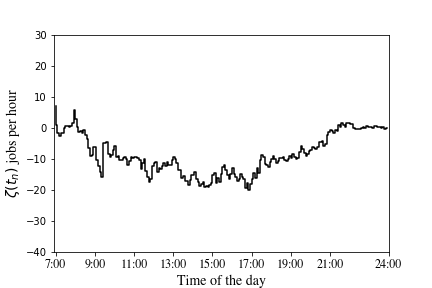}
         \captionsetup{font={footnotesize}}
         \caption*{(c) Class 3}
     \end{minipage}
     \caption{Hourly $\zeta_{k}(\cdot)$ terms}
     \label{graph_zeta_limit_3dim}
\end{figure}

\textbf{\hspace{-6mm}\textbf{Graphs of the second-order terms $\zeta_k(\cdot)$.}} Figure \ref{graph_zeta_limit_3dim} shows the hourly second-order terms, denoted by $\zeta_{k}(\cdot)$, which are calculated using Equation (\ref{eqn_limit_zeta}).

\subsection{Data used for 3-dimensional variant test problem} \label{3dim_variant_data}
\begin{table}[H]
	\centering
	\setlength\tabcolsep{4pt} 
	{\footnotesize 
        \scalebox{0.9}{
		\begin{tabular}{lccccccccc}
			\toprule
			Class &  Arrival & $\mu$ & $\theta$ & $p$  & $h$ & $c$ \\
			\rule{0pt}{3.5ex} & percentage (\%) & (per hr)& (per hr) & (per job) & (per hr) & (per hr) \\
                \midrule
			 1  & 33.90 & 15.74 & 8.14 & \$2.04 & \$24.48 & \$41.09 \\
			 2  & 33.29 & 15.77 & 6.03 & \$0.96 & \$11.46 & \$17.23 \\
                 3 & 32.81 & 14.68 & 6.64 & \$1.98 & \$23.75 & \$36.88\\
			\bottomrule 
		\end{tabular}
	}
 }
        \caption{Summary Statistics for the variant of the 3-dimensional problem.}
	\label{table_3dim_extend}
\end{table}
\subsection{Data used for high dimensional test problems}\label{appendix_higher}
Tables \ref{stats_30dim_path}, \ref{stats_50dim_path}, \ref{stats_100dim_path_part1} and \ref{stats_500dim_path} describe the system parameters used for the test problems of dimensions of 30, 50, 100, and 500 respectively. The second column of the tables lists the names of the classes from the main test problem that are used to set the arrival rate process $\{\tilde{\lambda}^{\tilde{r}}_{j}(\cdot),\, j = 1,\ldots, J\}$. Additionally, the tables provide information on the arrival percentages, denoted by $\tilde{q}_{j}$, common hourly service and abandonment rates, and abandonment penalty, denoted by $\tilde{\mu}$, $\tilde{\theta}$ and $\tilde{p}$, respectively. These common problem primitives are determined by taking a weighted average of $J$ randomly drawn service rates, abandonment rates, and abandonment penalties from the main test problem. The last two columns in each table present the hourly holding cost rates $h_{j}$ and total hourly cost rates $c_{j}$ for each class $j = 1,\ldots, J$. As explained in Section \ref{higher_dimensional}, the classes differ from each other only by their arrival, holding cost, and total cost rates. The holding cost rates $h_{j}$ are randomly drawn from a uniform grid between \$14 and \$34, with mesh sizes of $0.5, 0.25$, $0.125$ and $0.04$ for test problems of size $30, 50$, $100$ and $500$, respectively.

\subsubsection{The 30-dimensional test problem} \label{appendix_30dim}

\begin{figure}[H]
    \begin{minipage}{0.48\textwidth}
        \centering
        \includegraphics[scale=0.44]{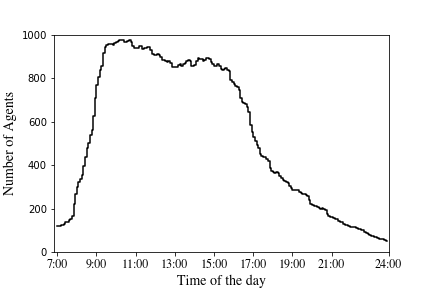}
    \end{minipage}
    \begin{minipage}{0.48\textwidth}
        \centering
        \includegraphics[scale = 0.44]{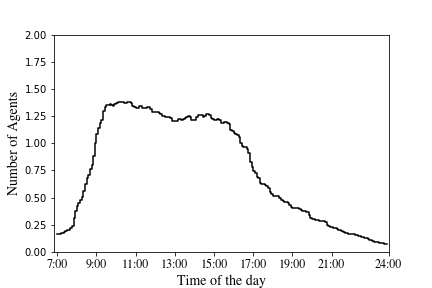}
    \end{minipage}
    \caption{The prelimit (left panel) and limiting (right panel) staffing levels $\tilde{N}^{\tilde{r}}(\cdot)$ and $\tilde{N}(\cdot)$}
    \label{agents_30dim}
\end{figure}

\textbf{The prelimit staffing levels $\tilde{N}^{\tilde{r}}(\cdot)$ and the limiting staffing levels $\tilde{N}(\cdot)$.} We use Equation (\ref{adjusted_agents}) to find the prelimit staffing levels for the 30-dimensional problem. Substituting the prelimit staffing levels $\tilde{N}^{\tilde{r}}(\cdot)$ and the system parameter $\tilde{r} = \lceil rJ/K\rceil = \lceil 400 \times 30/17\rceil = 706$ into Equation (\ref{eqn_limit_staff}), we find the limiting staffing levels $\tilde{N}(\cdot)$. Figure \ref{agents_30dim} displays the prelimit staffing levels $\tilde{N}^{\tilde{r}}(\cdot)$ and the limiting staffing levels $\tilde{N}(\cdot)$ throughout the day.

\subsubsection{The 50-dimensional test problem} \label{appendix_50dim}
\begin{figure}[H]
    \begin{minipage}{0.48\textwidth}
        \centering
        \includegraphics[scale=0.44]{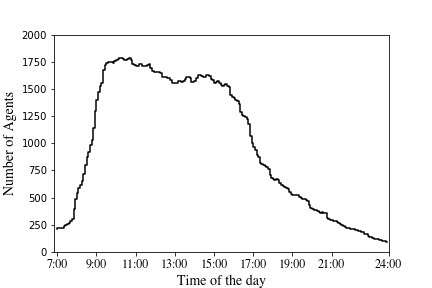}
    \end{minipage}
    \begin{minipage}{0.48\textwidth}
        \centering
        \includegraphics[scale = 0.44]{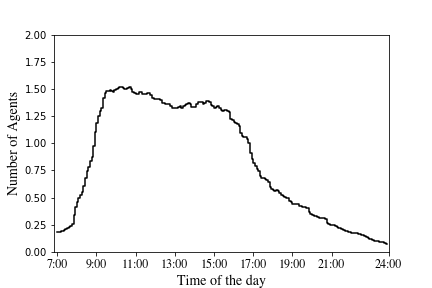}
    \end{minipage}
    \caption{The prelimit and limiting staffing levels $\tilde{N}^{\tilde{r}}(\cdot)$ and $\tilde{N}(\cdot)$}
    \label{agents_50dim}
\end{figure}

\textbf{The prelimit staffing levels $\tilde{N}^{\tilde{r}}(\cdot)$ and the limiting staffing levels $\tilde{N}(\cdot)$.} We use Equation (\ref{adjusted_agents}) to find the prelimit staffing levels for the 50-dimensional problem. Substituting the prelimit staffing levels $\tilde{N}^{\tilde{r}}(\cdot)$ and system parameter $\tilde{r} = \lceil rJ/K\rceil = \lceil 400 \times 50/17\rceil = 1177$ into Equation (\ref{eqn_limit_staff}), we find the limiting staffing levels $\tilde{N}(\cdot)$. The prelimit staffing levels $\tilde{N}^{\tilde{r}}(\cdot)$ and the limiting staffing levels $\tilde{N}(\cdot)$ throughout the day are shown in Figure \ref{agents_50dim}.

\subsubsection{The 100-dimensional test problem} \label{appendix_100dim}
\begin{figure}[H]
    \begin{minipage}{0.48\textwidth}
        \centering
        \includegraphics[scale=0.44]{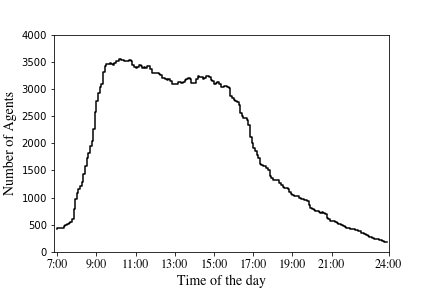}
    \end{minipage}
    \begin{minipage}{0.48\textwidth}
        \centering
        \includegraphics[scale = 0.44]{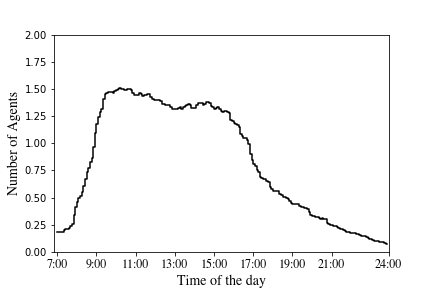}
    \end{minipage}
    \caption{The prelimit and limiting staffing levels $\tilde{N}^{\tilde{r}}(\cdot)$ and $\tilde{N}(\cdot)$}
    \label{agents_100dim}
\end{figure}

\textbf{The prelimit staffing levels $\tilde{N}^{\tilde{r}}(\cdot)$ and the limiting staffing levels $\tilde{N}(\cdot)$.} We use Equation (\ref{adjusted_agents}) to find the prelimit staffing levels for the 100-dimensional problem. Substituting the prelimit staffing levels $\tilde{N}^{\tilde{r}}(\cdot)$ and system parameter $\tilde{r} = \lceil rJ/K\rceil = \lceil 400 \times 100/17\rceil = 2353$ into Equation (\ref{eqn_limit_staff}), we find the limiting staffing levels $\tilde{N}(\cdot)$. The prelimit staffing levels $\tilde{N}^{\tilde{r}}(\cdot)$ and the limiting staffing levels $\tilde{N}(\cdot)$ throughout the day are shown in Figure \ref{agents_100dim}.

\subsubsection{The 500-dimensional test problem}\label{appendix_500dim}
\begin{figure}[H]
    \begin{minipage}{0.48\textwidth}
        \centering
        \includegraphics[scale=0.12]{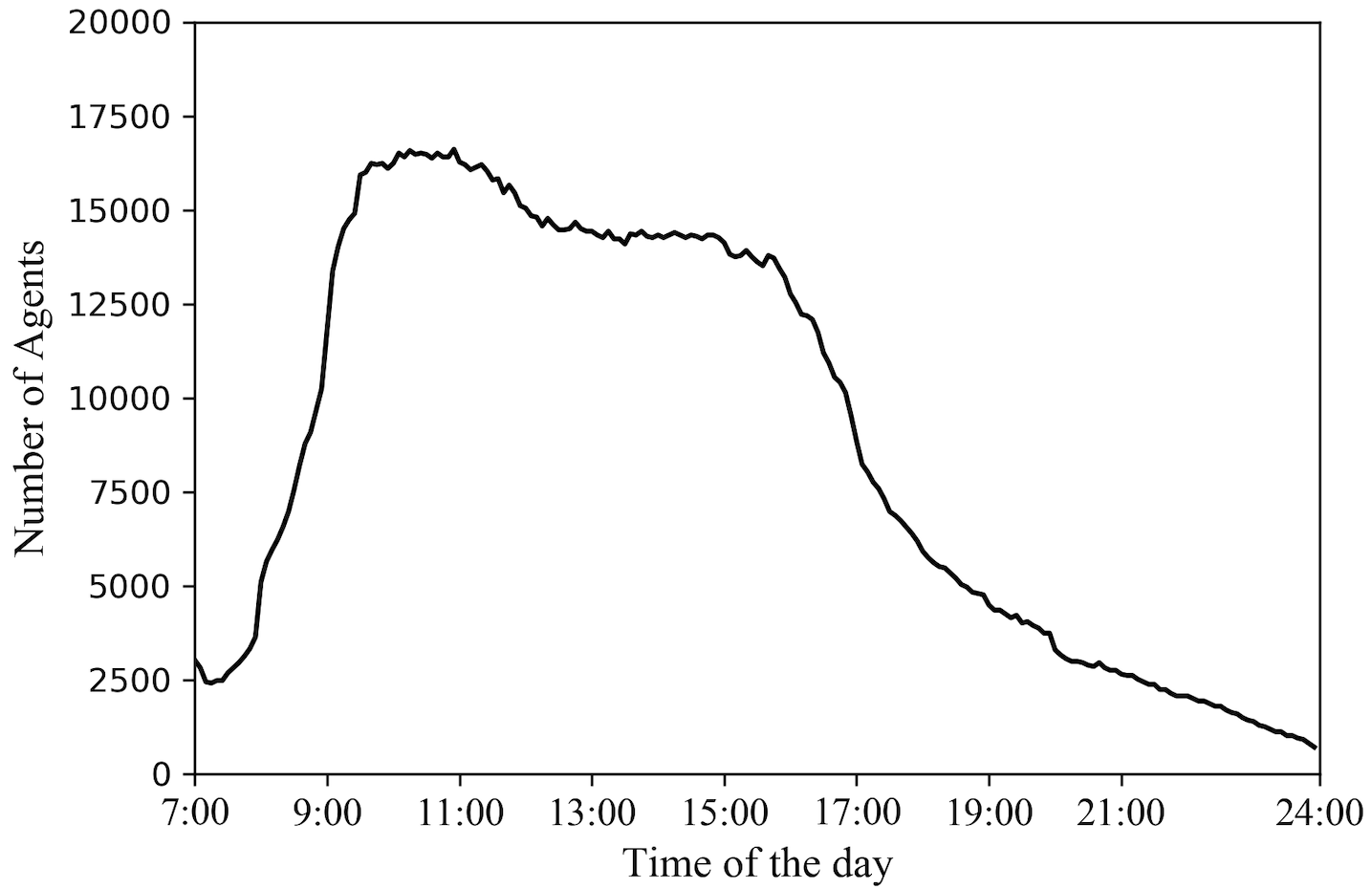}
    \end{minipage}
    \begin{minipage}{0.48\textwidth}
        \centering
        \includegraphics[scale = 0.12]{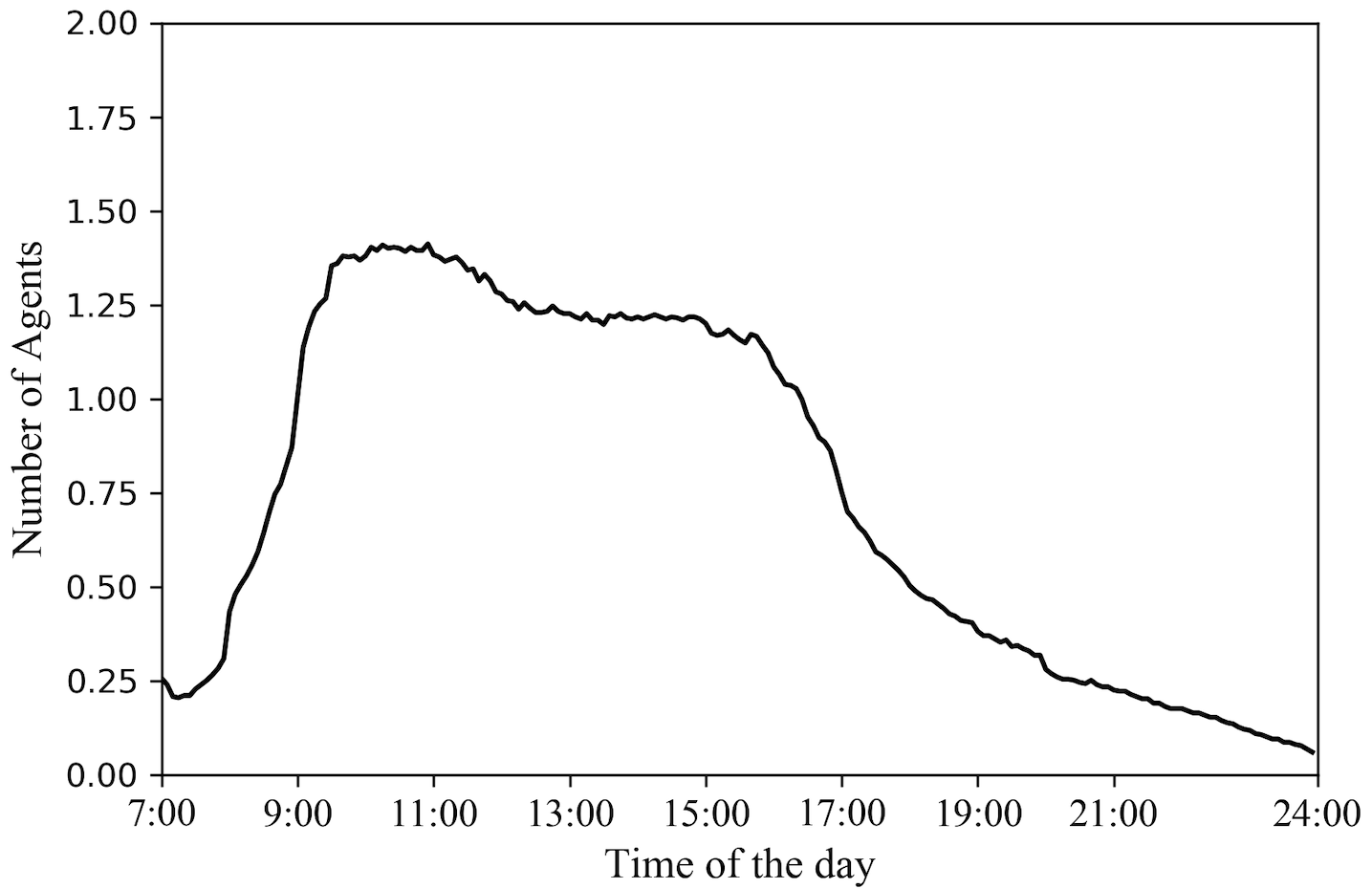}
    \end{minipage}
    \caption{The prelimit and limiting staffing levels $\tilde{N}^{\tilde{r}}(\cdot)$ and $\tilde{N}(\cdot)$}
    \label{agents_500dim}
\end{figure}

\textbf{The prelimit staffing levels $\tilde{N}^{\tilde{r}}(\cdot)$ and the limiting staffing levels $\tilde{N}(\cdot)$.} We use Equation (\ref{adjusted_agents}) to find the prelimit staffing levels for the 500-dimensional problem. Substituting the prelimit staffing levels $\tilde{N}^{\tilde{r}}(\cdot)$ and the system parameter $\tilde{r} = \lceil rJ/K\rceil = \lceil 400 \times 500/17\rceil = 11765$ into Equation (\ref{eqn_limit_staff}), we find the limiting staffing levels $\tilde{N}(\cdot)$. Figure \ref{agents_500dim} displays the prelimit staffing levels $\tilde{N}^{\tilde{r}}(\cdot)$ and the limiting staffing levels $\tilde{N}(\cdot)$ throughout the day.

\section{Computational benchmarks (for the pre-limit model)}\label{appendix_benchmarks}
\subsection{Optimal policies for MDP formulations of the low dimensional test problems}\label{appendix_mdp_low_dim}

The state process $X(t)$ is a continuous-time Markov chain (CTMC) on $\mathbb{Z}_{+}^{K}$.
The control is the $K$-dimensional process $u(t,x) \in \mathcal{U}(t,x)$, where $\mathcal{U}(t,x) = \{u \in \mathbb{R}^{K}_{+}: u\,(e\cdot x - N(t))^{+} \leq x, \,\, e\cdot u = 1\}$ and $u_{k}(t,x)$ is the fraction of the total backlog kept in class $k$ at time $t$ in state $x$.

The transition rate matrix $Q^{u} = (Q^{u}(x,y,t))$ under policy $u(t,x)$ is defined as follows\footnote{The case where $x_{k} = 0$ for $k = 1,\ldots,K$ requires special attention, as $x_{k} - e_{k}$ falls outside the bounds of the state space. However, the reader can verify that all transitions from $x_{k}$ to $x_{k} - e_{k}$, which result in the process leaving the state space, occur with a rate of zero.}: For $k = 1,\ldots, K$,
\begin{align}
    Q^{u}(x,x+e_{k},t) &=  \lambda_{k}(t), \label{eqn_generator1}\\
    Q^{u}(x,x-e_{k},t) &= \Big(x_{k} - u_{k}(t, x)(e \cdot x - N(t))^{+}\Big)\mu_{k} + u_{k}(t, x)(e\cdot x - N(t))^{+}\theta_{k}, \label{eqn_generator2}\\
    Q^{u}(x,x,t) \qquad &= - \Bigg(\sum_{k=1}^{K} \Big[\lambda_{k}(t) + \left(x_{k} - u_{k}(t, x)(e \cdot x - N(t))^{+}\right)\mu_{k} \\
    & \qquad \qquad \qquad \quad \, \,  + u_{k}(t, x)(e \cdot x - N(t))^{+}\theta_{k} \Big]\Bigg). \label{eqn_generator3}
\end{align}
Using Equation (\ref{eqn_$c_{k}$_of_policy}), we define the optimal value function of the cost-minimization problem in Section \ref{model_section} as follows:
\begin{equation}
    \tilde{V}(t,x) = \inf \, \mathbb{E}_{x}^{u}\left\{\int_{t}^{T} \left(e \cdot x - N(t)\right)^{+} c \cdot u(t, x)\,dt + g(X(T))\right\} \label{eqn_value_function_mdp},
\end{equation}
where the terminal $c_{k}$ is $g(x) = \bar{c}\,\left(e\cdot x - N(t)\right)^{+}$ for $x \in \mathbb{Z}_{+}^{K}$ and the infimum is taken over feasible policies. The associated Bellman equation, which helps us characterize the optimal value function $\tilde{V}$ and the corresponding optimal policy is given as follows: For $t \in [0,T]$ and $x \in \mathbb{Z}_{+}^{K}$,
\begin{equation}
    \frac{\partial \tilde{V}}{\partial t}(t,x) = - \min_{u \in \mathcal{U}(t,x)}\left\{(e \cdot x - N(t))^{+} c \cdot u + Q^{u}\tilde{V}(t,x)\right\}, \quad \tilde{V}(T,x) = g(x).\label{eqn_hjb_mdp}
\end{equation}

Substituting the definition of $Q^{u}$ from Equations (\ref{eqn_generator1}) - (\ref{eqn_generator3}) into Equation (\ref{eqn_hjb_mdp}) gives the following more explicit form:
\vspace{-2mm}
\begin{align}
 \frac{\partial \tilde{V}}{\partial t} (t,x) &=  -\sum_{k=1}^{K}\lambda_{k}(t)\Delta_{k}^{-}(t,x+e_{k}) + \sum_{k=1}^{K}x_{k}\mu_{k}\Delta_{k}^{-}(t,x) \nonumber \\
&\quad  - (e \cdot x - N(t))^{+} \min_{u\in \mathcal{U}(t,x)}  \Bigg\{ \sum_{k=1}^{K}\Big(c_{k} + (\mu_{k} - \theta_{k})\Delta_{k}^{-}(t,x)\Big)u_{k}\Bigg\},
\label{eqn_hjb_mdp2}
\vspace{10mm}
\end{align}
where $\Delta_k^{-}(t,\cdot)$ is an auxiliary function (used for notational simplicity) defined as follows: For $k = 1, \ldots, K$, and $(t,x) \in [0,T]\times \mathbb{Z}_+^K$,
\begin{equation}
    \Delta_{k}^{-}(t,x) = \tilde{V}(t,x) - \tilde{V}(t,x-e_{k}).\label{delta_x}
\end{equation}
In light of Equation (\ref{eqn_hjb_mdp2}), the optimal policy $u^{\star}(t,x)$ is as follows\footnote{Choosing a control $u$ is consequential only when $e\cdot x > N(t)$ because there are no customers waiting in the queue when $e \cdot x \leq N(t)$.}:
\vspace{2mm}
\begin{equation}
    u^{\star}(t,x) = \argmin_{u \in \mathcal{U}(t,x)}\Bigg\{\sum_{k=1}^{K} \Big(c_{k} + (\theta_{k} - \mu_{k})\Delta_{k}^{-}(t,x)\Big)u_{k}\Bigg\}, \quad (t,x) \in [0,T] \times \mathbb{Z}^{K}_{+}.
    \vspace{2mm}
\end{equation}

The optimal policy $u^{\star}(t,x)$ can be characterized equivalently using an effective holding cost function $\tilde{\phi}(\cdot)$ to prioritize different classes, cf. Section \ref{sect_hjb_equation}. For $(t,x) \in [0,T] \times \mathbb{Z}_{+}^{K}$ and $k = 1,\ldots,K$, we let
\begin{equation}
    \tilde{\phi}_{k}(t,x) = c_{k} + (\mu_{k} - \theta_{k})\Delta_{k}^{-}(t,x).
    \vspace{1mm}
\end{equation}
We then use the effective holding cost function $\tilde{\phi}(\cdot)$ to order classes from the most expensive to the cheapest. When assigning servers to calls, the system manager prioritizes classes with respect to their effective holding costs, where the classes with higher effective holding cost have higher priority. In other words, the system manager strives to keep the backlog in the buffers with lower effective holding costs.\\
\\
\noindent\textbf{Computational Method.} To numerically solve the Bellman equation, we consider a temporal discretization of Equation (\ref{eqn_hjb_mdp2}). Given a partition of the time interval $[0,T]$: $0 < t_{0} < t_{1} < \ldots < t_{N} = T$, we let $\Delta t_{n} = t_{n} - t_{n-1}$ and for simplicity set $\Delta t_{n} = T/N$ for $n = 1,\ldots,N$. Then we use the Euler method (see \citet{atkinson1991introduction}) as follows: For $n = 1, \ldots, N$ and $x \in \mathbb{Z}_{+}^{K}$
\begin{align}
    &\tilde{V}(t_{n-1},x) - \tilde{V}(t_{n},x) = \nonumber \\
    & \qquad \quad \quad   \vspace{-1mm} \quad + \Bigg(\sum_{k=1}^{K}\lambda_{k}(t_{n})\Delta_{k}^{-}(t_{n},x+e_{k})  - \sum_{k=1}^{K}x_{k}\mu_{k}\Delta_{k}^{-}(t_{n},x) \nonumber \\
    & \qquad \quad \quad \vspace{-1mm} \quad + (e \cdot x - N(t_{n}))^{+} \min_{u\in \mathcal{U}(t_{n},x)}  \left\{ \sum_{k=1}^{K}\Big(c_{k} + (\mu_{k} - \theta_{k})\Delta_{k}^{-}(t_n,x)\Big)u_{k}(t_n,x)\right\}\Bigg)\Delta t_n, \label{discrete_picard}\\
    &\tilde{V}(t_{N},x) = g(x). \label{discrete_picard_terminal}
\end{align}
We define an auxiliary function $f(t_n,x)$ to denote the right-hand side of Equation (\ref{discrete_picard}) and use Algorithm \ref{algorithm_forward_euler1} to recursively solve the value function for $n = 0,1,\ldots, N-1$.
\vspace{2mm}
\begin{algorithm}[H]    
	\caption{}
        \label{algorithm_forward_euler1}
        \begin{minipage}{\textwidth}
	\begin{algorithmic}[1]
        \Statex \textbf{Input:} The time horizon $T$, the number of intervals $N$, a discretization step-size $\Delta t_{n}$ (for simplicity, we assume $\Delta t_{n} \triangleq T/N$), terminal cost function $g(\cdot)$
        \Statex \textbf{Output:} The solution of the value function $\tilde{V}(t_{n}, \cdot)$ for $n = 0,1,\ldots, N-1$
        \State Initialize $\tilde{V}(t_{N}, \cdot) = g(\cdot)$.
        \While {$n \geq 1$}
        \State Calculate $\tilde{V}(t_{n-1},\cdot) = \tilde{V}(t_{n},\cdot) - \Delta t_{n} f(t_{n},\cdot)$.
        \State Update $t = t - \Delta t_{n}.$
        \State Update $n = n-1$.
	\EndWhile
    \State \textbf{return} $\tilde{V}(t_{n}, \cdot)$ for $n = 0,1,\ldots, N-1$ and $\Delta_{k}^{-}(t_n,\cdot)$ for $k = 1,\ldots, K$ and $n = 1,\ldots, N$.
	\end{algorithmic} 
        \end{minipage}
\end{algorithm}
\vspace{-5mm}
\noindent \textbf{Discretization step-size.} To ensure a close approximation of the continuous-time formulation by our discretization, we seek to ensure that $\Delta t_{n}$ is sufficiently small so that the probability of two or more arrivals during an interval of length $\Delta t_{n}$ is negligible, i.e., $o(t_{n})$. Similarly, we also want to ensure the probability of more than one service completion over such an interval is negligible. In doing so, we focus on the peak staffing level and assume all agents are busy. Ultimately, we set $\Delta t_{n} = 0.1$ seconds, for which the probability of more than one arrival or service completion is negligible, as shown in Table \ref{prob_arrivals}. Therefore, given $T = 17$ hours and $\Delta t_{n} = T/N$, the number of intervals is $N = 612,000.$ With our choice of $\Delta t_{n} = 0.1$ seconds, it takes about 6 days to solve the 3-dimensional test problems on a computer equipped with an AMD EPYC 7502 32c/64t CPU with 80 GB RAM using OpenMP to enable parallelization on the multicore CPU. Although one may consider an even smaller $\Delta t_{n}$, that would be even more time-consuming\footnote{Storing the optimal policy for every 0.1 seconds using double-precision floating-point numbers requires approximately 11 TB of storage space. In addition, simulating the pre-limit problem with optimal policies saved at 0.1-second intervals is not feasible due to memory constraints. Thus, we use OpenMP to parallelize our simulation code across 10 CPU cores and allocate 500 GB of memory to simulate the pre-limit system with optimal policies saved at one-minute intervals.}.
\vspace{5mm}
\begin{table}[H]
    \centering
    \renewcommand{\arraystretch}{1.2}
    \setlength\tabcolsep{4pt} 
	{\small 
             \scalebox{0.9}{
    \begin{tabular}{lllllcccccccc}
        \toprule
        Number of && Probability && Probability\\
        Events && of Arrivals && of Service Completions\\
        \midrule
         0 && 0.8311 &&  0.7912 \\
         1 && 0.1557 &&  0.1822 \\
         2 && 0.0125 &&  0.0249\\
         3 && 0.0007 &&  0.0017\\
        \bottomrule    
    \end{tabular}
    }
    \caption{Probabilities of Arrivals and Service Completions in $\Delta t_{n}$ = 0.1 seconds.}
     \label{prob_arrivals}
    }
\end{table}

\begin{figure}[H]
    \centering
\hspace{10mm}\includegraphics[scale=0.15]{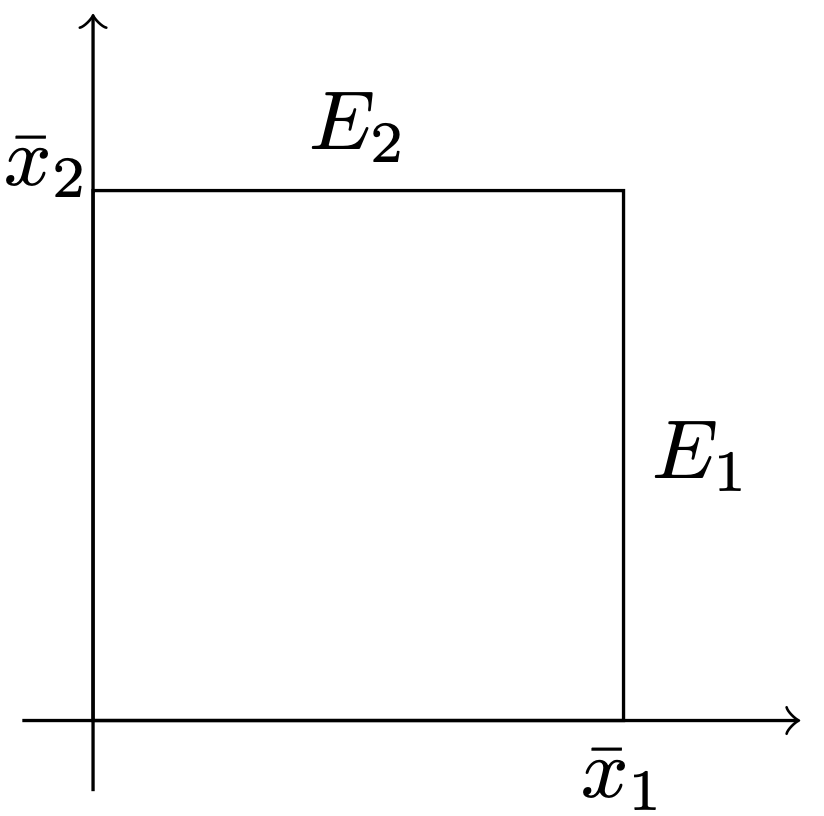}
    \caption{Truncated state space for $K = 2$}
    \label{state_space}
\end{figure}

\noindent \textbf{Truncating the state space.}
For computational feasibility, we truncate the state space by replacing it with $S_{\bar{x}}$ defined as follows:
\begin{equation*}
    S_{\bar{x}} = \{x \in \mathbb{Z}^{K}_{+}: 0 \leq x_{k} \leq \bar{x}_{k} \quad \text{for} \,\, k = 1, \ldots, K\}.
\end{equation*}
To define the behavior of the Markov chain at the boundary states, we modify the transition rate matrix $Q^{u}$. To be specific, when $K = 2$, we define the vertical boundary of the state space as $E_{1} = \{(\bar{x}_{1}, x_{2}): 0 \leq x_{2} < \bar{x}_{2}\}$. Similarly, we define the horizontal boundary of the state space as $E_{2} = \{(x_{1}, \bar{x}_{2}): 0 \leq x_{1} < \bar{x}_{1}\}$ as shown in Figure \ref{state_space}.
\noindent Then, we set 
$$ 
\lambda(t,x) = \begin{cases}
			(0,\lambda_{2}(t))^{\prime}, & \text{if $x \in E_{1}$},\\
            (\lambda_{1}(t),0)^{\prime}, & \text{if $x \in E_{2}$},\\
             (0,0)^{\prime}, & \text{if $x = (\bar{x}_{1}, \bar{x}_{2})$},\\
             (\lambda_{1}(t), \lambda_{2}(t))^{\prime}, & \text{otherwise}.
		 \end{cases}
$$
Similarly, for $K = 3$, we let
\begin{align}
   & F_{1} = \{(\bar{x}_{1}, x_{2}, x_{3}):\, 0 \leq  x_{2} < \bar{x}_{2}, \, 0 \leq x_{3} < \bar{x}_{3}\}, \nonumber \\
   & F_{2} = \{(x_{1}, \bar{x}_{2}, x_{3}):\, 0 \leq  x_{1} < \bar{x}_{1}, \, 0 \leq x_{3} < \bar{x}_{3}\}, \nonumber \\
   & F_{3} = \{(x_{1}, x_{2}, \bar{x}_{3}): \, 0 \leq x_{1} < \bar{x}_{1}, \, 0 \leq x_{2} < \bar{x}_{2}\}, \nonumber
\end{align}
and set
$$
\lambda(t,x) = \begin{cases}
			(0,\lambda_{2}(t), \lambda_{3}(t))^{\prime}, & \text{if $x \in F_{1}$},\\
            (\lambda_{1}(t),0, \lambda_{3}(t))^{\prime}, & \text{if $x \in F_{2}$},\\
            (\lambda_{1}(t), \lambda_{2}(t),0)^{\prime} & \text{if $x \in F_{3}$},\\
             (0,0, \lambda_{3}(t))^{\prime}, & \text{if $x \in F_{1} \cap F_{2}$},\\
             (0,\lambda_{2}(t), 0)^{\prime}, & \text{if $x \in F_{1} \cap F_{3}$},\\
             (\lambda_{1}(t),0, 0)^{\prime}, & \text{if $x \in F_{2} \cap F_{3}$},\\
             (0,0, 0)^{\prime}, & \text{if $x = (\bar{x}_{1}, \bar{x}_{2}, \bar{x}_{3})$},\\
             (\lambda_{1}(t), \lambda_{2}(t), \lambda_{3}(t))^{\prime}, & \text{otherwise}.
		 \end{cases}
$$

To set the upper bound vector $\bar{x}$ for the state space $S_{\bar{x}}$, we conduct a simulation study, considering all possible static priority policies for the lower-dimensional test cases $(K = 2,3)$. We observe the distribution of the maximum number of calls in the system for each class $k$ and set the 99$^{\text{th}}$ percentile of the observations as the upper bounds of the state space. To be specific, we establish $\bar{x} = (310, 310)$ for the 2-dimensional test problem and $\bar{x} = (210, 210, 210)$ for the 3-dimensional test problem.
\color{black}

\subsection{Further description and the performance of benchmark policies for the main test problem and its variants}
\label{benchmark_appendix}
\subsubsection{Numerical study of a 2-class queueing system using \citet{garnett2002designing} approximations}
\label{two_dim_numerical_study}
\begin{figure}[h]
\centering
    \includegraphics[scale = 0.4]{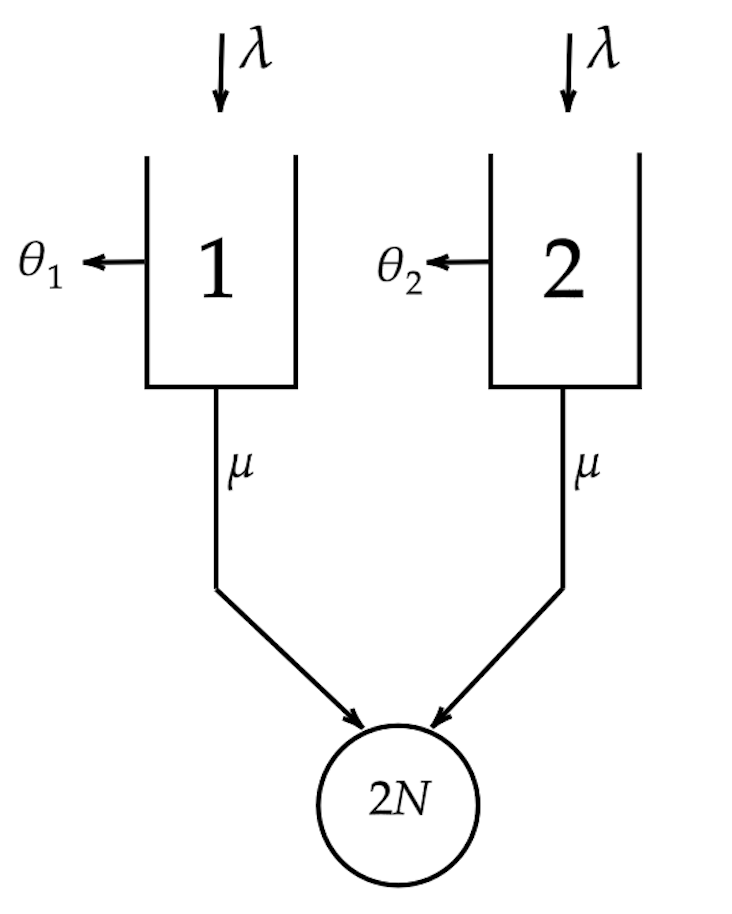}
    \caption{A Stylized 2-dimensional example.}
    \label{2dim_model}
\end{figure}

To build intuition, we first consider a 2-class Markovian queueing system with a single pool of homogeneous servers attending to both classes. A schematic description is given in Figure \ref{2dim_model}. Customers of class $k$ arrive at the system at rate $\lambda_{k}$ for $k = 1,2$. They leave the system either by receiving service or by abandoning while they wait in the queue. Servers process customers from class $k$ at a rate of $\mu_{k}$. Abandonment is represented by the horizontal arrows in Figure \ref{2dim_model}, with the associated abandonment rate $\theta_{k}$ for class $k$.

In particular, we focus on a system where $\lambda_{1} = \lambda_{2} = \lambda$,  $\mu_1 = \mu_{2} = \mu$ and without loss of generality, we assume $\theta_{2} > \theta_{1}$. There are a total of $2N$ servers staffing the system; see Figure \ref{2dim_model}. We assume $\lambda \approx N\mu$.
\vspace{-5mm}

\begin{figure}[h]
\centering
    \includegraphics[scale = 0.3]{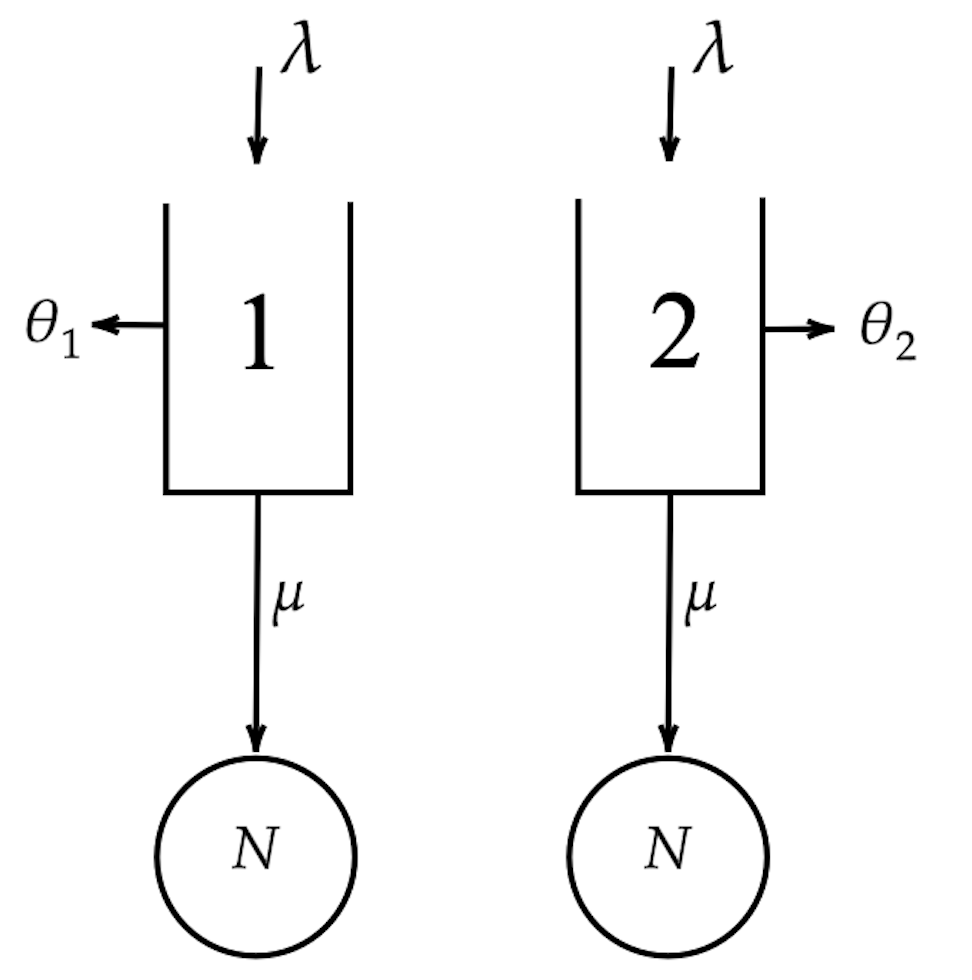}
    \caption{A separable system.}
    \label{2dim_model_separate}
\end{figure}

For future reference, we also consider the following system of two separate queues as shown in Figure \ref{2dim_model_separate}.

\begin{figure}[h]
\centering
    \includegraphics[scale = 0.4]{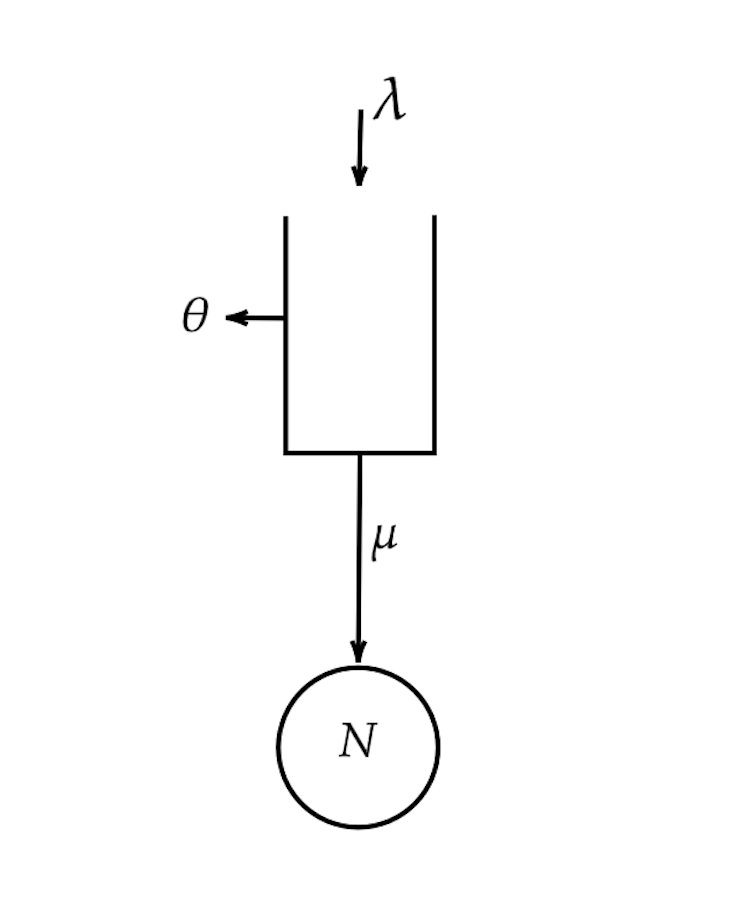}
    \caption{$M/M/N+M$ queuing model.}
    \label{1dim_model}
\end{figure}

For the system displayed in Figure \ref{2dim_model}, we restrict attention to static priority policies and assume only one of the queues (the one with lower priority) can be positive. For example, if class 1 is given priority, then we assume the queue length $Q_{1} \approx 0$, and all backlog is kept in buffer 2. (This approximation is based on the intuition from the heavy traffic literature.) To elaborate further, in this case, the left queue in Figure \ref{2dim_model_separate} is deleted, and we are left effectively with a single-server system. Therefore, we will study the single-server system displayed in Figure \ref{1dim_model} as a building block.

Figure \ref{1dim_model} shows the $M/M/N+M$ queue that is studied by \citet{garnett2002designing} in the Halfin-Whitt heavy traffic regime. Their study provides approximations for various performance measures, focusing on a regime where \( R = \lambda / \mu \) represents the (average) offered load and the appropriate staffing level is given by
\begin{equation}
N = R + \beta \sqrt{R}, \label{staffing_root}
\end{equation}
where the second term, \(\beta \sqrt{R}\), accounts for the excess capacity required beyond the nominal level (\(R\)) to meet a certain target service level under uncertainty.

Under this regime, \citet{garnett2002designing} approximate the steady-state average number of people in the queue as follows:
\begin{equation}
E[Q] \approx \frac{\lambda}{\theta} \left(1 - \frac{h\left(\beta \sqrt{\mu / \theta}\right)}{h\left(\beta \sqrt{\mu / \theta} + \sqrt{\theta / (N \mu)}\right)}\right) \cdot w(-\beta, \sqrt{\mu / \theta}), \label{eq_waiting_queue}
\end{equation}
where \( w(x, y) = \left(1 + \frac{h(-xy)}{y h(x)}\right)^{-1} \), and \( h(x) \) is the hazard rate function of the standard normal distribution.

\begin{figure}[h]
    \centering
    \includegraphics[scale=0.4]{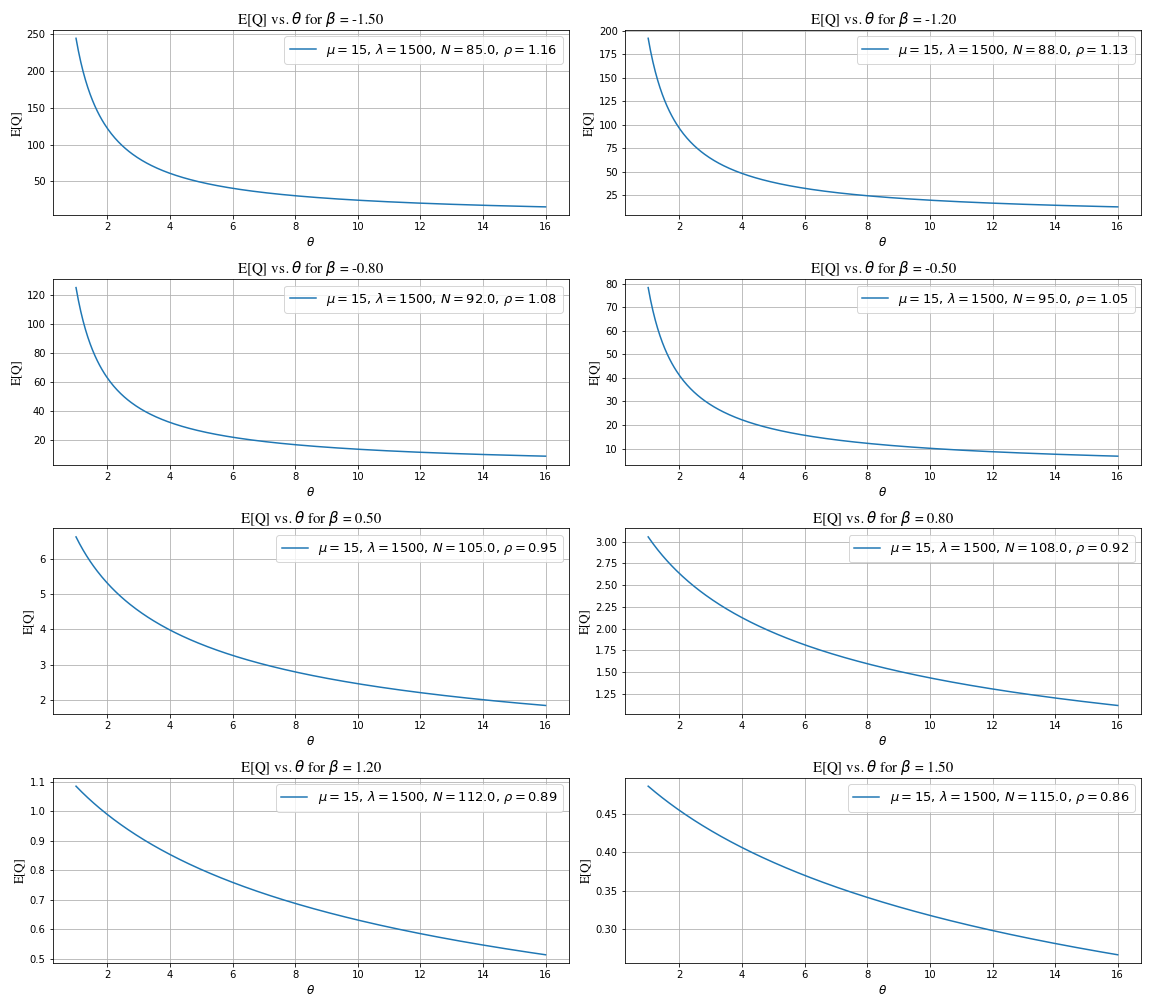}
    \caption{$\mathbb{E}[Q]$ vs. $\theta$ for overloaded, critically loaded, and underloaded systems in the 2-dimensional setting. We approximate the steady-state queue length by using the approximation given in Equation (\ref{eq_waiting_queue}).}
    \label{fig:expected_queue_theta}
\end{figure}

We use \citet{garnett2002designing}'s approximations in our analysis to identify when the $c\mu/\theta$ rule does not perform well. In addition, we verify our results for the 2-class system shown in Figure \ref{2dim_model} using simulation. We use $f(\theta)$ to denote \citet{garnett2002designing}'s approximation of the expected queue length in Equation (\ref{eq_waiting_queue}) and write 
\begin{equation}
\mathbb{E}[Q] \approx f(\theta),
\end{equation}
where $f(\theta)$ is given by the right-hand side of Equation (\ref{eq_waiting_queue}). We study how $f(\theta)$ varies with $\theta$ for different load factors, i.e., for different $\beta$ values. For the range of parameters we considered, we observe that \( f(\theta) \) decreases as \(\theta\) increases, as shown in Figure \ref{fig:expected_queue_theta}.

Recall that we restrict attention to the parameter regime $\theta_{2} > \theta_{1}$ throughout this section. (Of course, this can be viewed as a particular labeling of classes in the 2-class system of Figure \ref{2dim_model} without loss of generality). 

\begin{figure}[H]
  \centering
  \captionsetup{position=bottom} 
  \captionsetup[subfigure]{labelformat=empty}
  \subfloat[$f(\theta_{2})$ vs. $\theta_{2}$ (Underloaded)]{\includegraphics[width=0.45\linewidth]{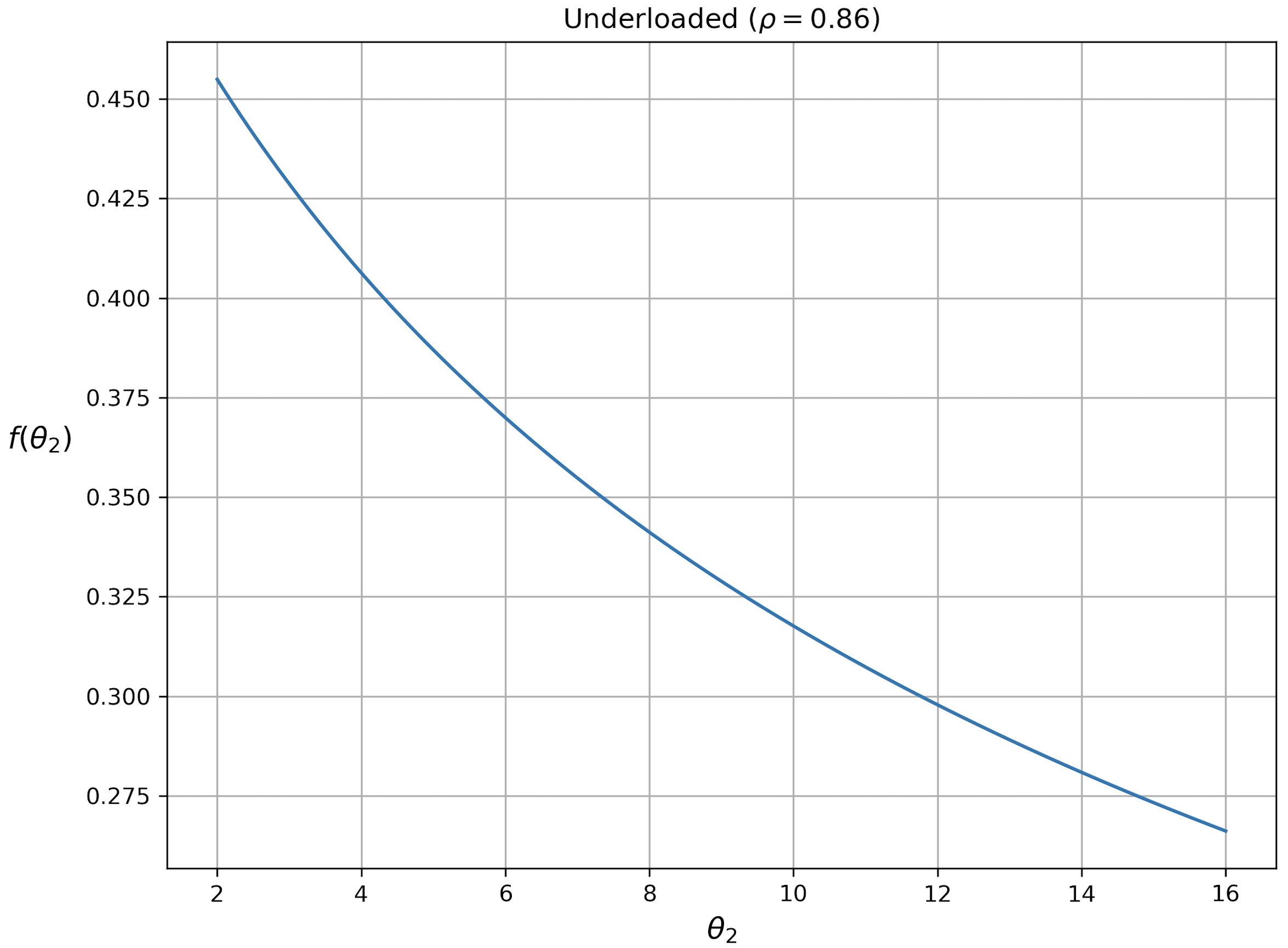}}\hfill
  \subfloat[$\bar{G}$ vs. $\theta_{2}$ (Underloaded)]{\includegraphics[width=0.45\linewidth]{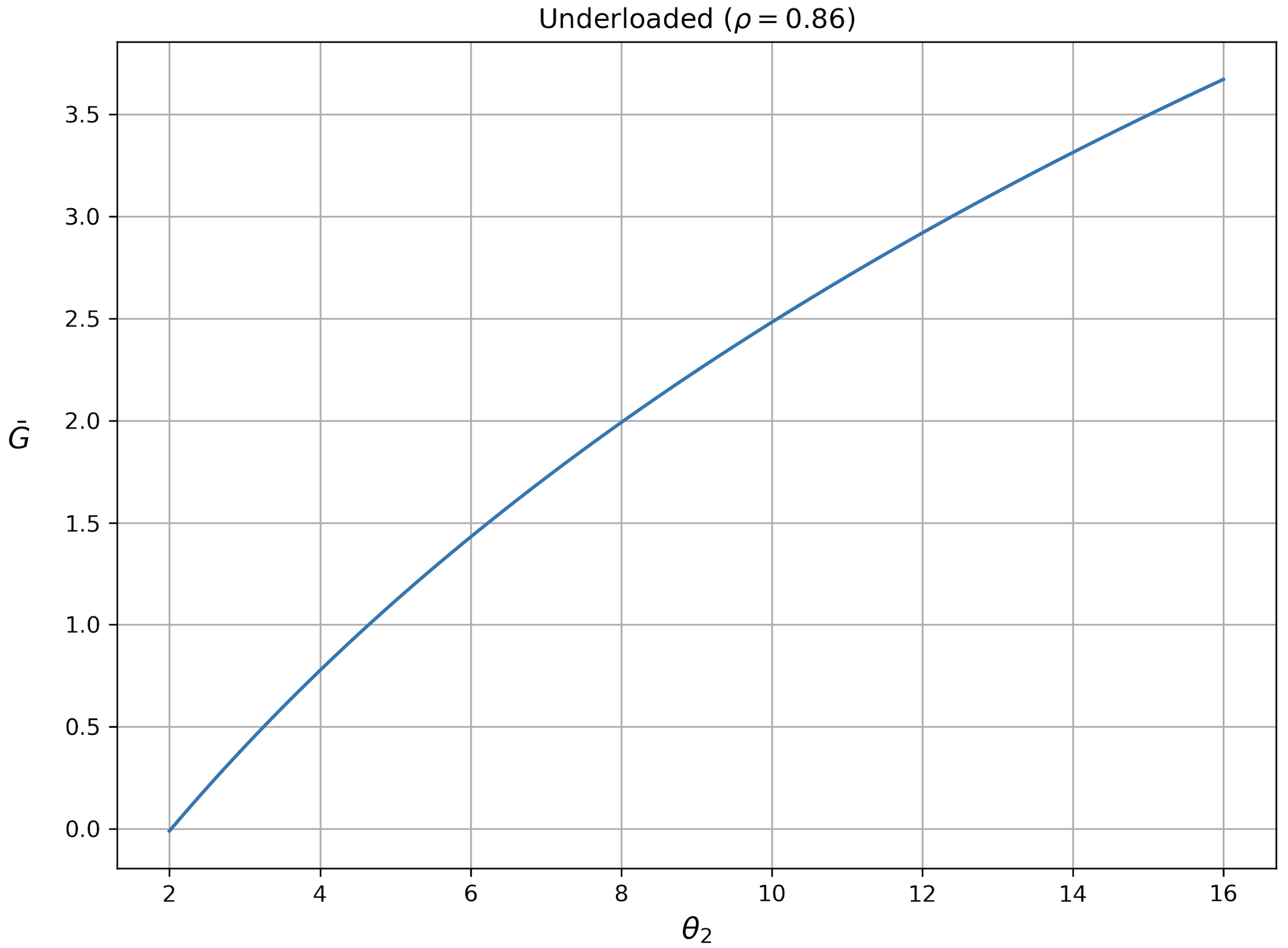}}\\
\caption{The graphs on the left show how $f(\theta_{2})$ decreases as $\theta_{2}$ increases for the overloaded, critically loaded and underloaded systems. The graphs on the right show how the optimality gap $\bar{G}$ increases as $\theta_{2}$ increases for the overloaded, critically loaded, and underloaded systems.}
 \label{f_underloaded}
\end{figure}

Let us reconsider the 2-class queueing system displayed in Figure \ref{2dim_model} and focus further on the parameter regime
\begin{equation}
    \frac{c_{1}}{\theta_{1}} > \frac{c_{2}}{\theta_{2}}.
\end{equation}
Because we assumed $\mu_{1} = \mu_{2}$, the $c\mu/\theta$ rule prioritizes class 1 in this case. Under the $c\mu/\theta$ rule, we attach the superscript of `$*$' to various quantities of interest to emphasize their dependence on the $c\mu/\theta$ rule. Because class 1 is prioritized, one expects to have
\begin{equation*}
    \mathbb{E}[Q^{*}_{1}] \approx 0 \,\, \text{ and } \,\, \mathbb{E}[Q^{*}_{2}] \approx f(\theta_{2}).
\end{equation*}
Thus, the total expected system cost, denoted by $\mathcal{C}^{*}$, is given by 
\begin{equation*}
    \mathcal{C}^{*} = c_{2}f(\theta_{2}).
\end{equation*}
Now, consider the alternative scheduling policy that prioritizes class 2 (the superscript `a' that will be attached to the quantities of interest under this alternative policy): We have
\begin{equation*}
    \mathbb{E}[Q_{2}^{a}] \approx 0 \,\, \text{ and } \,\, \mathbb{E}[Q_{1}^{a}] \approx f(\theta_{1}),
\end{equation*}
and the total expected system cost, denoted by $\mathcal{C}^{a}$, is given by 
\begin{equation*}
    \mathcal{C}^{a} = c_{1}f(\theta_{1}). 
\end{equation*}
If $\mathcal{C}^{a} < \mathcal{C}^{*}$ (i.e., $c_{1}f(\theta_{1}) < c_{2}f(\theta_{2})$), then the $c\mu/\theta$ rule isn't optimal, which we rewrite as 
\begin{equation}
    \frac{c_{2}}{c_{1}} > \frac{f(\theta_{1})}{f(\theta_{2})} > 1, \label{cost_compare}
\end{equation}
where the second inequality follows because $\theta_{2} > \theta_{1}$ and $f(\cdot)$ is decreasing. Of course, this is based on numerical observations; see Figure \ref{fig:expected_queue_theta}.
When Equation (\ref{cost_compare}) holds, the relative optimality gap of the $c\mu/\theta$ rule is 
\begin{equation}
    G(\theta_{1}, \theta_{2}) = \frac{\mathcal{C}^{*} - \mathcal{C}^{a}}{\mathcal{C}^{a}} = \frac{c_{2}f(\theta_{2}) - c_{1}f(\theta_{1})}{c_{1}f(\theta_{1})} = \frac{c_{2}/c_{1}}{f(\theta_{1})/f(\theta_{2})} - 1.
\end{equation}
Recall from (\ref{cost_compare}) that $c_{2}/c_{1} < \theta_{2}/\theta_{1}$. Thus, the optimality gap is given by 
\begin{equation}
    \bar{G} = \sup_{0 < \theta_{1} < \theta_{2}} \frac{\theta_{2}/\theta_{1}}{f(\theta_{1})/f(\theta_{2})} - 1.
\end{equation}
In words, the optimality gap is large when $\theta_{2}/\theta_{1}$ is large while $f(\theta_{1})/f(\theta_{2})$ is not. Our key observation is that for a fixed $\theta_{1}$, as $\theta_{2}$ increases so does 
\begin{equation}
\frac{\theta_{2}/\theta_{1}}{f(\theta_{1})/f(\theta_{2})} = \frac{\theta_{2}}{\theta_{1}}\cdot\frac{f(\theta_{2})}{f(\theta_{1})}. \label{bar_g}
\end{equation}
That is, as $\theta_{2}$ increases, the queue length $f(\theta_{2})$ does not decrease at the same rate across all regimes. Specifically, for all regimes, the rate of change in $\theta_{2}$ is greater than the rate of change in $f(\theta_{2})$, consistent with the fact that $f(\theta_{2})$ is convex in $\theta_{2}$. Consequently, as $\theta_{2}$ increases, $\theta_{2}f(\theta_{2})$ increases; see Figure \ref{f_underloaded} for how $f(\theta_{2})$ and $\bar{G}$ change as $\theta_{2}$ increases. Here, we fix $\theta_{1}$ = 2.
\newpage
\begin{figure}[!htb]
   \centering
  \captionsetup{position=bottom} 
  \captionsetup[subfigure]{labelformat=empty}
  \subfloat[$f(\theta_{2})$ vs. $\theta_{2}$ (Overloaded)]{\includegraphics[width=0.45\linewidth]{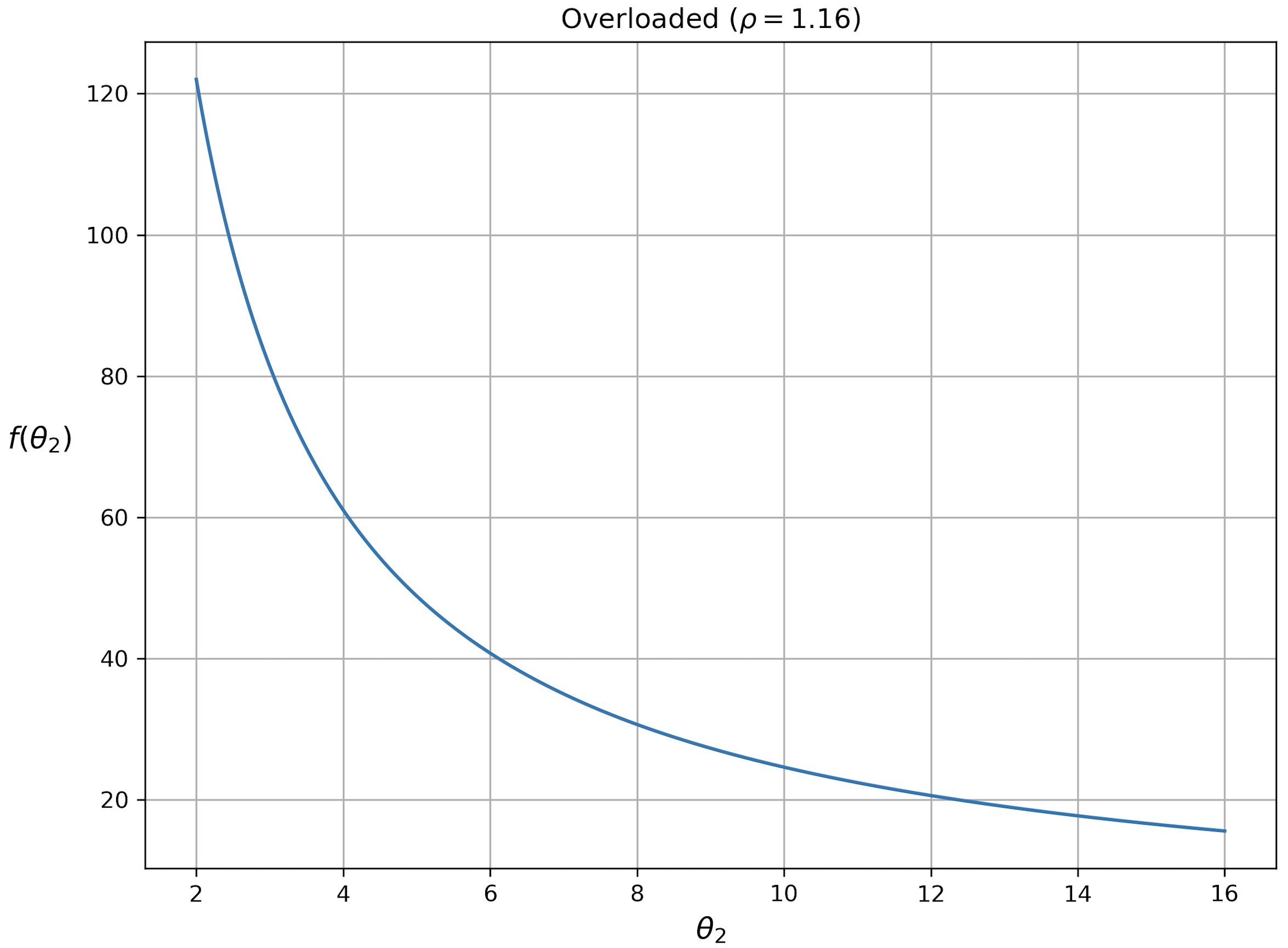}}\hfill
  \subfloat[$\bar{G}$ vs. $\theta_{2}$ (Overloaded)]{\includegraphics[width=0.45\linewidth]{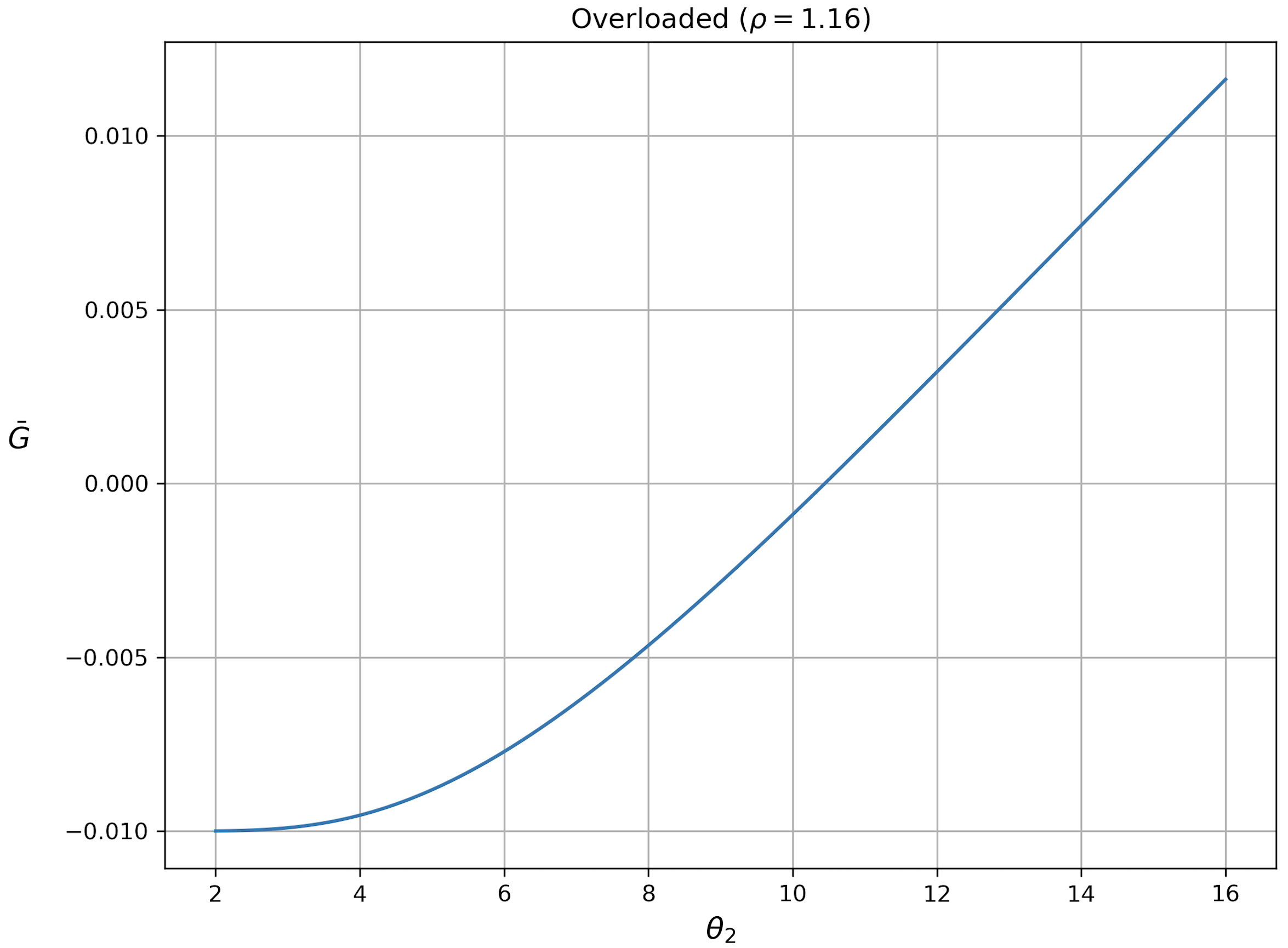}}\\
 \label{f_overloaded}
\end{figure}
\vspace{-7mm}
\begin{figure}[H]
  \centering
  \captionsetup{position=bottom} 
  \captionsetup[subfigure]{labelformat=empty}
  \subfloat[$f(\theta_{2})$ vs. $\theta_{2}$ (Critically Loaded)]{\includegraphics[width=0.45\linewidth]{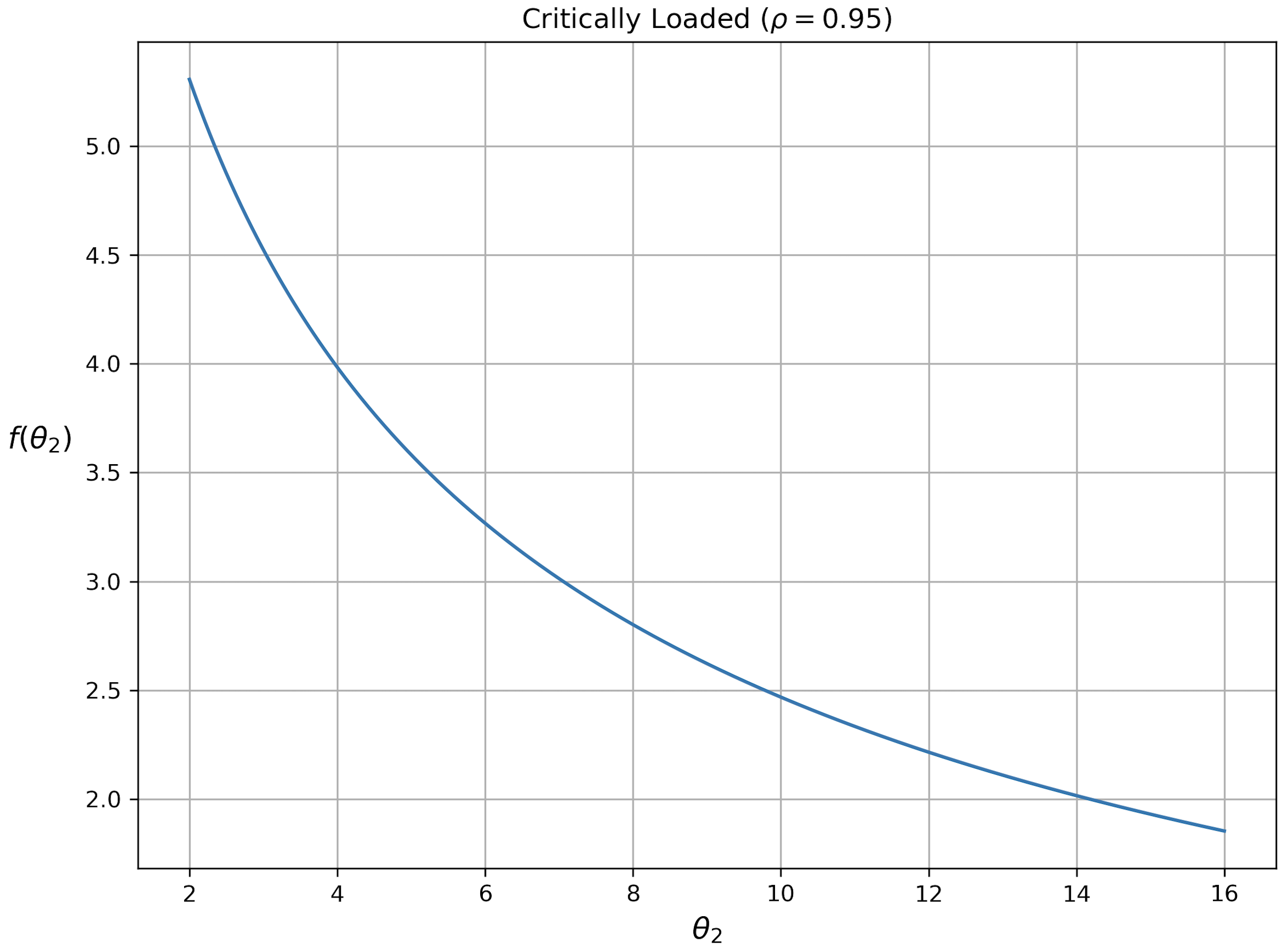}}\hfill
  \subfloat[$\bar{G}$ vs. $\theta_{2}$ (Critically Loaded)]{\includegraphics[width=0.45\linewidth]{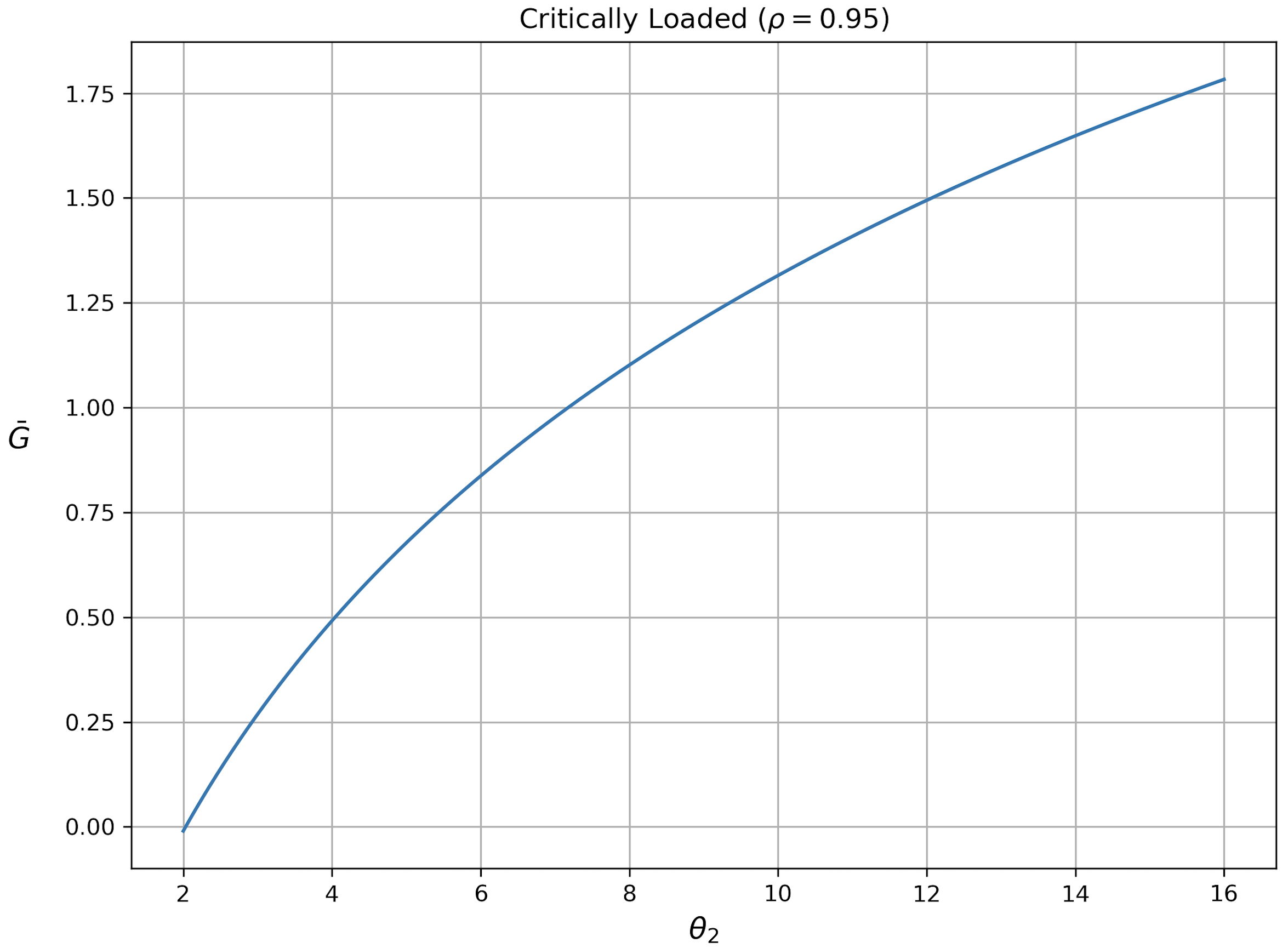}}\\
\label{f_critical}
\vspace{-10mm}
\end{figure}
\vspace{-4mm}
Figure \ref{f_underloaded} shows that the optimality gap $\bar{G}$ is negligible when the system is overloaded. However, this gap increases as the system utilization decreases. Notably, we observe substantial optimality gaps for critically loaded and underloaded systems as $\theta_{2}$ increases. To explore this observation further,
we plot heatmaps showing the optimality gap $\bar{G}$ for different values of $(\theta_{1}, \theta_{2}/\theta_{1})$ for the overloaded, critically loaded and underloaded systems calculated using \citet{garnett2002designing} and a simulation study.
\vspace{-6mm}
\begin{figure}[!htb]
   \centering
  \captionsetup{position=bottom} 
  \captionsetup[subfigure]{labelformat=empty}
  \subfloat[\citet{garnett2002designing}]{\includegraphics[width=0.48\linewidth]{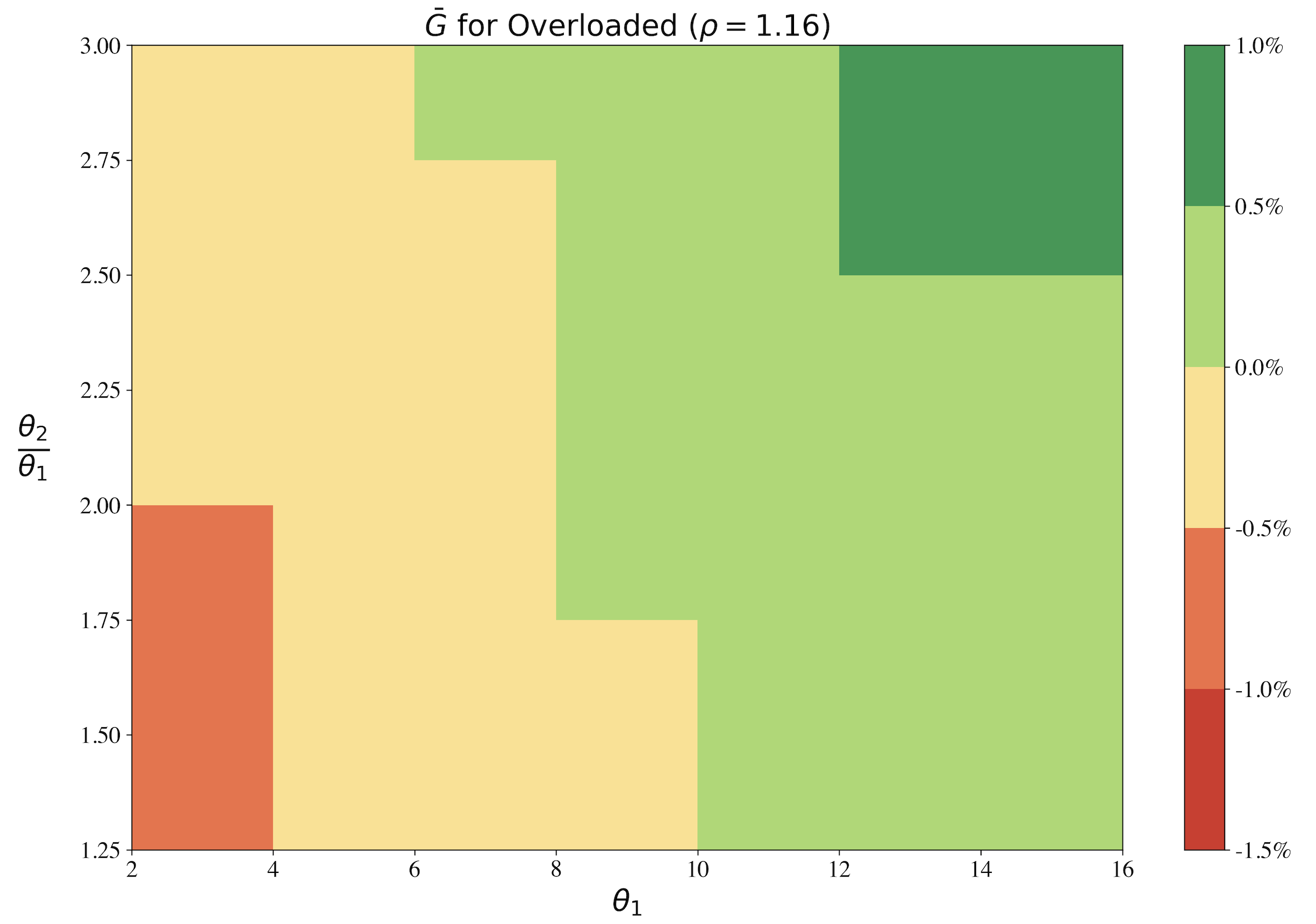}}\hfill
  \subfloat[Simulation]{\includegraphics[width=0.48\linewidth]{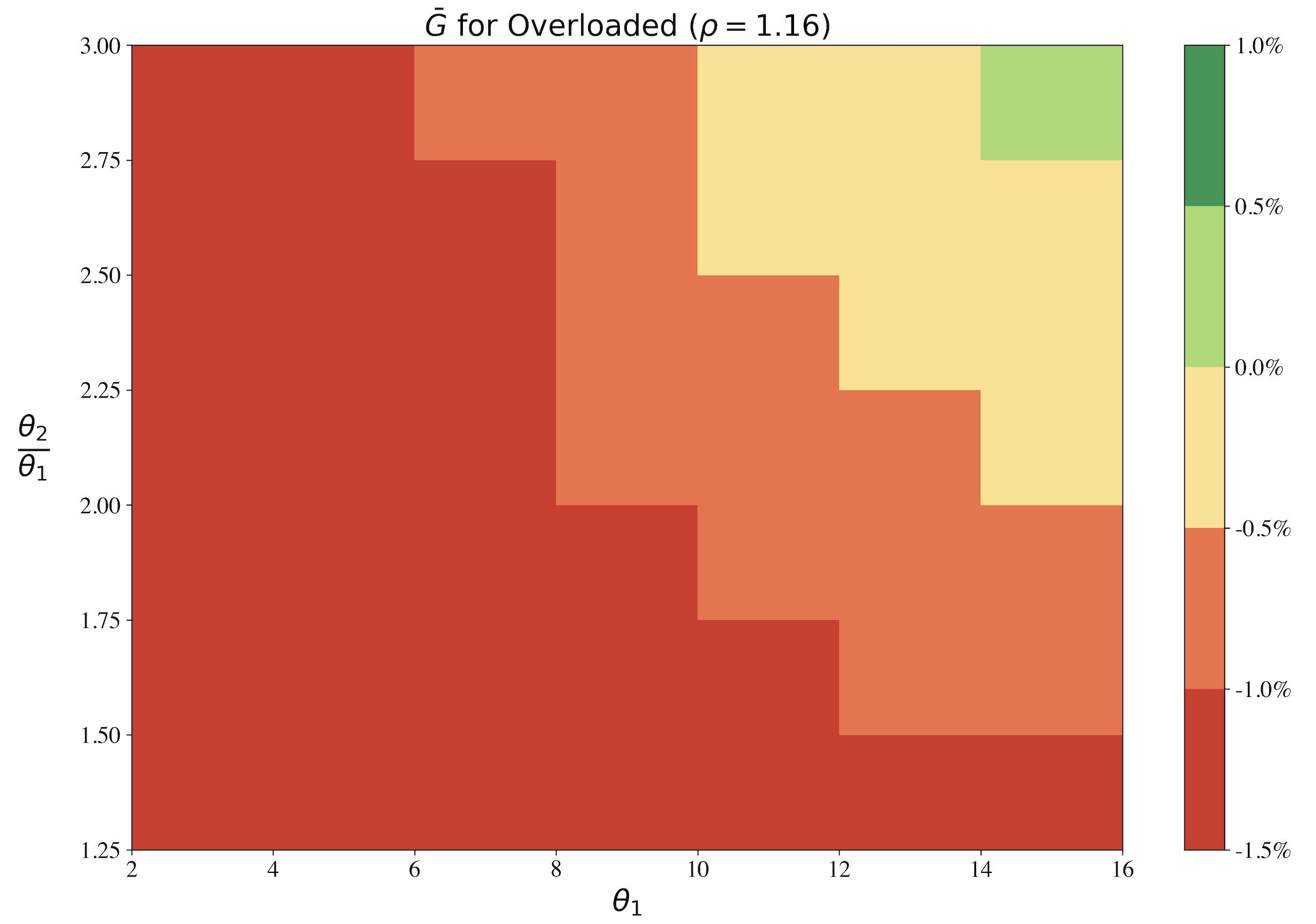}}\\
 \label{overloaded_2dim}
\end{figure}
\vspace{-5mm}
\begin{figure}[H]
   \centering
  \captionsetup{position=bottom} 
  \captionsetup[subfigure]{labelformat=empty}
  \subfloat[\citet{garnett2002designing}]{\includegraphics[width=0.48\linewidth]{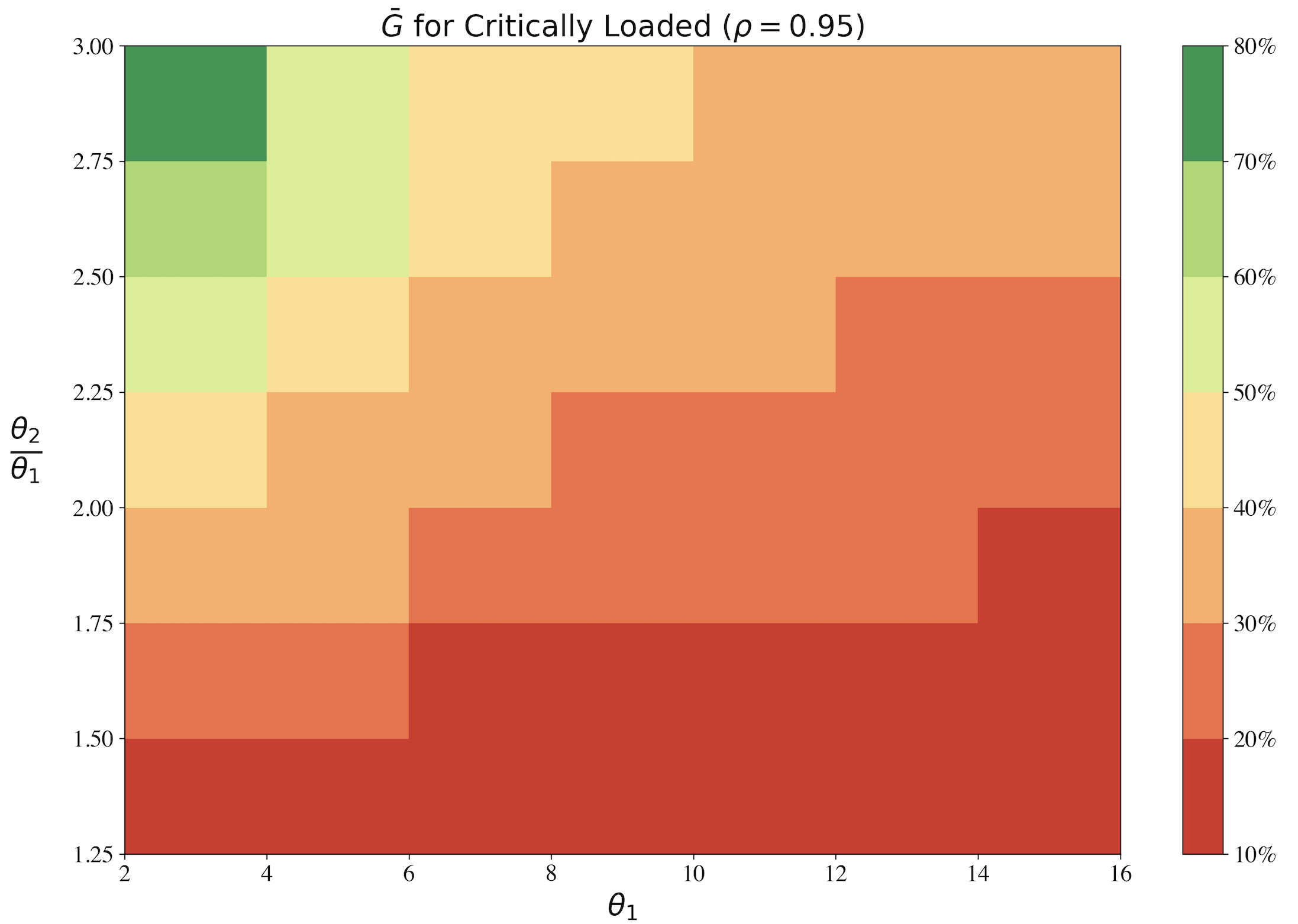}}\hfill
  \subfloat[Simulation]{\includegraphics[width=0.48\linewidth]{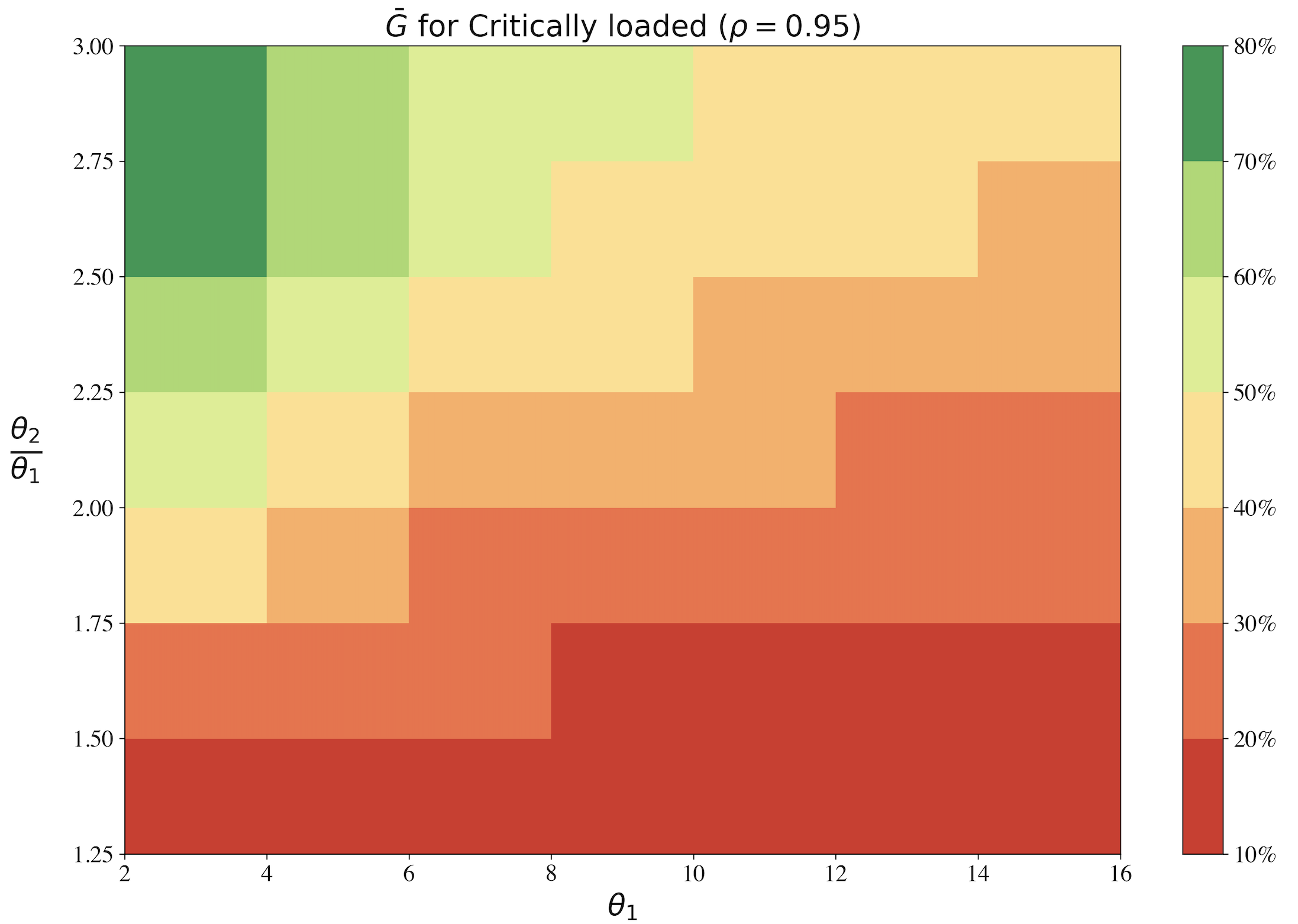}}\\
\label{heavy_2dim_heatmap}
\vspace{-7mm}
\end{figure}
\begin{figure}[H]
   \centering
  \captionsetup{position=bottom} 
  \captionsetup[subfigure]{labelformat=empty}
  \subfloat[\citet{garnett2002designing}]{\includegraphics[width=0.48\linewidth]{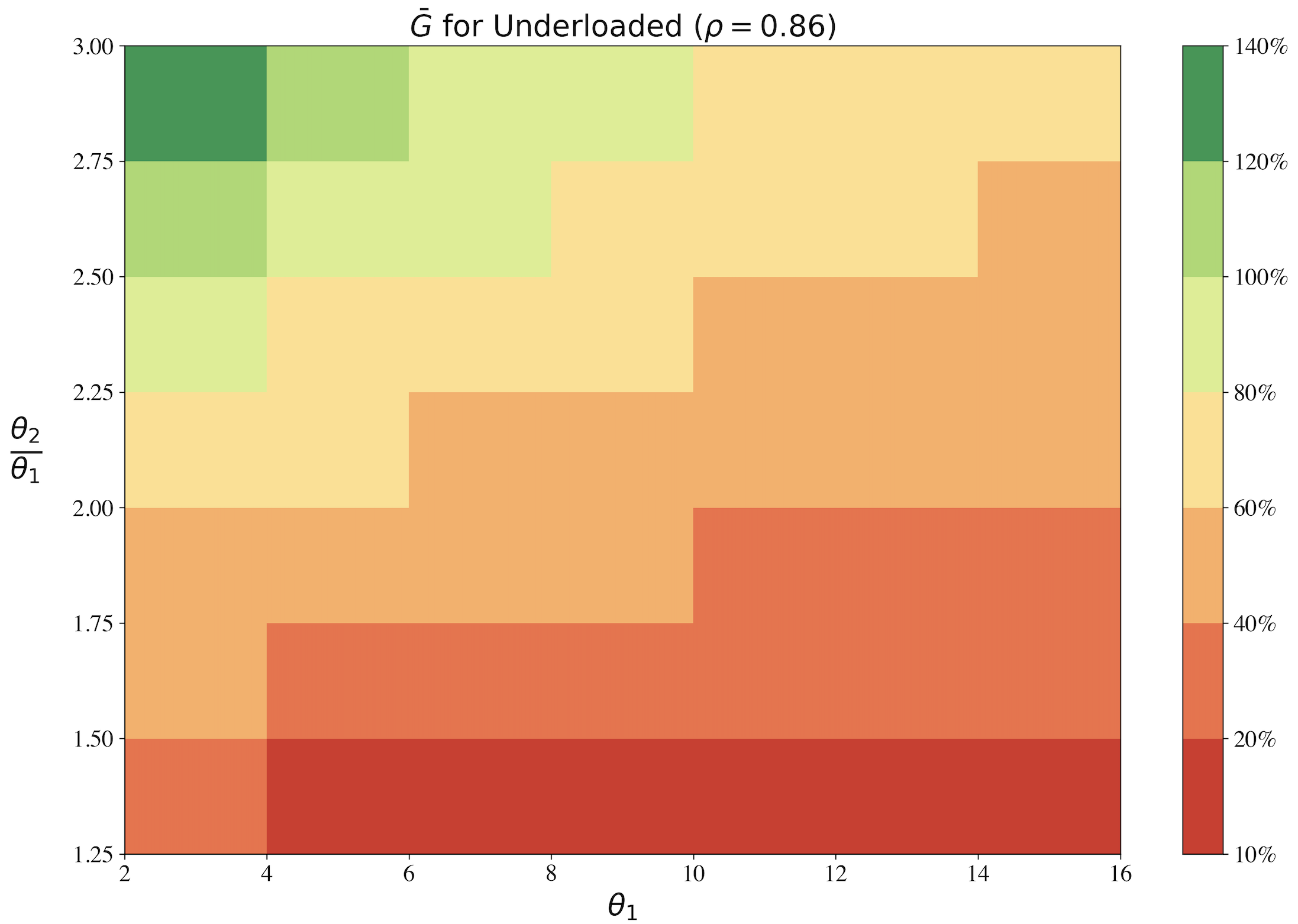}}\hfill
  \subfloat[Simulation]{\includegraphics[width=0.48\linewidth]{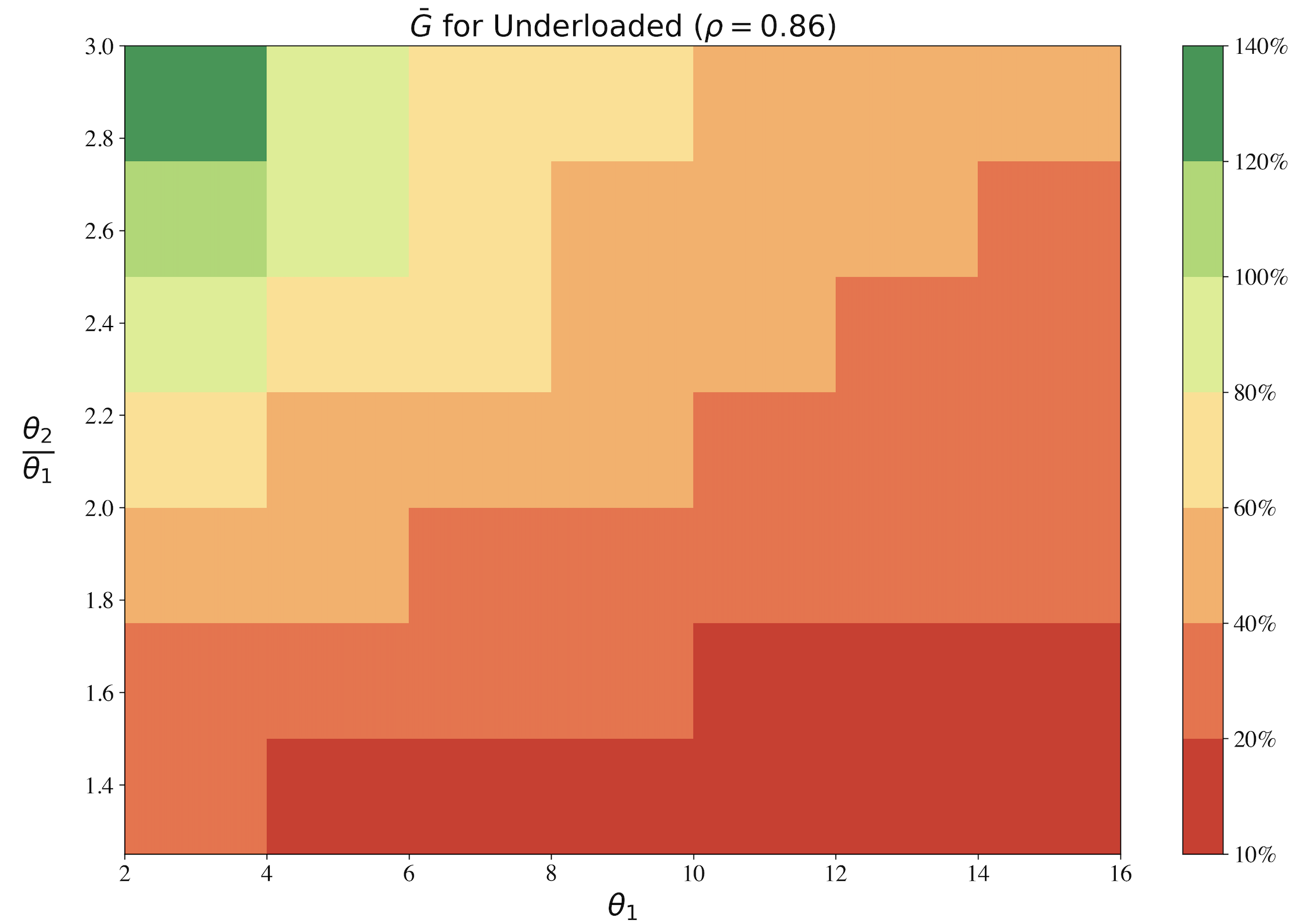}}\\
\caption{Heatmaps showing the optimality gap $\bar{G}$ for different values of $(\theta_{1}, \theta_{2}/\theta_{1})$ in the overloaded, critically loaded and underloaded systems.}
 \label{underloaded_2dim_heatmap}
\end{figure}
The graphs on the left column of Figure \ref{underloaded_2dim_heatmap} are based on the approximations of \citet{garnett2002designing}, whereas those on the right are obtained via simulation\footnote{We simulate the stationary system displayed in Figure \ref{2dim_model} using 10,000 instances and calculate the expected steady-state queue length.}. Figure \ref{underloaded_2dim_heatmap} shows that keeping $\theta_{1}$ fixed, the optimality gap increases in $\theta_{2}$. However, it is negligible in the overloaded regime as suggested by the asymptotic optimality result of \citet{atar2010cmu}; see the top row of Figure \ref{underloaded_2dim_heatmap}. (Recall that \citet{atar2010cmu} proves that the $c\mu/\theta$ rule is asymptotically optimal in the overloaded regime.) In contrast, if the system is critically loaded or underloaded, then we observe large optimality gaps when $\theta_{1}$ is small and $\theta_{2}/\theta_{1}$ is large; see the top-left corners of the four graphs in the middle and bottom rows of Figure \ref{underloaded_2dim_heatmap}.
The insights we glean from the simple two-class example considered above are twofold: 
\begin{enumerate}
    \begin{spacing}{1.5} 
    \item For $\theta_{1}$ small and $\theta_{2}$ large, the optimality gaps are large for critically loaded and underloaded systems.
    \item As the system utilization $(\rho)$ decreases, the optimality gap increases.
    \end{spacing}
\end{enumerate}
\vspace{-6mm}
Building on these insights, we design two variants of the main test problem introduced in Section \ref{sect_main_test}.

\subsubsection{Two variants of the main test problem}
\label{variants_design}

For the main test problem, we consider various benchmark policies introduced in Section \ref{static_benchmarks_main_variant} and compare their performance through a simulation study; see \ref{appendix_static_priority} for a performance summary. Notably, we observe the $c\mu/\theta$ performs well among all benchmark policies considered for this problem instance. To understand the reasons for this strong performance, we investigate the system dynamics throughout the day.

Using the estimated parameters from the data, we simulate the system and analyze the average traffic intensity; see Figure \ref{fig:traffic-intensity}. We observe that the traffic intensity exceeds 1 between 7 AM and 9 AM. For the rest of the day, it remains below 1: ranging between 0.9 and 1 from 9 AM to 2 PM and 9 PM to 12 AM, and dropping below 0.9 from 2 PM to 9 PM.

We also analyze when queueing costs occur and how they accumulate throughout the day. Figure \ref{fig:grh3} illustrates the percentage contributions to total costs over time. Drawing insights from Figures \ref{fig:traffic-intensity} and \ref{fig:grh3}, we segment the day into four time periods based on traffic intensity and cost contributions, as outlined in Table \ref{split}.

\begin{table}[H]
	\centering
	\setlength\tabcolsep{4pt} 
	{\small 
             \scalebox{0.9}{
		\begin{tabular}{lccccccc}
			\toprule
			Time Period && Avg. Traffic intensity && Avg. Contribution to total cost (\%)\\
			\midrule
			7 AM -- 9 AM && 1.17 && 76\%\\
                9 AM -- 2 PM && 0.93 && 9\%\\
			2 PM  -- 9 PM && 0.84 && 1\%\\
			9 PM -- 12 AM && 0.91 && 14\%\\
			\bottomrule 
		\end{tabular}
	}
 }
        \caption{The average traffic intensity and percentage contribution to the total cost aggregated into 4 time periods that show similar system dynamics.}
	\label{split}
\end{table}

We find that 76\% of the total cost occurs between 7 AM and 9 AM, when the system is overloaded. Meanwhile, 23\% of the total cost is incurred between 9 AM and 2 PM, and between 9 PM and 12 AM, when the system is critically loaded. The remaining 1\% of the cost is incurred between 2 PM and 9 PM, when the system is underloaded. We know from \citet{atar2010cmu} that the $c\mu/\theta$ rule is asymptotically optimal for overloaded systems. Because 76\% of the total cost is accumulated from 7 AM to 9 AM during which the system is overloaded, it is natural to expect the $c\mu/\theta$ rule to perform well for this system.

To better understand how $c\mu/\theta$ rule performs for a broader set of parameters, we consider a system with utilization level of 0.95. This utilization level is maintained throughout the day by adjusting the staffing levels of the original system.

To maintain the utilization at the target level of 0.95 consistently throughout the day, we denote the adjusted staffing level for the $n$-th interval as $\hat{N}(t_{n})$ and calculate it as  
\begin{equation}
\hat{N}(t_{n}) = \left\lceil \frac{\rho(t_{n})}{\rho^{T}} \, N(t_{n}) \right\rceil, \quad n = 0,\ldots, N-1, \label{adj_agents}
\end{equation}
where $\rho^{T}$ is the target system utilization and $\rho(t_{n})$ is the utilization of the original system during the $n$-th interval, as shown in Figure \ref{fig:traffic-intensity}. Figure \ref{fig:critically_loaded_agents} shows the adjusted staffing levels for the critically loaded system in comparison to the original system.

Building on the intuition from the analysis of 2-dimensional example in \ref{two_dim_numerical_study}, we focus on the classes that are expected to occupy the buffer more frequently. In doing so, we divide the classes into two groups: high-priority, denoted by $\mathcal{G}_{H}$ and low-priority, denoted by $\mathcal{G}_{L}$, based on their $c\mu/\theta$ order. We further divide the low-priority group into two subgroups, denoted by $\mathcal{G}_{L}^{\alpha}$ and $\mathcal{G}_{L}^{\beta}$, as shown in Table \ref{division}. These subgroups are determined such that the total call volume and mean service rates of the classes in low-priority subgroups $\mathcal{G}_{L}^{\alpha}$ and $\mathcal{G}_{L}^{\beta}$ are similar. Table \ref{division} lists the classes from left to right within each group according to their $c\mu/\theta$ order.

From the 2-dimensional example, we expect a large suboptimality gap for the $c\mu/\theta$ rule when it prioritizes classes with lower abandonment rates over those with higher abandonment rates. Therefore, we investigate how adjusting abandonment rates of classes across low-priority subgroups $\mathcal{G}_{L}^{\alpha}$ and $\mathcal{G}_{L}^{\beta}$ impacts the optimality gap. While doing so, we keep the $c\mu/\theta$ order of the classes the same as in the main test problem.

By Equation (\ref{eqn:defn:$c_{k}$:param}), the $c\mu/\theta$ ratio can be expressed as follows:
\begin{align}
    \frac{c_{k}\mu_{k}}{\theta_{k}} = \frac{h_{k}\mu_{k}}{\theta_{k}} + p_{k}\mu_{k}. \label{detached}
\end{align}
From Equation (\ref{detached}), it is clear that scaling both the abandonment rate $\theta_{k}$ and the holding cost rate $h_{k}$ by the same factor preserves the original $c\mu/\theta$ values shown in Table \ref{stats}, thereby maintaining the $c\mu/\theta$ order of the classes. Therefore, we consider the following parameter adjustments:
\begin{enumerate}
    \item We keep the parameters of the classes in the high-priority group $\mathcal{G}_{H}$ fixed.
    \item We decrease the abandonment rates $\theta_{k}$ and holding cost rates $h_{k}$ for $k \in$ $\mathcal{G}_{L}^{\alpha}$. To maintain the original $c\mu/\theta$ values from Table \ref{stats}, we scale $\theta_{k}$ and $h_{k}$ by a factor $\alpha \in (0,1]$.
    \item We increase the abandonment rates $\theta_{k}$ and holding cost rates $h_{k}$ for $k \in$ $\mathcal{G}_{L}^{\beta}$. To preserve the original $c\mu/\theta$
    values from Table \ref{stats}, we scale $\theta_{k}$ and $h_{k}$ by a factor $\beta \in [1, \infty)$.
\end{enumerate}
Table \ref{updated_stats} shows the adjustments to the original problem parameters for any $(\alpha, \beta)$ pair, ensuring that the original $c\mu/\theta$ values remain unchanged. 

As shown in Table \ref{updated_stats}, the $c\mu/\theta$ rule prioritizes the subgroups in the order \(\mathcal{G}_{H} > \mathcal{G}_{L}^{\alpha} > \mathcal{G}_{L}^{\beta}\). After scaling the parameters for the low-priority subgroups with $\alpha \in (0,1]$ and $\beta \in [1,\infty)$, classes in \(\mathcal{G}_{L}^{\alpha}\) are expected to have the lowest average abandonment rates and total cost rates. Drawing on insights from the 2-dimensional example, we anticipate that a static priority policy assigning the lowest priority to classes \(k \in \mathcal{G}_{L}^{\alpha}\) will outperform the $c\mu/\theta$ rule.

\begin{figure}[H]
    \centering
    \includegraphics[width=0.6\linewidth]{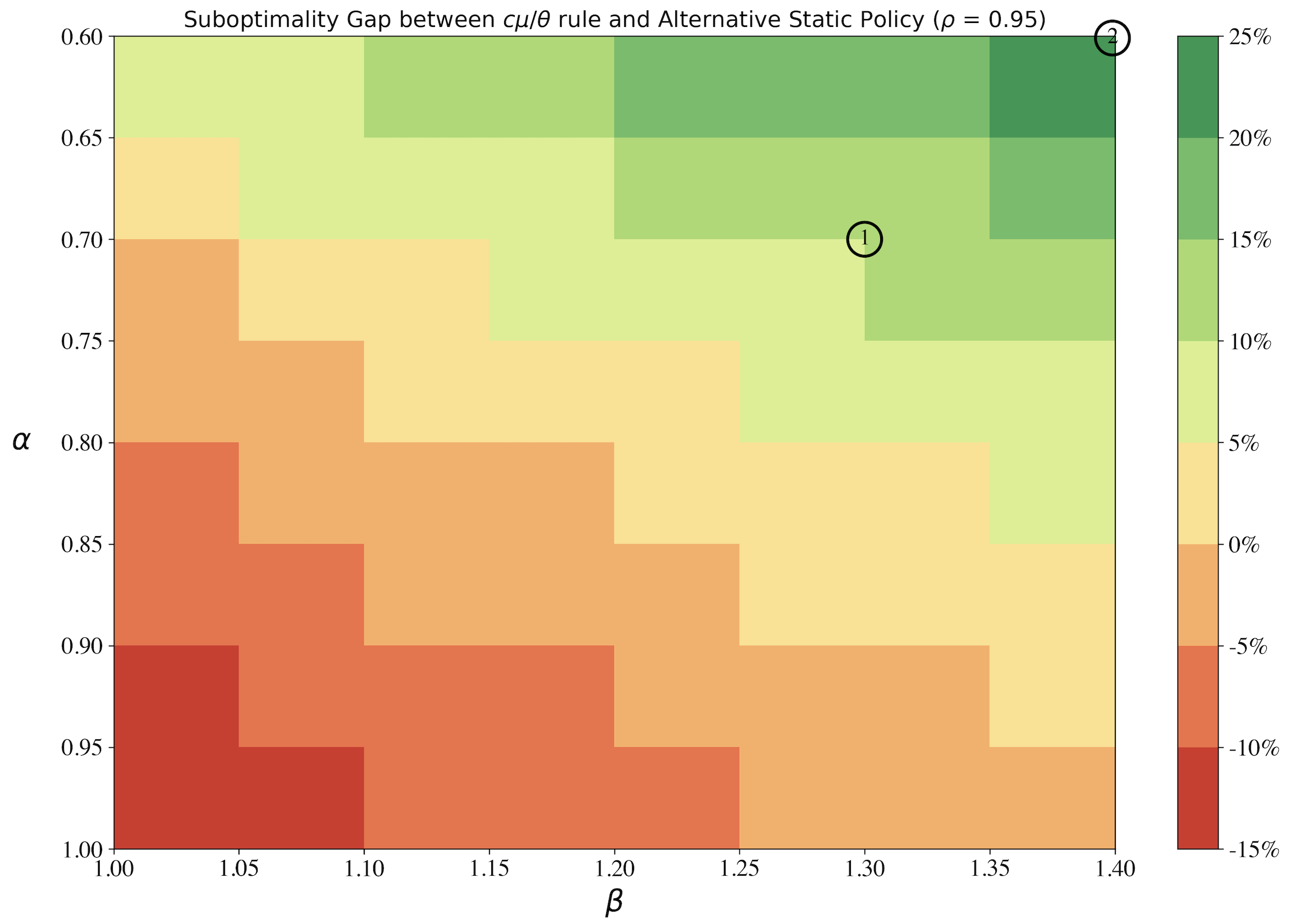}
    \caption{The performance gap between the $c\mu/\theta$ rule and the alternative static priority policy under a traffic intensity of 0.95, shown for different $(\alpha, \beta)$ pairs.}
    \label{fig:critically_loaded_instances}
\end{figure}

We then explore an alternative static priority policy that reverses the order of the low-priority subgroups, prioritizing them as \(\mathcal{G}_{H} > \mathcal{G}_{L}^{\beta} > \mathcal{G}_{L}^{\alpha}\). Within each subgroup, we maintain the $c\mu/\theta$ rule for class-level prioritization. Using a simulation study, we compare the performance of this alternative policy against the $c\mu/\theta$ rule. Figure \ref{fig:critically_loaded_instances} illustrates the performance gap between the two policies.
\vspace{-4.5mm}

As $\alpha$ decreases and $\beta$ increases, the abandonment rates ($\theta_{k}$) and total cost rates ($c_{k}$) of the classes in $\mathcal{G}_{L}^{\alpha}$ decrease while those in $\mathcal{G}_{L}^{\beta}$ increase. Figure \ref{fig:critically_loaded_instances} reveals a significant performance gap in critically loaded systems when $\alpha$ is small and $\beta$ is large, consistent with our insights from the 2-dimensional example.

Consequently, for the first variant of the main test problem, we choose \((\alpha, \beta) = (0.7, 1.3)\), marked as 1 in Figure \ref{fig:critically_loaded_instances}. In this case, the alternative priority policy, which reverses the priority order of the low-priority subgroups, outperforms the $c\mu/\theta$ rule by 10\%. For the second variant, we choose \((\alpha, \beta) = (0.6, 1.4)\), marked as 2 in Figure \ref{fig:critically_loaded_instances}. Here, the alternative priority policy exceeds the performance of the $c\mu/\theta$ rule by 20\%.

Next, building on the insights gained from designing the variant test examples, we introduce additional static priority benchmark policies for the main test problem and its two variants.

\subsubsection{Additional static priority policies for the main test problem and its two variants} \label{appendix_additional_static_benchmarks}

From Figure \ref{fig:critically_loaded_instances}, we observe that as \(\alpha\) decreases and \(\beta\) increases, abandonment rates (\(\theta_k\)) and total cost rates (\(c_k\)) decrease for classes in \(\mathcal{G}_{L}^{\alpha}\) and increase for those in \(\mathcal{G}_{L}^{\beta}\). Consequently, the alternative static priority rule, which ranks subgroups as \(\mathcal{G}_{H} > \mathcal{G}_{L}^{\beta} > \mathcal{G}_{L}^{\alpha}\), outperforms the $c\mu/\theta$ rule, which ranks subgroups as \(\mathcal{G}_{H} > \mathcal{G}_{L}^{\alpha} > \mathcal{G}_{L}^{\beta}\), in critically loaded systems. 

Building on this insight, we introduce two additional sets of static priority policies. In the first set, we rank the subgroups as $\mathcal{G}_{H} > \mathcal{G}_{L}^{\alpha} > \mathcal{G}_{L}^{\beta}$. In the second set, we reverse the ranking for the lower priority subgroups, ordering them as $\mathcal{G}_{H} > \mathcal{G}_{L}^{\beta} > \mathcal{G}_{L}^{\alpha}$. For these two sets of policies, we apply the five static priority policies introduced in Section \ref{static_benchmarks_main_variant} to each of the three subgroups, resulting in 120 additional benchmark policies for the main test problem and its two variants. See \ref{supp_tables_additional_static_benchmarks} for the performance tables of these additional static priority policies.

To summarize, the best-performing static priority policy ranks subgroups as 
$\mathcal{G}_{H} > \mathcal{G}_{L}^{\alpha} > \mathcal{G}_{L}^{\beta}$ 
for the main test problem and 
$\mathcal{G}_{H} > \mathcal{G}_{L}^{\beta} > \mathcal{G}_{L}^{\alpha}$ 
for its two variants. Table \ref{group_static_prioritization_additional_static} 
shows the static priority policies applied to each subgroup in the best-performing static benchmark. 
Notably, for the main test problem, the best-performing static policy considered here coincides with the $c\mu/\theta$ rule. 

\begin{table}[H]
	\centering
	\setlength\tabcolsep{4pt}
	{\footnotesize 
             \scalebox{1.0}{
		\begin{tabular}{lccccccccc}
			\toprule
			Group && Main  && First variant && Second variant \\
			\midrule
			$\mathcal{G}_{H}$ && $c\mu/\theta$ && $c\mu$ && $c\mu$\\[0.2em]
                $\mathcal{G}_{L}^{\alpha}$ && $c\mu/\theta$ && $c\mu$ && $c\mu$\\[0.2em]
                $\mathcal{G}_{L}^{\beta}$ && $c\mu/\theta$ && $c\mu/\theta$ && $c\mu/\theta$\\
			\bottomrule 
		\end{tabular}
	}
 }
\caption{Within-group static priority prioritization for the best-performing static priority benchmark.}
\label{group_static_prioritization_additional_static}
\end{table}

\subsubsection{Further benchmark policies derived from an auxiliary 3-dimensional MDP}\label{appendix_mdp_high_dim}
We design an auxiliary, low-dimensional MDP that focuses on the classes expected to occupy the buffer frequently. To construct this MDP, we first partition the classes into two groups: a high-priority group, denoted by $\mathcal{G}_{H}$, and a low-priority group, denoted by $\mathcal{G}_{L}$. For each test problem, this classification is determined by ranking the classes based on the best static priority policy, considering the five static priority policies introduced in Section \ref{computational_benchmarks} and their extensions in \ref{appendix_additional_static_benchmarks}. That is, for each test problem we consider 125 static priority policies. The top six ranked classes, which collectively account for approximately 65\% of the total call volume, form the high-priority group $\mathcal{G}_{H}$. Notably, these six classes remain consistent for the main test problem and its two variants, as shown in Table \ref{bench_subgroup_variant}. We assume these high-priority classes rarely occupy the queue so they are removed from the system. We also delete the service capacity needed to serve them. Therefore, we focus only on the remaining classes in the low-priority group $\mathcal{G}_{L}$.

Given our computational constraints, which limit us to solving at most three-dimensional MDPs, we further subdivide the low-priority group $\mathcal{G}_{L}$ into $I = 3$ subgroups, denoted by $\mathcal{G}_{L}^{i}$ for $i = 1,\ldots,I$. This procedure builds upon and extends the framework of the static priority policies outlined in \ref{appendix_additional_static_benchmarks}. In that framework, we divide the low-priority subgroup $\mathcal{G}_{L}$ into two subgroups $\mathcal{G}_{L}^{\alpha}$ and $\mathcal{G}_{L}^{\beta}$, as shown in Table \ref{division}. Since our MDP solution can accommodate three-dimensional systems, we expand this structure by introducing a third low-priority subgroup. In doing so, we ensure that each low-priority subgroup $\mathcal{G}_{L}^{i}$ contains at least three classes and serves a substantial call volume. To achieve this, we keep $\mathcal{G}_{L}^{\alpha}$ unchanged, as it already aggregates four classes and accounts for a significant call volume (\%16.36). Thus, we designate $\mathcal{G}_{L}^{\alpha}$ as the first low-priority subgroup, $\mathcal{G}_{L}^{1}$. 

Next, we divide $\mathcal{G}_{L}^{\beta}$ into two low-priority subgroups $\mathcal{G}_{L}^{2}$ and $\mathcal{G}_{L}^{3}$. This division follows the static priority rule used for within-group prioritization of $\mathcal{G}_{L}^{\beta}$ in \ref{appendix_additional_static_benchmarks}. Specifically, as shown in Table \ref{group_static_prioritization_additional_static}, the $c\mu/\theta$ rule is applied within $\mathcal{G}_{L}^{\beta}$, as it yields the best-performing static benchmark across the main test problem and its variants. Using this criterion, we assign the top-four ranked classes in group $\mathcal{G}_{L}^{\beta}$ to $\mathcal{G}_{L}^{2}$ and the remaining three classes to $\mathcal{G}_{L}^{3}$; see Table \ref{bench_subgroup_variant}.

This process defines the class groupings used to formulate and solve the MDP, which focuses on the low-priority classes in the main test problem and its two variants. Notably, the same grouping applies consistently for the main test problem and its two variants, as shown in Table \ref{bench_subgroup_variant}.

\vspace{2mm}
\begin{table}[H]
	\centering
	\setlength\tabcolsep{4pt} 
	{\footnotesize 
             \scalebox{0.9}{
		\begin{tabular}{llllcccccc}
			\toprule
			Group &    Classes \\
			\midrule
                 $\mathcal{G}_H$ & Retail (Node: 1, 2, 3), Business, Platinum, Consumer Loans\\[0.2em]
                 \midrule
                 $\mathcal{G}_{L}^{1}$  & Priority Service, CCO, Brokerage, BPS \\[0.2em]
                 $\mathcal{G}_{L}^{2}$ & Premier, Online Banking, AST, Subanco\\[0.2em]
                 $\mathcal{G}_{L}^{3}$  &  Telesales, EBO, Case Quality \\
			\bottomrule
		\end{tabular}
	}
 }
        \caption{The combination of original classes into high-priority group and low-priority subgroups for the auxiliary MDP designed as a benchmark policy for the main test problem and its two variant test examples.}
	\label{bench_subgroup_variant}
\end{table}

For the purpose of solving the 3-dimensional MDP, the mean service, abandonment, and cost rates of the low-priority subgroups $\mathcal{G}^{i}_{L}$ for $i = 1,2,3$ are calculated by taking the weighted averages of the rates associated with the corresponding classes within the subgroups; see Table \ref{stats_groups_main} for the main test problem, Table \ref{stats_groups_first_instance} for the first variant and Table \ref{stats_groups_second_instance} for the second variant. 
\begin{table}[H]
	\centering
	\setlength\tabcolsep{4pt}
	{\footnotesize 
             \scalebox{0.9}{
		\begin{tabular}{lccccccccc}
			\toprule
			Group & Arrival  & $\mu$ & $\theta$ & $c$ \\
			\rule{0pt}{3.5ex} & percentage(\%) &  (per hr) & (per hr) & (per hr)\\
			\midrule
			1 & 16.36 & 13.78 & 7.15 & \$36.56\\
                2 & 10.03 & 12.19 &  7.73 & \$38.52\\
                3 & 7.82 & 9.68 & 8.95 & \$37.72\\
			\bottomrule 
		\end{tabular}
	}
 }
\caption{Summary statistics for the three subgroups of the auxiliary MDP described as a benchmark policy for the main test example. The percentage (\%) column shows what fraction of the total arrivals to the call center belong to each of the groups.}
\label{stats_groups_main}
\end{table}
\begin{table}[H]
	\centering
	\setlength\tabcolsep{4pt}
	{\footnotesize 
             \scalebox{0.9}{
		\begin{tabular}{lccccccccc}
			\toprule
			Group & Arrival  & $\mu$ & $\theta$ & $c$ \\
			\rule{0pt}{3.5ex} & percentage(\%) &  (per hr) & (per hr) & (per hr)\\
			\midrule
                1 & 16.36 & 13.78 & 5.01 & \$25.59\\[0.2em]
			2 & 10.03 & 12.19 & 10.05 & \$50.08\\[0.2em]
                3 & 7.81 & 9.68 &  11.64 & \$49.05\\
			\bottomrule 
		\end{tabular}
	}
 }
\caption{Summary statistics for the three subgroups of the auxiliary MDP described as a benchmark policy for the first variant test example. The percentage (\%) column shows what fraction of the total arrivals to the call center belong to each of the groups.}
\label{stats_groups_first_instance}
\end{table}
\begin{table}[H]
	\centering
	\setlength\tabcolsep{4pt}
	{\footnotesize 
             \scalebox{0.9}{
		\begin{tabular}{lccccccccc}
			\toprule
			Group & Arrival  & $\mu$ & $\theta$ & $c$ \\
			\rule{0pt}{3.5ex} & percentage(\%) &  (per hr) & (per hr) & (per hr)\\
			\midrule
                1 & 16.36 & 13.78 & 4.29 & \$21.93\\[0.2em]
			2 & 10.03 & 12.19 & 10.82 & \$53.94\\[0.2em]
                3 & 7.81 & 9.68 &  12.53 & \$52.82\\
			\bottomrule 
		\end{tabular}
	}
 }
\caption{Summary statistics for the three subgroups of the auxiliary MDP described as a benchmark policy for the second variant test example. The percentage (\%) column shows what fraction of the total arrivals to the call center belong to each of the groups.}
\label{stats_groups_second_instance}
\end{table}
We define the system state of the auxiliary low-dimensional MDP at time $t$ as $X(t) = (X_{1}(t),X_{2}(t),X_{3}(t))$, where $X_{i}(t)$ denotes the number of calls from the subgroup $\mathcal{G}^{i}_{L}$ in the system. Recall that we assume the high-priority classes in the original problem rarely occupy the queue and are therefore removed from the low-dimensional system. Consequently, we remove the service capacity needed to serve these classes. Given the total number of agents serving all classes is $N(t)$, we define the number of agents serving only high-priority classes, denoted by $N_{H}(t)$ as follows (assuming high-priority calls are always in service): For $k = 1,\ldots, K$
\begin{equation*}
    N_{H}(t) = \left\lceil \sum_{k \in \mathcal{G}_{H}}\frac{\lambda_{k}(t)}{\mu_{k}(t)}\right\rceil, \quad t \in [0,T].
\end{equation*}
Then the number of agents serving only low-priority classes is defined as 
\begin{equation*}
    N_{L}(t) = \left(N(t) - N_{H}(t)\right)^{+}, \quad t \in [0,T].
\end{equation*}
Lastly, the control is $u(t,x) \in \tilde{\mathcal{U}}(t,x) = \{u \in \mathbb{R}_{+}^{I}: u\,(e\cdot x - N_{L}(t))^{+} \leq x, \, e \cdot u = 1\}$ where $u_{i}(t,x)$ is the fraction of the total backlog kept in subgroup $\mathcal{G}^{i}_{L}$ at time $t$ in state $x$. The associated Bellman equation is as follows:
\begin{align}
 \frac{\partial V}{\partial t} (t,x) &=  -\sum_{i=1}^{I}\lambda_{i}(t)\Delta_{i}^{-}(t,x+e_{i}) + \sum_{i=1}^{I}x_{i}\mu_{i}\Delta_{i}^{-}(t,x) \nonumber \\
&\quad  - (e \cdot x - N_{L}(t))^{+} \min_{u\in \tilde{\mathcal{U}}(t,x)}  \Bigg\{ \sum_{i=1}^{I}\Big(c_{i} + (\mu_{i} - \theta_{i})\Delta_{i}^{-}(t,x)\Big)u_{i}\Bigg\},
\label{eqn_hjb_high_dim}
\end{align}
where $c_{i}$ is the cost rate, $\mu_{i}$ is the service rate and $\theta_{i}$ is the abandonment rate of subgroup $\mathcal{G}^{i}_{L}$ for $i = 1,\ldots, I$; see Tables \ref{stats_groups_main} - \ref{stats_groups_second_instance}.

Given the solution to the Bellman equation (\ref{eqn_hjb_high_dim}), we define the approximate effective holding cost function for each subgroup $\mathcal{G}^{i}_{L}$ as follows:
\begin{equation*}
    \hat{\phi}_{i}(t,x) = c_{i} + (\mu_{i} - \theta_{i})\Delta_{i}^{-}(t,x), \quad i = 1,\ldots, I.\label{appendix_eff_$c_{k}$}
    \vspace{-2mm}
\end{equation*}
As in Section \ref{sect_hjb_equation}, we use the effective holding cost function $\hat{\phi}(\cdot)$ to order the subgroups from the most expensive to the cheapest. Specifically, the benchmark policy we propose is as follows: When assigning servers to callers, the system manager gives the highest priority to the high-priority group $\mathcal{G}_{H}$. Then if there is remaining service capacity, she prioritizes subgroups in the order induced by the approximate effective holding costs $\hat{\phi}(\cdot)$. While the (low-priority) subgroups are prioritized dynamically using the optimal policy given by the MDP solution. 

To establish a priority rule within each subgroup, we consider the five static priority policies described in Section \ref{static_benchmarks_main_variant}. Table \ref{priority_orders} provides the specific static policies used to prioritize classes within each subgroup for the best-performing auxiliary 3-dimensional MDP benchmark across the main test problem and its variants. This process generates 240 additional benchmark policies for the main test problem and its two variants. Their performance tables are provided in \ref{supp_tables_additional_mdp_benchmarks}.

\begin{table}[H]
	\centering
	\setlength\tabcolsep{4pt}
	{\footnotesize 
             \scalebox{1.0}{
		\begin{tabular}{lccccccccc}
			\toprule
			Group && Main && First variant && Second variant\\
			\midrule
                $\mathcal{G}_{H}$ && $c_{k}(\mu_{k}-\theta_{k})$ && $\mu_{k}-\theta_{k}$ && $c\mu/\theta$\\
                \\
			  $\mathcal{G}_{L}^{1}$ && $c\mu$ && $c\mu$ && $c\mu$ \\
                \\
                $\mathcal{G}_{L}^{2}$ && $c\mu/\theta$ && $c_{k}$ && $c_{k}$ \\
                \\
                $\mathcal{G}_{L}^{3}$ && $c\mu$ && $c\mu$ && $c\mu$ \\
			\bottomrule 
		\end{tabular}
	}
 }
\caption{The static priority policies used to prioritize classes within each subgroup for the best-performing auxiliary 3-dimensional MDP benchmark across the main test problem and its variants.}
\label{priority_orders}
\end{table}

\noindent \textbf{Truncating the state space.} To numerically solve the optimal policy for the low dimensional subgroups, we take the computational approach introduced in \ref{appendix_mdp_low_dim}. Therefore, to determine a suitable state-space truncation, we conducted a simulation study similar to the one described in \ref{appendix_mdp_low_dim}. Based on the simulation results, we chose 
$\{0,\ldots, 170\} \times \{0,\ldots,110\} \times \{0,\ldots, 100\}$ as the truncated state space for the main test problem and its two variants.

\subsubsection{Dynamic index policies}\label{appendix_dynamic_benchmarks}
Building on the intuition behind the policy proposed in Section \ref{sect_hjb_equation}, we propose three additional heuristics that approximate the effective holding cost function $\phi(\cdot)$ defined in Equation (\ref{eq_effective_$c_{k}$}). 

\noindent \textbf{Dynamic index heuristic 1.} We start by defining a partition $0 = t_{0} < t_{1} < \ldots < t_{N} = T$ of the time horizon $[0, T]$. For simplicity, we set $\Delta t_{n} = T/N$ for $n = 1,\ldots, N$, where we take $N = 204$, corresponding to 5-minute intervals. We then simulate the system using the best static priority rule considered in \ref{appendix_static_priority}\footnote{We use the $c\mu/\theta$ rule for the main test problem, the best static policy from Table \ref{first_instance} for the first variant and the best static policy from Table \ref{second_instance} for the second variant. These static policies are simulated using 10,000 replications.} and generate $L = 200$ discretized sample paths of the state process $X(t_{n}; l)$ for $n = 0,\ldots, N-1$ and $l = 1,\ldots,L$. 

Letting $x^{l} = $ $X(t_{n};l)$, we approximate the gradient of the value function as follows: For $l = 1,\ldots, L$, \, $k = 1,\ldots, K$ and $n = 0,\ldots, N-1$
\begin{equation}
    \frac{\partial V^{1}}{\partial x_{k}}(t_{n}, x^{l}) \approx \frac{1}{\mathcal{M}}\sum_{\mathcal{m}=1}^{\mathcal{M}} \left(\mathcal{C}^{m}(t_n, x_{k}^{l} + e_{k}) - \mathcal{C}^{m}(t_n, x_{k}^{l})\right), \label{gradient_approximation}
\end{equation}
where $\mathcal{M} = 50$ is the number of simulation replications using the best static priority rule that had the best performance in the earlier experiments and $\mathcal{C}^{\mathcal{m}}(t_{n},x)$ is the total cost starting in state $x$ at time $t_{n}$ for the $\mathcal{m}^{\text{th}}$ simulation replication. Equation (\ref{gradient_approximation}) provides an estimation of the gradient vector $\nabla_x V^{1}(t_{n}, x^{l})$ for $l = 1,\ldots,L$ and $n = 0,\ldots, N-1$. Given the discretized sample paths $X(t_{n};l)$ and the estimates of the gradient vector $\nabla_x V^{1} (t_{n}, x^{l})$, we approximate each gradient function $\nabla_{x}V(t_{n},\cdot)$ using a deep neural network $\hat{G}^{n}(\,\cdot \,;\hat{\nu}^{n})$ with parameter vector $\hat{\nu}_{n}$ for $n = 0,\ldots, N-1$. The hyperparameters used to train the deep neural network $\hat{G}^{n}(\,\cdot\,; \hat{\nu}^{n})$ for $n = 0,\ldots,N-1$ are shown in Table \ref{hyperparameter_dynamic_index}. The loss function for each neural network is defined as follows:
\begin{align*}
    \hat{\ell}^{n}(\hat{\nu}^{n}) = \frac{1}{L \cdot K}\sum_{l = 1}^{L}\sum_{k=1}^{K} \left(\frac{\partial V^{1}}{\partial x_{k}}(t_{n}, x^{l}) - \hat{G}_{k}^{n}(x^{l};\hat{\nu}^{n})\right)^{2}. 
\end{align*}
However, in order to avoid negative gradient approximations, we modify our loss function $\hat{\ell}^{n}(\hat{\nu}^{n})$ by adding a penalty term as shown in Algorithm \ref{method1}; also see \ref{appendix:penalty:neg:grad}. Given the neural network approximations $\hat{G}(\,\cdot \,,\hat{\nu}^{\star}_{n})$ for $n = 0,\ldots, N-1$, we approximate the effective holding cost function as follows: For $t \in [t_{n}, t_{n+1})$, $n = 0,\ldots, N-1$, $x\in \mathbb{R}_+^K$ and $k = 1,\ldots, K$,
\begin{equation}
    \hat{\phi}_{k}(t,x) = c_{k} + (\mu_{k} - \theta_{k}) \, \hat{G}_{k}^{n}(x;\hat{\nu}^{\star}_{n}).
\end{equation}
As done in Section \ref{sect_hjb_equation}, we use the approximate effective holding cost function $\hat{\phi}(\cdot)$ to order the classes, giving priority to classes with higher effective holding cost. 
\begin{algorithm}[H]    
	\caption{}
        \label{method1}
        \begin{minipage}{\textwidth}
	\begin{algorithmic}[1]
        \Statex \textbf{Input:} The number of iteration steps $M$, a learning rate $\alpha$, an optimization solver (SGD, ADAM, RMSProp, etc.), $L$ discretized sample paths of the state process $X(t_{n}; l)$ for $n = 0,\ldots, N-1$ and $l = 1,\ldots, L$ used as the training sample and penalty term $\Lambda$.
        \Statex \textbf{Output:} The approximation of the gradient function, $\hat{G}^{n}(\,\cdot\,;\,\hat{\nu}_{n})$, for $n = 0,\ldots, N-1$.
         \For {$m \leftarrow 0$ to $M -1$}
        \State Calculate the empirical loss \begin{align*}
    \hat{\ell}^{n}(\hat{\nu}^{n}) = \frac{1}{L \cdot K}\sum_{l = 1}^{L}\sum_{k=1}^{K} \left(\frac{\partial V^{1}}{\partial x_{k}}(t_{n}, x^{l}) - \hat{G}_{k}^{n}(x^{l};\hat{\nu}^{n})\right)^{2} + \Lambda \sum_{l =1}^{L}\sum_{k=1}^{K} \max\left(-\hat{G}_{k}^{n}(x^{l},\hat{\nu}^{n}),0\right).
\end{align*}
        \State Compute the gradient of the loss function $\nabla \hat{\ell}^{n}$ with respect to $\hat{\nu}^{n}$ and update $\hat{\nu}^{n}$
 \Statex \quad \, using the chosen optimizer solver. 
	\EndFor
    \State \textbf{return} Functions $\hat{G}^{n}(\,\cdot\,;\,\hat{\nu}_{n})$ for $n = 0,1,\ldots, N-1$.
	\end{algorithmic} 
        \end{minipage}
\end{algorithm}
\vspace{-2mm}
\noindent \textbf{Dynamic index heuristic 2.}
Using the samples $x^{l}$ and estimates of the gradient function $\frac{\partial V^{1}}{\partial x_{k}}(t_{n}, x^{l})$ for $l = 1,\dots,L$ and $n = 0,\ldots,N-1$ that are calculated for the preceding dynamic index heuristic, we have a total of $N\cdot L$ paired observations of time and state and denote them by $(t^{\mathcal{n}}, x^{\mathcal{n}})$ for $\mathcal{n} = 1,\ldots,\mathcal{N} = N\cdot L$. Using them, we train a deep neural network $\tilde{G}(t,x;\tilde{\nu})$ parameterized by $\tilde{\nu}$ to approximate the gradient function $\nabla_{x}V(t,x)$. We train the deep neural network $\tilde{G}(t,x;\tilde{\nu})$ using the hyperparameters listed in Table \ref{hyperparameter_dynamic_index}. Our loss function is defined as follows: 
\begin{equation}
    \tilde{\ell}(\tilde{\nu}) = \frac{1}{\mathcal{N}\cdot K}\sum_{\mathcal{n} = 1}^{\mathcal{N}}\sum_{k=1}^{K}\left(\frac{\partial V^{1}}{\partial x_{k}}(t^{\mathcal{n}}, x^{\mathcal{n}}) - \tilde{G}(t^{\mathcal{n}}, x^{\mathcal{n}};\tilde{\nu})\right)^{2}.
\end{equation}
However, to avoid negative gradient approximations, we add a penalty term as shown in Algorithm \ref{method2}. Given the optimal neural network parameters $\tilde{\nu}^{\star}$, we approximate the effective holding cost function as follows: For $(t,x) \in [0,T] \times \mathbb{R}_+^K$ and $k = 1,\ldots, K$,
\vspace{-2mm}
\begin{equation*}
    \hat{\phi}_{k}(t,x) = c_{k} + (\mu_{k} - \theta_{k})\,\tilde{G}_{k}(t, x;\tilde{\nu}^{\star}).
\end{equation*}
As in Section \ref{sect_hjb_equation}, we then use the approximate effective holding cost function $\hat{\phi}(\cdot)$ to order classes from the most expensive to the cheapest. 
\begin{algorithm}[H]    
	\caption{}
        \label{method2}
        \begin{minipage}{\textwidth}
	\begin{algorithmic}[1]
        \Statex \textbf{Input:} The number of iteration steps $M$, a learning rate $\alpha$, an optimization solver (SGD, ADAM, RMSProp, etc.) and $ \mathcal{N} = N \cdot L$ observations of the time and the state vector $X(t_{n}, x^{l})$ for $t = t_{n}$, $n = 0,\ldots, N-1$ and $l = 1,\ldots, L$ used as the training data, penalty term $\Lambda$.
        \Statex \textbf{Output:} The approximation of the gradient function, $\tilde{G}(\,\cdot\,;\,\tilde{\nu})$ for $(t,x) \in [0,T] \times \mathbb{Z}^{K}_{+}$
         \For {$m \leftarrow 0$ to $M -1$}
        \State Calculate the empirical loss \begin{equation*}
    \tilde{\ell}(\tilde{\nu}) = \frac{1}{\mathcal{N} \cdot K}\sum_{\mathcal{n} = 1}^{\mathcal{N}}\sum_{k=1}^{K}\left(\frac{\partial V^{1}}{\partial x_{k}}( t^{\mathcal{n}}, x^{\mathcal{n}}) - \tilde{G}(t^{\mathcal{n}}, x^{\mathcal{n}};\tilde{\nu})\right)^{2} + \Lambda \sum_{\mathcal{n}=1}^{\mathcal{N}} \sum_{k=1}^{K}\max\left(-\tilde{G}_{k}(t^{\mathcal{n}},x^{\mathcal{n}};\tilde{\nu}),0\right).
\end{equation*}
        \State Compute the gradient of the loss function $\tilde{\ell}(\tilde{\nu})$ with respect to $\tilde{\nu}$ and update $\tilde{\nu}$
 \Statex \quad \, using the chosen optimizer solver. 
	\EndFor
    \State \textbf{return} Function $\tilde{G}(\, \cdot\,;\,\tilde{\nu})$ for $(t,x) \in [0,T] \times \mathbb{Z}^{K}_{+}$.
	\end{algorithmic} 
        \end{minipage}
\end{algorithm}

\noindent \textbf{Implementation details of dynamic index heuristic 1 and 2.}

\noindent \textbf{Data generation.} To generate input for training the deep neural networks used in dynamic index heuristics 1 and 2, we set the number of intervals to $N = 204$ and generate $L = 200$ discretized sample paths of the state process $X(t_{n}, l)$ for $n = 0,\ldots, N-1$ and $l = 1,\ldots, L$. The generation of the state process $X(t_{n},l)$ for $n = 0,\ldots,N-1$ and $l = 1,\ldots,L$ takes approximately one hour on an AMD EPYC 7502 32c/64t CPU with 80GB RAM. 

\noindent \textbf{Gradient estimations.} To generate estimates of the gradient of the value function, denoted by $\frac{\partial V^{1}}{\partial x_{k}}(t_{n}, x^{l})$, we use Equation (\ref{gradient_approximation}) while setting the number of simulation replications to $\mathcal{M} = 50$. In total, this step requires $N \cdot L \cdot K \cdot \mathcal{M}$ discrete event simulations\footnote{The simulation runs are distributed across 5 CPUs.}, which take approximately 6 days to complete using 5 AMD EPYC 7502 32c/64t CPUs, each equipped with 80 GB RAM.\\
\\
\noindent \textbf{Neural network architecture.} We implement the deep neural networks for dynamic index heuristics 1 and 2 with ReLU activation function (see \citet{rasamoelina2020review}) and Adam optimizer in Tensorflow 2 (see \citet{abadi2016tensorflow}). For dynamic index heuristic 1, we have 200 observations for each time interval $n = 0,\ldots, N-1$. We use 95\% of the observations to train the deep neural network $\hat{G}^{n}(\,\cdot\,;\hat{\nu}_{n})$ for $n = 0,\ldots, N-1$ and reserve 5\% of the observations to calculate the validation loss after every iteration. Similarly, for dynamic index heuristic 2, we have 40,800 observations. Of these, 95\% are used for training the deep neural network $\tilde{G}(t,x;\tilde{\nu})$, and the remaining 5\% are reserved for validation loss calculations after each iteration. The training of the neural networks takes about 5 hours for the dynamic index heuristic 1 and about 30 minutes for the dynamic index heuristic 2 using an AMD EPYC 7502 32c/64t CPU with 80 GB RAM. 
\begin{minipage}{\textwidth}
\centering
\begin{table}[H]
	\centering
	\setlength\tabcolsep{4pt} 
	{\small 
             \scalebox{0.9}{
		\begin{tabular}{lllllll}
			\toprule
			Hyperparameters && Dynamic index heuristic 1 && Dynamic index heuristic 2 \\
			\midrule
                Number of hidden layers\footnotemark && 2 && 2 \\
                \\
                Number of neurons per layer && 100 && 100\\
                \\
                Number of networks && 204 && 1\\
                \\
                Batch size && 190 && 38760\\
                \\
                Number of iterations && 5000 && 10000\\
                \\
                Number of epochs && 5000 && 10000\\
                \\
                Learning rate (iteration range) && 1e-3 (0, 200) && 1e-3 (0,1000) \\
                && 1e-4 (200, 1000) && 1e-4 (1000, 4000)\\
                && 5e-5 (1000, 3000) && 1e-5 (4000, 5000)\\
                && 1e-5 (3000, 5000) && 5e-6 (5000, 8000)\\
                && && 1e-6 (8000, 10000)\\
                Negative gradient penalty\footnotemark $\Lambda$ && 0.5 && 0.5\\
                \\
                Runtime && about 5 hours && about 30 minutes \\
			\bottomrule 
		\end{tabular}
	}
 }
 \caption{Summary of the hyperparameters used for dynamic index heuristics 1 and 2.} 
 \label{hyperparameter_dynamic_index}
\end{table}
\end{minipage}
\footnotetext[30]{We observe overfitting when considering three- and four-layer neural networks.}
\footnotetext[31]{We perform a brute-force search over the grid [0, 2] with increments of 0.5 and choose $\Lambda$ based on the lowest final training loss values.}

\noindent \textbf{Dynamic index heuristic 3.} We divide the classes into $I = 5$ subgroups\footnote{The classes are divided into 5 subgroups such that each subgroup has at least three classes. Specifically, we have the three subgroups of the low-priority classes as shown in Table \ref{bench_subgroup_variant} for the main test problem and its two variants. In addition, we divide the high-priority classes into two subgroups.}, denoted by $\mathcal{D}_{i}$ for $i = 1,\ldots, I$; see Table \ref{dynamic_index3_subgroup}. For the classes in each subgroup, we approximate the value function gradient as follows: 
\begin{equation*}
    \frac{\partial V}{\partial x_{k}} \approx a_{i} x_{k} \ \textrm{for} \  k \in \mathcal{D}_{i} \ \textrm{and} \ a_{i} \in \mathbb{R}_{+}
\end{equation*}
Using this approximation, we further approximate the effective holding cost function $\hat{\phi}(\cdot)$ defined in Equation (\ref{eq_effective_$c_{k}$}) as follows: For $k = 1,\ldots, K$, and $n = 0,\ldots,N-1$
\begin{equation}
    \hat{\phi}_{k}(t_{n},x) = c_{k} + (\mu_{k} - \theta_{k})\,a_{i}x_{k}, \quad k \in \mathcal{D}_{i}, \,\,\, i = 1,\ldots, I.
    \label{eqn:app:eff:hold:$c_{k}$:dynamic:index3}
\end{equation}
Then we use the approximate effective holding cost function $\hat{\phi}(\cdot)$ to define a dynamic index policy, giving priority to classes with higher effective holding cost. Using the estimates of the gradient function given by Equation (\ref{gradient_approximation}), we search for $a_{i}$ values over a grid $[\underline{a}_{i}, \bar{a}_{i}]$. Specifically, we let $B_{i} = \bigcup_{k \in \mathcal{D}_{i}} E_{k}$, where 
\begin{equation*}
    E_{k} = \left\{  \frac{\frac{\partial V^{1}}{\partial x_{k}}(t_{0}, x^{1})}{x_{k}^{1}},\ldots, 
    \frac{\frac{\partial {V}^{1}}{\partial x_{k}}(t_{N-1}, x^{L})}{x_{k}^{L}}  
    \right\} \ \textrm{for} \ k \in \mathcal{D}_{i} \ \textrm{and} \ x_{k}^{l} \neq 0.
\end{equation*}
Let $P_{10}(B_{i})$ and $P_{90}(B_{i})$ denote the 10$^{\text{th}}$ and 90$^{\text{th}}$ percentiles of the values in set $B_{i}$. We use these percentiles to calculate the bounds of the search grid we use as follows: For $i = 1,\ldots, I$,
\begin{align}
    \underline{a}_{i} = P_{10}(B_{i}) \text{  and  } \bar{a}_{i} = P_{90}(B_{i}).
    \vspace{-2mm}
\end{align}
The resulting bounds are shown in Table \ref{grid_bounds}. Using them, we conduct a brute-force grid search on the 5-dimensional set $ \Pi_{i=1}^5 [\underline{a}_{i}, \bar{a}_{i}]$ via simulation. Specifically, we choose five equidistant points within the bounds $\underline{a}_{i}$ and $\bar{a}_{i}$ for $i = 1 \ldots 5$. These points are defined as $x_{ij} = \underline{a}_{i} + (j-1) \left(\frac{\bar{a}_{i} - \underline{a}_{i}}{4}\right)$ for $j = 1 \ldots 5$ and $i = 1 \ldots 5$. For the simulation, we prioritize each class $k = 1,\ldots,K$ using the approximate effective holding cost function in Equation (\ref{eqn:app:eff:hold:$c_{k}$:dynamic:index3}). This search takes about a day on an AMD EPYC 7502 32c/64t CPU with 80GB RAM. We use OpenMP to enable parallelization on the multicore CPU.
\vspace{6mm}
\begin{table}[h]
	\centering 
	\setlength\tabcolsep{4pt} 
	{\footnotesize 
             \scalebox{0.9}{
		\begin{tabular}{lccccccccc}
			\toprule
			Grids && Main && First variant && Second variant\\
			\midrule
                $[\underline{a}_{1}, \bar{a}_{1}]$ && [0.000101, 0.206033] && [0.031827, 1.250023] && [0.043587, 1.245195]\\
                $[\underline{a}_{2}, \bar{a}_{2}]$ && [0.000079, 0.224205] && [0.029197, 0.450270] && [0.033059, 0.654193]\\
                $[\underline{a}_{3}, \bar{a}_{3}]$ && [0.000262, 0.429432] && [0.070618, 1.240783] &&[0.105670, 1.528904]\\
                $[\underline{a}_{4}, \bar{a}_{4}]$ && [0.000145, 0.323169] && [0.032800, 1.313165] &&[0.047232, 1.482964]\\
                $[\underline{a}_{5}, \bar{a}_{5}]$ && [0.000134, 0.270906] && [0.057799, 0.657157] && [0.060852, 0.800778]\\
			\bottomrule 
		\end{tabular}
	}
 }      \caption{The upper and lower bounds of the grid $[\underline{a}_{i}, \bar{a}_{i}]$ for $i = 1,\ldots,5$}
	\label{grid_bounds}
\end{table}
\begin{table}[H]
	\centering
	\setlength\tabcolsep{4pt} 
	{\footnotesize 
             \scalebox{0.9}{
		\begin{tabular}{llllcccccc}
			\toprule
			Group &  Classes \\
			\midrule
                 $\mathcal{D}_{1}$  & Retail (Node: 1, 3), Platinum\\[0.2em]
                 $\mathcal{D}_{2}$ & Retail (Node: 2), Business, Consumer Loans\\[0.2em]
                 $\mathcal{D}_{3}$  & Premier, Online Banking, Subanco, AST \\[0.2em]
                 $\mathcal{D}_{4}$  & EBO, Telesales, Case Quality \\[0.2em]
                 $\mathcal{D}_{5}$  & Priority Service, CCO, Brokerage, BPS\\[0.2em]
			\bottomrule
		\end{tabular}
	}
 }
        \caption{The combination of original classes into five subgroups designed to search for a class of dynamic benchmark policies for main test problem and its two variants.}
	\label{dynamic_index3_subgroup}
\end{table}
While we divide the classes into five subgroups, it is important to note that we do not need to conduct an extensive search using five static priority policies to determine the relative priority of the classes within each subgroup, as was done for the benchmarks in \ref{appendix_additional_static_benchmarks} and \ref{appendix_mdp_high_dim}. Instead, we estimate $a_{i}$ for each subgroup $\mathcal{D}_{i}$, where $i = 1, \ldots, 5$. Using these estimates, we dynamically determine the priority of each class $k \in \mathcal{D}_{i}$ based on the effective holding cost function in Equation (\ref{eqn:app:eff:hold:$c_{k}$:dynamic:index3}).
\subsubsection{Summary of benchmark policy performance}\label{appendix_static_priority}
Table \ref{comp_benchmark} reports the simulated performance of all the benchmark policies discussed in Section \ref{static_benchmarks_main_variant} for the main test problem and its two variants, including those of the aforementioned static priority policies. To be specific, we report the average total cost for each policy\footnote{We simulate the policies with 10,000 replications, using the same seed for each policy.}. Among the benchmark policies considered, the auxiliary 3-dimensional MDP solution performs best for the main test problem and its two variants.
\begin{table}[H]
	\centering 
	\setlength\tabcolsep{4pt} 
	{\small 
             \scalebox{0.9}{
		\begin{tabular}{lcccccccccccccccccc}
			\toprule
			Benchmark Policies &&&& Main &&&& First variant &&&& Second variant\\[0.2em]
			\midrule
                $c\mu/\theta$ rule  &&&& 1157.85 $\pm$ 5.91  &&&& 1624.04 $\pm$ 7.50  &&&& 1678.40 $\pm$ 7.73 \\[0.3em]
                \midrule
                $c\mu$ rule  &&&& 1258.57 $\pm$ 6.78  &&&& 1531.84 $\pm$ 8.75  &&&& 1443.37 $\pm$ 8.39 \\[0.3em]
                \midrule
                $c_{k}$ rule &&&& 1612.20 $\pm$ 9.24  &&&& 1685.20 $\pm$ 9.87 &&&& 1587.54 $\pm$ 9.50  \\[0.3em]
                \midrule
                $\mu_{k} - \theta_{k}$ rule &&&& 1201.18 $\pm$ 6.08  &&&& 1653.54 $\pm$ 7.55  &&&& 1709.15 $\pm$ 7.85 \\[0.3em]
                \midrule
                $c_{k}(\mu_{k} - \theta_{k})$ rule &&&& 1183.95 $\pm$ 5.98  &&&& 1685.72 $\pm$ 7.70  &&&& 1736.33 $\pm$ 7.87  \\[0.3em]
                \midrule
                Auxiliary MDP solution &&&& 1157.23 $\pm$ 5.88  &&&& 1467.59 $\pm$ 7.70  &&&& 1394.72 $\pm$ 7.68 \\[0.3em]
                \midrule
                Dynamic index heuristic 1 &&&&  1184.94 $\pm$ 6.07  &&&& 1548.07 $\pm$ 8.01  &&&& 1467.26 $\pm$ 7.93 \\[0.3em]
                \midrule
                Dynamic index heuristic 2 &&&& 1172.06 $\pm$ 6.06  &&&& 1552.28 $\pm$ 8.18  &&&& 1425.77 $\pm$ 7.97 \\[0.3em]
                \midrule
                Dynamic index heuristic 3 &&&& 1183.40 $\pm$ 6.05   &&&& 1532.61 $\pm$ 8.58 &&&& 1471.06 $\pm$ 8.42 \\[0.3em]
			\bottomrule 
		\end{tabular}
	}
 }
        \caption{The performance comparison of the benchmark policies considered for the main test problem and its two variants. We show the total cost $\pm$ half-length of the 99\% confidence interval for the benchmark policies.}
	\label{comp_benchmark}
\end{table}
\vspace{-10mm}
\section{Comparison of proposed policy with the benchmarks at the 95\% confidence level}
\label{comparison_95}
\vspace{-2mm}
\subsection{Low dimensional test problems}
\begin{table}[H]
    \centering
    \renewcommand{\arraystretch}{1.2}
    \setlength\tabcolsep{4pt} 
	{\small 
             \scalebox{0.9}{
    \begin{tabular}{lcccccccccccccccc}
        \toprule
        Method &&&& 2-Dimensional &&&& 3-Dimensional &&&& 3-Dimensional variant\\
        \midrule
        Our Policy &&&& 1682.84 $\pm$ 6.28  &&&& 1690.25 $\pm$ 6.38 &&&& 934.02 $\pm$ 3.60 \\
        Benchmark &&&& 1681.19 $\pm$ 6.26  &&&& 1682.67 $\pm$ 6.31  &&&& 934.02 $\pm$ 3.60 \\
        \midrule
        Optimality Gap &&&&  0.10\% $\pm$ 0.53\% &&&& 0.45\% $\pm$ 0.53\% &&&& 0.00\% $\pm$ 0.55\%\\
        \bottomrule
    \end{tabular}
    }
    }
    \caption{Performance comparison of our proposed policy with the benchmark policy in the low dimensional test problems. The first two rows show the total cost $\pm$ half-length of the 95\% confidence interval. Similarly, the last row shows the percentage optimality gap $\pm$ half-length of the 95\% confidence interval.}
    \label{results_lower_test_95}
\end{table}

\subsection{Main test problem and its variants}
\begin{table}[H]
    \centering
    \renewcommand{\arraystretch}{1.2}
    \setlength\tabcolsep{4pt} 
	{\small 
             \scalebox{0.9}{
    \begin{tabular}{lcccccccccccccccc}
        \toprule
          Method &&&& Main &&&& First variant &&&& Second variant\\
        \midrule
          Our Policy &&&& 1154.41 $\pm$ 4.50 &&&& 1470.06 $\pm$ 5.98 &&&& 1396.31 $\pm$ 5.93\\
          Benchmark &&&& 1157.23 $\pm$ 4.47 &&&& 1467.59 $\pm$ 5.86 &&&& 1394.72 $\pm$ 5.85\\
          \midrule
          Performance Gap &&&& -0.24\% $\pm$ 0.55\% &&&& 0.17\% $\pm$ 0.57\% &&&& 0.11\% $\pm$ 0.60\%\\
        \bottomrule
    \end{tabular}
    }
    \caption{Performance comparison of our proposed policy with the benchmark policy in the main test problem and its two variants. The first two rows show the total cost $\pm$ half-length of the 95\% confidence interval for our proposed policy and the best benchmark policy. The last row shows the percentage performance gap $\pm$ half-length of the 95\% confidence interval.}
    \label{results_main_test_95}
    }
\end{table}

\subsection{High dimensional test problems that have a pathwise optimal policy}
\begin{table}[H]
    \centering
    \renewcommand{\arraystretch}{1.2}
    \setlength\tabcolsep{4pt} 
	{\small 
             \scalebox{0.9}{
    \begin{tabular}{lcccccccccccccccccc}
        \toprule
        Method &&&& 30-Dimensional &&&& 50-Dimensional &&&& 100-Dimensional &&&& 500-Dimensional\\
        \midrule
        Our Policy &&&& 911.23 $\pm$ 4.88 &&&& 2428.70 $\pm$ 9.31 &&&& 11067.63 $\pm$ 16.51 &&&& 53340.77 $\pm$ 60.18\\
        Benchmark &&&& 908.05 $\pm$ 4.87 &&&& 2422.18 $\pm$ 9.29 &&&& 11062.99 $\pm$ 16.66 &&&& 53309.68 $\pm$ 59.50\\
        \midrule
        Optimality Gap &&&&  0.35\% $\pm$ 0.76\% &&&& 0.27\% $\pm$ 0.54\% &&&& 0.04\% $\pm$ 0.21\% &&&& 0.06\% $\pm$ 0.16\%\\
        \bottomrule
    \end{tabular}
    }
    \caption{Performance comparison of our proposed policy with the benchmark policy in the high dimensional test problems that have a pathwise optimal solution. The first two rows show the total cost $\pm$ half-length of the 95\% confidence interval for each case. Similarly, the last row shows the percentage optimality gap $\pm$ half-length of the 95\% confidence interval for each case.}
    \label{results_high_test_95}
    }
\end{table}
\subsection{High dimensional problems that do not admit a pathwise optimal policy}
\begin{table}[H]
    \centering
    \renewcommand{\arraystretch}{1.2}
    \setlength\tabcolsep{4pt} 
	{\small 
             \scalebox{0.84}{
    \begin{tabular}{lcccccccccccccccc}
        \toprule
           Method &&&& 100-Dimensional Main &&&& 100-Dimensional First Variant &&&& 100-Dimensional Second Variant\\
        \midrule
            Our Policy &&&& 1382.14 $\pm$ 6.77 &&&& 1342.77 $\pm$ 6.37 &&&& 1337.13 $\pm$ 6.23\\
            $c\mu/\theta$ rule  &&&& 1490.02 $\pm$ 6.72 &&&& 1458.34 $\pm$ 6.27 &&&& 1454.49 $\pm$ 6.16\\
            $c\mu$ rule  &&&& 1409.68 $\pm$ 7.05 &&&& 1415.07 $\pm$ 6.82 &&&& 1466.97 $\pm$ 7.31 \\
            $c_{k}$ rule &&&&  1441.79 $\pm$ 7.28 &&&&  1647.21 $\pm$ 8.68 &&&&  1812.00 $\pm$ 9.83\\
            $\mu_{k} - \theta_{k}$ rule &&&& 2009.27 $\pm$ 8.98 &&&&  1958.73 $\pm$ 8.40 &&&&  1953.13 $\pm$ 8.26  \\
            $c_{k}(\mu_{k} - \theta_{k})$ rule &&&& 1537.11 $\pm$ 6.95 &&&&  2308.48 $\pm$ 9.67 &&&&  2298.48 $\pm$ 9.54 \\
            \midrule
            Performance Gap &&&& -1.95\% $\pm$ 0.69\% &&&& -5.11\% $\pm$ 0.64\% &&&& -8.07\% $\pm$ 0.58\% \\
            \bottomrule
    \end{tabular}
    }
    \caption{Performance comparison of our proposed policy and the benchmark policies considered for the three 100-dimensional problems that do not admit a pathwise optimal solution. The first two rows show the total cost $\pm$ half-length of the 95\% confidence interval for our proposed policy and the best benchmark policy. The last row shows the percentage performance gap $\pm$ half-length of the 95\% confidence interval.}
    \label{results_100dim_non_pathwise_95confidence}
    }
\end{table}
\subsection{Summary of benchmark policy performance for the main test problem and its variants}
\begin{table}[H]
	\centering 
	\setlength\tabcolsep{4pt} 
	{\small 
             \scalebox{0.9}{
		\begin{tabular}{lcccccccccccccccccc}
			\toprule
			Benchmark Policies &&&& Main &&&& First variant &&&& Second variant\\[0.2em]
			\midrule
                $c\mu/\theta$ rule  &&&& 1157.85 $\pm$ 4.49  &&&& 1624.04 $\pm$ 5.71 &&&& 1678.40 $\pm$ 5.88\\[0.3em]
                \midrule
                $c\mu$ rule  &&&& 1258.57 $\pm$ 5.16 &&&& 1531.84 $\pm$ 6.66 &&&& 1443.37 $\pm$ 6.38\\[0.3em]
                \midrule
                $c_{k}$ rule &&&& 1612.20 $\pm$ 7.03 &&&& 1685.20 $\pm$ 7.51 &&&& 1587.54 $\pm$ 7.23 \\[0.3em]
                \midrule
                $\mu_{k} - \theta_{k}$ rule &&&& 1201.18 $\pm$ 4.62 &&&& 1653.54 $\pm$ 5.75 &&&& 1709.15 $\pm$ 5.97\\[0.3em]
                \midrule
                $c_{k}(\mu_{k} - \theta_{k})$ rule &&&& 1183.95 $\pm$ 4.55 &&&& 1685.72 $\pm$ 5.89 &&&& 1736.33 $\pm$ 5.99 \\[0.3em]
                \midrule
                Auxiliary MDP solution &&&& 1157.23 $\pm$ 4.47 &&&& 1467.59 $\pm$ 5.86 &&&& 1394.72 $\pm$ 5.85\\[0.3em]
                \midrule
                Dynamic index heuristic 1 &&&&  1184.94 $\pm$ 4.62 &&&& 1548.07 $\pm$ 6.09 &&&& 1467.26 $\pm$ 6.03\\[0.3em]
                \midrule
                Dynamic index heuristic 2 &&&& 1172.06 $\pm$ 4.61 &&&& 1552.28 $\pm$ 6.22 &&&& 1425.77 $\pm$ 6.07\\[0.3em]
                \midrule
                Dynamic index heuristic 3 &&&& 1183.40 $\pm$ 4.60 &&&& 1532.61 $\pm$ 6.53 &&&& 1471.06 $\pm$ 6.41\\[0.3em]
			\bottomrule 
		\end{tabular}
	}
 }
        \caption{The performance comparison of the benchmark policies considered for the main test problem and its two variants. We show the total cost $\pm$ half-length of the 95\% confidence interval for the benchmark policies.}
	\label{comp_benchmark_95}
\end{table}
\section{Implementation details of our computational method}\label{appendix_computational_method}

We implement our method using a fully connected deep neural network. The number of subnetworks ranges from 204 to 3060, depending on the specific test instance and the required level of discretization for training. We utilize Leaky ReLU and GELU as choices for the activation function, adapting code from the work of \citet{beck2021deep}. The implementation is carried out in Python (see \citet{van1995python}) using the PyTorch package (see \citet{paszke2019pytorch}).\\
\noindent \textbf{Neural network architecture.} Fully connected neural networks with 100 neurons per layer. The number of hidden layers ranges from two to five, depending on the test instance.

\noindent\textbf{Common Hyperparameters.} Initial state $x_{0} \sim U[-10,10]$; Batch size = 256; time horizon $T = 17$; the number of epochs is 1 and the number of iteration steps varies across different test cases, please see below for further detail.

\noindent\textbf{Optimizer.} We use the Adam optimizer and apply gradient clipping for four high-dimensional test instances that admit pathwise optimal solutions.

\noindent \textbf{Weighted-split reference policy.} For the main test problem and its two variants, we use the weighted-split reference policy. For the main test problem, we let $\mathcal{C} = \{ \textrm{Retail (Node: 1,2,3)} \}$ which constitutes $53.71 \%$ of all incoming calls. To reflect this percentage, we set $(w_1,w_2) = (0.18, 0.03285)$. On the other hand, setting $(w_1,w_2) = (1/17, 1/17)$ corresponds to the evenly split reference policy. In addition to these two, we consider four additional weight vectors defined as their convex combinations:
\vspace{-5mm}
\begin{equation*}
    \alpha \,(0.18,0.03285) + (1- \alpha) (1/17,1/17) \ \textrm{for} \ \alpha \in \{0, 1/6, 2/6, 3/6, 4/6, 5/6, 1 \}.
    \vspace{-5mm}
\end{equation*}
For the variant test problems, we redefine $\mathcal{C} = \{ \textrm{Retail (Nodes: 1,2,3), Consumer Loans, CCO} \}$ which constitutes $68.97 \%$ of all incoming calls. To represent this proportion, we set $(w_1,w_2) = (0.14, 0.025)$. Similarly, setting $(w_1,w_2) = (1/17, 1/17)$ corresponds to the evenly split reference policy. As in the main test problem, we include four additional weight vectors defined as convex combinations:
\vspace{-5mm}
\begin{equation*}
    \alpha \,(0.14,0.025) + (1- \alpha) (1/17,1/17) \ \textrm{for} \ \alpha \in \{0, 1/6, 2/6, 3/6, 4/6, 5/6, 1 \}.
    \vspace{-5mm}
\end{equation*}

\subsection{Hyperparameters used for the test problems}\label{hyperparameters}
All the neural networks in this section were executed on two high-performance computing clusters. The first cluster was equipped with NVIDIA V100 GPUs, each with 16 GB of memory, and Intel Gold 6248R CPUs. The second cluster featured NVIDIA A100 HGX GPUs, each with 160 GB memory, and AMD EPYC Milan CPUs. The job was allocated 50 GB of RAM and 8 CPU cores per task. The runtime results reported in this section correspond to the V100 system. However, the A100 GPUs reduced runtime by about 10-20\% due to the enhanced performance of their architecture.
\begin{table}[H]
\centering
\setlength\tabcolsep{6pt}
\renewcommand{\arraystretch}{1.2}
{\small
\scalebox{0.8}{
\begin{tabular}{lccc}
\toprule
\textbf{Hyperparameters} & \textbf{2D} & \textbf{3D} & \textbf{3D Variant} \\
\midrule
Number of hidden layers         & 2              & 3              & 4 \\
Number of neurons per layer     & 100            & 100            & 100 \\
Number of networks $N$        & 204            & 204            & 1020 \\
$\Delta T = T/N$  & 5 min         & 5 min         & 1 min \\
Batch size                      & 256            & 256            & 256 \\
\midrule
Number of iterations (final interval) & 2000           & 4000           & 5000 \\
Number of iterations (rest of intervals)      & 2000           & 2000           & 2000 \\
\midrule
Reference policy                & $u(t,x) = (0,1)$   & $u(t,x) = (0,1,0)$   & $u(t,x) = (1,0,0)$   \\
Decision frequency & 5 min         & 5 min         & 1 min \\
Learning rate (final interval)  & $1\text{e-3}[0,1000], 1\text{e-4}[1000,2000]$ 
                                & $1\text{e-3}[0,1000], 1\text{e-4}[1000,4000]$ 
                                & $1\text{e-3}[0,1000], 1\text{e-4}[1000,5000]$ \\
Learning rate (rest of intervals) & $1\text{e-3}[0,1000], 1\text{e-4}[1000,2000]$ 
                                & $1\text{e-3}[0,1000], 1\text{e-4}[1000,1500],$ 
                                & $1\text{e-3}[0,1000], 1\text{e-4}[1000,2000]$ \\
                                && $1\text{e-5}[1500,2000]$ & \\
Negative gradient penalty $\Lambda$ & 0.1            & 0.001            & 0.5 \\
\midrule
Early stopping                  & Yes (100 patience iter)   & Yes (250 patience iter)   & Yes (100 patience iter) \\
Gradient clipping               & Not applied          & Not applied          & Not applied \\
Activation function             & Leaky ReLU $(\alpha = 0.2)$ & Leaky ReLU ($\alpha = 0.01$) & Leaky ReLU ($\alpha = 0.2$) \\
Initialization                  & Kaiming              & Kaiming              & Kaiming \\
Runtime                         & $\sim$10 min              &  $\sim$10 min          & $\sim$15 min \\
\bottomrule
\end{tabular}
}}
\caption{Summary of the hyperparameters used for low-dimensional test problems.}
\label{hyperparameter_low}
\end{table}

\begin{table}[H]
\centering
\setlength\tabcolsep{6pt}
\renewcommand{\arraystretch}{1.2}
{\small
\scalebox{0.75}{
\begin{tabular}{lccc}
\toprule
\textbf{Hyperparameters} & \textbf{Main} & \textbf{First Variant} & \textbf{Second Variant} \\
\midrule
Number of hidden layers         & 4              & 5              & 5 \\
Number of neurons per layer     & 100            & 100            & 100 \\
Number of networks $N$        & 1020            & 3060            & 3060 \\
$\Delta T = T/N$  & 1 min        & 20 sec         & 20 sec \\
Batch size                      & 256            & 256            & 256 \\
\midrule
Number of iterations (final interval) & 5000           & 5000           & 5000 \\
Number of iterations (rest of intervals)      & 2000           & 2000           & 2000 \\
\midrule
Reference policy                & weighted split   & weighted split   & weighted split   \\
Decision frequency & 1 min        & 5 min         & 5 min \\
Learning rate (final interval)  & $1\text{e-3}[0,1000], 1\text{e-4}[1000,5000]$ 
                                & $1\text{e-3}[0,550], 1\text{e-4}[550,1200],$ 
                                & $1\text{e-3}[0,250], 1\text{e-4}[250,500],$ \\
& & $1\text{e-5}[1200,2500], 1\text{e-6}[2500,5000]$ & $1\text{e-5}[500,1000], 1\text{e-6}[1000,5000]$\\
\\
Learning rate (rest of intervals) & $1\text{e-3}[0,1000], 1\text{e-4}[1000,2000]$ 
                                & $1\text{e-3}[0,250], 1\text{e-4}[250,500],$
                                & $5\text{e-4}[0,1000], 5\text{e-5}[1000,2000]$\\
                                && $1\text{e-5}[500,2000]$&\\
                                \\
Negative gradient penalty \(\Lambda\) & 0.5            & 0.75            & 0.5 \\
\midrule
Early stopping                  & Yes (100 patience iter)   & Yes (120 patience iter)   & Yes (100 patience iter) \\
Gradient clipping               & Not applied          & Not applied          & Not applied \\
Activation function             & Leaky ReLU ($\alpha = 0.01$) & GELU & Leaky ReLU ($\alpha = 0.4$) \\
Initialization                  & Kaiming              & Xavier              & Kaiming \\
Runtime                         & $\sim$15 min              & $\sim$1 hour           & $\sim$45 min \\
\bottomrule
\end{tabular}
}}
\caption{Summary of the hyperparameters used for main test problem and its two variants.}
\label{hyperparameter_main}
\end{table}

\begin{table}[H]
\centering
\setlength\tabcolsep{6pt}
\renewcommand{\arraystretch}{1.2}
{\small
\scalebox{0.75}{
\begin{tabular}{lcccc}
\toprule
\textbf{Hyperparameters} & \textbf{30D} & \textbf{50D} & \textbf{100D} & \textbf{500D} \\
\midrule
Number of hidden layers         & 4              & 4              & 4 & 4 \\
Number of neurons per layer     & 100            & 100            & 100 & 100 \\
Number of networks $N$        & 1020            & 1020            & 1020 & 1020\\
$\Delta T = T/N$  & 1 min         & 1 min         & 1 min & 1 min \\
Batch size                      & 256            & 256            & 256 & 256 \\
\midrule
Number of iterations (final interval) & 5000           & 5000           & 5000 & 5000 \\
Number of iterations (rest of intervals)      & 2000           & 2000           & 2000 & 2000 \\
\midrule
Reference policy                & minimal   & minimal   & minimal   & minimal  \\
Decision frequency & 1 min         & 1 min         & 5 min & 5 min \\
Learning rate (final interval)  & $1\text{e-3}[0,1000],$ 
                                & $1\text{e-3}[0,1000],$ 
                                & $1\text{e-3}[0,1000],$ & $1\text{e-3}[0,1000],$ \\
& $1\text{e-4}[1000,5000]$ &  $1\text{e-4}[1000,5000]$ &  $1\text{e-4}[1000,5000]$ &  $1\text{e-4}[1000,5000]$\\
\\
                                
Learning rate (rest of intervals) & $1\text{e-3}[0,1000],$ 
                                & $1\text{e-4}[0,1000],$ 
                                & $1\text{e-4}[0,1000],$ & $1\text{e-3}[0,1000],$ \\
& $1\text{e-4}[1000,2000]$ & $1\text{e-5}[1000,2000]$ & $1\text{e-5}[1000,2000]$  & $1\text{e-4}[1000,2000]$ \\
\\
Negative gradient penalty $\Lambda$ & 0.5            & 0.5            & 0.5  & 0.5 \\
\midrule
Early stopping                  & Yes (100 patience iter)   & Yes (200 patience iter)   & Yes (200 patience iter) & Yes (100 patience iter) \\
Gradient clipping               & applied (norm: 5)          & applied (norm: 5)          & applied (norm: 5) & applied (norm: 5)  \\
Activation function             & Leaky ReLU ($\alpha = 0.2$) & Leaky ReLU ($\alpha = 0.2$) & Leaky ReLU ($\alpha = 0.2$) & Leaky ReLU ($\alpha = 0.2$) \\
Initialization                  & Kaiming              & Kaiming              & Kaiming & Kaiming \\
Runtime                         & $\sim$20 min              &  $\sim$40 min          & $\sim$1 hour  & $\sim$1 hour \\
\bottomrule
\end{tabular}
}}
\caption{Summary of the hyperparameters used for high-dimensional test problems that admit pathwise optimal solutions.}
\label{hyperparameter_high}
\end{table}
\vspace{-5mm}
\begin{table}[H]
\centering
\setlength\tabcolsep{6pt}
\renewcommand{\arraystretch}{1.2}
{\small
\scalebox{0.78}{
\begin{tabular}{lcccc}
\toprule
\textbf{Hyperparameters} & \textbf{100D Main}  & \textbf{100D First Variant}  & \textbf{100D Second Variant }\\
\midrule
Number of hidden layers         & 4 & 4 & 4 \\
Number of neurons per layer     & 100 & 100 & 100\\
Number of networks \(N\)        & 2040 & 1020 & 1020\\
\(\Delta T = T/N\)  & 30 sec & 1 min & 1 min\\
Batch size                      & 256 & 256 & 256\\
\midrule
Number of iterations (final interval) & 5000 & 5000 & 5000\\
Number of iterations (rest of intervals)      & 2000 & 2000 & 2000\\
\midrule
Reference policy                & even  & even & even\\
Decision frequency & 5 min & 1 min & 1 min \\
Learning rate (final interval)  & \(1\text{e-3}[0,1000],\) \(1\text{e-4}[1000,5000]\) & \(1\text{e-3}[0,1000],\) \(1\text{e-4}[1000,5000]\) & \(1\text{e-3}[0,1000],\) \(1\text{e-4}[1000,5000]\) \\
                                
Learning rate (rest of intervals) & \(1\text{e-3}[0,1000],\) \(1\text{e-4}[1000,2000]\) &  \(1\text{e-3}[0,1000],\) \(1\text{e-4}[1000,2000]\) & \(1\text{e-3}[0,1000],\) \(1\text{e-4}[1000,2000]\) \\
                             
\\
Negative gradient penalty \(\Lambda\) & 0.5 & 0.5 & 0.5 \\
\midrule
Early stopping                  & Yes (100 patience iter)  & Yes (100 patience iter) & Yes (100 patience iter)\\
Gradient clipping               & Not applied & Not applied & Not applied\\
Activation function             & Leaky ReLU (\(\alpha = 0.4\)) & Leaky ReLU (\(\alpha = 0.4\)) & Leaky ReLU (\(\alpha = 0.4\)) \\
Initialization                  & Kaiming  & Kaiming & Kaiming \\
Runtime                         & $\sim$1.5 hours & $\sim$1 hour & $\sim$1 hour \\
\bottomrule
\end{tabular}
}}
\caption{Summary of the hyperparameters used for the three 100-dimensional test problems that do not admit a pathwise optimal solution.}
\label{hyperparameter_high}
\end{table}

\subsection{Technical modifications}
Our code follows the structure outlined by \citet{beck2021deep}. In addition, we implemented two features discussed in the following two subsections.

\subsubsection{Decay loss}
\begin{figure}[H]
    \centering
    \includegraphics[scale=0.38]{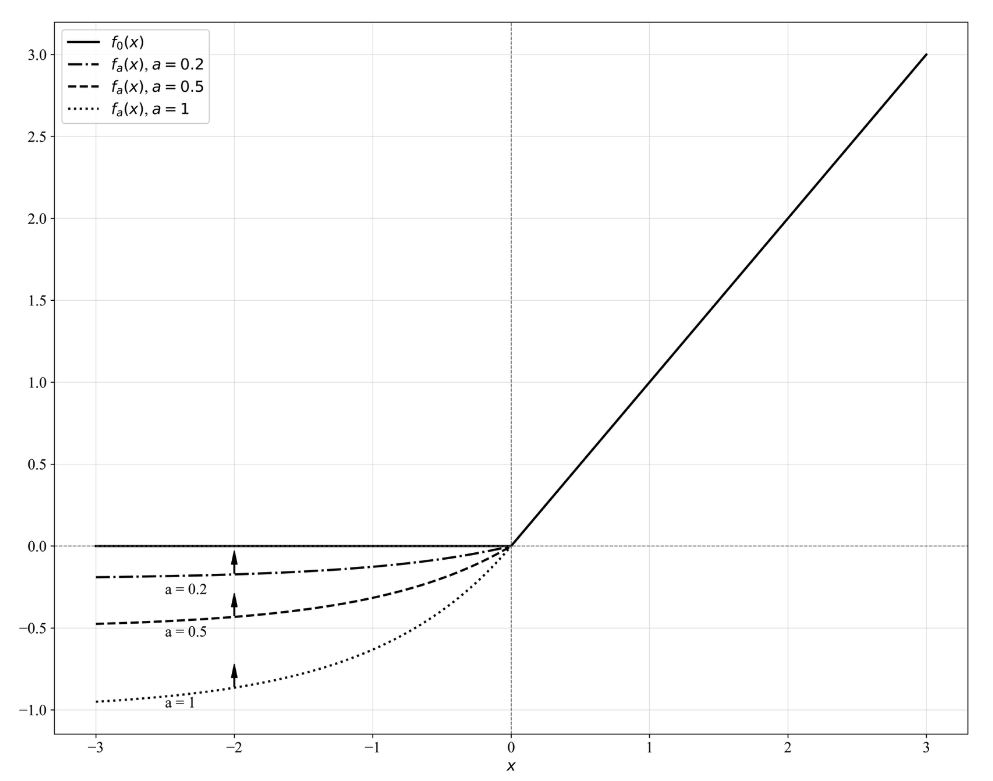}
    \caption{Plot of functions $f_0$ and $f_{a}$ for different values of $a$.}
    \label{fig:modified_relu}
\end{figure}

Recall from Section \ref{computational_method} that, we define the loss function for the final interval as follows:
\begin{align*}
      \ell_{N-1}(\omega_{N-1}, \nu_{N-1}) = \mathbb{E}\Big[\Big(\hat{g}(\tilde{X}(t_{N})) &- H^{N-1}(\tilde{X}(t_{N-1})\,;\,\omega_{N-1}) \\
      & - F\big(t_{N-1}, \tilde{X}(t_{N-1}),  G^{N-1}(\tilde{X}(t_{N-1})\,;\,\nu_{N-1})\big)\Delta t_{N-1} \Big)^{2}\Big].
\end{align*}
Similarly for $n = N-2, \ldots,1,0$, we define the loss function $\ell_{n}(\omega_{n}, \nu_{n})$ as follows: 
\begin{align*}
      \ell_{n}(\omega_{n}, \nu_{n}) = \mathbb{E}\Big[\Big( H^{n+1}(\tilde{X}(t_{n+1})\,;\,\omega^{\star}_{n+1}) - H^{n}(\tilde{X}(t_{n})\,;\,\omega_{n}) 
      - F\big(t_{n}, \tilde{X}(t_{n}),  G^{n}(\tilde{X}(t_{n})\,;\,\nu_{n})\big)\Delta t_{n} \Big)^{2}\Big],
\end{align*}
where 
\begin{flalign}
F(t_{n}, \tilde{X}(t_{n}), G^{n}(\tilde{X}(t_{n});\nu_{n})) &= (e \cdot \tilde{X}(t_{n}))^{+}\left(\sum_{k=1}^{K}G_{k}^{n}(\tilde{X}(t_{n});\nu_{n})\,(\mu_{k} - \theta_{k})\,\tilde{u}_{k}(t_{n}, \tilde{X}(t_{n})) \right. \nonumber\\
&\quad\quad\quad\quad\quad\quad - \left. \min_{k=1,\ldots, K}\Big[c_{k} + (\mu_{k} - \theta_{k})\,G_{k}^{n}(\tilde{X}(t_{n});\nu_{n})\Big]\right). \label{f_function}
\end{flalign}
Note that if $(e \cdot \tilde{X}(t_{n}))^{+} = 0$, we have for $n = 0,1,\ldots,N-1$
\begin{align*}
    \frac{\partial F(t_{n}, \tilde{X}(t_{n}), G^{n}(\tilde{X}(t_{n});\nu_{n}))}{\partial \nu_{n}} = 0,
\end{align*}
which may lead to the vanishing gradient problem (see \citet{hochreiter1998vanishing}). To address this, we introduce two functions:
\newpage
\begin{enumerate}
    \item $f_{0}(x) = x^{+}$,
    \item $f_{a}(x)$, inspired by the structure of the eLU function \citep{clevert2016elu}, defined as: 
    \begin{align} & f_{a}(x) = x^{+} + a \, \big(e^{x^{-}} - 1\big), \end{align}
    where $a$ is a decaying parameter with respect to the training iteration as follows:
    \begin{align*}
    a = \left(c_{0} - \frac{\text{iteration}}{c_{1}}\right)^{+}\hspace{-2mm},\,\,\, c_{0} = 1, \text{ and } c_{1} = 1000.
\end{align*}
\end{enumerate}
As shown in Figure \ref{fig:modified_relu}, $f_{a}$ converges to $f_{0}$ as training progresses. The arrows indicate this convergence as $a$ decreases. Consequently, to overcome the problem of the vanishing gradients, we replace the $f_{0}(e \cdot \tilde{X}(t_{n})) = (e \cdot \tilde{X}(t_{n}))^{+}$ term in Equation (\ref{f_function}) with $f_{a}(e \cdot \tilde{X}(t_{n}))$. 

\subsubsection{Penalty function for enforcing nonnegative gradients}\label{appendix:penalty:neg:grad}
To avoid negative neural network approximations of $V(t_{n},\cdot)$ and $\nabla_{x} V(t_{n},\cdot)$ for $n = 0, \ldots, N-1$, we introduce the following two penalty terms:
\begin{align*}
p_{1}\Big(\,\tilde{X}(t_{n}),\, H^{n}(\tilde{X}(t_{n}); \omega_n)\,\Big)  &= \min\Big(0, H^{n}(\tilde{X}(t_{n}); \omega_n)\Big),\\
p_{2}\Big(\tilde{X}(t_{n}),\, G^{n}(\tilde{X}(t_{n});\nu_{n})\Big) &= \min\Big(0, G_{1}^{n}(\tilde{X}(t_{n});\nu_{n}), \ldots, G_{K}^{n}(\tilde{X}(t_{n});\nu_{n})\Big),
\end{align*}
We then redefine the loss function for the final interval as follows:
\begin{align*}
      \ell_{N-1}&(\omega_{N-1}, \nu_{N-1}) = \\
      &\mathbb{E}\Bigg[\Bigg(\hat{g}(\tilde{X}(t_{N})) - H^{N-1}(\tilde{X}(t_{N-1})\,;\,\omega_{N-1}) - F\big(t_{N-1}, \tilde{X}(t_{N-1}),  G^{N-1}(\tilde{X}(t_{N-1})\,;\,\nu_{N-1})\big)\Delta t_{N-1} \Bigg)^{2} \\
      & \quad+  \Lambda\Bigg(\left| p_1\left(\tilde{X}(t_{N-1}), H^{N-1}(\tilde{X}(t_{N-1});\omega_{N-1})\right)\right|^{2} + \left| p_2\left(\tilde{X}(t_{N-1}), G^{N-1}(\tilde{X}(t_{N-1});\nu_{N-1})\right)\right|^{2}\Bigg)\Bigg].
      \label{eqn_loss_redefined}
\end{align*}
Similarly for $n = N-2, \ldots,1,0$, we redefine the loss function $\ell_{n}(\omega_{n}, \nu_{n})$ as follows: 
\begin{align*}
      \ell_{n}(\omega_{n}, \nu_{n}) = &\mathbb{E}\Bigg[\Big( H^{n+1}(\tilde{X}(t_{n+1})\,;\,\omega^{\star}_{n+1}) - H^{n}(\tilde{X}(t_{n})\,;\,\omega_{n}) - F\big(t_{n}, \tilde{X}(t_{n}),  G^{n}(\tilde{X}(t_{n})\,;\,\nu_{n})\big)\Delta t_{n} \Big)^{2}\\
      & \quad +  \Lambda\left(\left| p_1\left(\tilde{X}(t_{n}), H^{n}(\tilde{X}(t_{n});\omega_{n})\right)\right|^{2} + \left| p_2\left(\tilde{X}(t_{n}), G^{n}(\tilde{X}(t_{n});\nu_{n})\right)\right|^{2}\right)\Bigg],
\end{align*}
for some positive constant $\Lambda$\footnote{We consider a range from 0 to 2 and pick $\Lambda$ based on the best policy performance.}. 
\section{Derivation of the deep splitting algorithm}
\label{alg_derivation_appendix}

In what follows, we provide a purely formal derivation mainly to provide the intuition behind the algorithm. As such, we suppress certain technical details while putting the emphasis on the features of the algorithm; the reader is referred to \citet{beck2021deep} for further details.

To derive the algorithm, we focus on the following equivalent characterization of the HJB equation: For $(t,x) \in [0,T] \times \mathbb{R}^{K},$
\begin{align}
    &\frac{\partial V}{\partial t}(t,x) + \sum_{k=1}^{K}\lambda_{k}(t)\frac{\partial^{2}V}{\partial x_{k}^{2}}(t,x) + \sum_{k=1}^{K}b_{k}\big(t, x, u(t,x)\big)\frac{\partial V}{\partial x_{k}}(t,x) - F(t,x,\nabla_{x}V(t,x)) = 0,\\ \label{modified_HJB_appendix}
    & V(T,x) = \hat{g}(x), \nonumber
\end{align}
where $b_{k}(\cdot)$ and $F(\cdot)$ are defined as in Equations (\ref{eqn_drift}) and (\ref{eqn_f_function}), respectively. 

First, we fix a partition of $[0,T]$ such that $0 = t_{0} < t_{1} < \ldots < t_{N} = T$. Under suitable integrability assumptions, for $n = 0,1,\ldots,N-1$, $t \in [t_{n}, t_{n+1}]$ and $x \in \mathbb{R}^{K}$, we have
\begin{align}
V(t,x) = V(t_{n}, x) &- \int_{t_n}^{t}\sum_{k = 1}^{K} \lambda_{k}(s)\frac{\partial^{2}V}{\partial x_{k}^{2}}(s,x)\,ds - \int_{t_n}^{t} \sum_{k=1}^{K}b_{k}(s,x,u(s,x))\frac{\partial V}{\partial x_{k}}(s,x)\,ds \nonumber\\
& + \int_{t_{n}}^{t} F(s,x, \nabla_{x} V(s,x))\,ds.
\end{align}
\color{black}
For a sufficiently fine partition, one can write
\begin{align}
V(t,x) \approx V(t_{n}, x) &- \int_{t_n}^{t}\sum_{k = 1}^{K} \lambda_{k}(s)\frac{\partial^{2}V}{\partial x_{k}^{2}}(s,x)\,ds - \int_{t_n}^{t} \sum_{k=1}^{K}b_{k}(s,x,u(s,x))\frac{\partial V}{\partial x_{k}}(s,x)\,ds \nonumber\\
& + F(t_{n},x, \nabla_{x} V(t_{n},x))\,(t_{n+1} - t_{n}). \label{value_function_approx_appendix}
\end{align}
We assume there exists a function \( \mathscr{V} : [0,T] \times \mathbb{R}^K \to \mathbb{R} \) satisfying $\mathscr{V}(T, x) = \hat{g}(x), \,\, x \in \mathbb{R}^K,$ such that for every \( n \in \{0, 1, \dots, N-1\} \), $\mathscr{V}\mid_{(t_n, t_{n+1}] \times \mathbb{R}^K }$ belongs to \( C^{1,2}((t_n, t_{n+1}] \times \mathbb{R}^K, \mathbb{R}) \) with at most polynomially growing partial derivatives and for all \( n \in \{0, 1, \dots, N-1\} \) and $t \in (t_n, t_{n+1}]$, and $x \in \mathbb{R}^K$  satisfying
\begin{align}
\mathscr{V}(t,x) = \mathscr{V}(t_{n}, x) &- \int_{t_{n}}^{t}\sum_{k = 1}^{K} \lambda_{k}(s)\frac{\partial^{2}\mathscr{V}}{\partial x_{k}^{2}}(s,x)\,ds - \int_{t_{n}}^{t} \sum_{k=1}^{K}b_{k}(s,x,u(s,x))\frac{\partial \mathscr{V}}{\partial x_{k}}(s,x)\,ds \nonumber\\
& + F(t_{n},x, \nabla_{x} \mathscr{V}(t_{n},x))\,(t_{n+1} - t_{n}). \label{deep_split_approx_value}
\end{align}
The preceding equation is essentially the aforementioned linear approximation of the HJB equation over small time intervals $(t_{n}, t_{n+1}]$ for $n = 0,1,\ldots,N-1$. Comparing Equation (\ref{deep_split_approx_value}) with Equation (\ref{value_function_approx_appendix}), one expects that
\begin{equation}
    \mathscr{V}(t_{n},x) \approx V(t_{n},x) \quad \text{for} \quad n = 0,1,\ldots,N.
\end{equation}
It follows from Equation (\ref{deep_split_approx_value}) that for $n = 0,1,\ldots,N$ 
\begin{equation}
    \frac{\partial \mathscr{V}}{\partial t}(t, x) + \sum_{k = 1}^{K} b_{k}(t,x,u(t,x))\frac{\partial \mathscr{V}}{\partial x_{k}}(t,x) + \sum_{k=1}^{K}\lambda_{k}(t)\frac{\partial^{2}\mathscr{V}}{\partial x_{k}^{2}}(t,x) = 0, \quad (t,x) \in (t_{n}, t_{n+1}] \times \mathbb{R}^{K}. \label{eq1_appendix}
\end{equation}
From Ito's formula, we have 
\begin{align}
    d\mathscr{V}(t,\tilde{X}(t)) &= \frac{\partial \mathscr{V}}{\partial t}(t, \tilde{X}(t)) + \sum_{k=1}^{K}\frac{\partial \mathscr{V}}{\partial x_{k}}(t, \tilde{X}(t))\,b_{k}\big(t,\tilde{X}(t), \tilde{u}(t,\tilde{X}(t))\big) + \sum_{k=1}^{K} \lambda_{k}(t)\frac{\partial^{2}\mathscr{V}}{\partial x_{k}^{2}}(t, \tilde{X}(t)) \nonumber \\
    & \quad + \sum_{k=1}^{K} \frac{\partial \mathscr{V}}{\partial x_{k}}(t, \tilde{X}(t))\sqrt{2\lambda_{k}(t)}dB_{k}(t). \label{eqn_ito_value_appendix}
\end{align}
Substituting Equation (\ref{eq1_appendix}) into Equation (\ref{eqn_ito_value_appendix}) yields
\begin{align}
    d\mathscr{V}(t,\tilde{X}(t)) 
    &=  \sum_{k=1}^{K} \frac{\partial \mathscr{V}}{\partial x_{k}}(t, \tilde{X}(t))\sqrt{2\lambda_{k}(t)}dB_{k}(t).
\end{align}
Then integrating both sides on $(t,t_{n+1}]$ for $t \in (t_{n}, t_{n+1}]$ gives that $\mathbb{P}$-a.s.
\begin{align}
\mathscr{V}(t_{n+1}, \tilde{X}(t_{n+1})) - \mathscr{V}(t, \tilde{X}(t)) &=  \int_{t}^{t_{n+1}}\sum_{k=1}^{K} \frac{\partial \mathscr{V}}{\partial x_{k}}(s, \tilde{X}(s))\sqrt{2\lambda_{k}(s)}dB_{k}(s). \label{appendix_value_function_change}
\end{align}
Let $(\Omega, \mathcal{F}, (\mathcal{F}_{t})_{t \in [0,T]}, \mathbb{P})$ denote the filtered probability space that supports the reference process and its driving Brownian motion $B$; both $B$ and $\tilde{X}$ are adapted to the filtration $(\mathcal{F}_{t})_{t \in [0,T]}$. Then under suitable integrability assumptions, it follows that 

\begin{equation}
\mathbb{E}\left[\int_{t}^{t_{n+1}} \sum_{k=1}^{K} \frac{\partial V}{\partial x_{k}}(s, \tilde{X}(s))\sqrt{2\lambda_{k}(s)} dB_{k}(s) \Bigm| \mathcal{F}_{t}\right] = 0 \qquad \text{a.s.} \label{expectation_zero}
\end{equation}

Combining Equations (\ref{appendix_value_function_change}) and (\ref{expectation_zero}), one arrives at the following for $n = 0,1,\ldots,N-1$ and $t \in (t_{n}, t_{n+1}]$.
\begin{align}
\mathscr{V}(t,\tilde{X}(t)) = \mathbb{E}\left[\mathscr{V}(t_{n+1}, \tilde{X}(t_{n+1})) \Bigm| \mathcal{F}_{t} \right] \quad \quad \text{a.s.} 
\end{align}

From the tower property of conditional expectations, we can write for $t \in (t_{n}, t_{n+1}]$ and $n = 0,1,\ldots,N-1$
\begin{equation}
    \mathbb{E}\left[\mathscr{V}(t, \tilde{X}(t)) \Bigm| \mathcal{F}_{t_{n}}\right] = \mathbb{E}\left[\mathscr{V}(t_{n+1}, \tilde{X}(t_{n+1})) \Bigm| \mathcal{F}_{t_{n}}\right] \quad \text{a.s.} \label{identity_part_one}
\end{equation}
Now, note from Equation (\ref{deep_split_approx_value}) that for $n \in \{0,1,\ldots,N-1\}$ and $t \in (t_{n}, t_{n+1}]$ the following holds: 
\begin{align}
\mathscr{V}(t,\tilde{X}(t)) = \mathscr{V}(t_{n}, \tilde{X}(t)) &- \int_{t_{n}}^{t} \sum_{k=1}^{K} \lambda_{k}(s) \frac{\partial^{2}\mathscr{V}}{\partial x_{k}^{2}} (s,\tilde{X}(s))ds \nonumber\\
& - \int_{t_{n}}^{t} \sum_{k=1}^{K} b_{k}(s, \tilde{X}(s), u(s, \tilde{X}(s)))\frac{\partial \mathscr{V}}{\partial x_{k}}(s, \tilde{X}(s))ds \nonumber \\
& + F(t_{n}, \tilde{X}(t),\nabla_{x}\mathscr{V}(t_{n}, \tilde{X}(t)))(t_{n+1} - t_{n}). 
\end{align}
Now, letting $t \searrow t_{n}$ yields the following
\begin{equation}
    \lim_{t \searrow t_{n}} \mathscr{V}(t, \tilde{X}(t)) = \mathscr{V}(t_{n}, \tilde{X}(t_{n})) + F\big(t_{n}, \tilde{X}(t_{n}), \nabla_{x}\mathscr{V}(t_{n}, \tilde{X}(t_{n}))\big)(t_{n+1} - t_{n}). \label{limit_identity}
\end{equation}
Then, under suitable integrability assumptions, passing to the limit in (\ref{identity_part_one}) as $t \searrow t_{n}$ and substituting (\ref{limit_identity}) yields the following: 
\begin{align}
    \mathbb{E}\left[\mathscr{V}(t_{n+1}, \tilde{X}(t_{n+1})) \Bigm| \mathcal{F}_{t_{n}} \right] &= \lim_{t \searrow t_{n}}\mathbb{E}\left[\mathscr{V}(t, \tilde{X}(t)) \Bigm| \mathcal{F}_{t_{n}}\right] \nonumber \\
    & = \mathbb{E}\left[\lim_{t \searrow t_{n}} \mathscr{V}(t,\tilde{X}(t)) \Bigm| \mathcal{F}_{t_{n}}\right] \nonumber \\
    & = \mathscr{V}(t_{n}, \tilde{X}(t_{n})) + F\big(t_{n}, \tilde{X}(t_{n}), \nabla_{x}\mathscr{V}(t_{n}, \tilde{X}(t_{n}))\big)(t_{n+1} - t_{n}).
\end{align}
Then taking the conditional expectations of both sides with respect to the $\sigma$-algebra generated by $\tilde{X}(t_{n})$ and using the tower property yields
\begin{align}
    \mathbb{E}\left[\mathscr{V}(t_{n+1}, \tilde{X}(t_{n+1})) \Bigm| \tilde{X}(t_{n}) \right] = \mathscr{V}(t_{n}, \tilde{X}(t_{n})) + F\big(t_{n}, \tilde{X}(t_{n}), \nabla_{x}\mathscr{V}(t_{n}, \tilde{X}(t_{n}))\big)(t_{n+1} - t_{n}).
\end{align}
\section{Supplementary tables for the high-dimensional problems}
\label{high_dimensional_supplementary_tables}
\subsection{High-dimensional test problems that have a pathwise optimal policy}
\subsubsection{30-dimensional problem}
\renewcommand{\arraystretch}{0.7} 
\setlength{\extrarowheight}{-2pt} 
\footnotesize
\begin{longtable}{llcllcc}
        \caption{Summary statistics for the 30-dimensional test problem that have a pathwise optimal solution.}
        \label{stats_30dim_path}\\
	\toprule
	Class & Arrivals & Arrival & $\tilde{\mu}$ & $\tilde{\theta}$ & $h$ & $c$\\
	\rule{0pt}{3.5ex} & sourced from & percentage (\%) & (per hr) & (per hr) & (per hr) & (per hr)\\
	\midrule
        \endfirsthead
        \endhead
        \endfoot
        \bottomrule
        \endlastfoot
1 & Consumer Loans & 3.77 & EBO & EBO & \$15.50 & \$30.86\\
2 & CCO & 4.54 & Retail (Node: 2) & Retail (Node: 2) & \$20.50 & \$35.86\\
3 & Subanco &  0.28 & CCO & CCO & \$21.50 & \$36.86\\
4 & Online Banking & 1.44 & Consumer Loans & Consumer Loans & \$24.00 & \$39.36\\
5 & Consumer Loans & 3.77 & Case Quality & Case Quality & \$30.50 & \$45.86\\
6 & Subanco & 0.28 & Online Banking & Online Banking & \$16.50 & \$31.86\\
7 & Subanco &  0.28 & CCO & CCO & \$20.00 & \$35.36\\
8 & Premier & 1.88 & Retail (Node: 3) & Retail (Node: 3) & \$33.00 & \$48.36\\
\end{longtable}

\newpage
\subsubsection{50-dimensional problem}
\renewcommand{\arraystretch}{0.7} 
\setlength{\extrarowheight}{-2pt} 
\footnotesize
\begin{longtable}{llcllcc}
    \caption{Summary statistics for the 50-dimensional test problem that have a pathwise optimal solution.}
    \label{stats_50dim_path}\\
    \toprule
    Class & Arrivals & Arrival & $\tilde{\mu}$ & $\tilde{\theta}$ & $h$ & $c$\\
    \rule{0pt}{3.5ex} & sourced from & percentage (\%) & (per hr) & (per hr) & (per hr) & (per hr)\\
    \midrule
    \endfirsthead
    \endhead
    \endfoot
    \bottomrule
    \endlastfoot
    1 & Consumer Loans & 2.12 & Premier & Premier & \$17.75 & \$32.58\\
    2 & CCO & 2.56 & AST & AST & \$16.25 & \$31.08\\
    3 & Subanco & 0.16 & Brokerage & Brokerage & \$31.75 & \$46.58\\
    4 & Online Banking & 0.81 & CCO & CCO & \$33.75 & \$48.58\\
    5 & Consumer Loans & 2.12 & Online Banking & Online Banking & \$30.75 & \$45.58\\
    6 & Subanco & 0.16 & AST & AST & \$25.50 & \$40.33\\
    7 & Subanco & 0.16 & Online Banking & Online Banking & \$24.00 & \$38.83\\
    8 & Premier & 1.06 & Brokerage & Brokerage & \$18.25 & \$33.08\\
\end{longtable}

\newpage
\subsubsection{100-dimensional problem}
\renewcommand{\arraystretch}{0.7} 
\setlength{\extrarowheight}{-2pt} 
\footnotesize
\begin{longtable}{llcllcc}
    \caption{Summary statistics for the 100-dimensional test problem that have a pathwise optimal solution.}
    \label{stats_100dim_path_part1}\\
    \toprule
    Class & Arrivals & Arrival & $\tilde{\mu}$ & $\tilde{\theta}$ & $h$ & $c$\\
    \rule{0pt}{3.5ex} & sourced from & percentage (\%) & (per hr) & (per hr) & (per hr) & (per hr)\\
    \midrule
    \endfirsthead
    \endhead
    \endfoot
    \bottomrule
    \endlastfoot
    1 & Consumer Loans & 1.03 & AST & AST & \$26.75 & \$40.39\\
    2 & CCO & 1.24 & Consumer Loans & Consumer Loans & \$23.38 & \$37.02\\
    3 & Subanco & 0.08 & EBO & EBO & \$15.25 & \$28.89\\
    4 & Online Banking & 0.39 & CCO & CCO & \$17.75 & \$31.39\\
    5 & Consumer Loans & 1.03 & CCO & CCO & \$27.12 & \$40.77\\
    6 & Subanco & 0.08 & Telesales & Telesales & \$30.62 & \$44.27\\
    7 & Subanco & 0.08 & Priority Service & Priority Service & \$32.50 & \$46.14\\
    8 & Premier & 0.51 & Consumer Loans & Consumer Loans & \$33.12 & \$46.77\\
\end{longtable}

\newpage
\subsubsection{500-dimensional problem}
{
\renewcommand{\arraystretch}{0.7} 
\setlength{\extrarowheight}{-2pt} 
\footnotesize
\begin{longtable}{llcllccccc}
	\caption{Summary statistics for the 500-dimensional test problem that have a pathwise optimal solution.}
        \label{stats_500dim_path}\\
	\toprule
	Class & Arrivals & Arrival & $\tilde{\mu}$ & $\tilde{\theta}$  & $h$ & $c$\\
	\rule{0pt}{3.5ex} & sourced from & percentage (\%) & (per hr) & (per hr) & (per hr) & (per hr)\\
	\midrule
        \endfirsthead
        \endhead
        \endfoot
        \bottomrule
        \endlastfoot
1 & Consumer Loans & 0.22 & EBO & EBO & \$31.32 & \$46.12 \\
2 & CCO & 0.27 & Premier & Premier & \$22.08 & \$36.88 \\
3 & Subanco & 0.02 & BPS & BPS & \$32.40 & \$47.20 \\
4 & Online Banking & 0.08 & EBO & EBO & \$30.08 & \$44.88 \\
5 & Consumer Loans & 0.22 & Retail (Node: 1) & Retail (Node: 1) & \$30.32 & \$45.12 \\
6 & Subanco & 0.02 & Priority Service & Priority Service & \$28.40 & \$43.20 \\
7 & Subanco & 0.02 & Brokerage & Brokerage & \$23.56 & \$38.36 \\
8 & Premier & 0.11 & Priority Service & Priority Service & \$20.56 & \$35.36 \\
\end{longtable}
}
\newpage
\subsection{High-dimensional test problems that do not admit a pathwise optimal policy}
\label{appendix_non_pathwise_optimal_100dim}
\subsection{100-dimensional main test problem}
\renewcommand{\arraystretch}{0.7} 
\setlength{\extrarowheight}{-2pt} 
\footnotesize
\begin{longtable}{llcllcc}
    \caption{Summary statistics for the 100-dimensional test problem, where the abandonment rates are sampled with replacement from the main test problem.}
    \label{stats_100dim_non_path_main}\\
    \toprule
    Class & Arrivals & Arrival & $\mu$ & $\theta$ & $h$  & $c$ \\
    \rule{0pt}{3.5ex} & sourced from & percentage (\%) & (per hr) & (per hr) & (per hr) & (per hr)\\
    \midrule
    \endfirsthead
    \endhead
    \endfoot
    \bottomrule
    \endlastfoot
    1 & Premier & 0.66 & Online Banking & Online Banking & \$16.00 & \$23.44\\
    2 & AST & 0.65 & Premier & Premier & \$33.75 & \$61.28\\
    3 & Premier & 0.66 & Retail (Node: 1) & Retail (Node: 1) & \$30.38 & \$45.71\\
    4 & Retail (Node: 1) & 2.92 & Retail (Node: 2) & Retail (Node: 2) & \$24.88 & \$41.06\\
    5 & Premier & 0.66 & Subanco & Subanco & \$18.75 & \$28.73\\
    6 & Brokerage & 1.08 & Consumer Loans & Consumer Loans & \$23.50 & \$33.04\\
    7 & Subanco & 0.10 & EBO & EBO & \$22.88 & \$38.58\\
    8 & Premier & 0.66 & Subanco & Subanco & \$31.12 & \$47.70\\
\end{longtable}

\subsection{100-dimensional first variant test problem}
\renewcommand{\arraystretch}{0.7} 
\setlength{\extrarowheight}{-2pt} 
\footnotesize
\begin{longtable}{llcllcc}
    \caption{Summary statistics for the 100-dimensional test problem, where the abandonment rates are sampled with replacement from the first variant of the main test problem.}
    \label{stats_100dim_non_path_var1}\\
    \toprule
    Class & Arrivals & Arrival & $\mu$ & $\theta$ & $h$  & $c$ \\
    \rule{0pt}{3.5ex} & sourced from & percentage (\%) & (per hr) & (per hr) & (per hr) & (per hr)\\
    \midrule
    \endfirsthead
    \endhead
    \endfoot
    \bottomrule
    \endlastfoot
    1 & Premier & 0.66 & Online Banking & Online Banking & \$16.00 & \$25.67\\
    2 & AST & 0.65 & Premier & Premier & \$33.75 & \$69.55\\
    3 & Premier & 0.66 & Retail (Node: 1) & Retail (Node: 1) & \$30.38 & \$45.71\\
    4 & Retail (Node: 1) & 2.92 & Retail (Node: 2) & Retail (Node: 2) & \$24.88 & \$41.06\\
    5 & Premier & 0.66 & Subanco & Subanco & \$18.75 & \$31.73\\
    6 & Brokerage & 1.08 & Consumer Loans & Consumer Loans & \$23.50 & \$33.04\\
    7 & Subanco & 0.10 & EBO & EBO & \$22.88 & \$43.29\\
    8 & Premier & 0.66 & Subanco & Subanco & \$31.12 & \$52.68\\
\end{longtable}
\newpage
\subsection{100-dimensional second variant test problem}
\renewcommand{\arraystretch}{0.7} 
\setlength{\extrarowheight}{-2pt} 
\footnotesize
\begin{longtable}{llcllcc}
    \caption{Summary statistics for the 100-dimensional test problem, where the abandonment rates are sampled with replacement from the second variant of the main test problem.}
    \label{stats_100dim_non_path_var2}\\
    \toprule
    Class & Arrivals & Arrival & $\mu$ & $\theta$ & $h$  & $c$ \\
    \rule{0pt}{3.5ex} & sourced from & percentage (\%) & (per hr) & (per hr) & (per hr) & (per hr)\\
    \midrule
    \endfirsthead
    \endhead
    \endfoot
    \bottomrule
    \endlastfoot
    1 & Premier & 0.66 & Online Banking & Online Banking & \$16.00 & \$26.41\\
    2 & AST & 0.65 & Premier & Premier & \$33.75 & \$72.31\\
    3 & Premier & 0.66 & Retail (Node: 1) & Retail (Node: 1) & \$30.38 & \$45.71\\
    4 & Retail (Node: 1) & 2.92 & Retail (Node: 2) & Retail (Node: 2) & \$24.88 & \$41.06\\
    5 & Premier & 0.66 & Subanco & Subanco & \$18.75 & \$32.73\\
    6 & Brokerage & 1.08 & Consumer Loans & Consumer Loans & \$23.50 & \$33.04\\
    7 & Subanco & 0.10 & EBO & EBO & \$22.88 & \$44.87\\
    8 & Premier & 0.66 & Subanco & Subanco & \$31.12 & \$54.34\\
\end{longtable}

\newpage
\section{Benchmark policy performance tables}
\subsection{Supplementary tables for \ref{appendix_additional_static_benchmarks}}
\label{supp_tables_additional_static_benchmarks}
\subsubsection{Main test problem} 
{
\renewcommand{\arraystretch}{0.8} 
\setlength{\extrarowheight}{-2pt} 
\footnotesize
\begin{longtable}{lllllllccc}
    \caption{The performance comparison of all combinations of benchmark policies considered for the main test problem. We consider static priority policies that rank the subgroups in the order $\mathcal{G}_{H} > \mathcal{G}_{L}^{\alpha} > \mathcal{G}_{L}^{\beta}$. The first three columns show the priority policy used to schedule classes within each subgroup. The last column shows the total cost $\pm$ half-length of the 99\% confidence interval.} \label{main_instance1}\\
            \toprule
            $\mathcal{G}_{H}$ & & & $\mathcal{G}^{\alpha}_{L}$ & & & $\mathcal{G}^{\beta}_{L}$ & & & Total cost $\pm$ 99\% CI Half-length \\
            \midrule
            \endfirsthead

            \endhead

            \endfoot

            \bottomrule
            \endlastfoot
	    $c\mu$ &&& $c\mu$ &&& $c\mu$ &&& 1214.23 $\pm$ 6.33 \\
    $c\mu$ &&& $c\mu$ &&& $c\mu/\theta$ &&& 1190.49 $\pm$ 6.08 \\
    $c\mu$ &&& $c\mu$ &&& $c_{k}(\mu_{k} - \theta_{k})$ &&& 1213.54 $\pm$ 6.19 \\
    $c\mu$ &&& $c\mu$ &&& $c_{k}$ &&& 1225.83 $\pm$ 6.36 \\
    $c\mu$ &&& $c\mu$ &&& $\mu_{k} - \theta_{k}$ &&& 1222.68 $\pm$ 6.27 \\
    $c\mu$ &&& $c\mu/\theta$ &&& $c\mu$ &&& 1212.53 $\pm$ 6.36 \\
    $c\mu$ &&& $c\mu/\theta$ &&& $c\mu/\theta$ &&& 1190.71 $\pm$ 6.09 \\
    $c\mu$ &&& $c\mu/\theta$ &&& $c_{k}(\mu_{k} - \theta_{k})$ &&& 1214.41 $\pm$ 6.19 \\
\end{longtable}
}
\newpage
{
\renewcommand{\arraystretch}{0.8} 
\setlength{\extrarowheight}{-2pt} 
\footnotesize
\begin{longtable}{lllllllccc}
    \caption{The performance comparison of all combinations of benchmark policies considered for the main test problem. We consider static priority policies that rank the subgroups in the order $\mathcal{G}_{H} > \mathcal{G}_{L}^{\beta} > \mathcal{G}_{L}^{\alpha}$. The first three columns show the priority policy used to schedule classes within each subgroup. The last column shows the total cost $\pm$ half-length of the 99\% confidence interval.} \label{main_instance1}\\
            \toprule
            $\mathcal{G}_{H}$ & & & $\mathcal{G}^{\beta}_{L}$ & & & $\mathcal{G}^{\alpha}_{L}$ & & & Total cost $\pm$ 99\% CI Half-length \\
            \midrule
            \endfirsthead

            \endhead

            \endfoot

            \bottomrule
            \endlastfoot
            $c\mu$ &&& $c\mu$ &&& $c\mu$ &&& 1295.64 $\pm$ 6.88 \\
    $c\mu$ &&& $c\mu$ &&& $c\mu/\theta$ &&& 1300.40 $\pm$ 6.90 \\
    $c\mu$ &&& $c\mu$ &&& $c_{k}(\mu_{k} - \theta_{k})$ &&& 1346.37 $\pm$ 7.04 \\
    $c\mu$ &&& $c\mu$ &&& $c_{k}$ &&& 1295.64 $\pm$ 6.88 \\
    $c\mu$ &&& $c\mu$ &&& $\mu_{k} - \theta_{k}$ &&& 1346.37 $\pm$ 7.04 \\
    $c\mu$ &&& $c\mu/\theta$ &&& $c\mu$ &&& 1275.83 $\pm$ 6.72 \\
    $c\mu$ &&& $c\mu/\theta$ &&& $c\mu/\theta$ &&& 1288.53 $\pm$ 6.80 \\
    $c\mu$ &&& $c\mu/\theta$ &&& $c_{k}(\mu_{k} - \theta_{k})$ &&& 1329.46 $\pm$ 6.94 \\
\end{longtable}
}
\newpage
\subsubsection{First variant problem}
{
\renewcommand{\arraystretch}{0.8} 
\setlength{\extrarowheight}{-2pt} 
\footnotesize
\begin{longtable}{lllllllccc}
    \caption{The performance comparison of all combinations of benchmark policies considered for the first variant. We consider static priority policies that rank the subgroups in the order $\mathcal{G}_{H} > \mathcal{G}_{L}^{\alpha} > \mathcal{G}_{L}^{\beta}$. The first three columns show the priority policy used to schedule classes within each subgroup. The last column shows the total cost $\pm$ half-length of the 99\% confidence interval.} \label{first_instance_sub}\\
            \toprule
            $\mathcal{G}_{H}$ & & & $\mathcal{G}^{\alpha}_{L}$ & & & $\mathcal{G}^{\beta}_{L}$ & & & Total cost $\pm$ 99\% CI Half-length \\
            \midrule
            \endfirsthead

            \endhead

            \endfoot

            \bottomrule
            \endlastfoot
		 $c\mu$ &&& $c\mu$ &&& $c\mu$ &&& 1727.77 $\pm$ 8.42 \\
    $c\mu$ &&& $c\mu$ &&& $c\mu/\theta$ &&& 1624.82 $\pm$ 7.52 \\
    $c\mu$ &&& $c\mu$ &&& $c_{k}(\mu_{k} - \theta_{k})$ &&& 1680.82 $\pm$ 7.78 \\
    $c\mu$ &&& $c\mu$ &&& $c_{k}$ &&& 1773.82 $\pm$ 8.71 \\
    $c\mu$ &&& $c\mu$ &&& $\mu_{k} - \theta_{k}$ &&& 1653.91 $\pm$ 7.52 \\
    $c\mu$ &&& $c\mu/\theta$ &&& $c\mu$ &&& 1727.89 $\pm$ 8.42 \\
    $c\mu$ &&& $c\mu/\theta$ &&& $c\mu/\theta$ &&& 1624.80 $\pm$ 7.52 \\
    $c\mu$ &&& $c\mu/\theta$ &&& $c_{k}(\mu_{k} - \theta_{k})$ &&& 1680.65 $\pm$ 7.78 \\
\end{longtable}
}
\newpage
{
\renewcommand{\arraystretch}{0.8} 
\setlength{\extrarowheight}{-2pt} 
\footnotesize
\begin{longtable}{lllllllccc}
    \caption{The performance comparison of all combinations of benchmark policies considered for the first variant. We consider static priority policies that rank the subgroups in the order $\mathcal{G}_{H} > \mathcal{G}_{L}^{\beta} > \mathcal{G}_{L}^{\alpha}$. The first three columns show the priority policy used to schedule classes within each subgroup. The last column shows the total cost $\pm$ half-length of the 99\% confidence interval.} \label{first_instance}\\
            \toprule
            $\mathcal{G}_{H}$ & & & $\mathcal{G}^{\beta}_{L}$ & & & $\mathcal{G}^{\alpha}_{L}$ & & & Total cost $\pm$ 99\% CI Half-length \\
            \midrule
            \endfirsthead

            \endhead

            \endfoot

            \bottomrule
            \endlastfoot
		 $c\mu$ &&& $c\mu$ &&& $c\mu$ &&& 1527.27 $\pm$ 8.65 \\
    $c\mu$ &&& $c\mu$ &&& $c\mu/\theta$ &&& 1550.73 $\pm$ 8.65 \\
    $c\mu$ &&& $c\mu$ &&& $c_{k}(\mu_{k} - \theta_{k})$ &&& 1677.93 $\pm$ 9.08 \\
    $c\mu$ &&& $c\mu$ &&& $c_{k}$ &&& 1527.27 $\pm$ 8.65 \\
    $c\mu$ &&& $c\mu$ &&& $\mu_{k} - \theta_{k}$ &&& 1677.93 $\pm$ 9.08 \\
    $c\mu$ &&& $c\mu/\theta$ &&& $c\mu$ &&& 1497.78 $\pm$ 8.41 \\
    $c\mu$ &&& $c\mu/\theta$ &&& $c\mu/\theta$ &&& 1523.57 $\pm$ 8.59 \\
    $c\mu$ &&& $c\mu/\theta$ &&& $c_{k}(\mu_{k} - \theta_{k})$ &&& 1656.15 $\pm$ 8.82 \\
\end{longtable}
}
\newpage
\subsubsection{Second variant problem}
{
\renewcommand{\arraystretch}{0.8} 
\setlength{\extrarowheight}{-2pt} 
\footnotesize
\begin{longtable}{lllllllccc}
\caption{The performance comparison of all combinations of benchmark policies considered for the second variant. We consider static priority policies that rank the subgroups in the order $\mathcal{G}_{H} > \mathcal{G}_{L}^{\alpha} > \mathcal{G}_{L}^{\beta}$. The first three columns show the priority policy used to schedule classes within each subgroup. The last column shows the average total cost $\pm$ half-length of the 99\% confidence interval.} \label{second_instance_sub}\\
            \toprule
            $\mathcal{G}_{H}$ & & & $\mathcal{G}^{\alpha}_{L}$ & & & $\mathcal{G}^{\beta}_{L}$ & & & Total cost $\pm$ 99\% CI Half-length \\
            \midrule
            \endfirsthead

            \endhead

            \endfoot

            \bottomrule
            \endlastfoot
               $c\mu$ &&& $c\mu$ &&& $c\mu$ &&& 1780.84 $\pm$ 8.47 \\
    $c\mu$ &&& $c\mu$ &&& $c\mu/\theta$ &&& 1674.46 $\pm$ 7.63 \\
    $c\mu$ &&& $c\mu$ &&& $c_{k}(\mu_{k} - \theta_{k})$ &&& 1736.67 $\pm$ 7.82 \\
    $c\mu$ &&& $c\mu$ &&& $c_{k}$ &&& 1829.92 $\pm$ 8.89 \\
    $c\mu$ &&& $c\mu$ &&& $\mu_{k} - \theta_{k}$ &&& 1710.57 $\pm$ 7.73 \\
    $c\mu$ &&& $c\mu/\theta$ &&& $c\mu$ &&& 1780.77 $\pm$ 8.48 \\
    $c\mu$ &&& $c\mu/\theta$ &&& $c\mu/\theta$ &&& 1674.47 $\pm$ 7.62 \\
    $c\mu$ &&& $c\mu/\theta$ &&& $c_{k}(\mu_{k} - \theta_{k})$ &&& 1736.39 $\pm$ 7.83 \\
\end{longtable}
}
\newpage
{
\renewcommand{\arraystretch}{0.8} 
\setlength{\extrarowheight}{-2pt} 
\footnotesize
\begin{longtable}{lllllllccc}
\caption{The performance comparison of all combinations of benchmark policies considered for the second variant. We consider static priority policies that rank the subgroups in the order $\mathcal{G}_{H} > \mathcal{G}_{L}^{\beta} > \mathcal{G}_{L}^{\alpha}$. The first three columns show the priority policy used to schedule classes within each subgroup. The last column shows the average total cost $\pm$ half-length of the 99\% confidence interval.} \label{second_instance}\\
            \toprule
            $\mathcal{G}_{H}$ & & & $\mathcal{G}^{\beta}_{L}$ & & & $\mathcal{G}^{\alpha}_{L}$ & & & Total cost $\pm$ 99\% CI Half-length \\
            \midrule
            \endfirsthead

            \endhead

            \endfoot

            \bottomrule
            \endlastfoot
		  $c\mu$ &&& $c\mu$ &&& $c\mu$ &&& 1438.51 $\pm$ 8.36 \\
    $c\mu$ &&& $c\mu$ &&& $c\mu/\theta$ &&& 1451.56 $\pm$ 8.36 \\
    $c\mu$ &&& $c\mu$ &&& $c_{k}(\mu_{k} - \theta_{k})$ &&& 1577.60 $\pm$ 8.67 \\
    $c\mu$ &&& $c\mu$ &&& $c_{k}$ &&& 1438.51 $\pm$ 8.36 \\
    $c\mu$ &&& $c\mu$ &&& $\mu_{k} - \theta_{k}$ &&& 1577.60 $\pm$ 8.67 \\
    $c\mu$ &&& $c\mu/\theta$ &&& $c\mu$ &&& 1410.87 $\pm$ 8.13 \\
    $c\mu$ &&& $c\mu/\theta$ &&& $c\mu/\theta$ &&& 1424.63 $\pm$ 8.19 \\
    $c\mu$ &&& $c\mu/\theta$ &&& $c_{k}(\mu_{k} - \theta_{k})$ &&& 1555.69 $\pm$ 8.50 \\
\end{longtable}
}
\newpage
\subsection{Supplementary tables for \ref{appendix_mdp_high_dim}}
\label{supp_tables_additional_mdp_benchmarks}
\subsubsection{Main test problem}
{
\renewcommand{\arraystretch}{0.8} 
\setlength{\extrarowheight}{-2pt} 
\footnotesize
\begin{longtable}{lllllllccccccccccccccc}
\caption{The performance comparison of all combinations of static priority policies considered for auxiliary 3-dimensional MDP solution of the main test problem. The first four columns show the priority policy used to schedule classes within each subgroup. The last column shows the average total cost $\pm$ half-length of the 99\% confidence interval.} \label{aux_mdp_main_test_static}\\
            \toprule
            $\mathcal{G}_{H}$ && $\mathcal{G}^{1}_{L}$ && $\mathcal{G}^{2}_{L}$ && $\mathcal{G}^{3}_{L}$ && Total cost $\pm$ 99\% CI Half-length \\
            \midrule
            \endfirsthead
            \endhead

            \endfoot

            \bottomrule
            \endlastfoot
    $c\mu$ && $c\mu$ && $c\mu$ && $c\mu$ && 1194.23 $\pm$ 6.16 \\
    $c\mu$ && $c\mu$ && $c\mu$ && $c_{k}$ && 1194.23 $\pm$ 6.16 \\
    $c\mu$ && $c\mu$ && $c\mu$ && $c\mu/\theta$ && 1193.63 $\pm$ 6.14 \\
    $c\mu$ && $c\mu$ && $c\mu$ && $c_{k}(\mu_{k}-\theta_{k})$ && 1202.42 $\pm$ 6.21 \\
    $c\mu$ && $c\mu$ && $c\mu$ && $\mu_{k}-\theta_{k}$ && 1205.85 $\pm$ 6.26 \\
    $c\mu$ && $c\mu$ && $c_{k}$ && $c\mu$ && 1194.49 $\pm$ 6.25 \\
    $c\mu$ && $c\mu$ && $c_{k}$ && $c_{k}$ && 1194.49 $\pm$ 6.25 \\
    $c\mu$ && $c\mu$ && $c_{k}$ && $c\mu/\theta$ && 1194.74 $\pm$ 6.18 \\
\end{longtable}
}
\newpage
\subsubsection{First variant problem}
{
\renewcommand{\arraystretch}{0.8} 
\setlength{\extrarowheight}{-2pt} 
\footnotesize
\begin{longtable}{lllllllccccccccccccccc}
\caption{The performance comparison of all combinations of static priority policies considered for the auxiliary 3-dimensional MDP solution of the first variant problem. The first four columns show the priority policy used to schedule classes within each subgroup. The last column shows the average total cost $\pm$ half-length of the 99\% confidence interval.} \label{aux_mdp_first_variant_static}\\
            \toprule
            $\mathcal{G}_{H}$ && $\mathcal{G}^{1}_{L}$ && $\mathcal{G}^{2}_{L}$ && $\mathcal{G}^{3}_{L}$ && Total cost $\pm$ 99\% CI Half-length \\
            \midrule
            \endfirsthead
            \endhead

            \endfoot

            \bottomrule
            \endlastfoot
$c\mu$ && $c\mu$ && $c\mu$ && $c\mu$ && 1476.49 $\pm$ 7.85 \\
$c\mu$ && $c\mu$ && $c\mu$ && $c\mu/\theta$ && 1476.96 $\pm$ 7.85 \\
$c\mu$ && $c\mu$ && $c\mu$ && $c_{k}(\mu_{k} - \theta_{k})$ && 1475.01 $\pm$ 7.79 \\
$c\mu$ && $c\mu$ && $c\mu$ && $c_{k}$ && 1476.49 $\pm$ 7.85 \\
$c\mu$ && $c\mu$ && $c\mu$ && $\mu_{k} - \theta_{k}$ && 1475.32 $\pm$ 7.84 \\
$c\mu$ && $c\mu$ && $c\mu/\theta$ && $c\mu$ && 1476.13 $\pm$ 7.87 \\
$c\mu$ && $c\mu$ && $c\mu/\theta$ && $c\mu/\theta$ && 1474.24 $\pm$ 7.81 \\
$c\mu$ && $c\mu$ && $c\mu/\theta$ && $c_{k}(\mu_{k} - \theta_{k})$ && 1477.47 $\pm$ 7.85 \\
\end{longtable}
}
\newpage
\subsubsection{Second variant problem}
{
\renewcommand{\arraystretch}{0.8} 
\setlength{\extrarowheight}{-2pt} 
\footnotesize
\begin{longtable}{lllllllccccccccccccccc}
\caption{The performance comparison of all combinations of static priority policies considered for auxiliary 3-dimensional MDP solution of the second variant problem. The first four columns show the priority policy used to schedule classes within each subgroup. The last column shows the average total cost $\pm$ half-length of the 99\% confidence interval.} \label{aux_mdp_second_variant_static}\\
            \toprule
            $\mathcal{G}_{H}$ && $\mathcal{G}^{1}_{L}$ && $\mathcal{G}^{2}_{L}$ && $\mathcal{G}^{3}_{L}$ && Total cost $\pm$ 99\% CI Half-length \\
            \midrule
            \endfirsthead
            \endhead

            \endfoot

            \bottomrule
            \endlastfoot
$c\mu$ && $c\mu$ && $c\mu$ && $c\mu$ && 1401.10 $\pm$ 7.66 \\
$c\mu$ && $c\mu$ && $c\mu$ && $c\mu/\theta$ && 1401.32 $\pm$ 7.73 \\
$c\mu$ && $c\mu$ && $c\mu$ && $c_{k}(\mu_{k} - \theta_{k})$ && 1403.80 $\pm$ 7.71 \\
$c\mu$ && $c\mu$ && $c\mu$ && $c_{k}$ && 1401.10 $\pm$ 7.66 \\
$c\mu$ && $c\mu$ && $c\mu$ && $\mu_{k} - \theta_{k}$ && 1403.34 $\pm$ 7.71 \\
$c\mu$ && $c\mu$ && $c\mu/\theta$ && $c\mu$ && 1400.04 $\pm$ 7.67 \\
$c\mu$ && $c\mu$ && $c\mu/\theta$ && $c\mu/\theta$ && 1401.44 $\pm$ 7.74 \\
$c\mu$ && $c\mu$ && $c\mu/\theta$ && $c_{k}(\mu_{k} - \theta_{k})$ && 1401.36 $\pm$ 7.73 \\
\end{longtable}
}

\end{document}